\documentclass[aps,rmp,reprint,twocolumn,superscriptaddress,floatfix,nofootinbib,longbibliography]{revtex4-2}
\usepackage{graphicx,amsmath,amsfonts,amssymb,amsthm,xr}
\usepackage{epsfig,amsmath,amssymb,color,dsfont,upgreek,physics}
\usepackage{mathrsfs}
\usepackage{mathtools}
\usepackage{bbold}
\usepackage{comment}
\usepackage{float}
\usepackage{comment}
\usepackage{bm}

\usepackage[bookmarks=true,colorlinks,linkcolor=OrangeRed,urlcolor=NavyBlue,citecolor=RoyalBlue]{hyperref}
\usepackage[dvipsnames]{xcolor}

 \newcommand{\Eq}[1]{Eq.~(\ref{#1})}

\newcommand{\Eqs}[2]{Eqs.~(\ref{#1}-\ref{#2})}
\newcommand{\Fig}[1]{Fig.~\ref{#1}}

\begin{document}

\title{Quantum simulation of out-of-equilibrium dynamics in gauge theories}

\author{Jad~C.~Halimeh${}^{*}{}^{\dagger}$}
\affiliation{Department of Physics and Arnold Sommerfeld Center for Theoretical Physics (ASC), Ludwig Maximilian University of Munich, 80333 Munich, Germany}
\affiliation{Max Planck Institute of Quantum Optics, 85748 Garching, Germany}
\affiliation{Munich Center for Quantum Science and Technology (MCQST), 80799 Munich, Germany}
\affiliation{Department of Physics, College of Science, Kyung Hee University, Seoul 02447, Republic of Korea}

\author{Niklas~Mueller${}^{*}$}
\affiliation{Center for Quantum Information and Control, University of New Mexico, Albuquerque, NM 87106, USA}
\affiliation{Department of Physics and Astronomy, University of New Mexico, Albuquerque, NM 87106, USA}

\author{Johannes~Knolle}
\affiliation{Department of Physics, Technische Universit\"at M\"unchen, 85748 Garching, Germany}
\affiliation{Munich Center for Quantum Science and Technology (MCQST), 80799 Munich, Germany}
\affiliation{Blackett Laboratory, Imperial College London, London SW7 2AZ, UK}

\author{Zlatko~Papi\'c}
\affiliation{School of Physics and Astronomy, University of Leeds, Leeds LS2 9JT, UK}

\author{Zohreh~Davoudi${}^{\dagger}$}
\affiliation{Maryland Center for Fundamental Physics and Department of Physics, University of Maryland, College Park, MD 20742, USA}
\affiliation{Joint Center for Quantum Information and Computer Science, NIST and University of Maryland, College Park, MD 20742, USA}
\affiliation{The NSF Institute for Robust Quantum Simulation, University of Maryland, College Park, MD 20742, USA}

% Equal contribution footnote with *
\begingroup
\renewcommand\thefootnote{*}
\footnotetext{These authors contributed equally to this work.}
\endgroup

% Corresponding authors footnote with †
\begingroup
\renewcommand\thefootnote{$\dagger$}
\footnotetext{Corresponding authors: \href{mailto:jad.halimeh@lmu.de}{jad.halimeh@lmu.de}, \href{mailto:davoudi@umd.edu}{davoudi@umd.edu}}
\endgroup

% Reset counter so normal numbering starts at 1
\setcounter{footnote}{0}

%%%%%%%%%%%%%%%%%%%%%%%%%%%%%%%%%%%%%%%%%%%%%%%%%%%%%%%%%%%%%%%%%%%%%%%%%%%%%%%%%%%%%%%%%%%%%%%%%%%%%%%%%%%%%%%%%%%%%%%%%%%%%%%%%%%%%%%%%%%%%%%%%%%%%%%%%%%%%%%
\begin{abstract}
Recent advances in quantum technologies have enabled quantum simulation of gauge theories---some of the most fundamental frameworks of nature---in regimes far from equilibrium, where classical computation is severely limited. These simulators, primarily based on neutral atoms, trapped ions, and superconducting circuits, hold the potential to address long-standing questions in nuclear, high-energy, and condensed-matter physics, and may ultimately allow first-principles studies of matter evolution in settings ranging from the early universe to high-energy collisions.
Research in this rapidly growing field is also driving the convergence of concepts across disciplines and uncovering new phenomena.
In this Review, we highlight recent experimental and theoretical developments, focusing on phenomena accessible in current and near-term quantum simulators, including particle production and string breaking, collision dynamics, thermalization, ergodicity breaking, and dynamical quantum phase transitions. We conclude by outlining promising directions for future research and opportunities enabled by available quantum hardware.
\end{abstract}

%\date{\today}
\maketitle

\tableofcontents

%%%%%%%%%%%%%%%%%%%%%%%%%%%%%%%%%%%%%%%%%%%%%%%%%%%%%%%%%%%%%%%%%%%%%%%%%%%%%%%%%%%%%%%%%%%%%%%%%%%%%%%%%%%%%%%%%%%%%%%%%%%%%%%%%%%%%%%%%%%%%%%%%%%%%%%%%%%%%%%
\section{Introduction}
The universe around us, for the most part, appears steady and in equilibrium. In fact, we have learned a great deal about fundamental interactions underlying natural phenomena by observing and analyzing low-energy states in nature, and states in thermal equilibrium. Nonetheless, the universe did not start in equilibrium, rather it evolved from a far-from-equilibrium state after the Big Bang, undergoing inflation, the electroweak phase transition, baryogenesis, and nucleosynthesis. Indeed, nonequilibrium conditions were necessary for generating a matter-dominated universe~\cite{Sakharov:1967dj}, may have led to unnatural parameter values in the Standard Model and cosmology~\cite{peccei1977cp}, or may explain today's dark-matter abundance (assuming the presence of dark matter during a metastable phase of the universe)~\cite{baker2020filtered,hong2020fermi}. The purpose of high-energy particle colliders~\cite{evans2008lhc,aamodt2008alice,florkowski2010phenomenology,lisa2005femtoscopy} is to recreate such nonequilibrium conditions in experiment, and to reach densities and temperatures necessary for generating some of the short-lived states of matter in the early universe~\cite{mclerran1986physics,harris1996search,rischke2004quark,braun2007quest}. Unfortunately, probing such states of matter is extremely challenging, often requiring indirect means~\cite{chatterjee2010electromagnetic,van2006heavy,foka2016overview,guangyou2020soft,luo2017search,metag1993near,lovato2022long}. First-principles theoretical tools fall short, as studying real-time dynamics of quantum many-body systems in general, and of complex gauge theories of the Standard Model and beyond in particular, have proven computationally intractable to date. Can large-scale, reliable, and powerful quantum simulators ultimately attack such fundamental nonequilibrium questions from first principles? 

Let us take a step back and recall that gauge theories are fundamental frameworks describing the Standard Model of particle physics. Concretely, the Standard Model is a gauge theory with the symmetry group U$(1)$ $\times$ SU$(2)$ $\times$ SU$(3)$, describing the electroweak and strong interactions~\cite{weinberg2004making,aitchison2012gauge}. More generally, gauge theories of various form, and other quantum field theories, have been conjectured to govern physics beyond the Standard Model~\cite{glashow1980towards,cveti1987gauge,brivio2019standard,langacker2017standard,svetitsky2018looking,usqcd2019lattice}, 
and the physics of early universe~\cite{linde1979phase,bailin2004cosmology,narlikar2012gravity,kumar2018heavy,blagojevic2012gauge}. 
The primary focus is often on the theory of the strong force, quantum chromodynamics (QCD)~\cite{gross202350}, since predictions based in QCD generally require nonperturbative numerical methods, even in equilibrium settings. QCD is most prevalent in the physics of colliding hadrons and nuclei in experiment. In these processes, the equilibrated matter after the collision involves abundant sprays of hadrons. These hadrons are produced from the hadronization of quarks and gluons formed shortly after the collision. Extracting information about this transient exotic phases of matter is extremely challenging, requiring tracing the equilibrated final state back to the nonequilibrium stages of evolution.

Monte-Carlo-based computing of the QCD expectation values on a Euclidean finite discretized space time,  namely lattice-QCD methods~\cite{wilson1974confinement,creutz1983monte,montvay1994quantum,rothe2012lattice,gattringer2009quantum}, have enabled successful first-principles determination of a range of static and equilibrium properties of hadrons, nucleons, and dilute thermal matter in recent years~\cite{aoki2024flag,usqcd2019hot,davoudi2021nuclear,davoudi2022report,kronfeld2022lattice}. 
However, these methods run into a sign problem when applied to dense baryonic systems~\cite{troyer2005computational,goy2017sign,nagata2022finite},
and to dynamical, out-of-equilibrium scenarios~\cite{gattringer2016approaches,cohen2015taming}, which occur in collision processes and in the early universe. Another powerful classical method is based on tensor networks~\cite{perez2006matrix,verstraete2008matrix,verstraete2006criticality,shi2006classical,murg2010simulating,orus2014practical,orus2019tensor,cirac2021matrix}. Applications of tensor networks  to LGT dynamics have seen a surge in recent years~\cite{banuls2020review,meurice2022tensor,banuls2023tensor}. However, tensor networks still fall short in high-energy processes yielding abundant entanglement generation, and further become increasingly ineffective beyond $(1+1)$ dimensions (D)~\cite{zohar2022quantum,magnifico2024tensor}. 

The rise of quantum-simulation and quantum-computing technologies~\cite{buluta2009quantum,altman2021quantum} has brought hope for potential breakthroughs in several scientific disciplines, including quantum chemistry~\cite{cao2019quantum,lanyon2010towards,mcardle2020quantum,bauer2020quantum,liu2022prospects}, material science~\cite{bauer2020quantum,alexeev2024quantum}, fusion and energy research~\cite{joseph2023quantum,paudel2022quantum}, and high-energy~\cite{bauer2023quantum,di2024quantum} and nuclear physics~\cite{cloet2019opportunities,beck2023quantum}. Quantum technologies are ideally suited for simulating complex quantum systems, as they leverage intrinsically quantum storage and processing units, providing exponentially more efficient encoding of quantum states compared to their classical counterparts~\cite{feynman2018simulating,simon1997power,lloyd1996universal}. Real-time nonequilibrium problems also constitute a natural use case for these computing systems, since encoding unitary time evolution is known to be an efficient operation in quantum hardware. In fact, time evolution can be implemented most naturally in an analog quantum simulator, upon engineering a Hamiltonian closely resembling the target Hamiltonian~\cite{cirac2012goals,georgescu2014quantum,daley2022practical}. In a digital gate-based quantum computer, evolution is digitized, via, e.g., product formulas~\cite{lie1880theorie,trotter1959product,suzuki1976generalized,lloyd1996universal,childs2021theory} or other methods~\cite{childs2012hamiltonian,berry2014exponential,berry2015simulating,berry2015hamiltonian,low2017optimal,low2019hamiltonian,low2018hamiltonian,haah2021quantum}, and can be decomposed into a set of universal elementary gates, with resources that scale at worst polynomially, and at best logarithmically, in inverse precision. In both analog and digital cases, the computational cost is polynomial in system size, contrary to path-integral Monte Carlo-based simulations of real-time observables whose cost grows exponentially in system size. State preparation and observable estimation constitute more nontrivial tasks on a quantum computer, but efficient, sometimes heuristic methods, have been developed in recent years to facilitate these steps~\cite{davoudi2025tasi}. Among popular quantum-simulation and quantum-computing architectures are optical lattices, neutral-atom arrays, trapped ions, and superconducting circuits; see Refs.~\cite{buluta2009quantum,altman2021quantum} for reviews. As these architectures are commonly used in gauge-theory quantum simulations, we briefly review the basic properties of these architectures and their underlying physics in Sec.~\ref{sec:platforms}.

Unfortunately, many nonequilibrium aspects of gauge theories of the Standard Model and beyond may need to await the arrival of large-scale fault-tolerant quantum computers; see, e.g., digital-algorithm developments for gauge theories and the estimated costs in Refs.~\cite{byrnes2006simulating,Shaw:2020udc, ciavarella2021trailhead, kan2021lattice,Lamm:2019bik,haase2021resource,Davoudi:2022xmb,Murairi:2022zdg,rhodes2024exponential,Lamm:2024jnl,Balaji:2025afl}. Nonetheless, a plethora of explorations and results can be achieved in the meantime in the context of simpler prototype models, which are amenable to present-time quantum simulators. For example, some defining features of QCD, such as confinement, chiral symmetry breaking, and a nontrivial CP-violating term, feature in a 1+1 dimensional U$(1)$ gauge theory, or the Schwinger model~\cite{schwinger1962gauge,coleman1976more}. Other testing grounds include $\mathbb{Z}_2$ lattice gauge theories (LGTs) and their spin-model equivalents, various quantum link models of gauge theories~\cite{chandrasekharan1997quantum}, and highly truncated Kogut--Susskind LGTs~\cite{kogut1975hamiltonian} in lower dimensions. These formulations are briefly introduced in Sec.~\ref{sec:formulations}. While reliable and quantitative QCD predictions will not be possible by these early studies, a great deal of insight will be gained along the way on problems that will essentially be entirely inaccessible  via classical simulations. Another aspect is the role of these simpler LGTs not only as fundamental but also as effective descriptions, e.g., in condensed matter systems~\cite{wen2004quantum,fradkin2013field,wen1990topological,levin2005string}, as ingredient for quantum-error-correction schemes~\cite{kitaev2003fault,kitaev2006anyons,das2006topological}, or in fermion-to-boson mappings~\cite{chen2018exact,chen2020exact,chen2023equivalence}. Shining light on the nonequilibrium physics of these models via quantum simulation, therefore, is a valuable, and more achievable objective, in the near term.

This Review covers quantum-simulation proposals and implementation for a range of nonequilibrium phenomena in gauge theories. These will ultimately pave the way toward studying the same or related phenomena in nuclear and particle physics, cosmology, as well as condensed-matter and quantum-information theory. With an emphasis on nuclear- and high-energy-physics motivations, we summarize the phenomena we cover in this Review. For further context and references, we refer the reader to the corresponding Sections.
\begin{itemize}
\item[$\diamond$] Confinement is a hallmark of QCD. It forbids  color charged objects, i.e., quarks and gluons, to exist in isolation, forcing them to form color-neutral bound states. In a thought experiment, one pulls apart the color charges, which results in the increase in the electric energy stored in the ``flux tube'' connecting them. Eventually, the creation of additional charges from the interacting vacuum becomes energetically favored. Hence, the string breaks to smaller fragments, and the original charges get screened by the new ones. Deciphering mechanisms of \emph{particle production and string breaking} in real-time, out-of-equilibrium phenomena, is of paramount importance in nuclear and particle physics. Particle production in quantum quenches of trivial vacuum, and string-breaking dynamics in quench, diabatic, and adiabatic processes, and closely related false-vacuum decay phenomenon, have received significant attention in recent years, albeit in simpler lower-dimensional gauge-theory models. In fact, several experimental demonstrations have become reality only recently. This progress will be reviewed in Sec.~\ref{sec:particle-string}.

\item[$\diamond$] Producing nonequilibrium conditions in particle colliders necessitates going beyond simple quench, diabatic, and adiabatic processes. Instead of starting from a simple initial state, nontrivial initial states, such as high-energy colliding hadrons are needed to study the nonequilibrium post-collision dynamics. As a result, state preparation is a crucial, and often costly step, in quantum simulation of \emph{collision processes}. Particle production and hadronization processes, such as string breaking, remain of relevance in these realistic scenarios. Asymptotic scattering amplitudes, or nonperturbative contributions to them, are further sought after in these simulations. Accessing these quantities demands efficient quantum algorithms, oftentimes only accessible in digital quantum computers. Considerable progress is made in recent years to establish routes to studies of collision processes on quantum simulators, in both analog and digital fashions, and the first hardware studies, albeit in simpler yet confining gauge theories, have emerged. The progress in quantum simulation of particle collisions will be reviewed in Sec.~\ref{sec:scattering}. 

\item[$\diamond$] Developing a first-principles understanding of the evolution of matter toward equilibrium under extreme densities and temperatures post Big Bang and in heavy-ion colliders is a major undertaking. In general, \emph{thermalization} in quantum many-body dynamics, including in gauge theories, is a common, yet complex phenomenon. Many questions remain under debate, particularly in the quantum regime: Is there a universal thermalization paradigm? What are the thermalization stages and are they similar among radically different systems and initial conditions? What are the time scales associated with thermalization compared to system's characteristic times? What is the role of entanglement in thermalizing systems and can it serve as a universal probe of thermalization? Theoretical tools developed in recent years to study QCD thermalization require working in limited parameter regimes. Unfortunately, probing gauge-theory thermalization in all coupling regimes is challenging, hence providing another window of opportunity for quantum simulators. A wealth of thermalization studies with quantum-simulation and quantum-information perspectives has appeared in recent years. We review the developments of relevance to gauge theories in Sec.~\ref{sec:thermalization}.

\item[$\diamond$] While many quantum many-body systems evolved under nonintegrable Hamiltonians eventually thermalize, deviation from ergodic behavior can still occur. Understanding the type and origin of such \emph{ergodicity-breaking} situations has constituted an exciting area of investigation in recent years. Various paradigms include Hilbert-space fragmentation, many-body localization, disorder-free localization, and quantum many-body scarring. While these studies originated in other disciplines, such as in quantum many-body physics and condensed-matter physics, their implications for lattice-gauge-theory dynamics have been recently brought to spotlight. Many questions remain unanswered, including the fate of nonergodic behavior toward the continuum limit of lattice field theories, and the relevance, and potential signatures, of nonergodic dynamics and initial conditions in particle colliders, early universe, and cosmological settings. Quantum simulators have begun to explore these open questions, albeit via simpler prototype models. These developments will be reviewed in Sec.~\ref{sec:ergodicitybreaking}.

\item[$\diamond$] A framework for describing phase transitions out of equilibrium, and to relate the phases and phase transitions in equilibrium states to their nonequilibrium counterparts, will be of tremendous value. It would also be worthwhile to search for universalities in out-of-equilibrium settings, e.g., using \emph{dynamical quantum phase transitions}, to understand relations to topological phenomena, or to devise entanglement and other information-theoretic probes of such nonequilibrium transitions. These goals have resulted in an active frontier of research in quantum many-body physics, which has further been applied to simple gauge theories in recent years. From a fundamental standpoint, potential nonequilibrium dynamics may have changed the value of the topological $\theta$ angle in the early universe, hence providing a potential solution to the strong CP problem. Confirming such a possibility, nonetheless, requires a first-principle study of quantum field theories of nature out of equilibrium, which will be accessible using quantum simulations. In Sec.~\ref{sec:DQPT}, we review progress in quantum simulating dynamical quantum phase transitions in gauge theories.
\end{itemize}
The scope and diversity of the topics covered in this Review, and the extent and course of developments reviewed,\footnote{Disclaimer: Only developments released by July 31, 2025 are covered in this Review.} demonstrate that this field is only at the beginning of a long, but exciting and clear path toward addressing a range of questions in nuclear physics, particle physics, cosmology, and quantum many-body dynamics. Along the way, this frontier will leverage tools and expertise in multiple disciplines. It will provide insights into the unexplored nonequilibrium physics of gauge theories for other applications, from condensed-matter physics to quantum information science. We will remark on the potential impact of this frontier, and on some of the future directions, in Sec.~\ref{sec:outlook}. 

%%%%%%%%%%%%%%%%%%%%%%%%%%%%%%%%%%%%%%%%%%%%%%%%%%%%%%%%%%%%%%%%%%%%%%%%%%%%%%%%%%%%%%%%%%%%%%%%%%%%%%%%%%%%%%%%%%%%%%%%%%%%%%%%%%%%%%%%%%%%%%%%%%%%%%%%%%%%%%%
\section{Overview of relevant theoretical frameworks and experimental platforms}
Before diving into the topic of nonequilibrium gauge-theory phenomena on quantum simulators, we introduce some of the theoretical frameworks and experimental platforms that are used throughout this Review. As experimental progress has happened in simpler, lower-dimensional gauge theories, we put the focus on such models, and refrain from a comprehensive overview of Hamiltonian gauge theories of relevance to nature. For more complete discussions, see Refs.~\cite{bauer2023quantum,di2024quantum,klco2022standard,bauer2024quantum}. Similarly, while the field of simulation-hardware development is rather diverse, and a broad range of hardware architectures are being explored in the context of quantum simulation, we refrain from a comprehensive overview, and instead refer the reader to an excellent recent review in Ref.~\cite{altman2021quantum}. This Section further sets the notation and conventions for the rest of the Review. As a first convention, we set the natural constants to unity: $c=\hbar=k_\text{B}=1$. As a first notation, we use $(d+1)$D or $d$d to denote a system or phenomenon in $d$ spatial dimensions.

%%%%%%%%%%%%
%%%%%%%%%%%%
\subsection{Gauge theories and their finite-dimensional descriptions}\label{sec:formulations} 

The gauge theories of relevance to the Standard Model are described by continuous groups, and are hence infinite-dimensional. There are many ways to truncate the infinite-dimensional Hilbert space of a gauge theory to a finite-dimensional one, which is a requirement for classical and quantum simulation; see, e.g., Sec.~VI in Ref.~\cite{bauer2023quantum}. A common approach is to truncate the space and discretize it, i.e., define the fields only on a finite lattice. However, even then, for continuous groups, a truncation and/or digitization of the gauge-boson degrees of freedom is required at even a single lattice site. This amounts to first choosing a set of basis states, with the requirement that they are countable and finite. A common choice is the electric-field basis, which is equivalent to the irreducible representation (irrep) of the group. Nonetheless, simulation in group-element basis, magnetic basis, and other bases have also been considered in recent years. To put these discussions on a concrete footing, we proceed with introducing two paradigmatic LGTs based on the $\mathrm{U}(1)$ (continuous) and $\mathbb{Z}_2$ (discrete) gauge groups. These models will accompany us throughout the majority of this Review. For simplicity, we will first focus on the $(1+1)$D case, but we will also discuss two and higher spatial dimensions briefly. A general form of a U$(1)$ and SU$(N)$ LGTs in any dimensions within the Kogut--Susskind framework will be presented later in this Section.

%%%
%%%
\subsubsection{$\mathrm{U}(1)$ lattice gauge theory and its quantum link models}\label{sec:U1LGT_formulations}
Let us first begin with the provenance of $\mathrm{U}(1)$ gauge theories, namely the massive Schwinger model~\cite{schwinger1962gauge}, which is quantum electrodynamics in $(1+1)$D, with a topological $\theta$-term. In the temporal gauge ($A_0=0$), its continuous Hamiltonian can be written as
\begin{align}\label{eq:QED_continuum}
\hat{H}_\text{QED}^{(1+1)\text{D}}=\int dx \, \bigg[\hat{\Psi}^\dagger(x)\Big({-}i\gamma^1\hat{D}_1+
& m\Big)\hat{\Psi}(x)+\frac{1}{2}\hat{E}(x)^2
\nonumber\\
& +\frac{g\theta}{2\pi}\hat{E}(x)\bigg],
\end{align}
where $\hat{\psi}(x)$ is the two-component Dirac-fermion operators at spatial position $x$, $\hat{E}(x)$ is the electric field, $g$ is the gauge coupling, $\gamma^{0,1}$ are the Dirac matrices in one spatial dimension, $\hat{D}_1 \coloneq \partial_1+ig\hat{A}(x)$ is the covariant derivative, $\hat{A}(x)$ is the gauge-vector potential, $m$ is the fermionic rest mass, and $\theta$ is the topological angle. The first term of the Hamiltonian in Eq.~\eqref{eq:QED_continuum} is the kinetic energy of the fermions, which couple to the gauge potential $\hat{A}(x)$ through $\hat{D}_1$, the second term describes the electric-field energy, while the third is the topological $\theta$-term. In this model, the latter is equivalent to a background charge $E_\theta \coloneq g\theta/(2\pi)$~\cite{coleman1976more}. $\theta$ is an angular parameter, and physics is invariant under $\theta \to \theta+2\pi$. In the following, we shift $\theta$ to lie within $[-\pi,\pi]$ by convention (such that $\theta = \pi$ would correspond to no linear electric-field term)~\cite{coleman1976more}.

\vspace{0.125 cm}
\noindent
\emph{The bosonized Schwinger model:} Consider the continuum QED Hamiltonian in Eq.~\eqref{eq:QED_continuum}. Via a standard bosonization procedure~\cite{von1998bosonization,senechal2004introduction}, which applies to fermionic theories in $(1+1)$D, this theory can be shown to be dual to a bosonic scalar field theory with the Hamiltonian~\cite{coleman1975charge,coleman1976more}
\begin{align}
	\label{eq:Schwinger-bose-ham}
	\hat H^{(1+1)\text{D}}_{\text{bos.}} =  \int dx & :\bigg[\frac{\hat \Pi^2(x)}{2} +\frac{(\partial_x\hat \phi(x))^2}{2}+\frac{g^2\hat\phi^2(x)}{2\pi}\nonumber\\
    &-\frac{bmg\cos(\sqrt{4\pi}\hat \phi(x)-\theta)}{2\pi^{3/2}} \bigg]:.
\end{align}
Here, $\hat \phi(x)$ and $\hat \Pi(x)$ are the scalar field and its conjugate momentum, respectively, and $b \coloneq e^\gamma$ with $\gamma$ being Euler's constant. The lattice-regularized form of this theory,
\begin{align}
	\label{eq:Schwinger-bose-lattice}
	\hat H^{(1+1)\text{D}}_{\text{bos.}} = \xi\sum_i \bigg[\frac{\hat \pi_i^2}{2}&+\frac{(\hat \phi_{i}-\hat \phi_{i-1})^2}{2} +\frac{\mu^2\hat\phi_i^2}{2}
    \nonumber\\
    &-\lambda\cos(\beta\hat\phi_i-\theta)\bigg],
\end{align}
can be mapped to bosonic quantum simulators. Here, $i$ labels lattice sites, $\comm{\hat \phi_i}{\hat \pi_j}=i\delta_{i,j}$, $\xi \coloneq 1/a$, $\beta \coloneq \sqrt{4\pi}$, $\mu^2 \coloneq a^2g^2/\pi$, $\lambda \coloneq a^2b mg \exp[2\pi\Delta(a)]/2\pi^{3/2}$, $a$ is the lattice spacing, and $\Delta(a)$ is the lattice Feynman propagator at the origin \cite{coleman1975quantum,ohata2023monte}. $a$ will be set to 1 (i.e., quantities will be expressed in units of lattice spacing). With this choice, continuum limit corresponds to $\mu,\lambda \to 0$. The model has two dimensionless parameters, the ratio $g/m$, corresponding to $\mu/\lambda$ in the lattice bosonized form, and the angle $\theta$ corresponding to the constant background electric field $E_\theta$. Total electric field in the original form maps to $\hat E_T=\hat E_\theta+\hat E=\frac{g}{\sqrt{\pi}} \hat \phi$ in the bosonic form~\cite{shankar2005deconfinement}. The electric field $\hat E$ is related to dynamical charges via Gauss's law, $\partial_x \hat E(x)=g\hat\Psi^\dagger(x)\hat\Psi(x)$.

Two regimes are of interest: i) A deconfined phase at $\theta=\pi$ (terminating at the Ising critical point), where the ground state is two-fold degenerate. The fundamental excitations are  ``half-asymptotic" fermions (``quarks"), or topological kinks in the bosonic dual~\cite{coleman1976more}. ii) A confined phase, with a unique ground state, exhibiting quark-antiquark bound-state (``meson") excitations.

\vspace{0.125 cm}
\noindent
\emph{The Kogut--Susskind formulation:} 
A popular choice to discretize fermionic field theories is the Kogut--Susskind formulation~\cite{kogut1975hamiltonian}, in which the two-component Dirac fermions $\Psi$ in $(1+1)$D are replaced with one-component staggered fermions $\psi$ on a lattice. Then upon a field rescaling $\psi \to \sqrt{a} \psi$ where $a$ is the lattice spacing, the Schwinger Hamiltonian takes the form:
\begin{align}\nonumber
    \hat{H}_\text{KS}^{(1+1)\text{D}}=&-\frac{\kappa}{2a}\sum_{\ell=1}^{L-1}\Big(\hat{\psi}_\ell^\dagger\hat{U}_{\ell,\ell+1}\hat{\psi}_{\ell+1}+\text{H.c.}\Big)\\\nonumber
    &+m\sum_{\ell=1}^{L}(-1)^\ell\hat{\psi}_\ell^\dagger\hat{\psi}_\ell+\frac{a}{2}\sum_{\ell=1}^{L-1}\hat{E}_{\ell,\ell+1}^2\\
    &-\frac{ag\big(\theta-\pi\big)}{2\pi}\sum_{\ell=1}^{L-1}\hat{E}_{\ell,\ell+1}.
    \label{eq:SchwingerKS}
\end{align}
Here, $\kappa$ is the minimal coupling or hopping strength (which should be set to one when matched to the continuum U$(1)$ Hamiltonian), $L$ is the number of lattice sites, and the fields' subscripts are the site indices in units of lattice spacing. Equation~\eqref{eq:SchwingerKS} assumes open boundary conditions on the fields; periodic boundary conditions can be described by the same Hamiltonian upon adding the hopping and electric Hamiltonians on the link connecting sites $1$ and $L$. $\hat{U}_{\ell,\ell+1}=e^{iga\hat{A}_{\ell,\ell+1}}$ is the parallel transporter at the link between sites $\ell$ and $\ell+1$. The parallel transporter and electric-field operator satisfy the commutation relations
\begin{subequations}
    \begin{align}\label{eq:UEcommA}
        \Big[\hat{E}_{\ell,\ell+1},\hat{U}_{\ell',\ell'+1}\Big]&=g\delta_{\ell,\ell'}\hat{U}_{\ell,\ell+1},\\\label{eq:UEcommB}
        \Big[\hat{U}_{\ell,\ell+1},\hat{U}_{\ell',\ell'+1}^\dagger\Big]&=0.
    \end{align}
\end{subequations}
The generator of Gauss's law is
\begin{align}\label{eq:GaussLawKS}
    \hat{G}_\ell=\hat{E}_{\ell,\ell+1}-\hat{E}_{\ell-1,\ell}-g\left[\hat{\psi}_\ell^\dagger\hat{\psi}_\ell+\frac{(-1)^\ell-1}{2}\right].
\end{align}
Physical states are those satisfying the Gauss's law:
\begin{align}
\hat{G}_\ell \ket{\Psi}_{\rm phys}=0,~\forall \ell.
\end{align}

Although space is now discretized, the operators $\hat{U}_{\ell,\ell+1}$ and $\hat{E}_{\ell,\ell+1}$ still have an infinite-dimensional local Hilbert space. Thus, unless the quantum simulator has a continuous-variable or bosonic degrees of freedom, encoding the gauge fields demands further truncation and/or digitization. There are several ways around this issue. Here, we enumerate two schemes that have been frequently adopted in experimental realizations in discrete-variable platforms. 

\vspace{0.125 cm}
\noindent
\emph{Integrating out the gauge fields in the Kogut--Susskind formulation:} 
Since the gauge fields are locally infinite-dimensional, it would be useful to employ Gauss's law in Eq.~\eqref{eq:GaussLawKS} to integrate them out. This procedure, which is applicable in $(1+1)$D with open boundary conditions, obtains a Hamiltonian containing only the matter fields~\cite{hamer1997series}. First, the gauge fields can be eliminated by the gauge transformations $\hat{\psi}_\ell\to\left(\prod_{\ell'=1}^{\ell-1}\hat{U}_{\ell',\ell'+1}\right)\hat{\psi}_\ell$, which render the gauge links equal to identity, $\hat{U}_{\ell,\ell+1}=\mathds{1},~\forall \ell$. Second, from the Gauss's law, the electric field at each link can be replaced by the sum of the all electric charges prior to the link plus an incoming electric field to the one-dimensional lattice, which we set to zero, i.e., $E_{0,1}=0$. Explicitly, $\hat{E}_{\ell,\ell+1}=g\sum_{\ell'=1}^{\ell}\left[\hat{\psi}_{\ell'}^\dagger\hat{\psi}_{\ell'}+\frac{(-1)^{\ell'}-1}{2}\right]$. Furthermore, one can employ the Jordan--Wigner transformation $\hat{\psi}_\ell=\left(\prod_{\ell'=1}^{\ell-1}\hat{\sigma}^z_{\ell'}\right)\hat{\sigma}^-_\ell$ to express fermionic fields in terms of Pauli spin operators.\footnote{The convention $\hat{\psi}_\ell=\prod_{\ell'=1}^{\ell-1}\hat{\sigma}^z_{\ell'}\hat{\sigma}^+_\ell$ is also commonly used, which upon proper identification of computational basis states with the fermion occupation basis, yields an identical formulation.}

Substituting these relations into the Hamiltonian in Eq.~\eqref{eq:SchwingerKS} gives rise to
\begin{align}
    \label{eq:SchwingerGaugeFieldsIntegratedOut}
    &\hat{H}_\text{KS,f}^{(1+1)\text{D}}=\hat{H}_\pm+\hat{H}_{ZZ}+\hat{H}_Z,
\end{align}
where~\cite{martinez2016real}
\begin{subequations}
\label{eq:SchwingerGaugeFieldsIntegratedOut-terms}
\begin{align}
    &\hat{H}_\pm=-\frac{\kappa}{2a}\sum_{\ell=1}^{L-1}\Big(\hat{\sigma}_\ell^+\hat{\sigma}_{\ell+1}^-+\hat{\sigma}_\ell^-\hat{\sigma}_{\ell+1}^+\Big),\\
    &\hat{H}_{ZZ}=\frac{g^2a}{2}\sum_{\ell=1}^{L-2}\sum_{\ell'=\ell+1}^{L-1}(L-\ell')\hat{\sigma}^z_\ell\hat{\sigma}^z_{\ell'},\\\nonumber
    &\hat{H}_Z=\frac{g^2a}{    8}\sum_{\ell=1}^{L-1}\big[(-1)^\ell-1\big]\sum_{\ell'=1}^\ell\hat{\sigma}^z_{\ell'}\\
    &\hspace{0.95 cm}-\frac{g^2a\big(\theta-\pi\big)}{4\pi}\sum_{\ell=1}^{L-1}\sum_{\ell'=1}^\ell\hat{\sigma}^z_{\ell'}+\frac{m}{2}\sum_{\ell=1}^{L}(-1)^\ell\hat{\sigma}^z_\ell.
\end{align}
\end{subequations}

\vspace{0.125 cm}
\noindent
\emph{The quantum-link-model formulation:} 
The quantum link model (QLM) formulation~\cite{chandrasekharan1997quantum,brower1999qcd,Wiese_review} maps the gauge and electric field operators onto spin-$S$ operators according to
\begin{subequations}\label{eq:QLMmapping}
    \begin{align}
        \hat{U}_{\ell,\ell+1}&\to\frac{\hat{S}^+_{\ell,\ell+1}}{\sqrt{S(S+1)}},\\
        \hat{E}_{\ell,\ell+1}&\to g\hat{S}^z_{\ell,\ell+1}.
    \end{align}
\end{subequations}
This formulation allows for discretizing these operators into a $(2S+1)$-dimensional local Hilbert space. This map automatically satisfies the commutation relation in Eq.~\eqref{eq:UEcommA} for any $S$, while Eq.~\eqref{eq:UEcommB} is satisfied only in the limit $S\to\infty$. Plugging Eqs.~\eqref{eq:QLMmapping} into the Hamiltonian in Eq~\eqref{eq:SchwingerKS}, and upon a Jordan--Wigner transformation of the fermion fields, one obtains the QLM Hamiltonian
\begin{align}\nonumber
    \hat{H}^{(1+1)\text{D}}_\text{QLM}=&-
    \tilde{\kappa} \sum_{\ell=1}^{L-1}\Big(\hat{\sigma}^+_\ell\hat{S}^+_{\ell,\ell+1}\hat{\sigma}^-_{\ell+1}+\text{H.c.}\Big)\\\nonumber
    &+\frac{m}{2}\sum_{\ell=1}^{L}(-1)^\ell\hat{\sigma}^z_\ell+\frac{g^2a}{2}\sum_{\ell=1}^{L-1}\big(\hat{S}^z_{\ell,\ell+1}\big)^2\\\label{eq:U1QLM}
    &-
    a\chi \sum_{\ell=1}^{L-1}\hat{S}^z_{\ell,\ell+1},
\end{align}
with 
\begin{subequations}
    \begin{align}\label{eq:tildekappa}
    &\tilde{\kappa} \coloneq\frac{\kappa}{2a\sqrt{S(S+1)}}, \\\label{eq:chi}
    &\chi \coloneq \frac{g^2(\theta-\pi)}{2\pi}.
    \end{align}
\end{subequations}
The generator of the $\mathrm{U}(1)$ gauge symmetry, i.e., the Gauss's law operator, takes the form $\hat{G}_\ell=\hat{S}^z_{\ell,\ell+1}-\hat{S}^z_{\ell-1,\ell}-\hat{\psi}_\ell^\dagger\hat{\psi}_\ell+[1-(-1)^\ell]/2=\hat{S}^z_{\ell,\ell+1}-\hat{S}^z_{\ell-1,\ell}-[\sigma^z_\ell+(-1)^\ell]/2$. The physical gauge sector constitutes states that are annihilated by this operator. The continuum limit is recovered when $S \rightarrow \infty$ and $ag \rightarrow 0$ \cite{buyens2017finite, zache2022toward,halimeh2022achieving}. As we will see later in this Review, even at rather small values such as $S=1/2$, the Hamiltonian in Eq.~\eqref{eq:U1QLM} captures many of the salient features of the Schwinger model.

For example, the model in Eq.~(\ref{eq:U1QLM}) hosts a $\mathbb{Z}_2$ symmetry-breaking transition as a function of $m/\tilde{\kappa}$. Consider the spin-$\frac{1}{2}$ variant of the model with $\chi=0$ (i.e., $\theta = \pi$). The ground state of this model can be deduced easily in two extreme cases. First, for $m/\tilde{\kappa} \to +\infty$, the Hamiltonian has two degenerate ground states. These are states that host no particle or antiparticle while hosting one of the two possible electric-field configurations consistent with Gauss's law:
\begin{align}
\ket{\cdots \triangleleft,\varnothing,\triangleleft,\varnothing,\triangleleft \cdots}~\text{and}~\ket{\cdots \triangleright,\varnothing,\triangleright,\varnothing,\triangleright \cdots}.
\label{eq:vacuua}
\end{align}
Here, $\{ \ket{\triangleleft},\ket{\triangleright} \}$ denote two eigenstates of the spin-$\frac{1}{2}$ operator $\hat{S}_{i}^z$ and $\varnothing$ represents the absence of matter at a site. The vacuum states break charge and parity symmetry and exhibit nonzero electric flux. The model is nonconfining in this limit. Second, if $m/\tilde{\kappa} \to -\infty$, the vacuum is unique; all particle and antiparticle sites are occupied and the electric-field configuration is a staggered one according to Gauss's law:
\begin{align}
\ket{\cdots \triangleleft,e^+,\triangleright,e^-,\triangleleft \cdots}.
\end{align}
This phase is C and P symmetric and hosts a net zero electric flux. It turned out that the model exhibits an equilibrium quantum phase transition at $(m/\tilde{\kappa})_c=0.655$, within the $2$d Ising universality class, 
which separates a symmetry-broken phase with two degenerate ground states [for $m/\tilde{\kappa}>(m/\tilde\kappa)_c$] from a paramagnetic one [for $m/\tilde{\kappa}<(m/\tilde\kappa)_c$]~\cite{byrnes2002density,Rico2014,huang2019dynamical,yang2020observation}. This is the Coleman phase transition also observed in the Schwinger model~\cite{coleman1976more,yang2020observation}.

For $\chi \neq 0$ (i.e., $\theta \neq \pi$), a background electric field is generated that explicitly breaks the global $\mathbb{Z}_2$ symmetry ($\ket{\triangleleft} \to \ket{\triangleright}$), creating an energy difference between the two electric fluxes $\{ \ket{\triangleleft},\ket{\triangleright} \}$.\footnote{Note that the term proportional to $\big(\hat{S}^z_\ell\big)^2$ in Eq.~\eqref{eq:U1QLM} becomes an irrelevant energy constant when $S=\frac{1}{2}$.} As a result, a particle-antiparticle pair connected by a string of electric fluxes experiences a string energy that increases linearly with the string length. Subsequently, the spin-$\frac{1}{2}$ QLM becomes a confining theory.

\vspace{0.125 cm}
\noindent
\emph{Integrating out the matter fields in the QLM formulation:} 
Another useful approach is to integrate out the matter fields, also through Gauss's law, which renders a pure spin model representing the QLM in the physical sector. As a first step, one can perform a convenient particle-hole transformation, that is performing a staggered rotation about the Bloch x axis: $\hat{\sigma}^z_\ell\to(-1)^\ell\hat{\sigma}^z_\ell$, $\hat{S}^z_{\ell,\ell+1}\to(-1)^{\ell+1}\hat{S}^z_{\ell,\ell+1}$, $\hat{\sigma}^y_\ell\to(-1)^\ell\hat{\sigma}^y_\ell$, $\hat{S}^y_{\ell,\ell+1}\to(-1)^{\ell+1}\hat{S}^y_{\ell,\ell+1}$, which leaves the Hamiltonian in Eq.~\eqref{eq:U1QLM} in the form \cite{Hauke2013quantumsimulation}
\begin{align}
    \hat{H}^{(1+1)\text{D}}_\text{QLM}=&-
    \tilde{\kappa}\sum_{\ell=1}^{L-1}\Big(\hat{\sigma}^-_\ell\hat{S}^+_{\ell,\ell+1}\hat{\sigma}^-_{\ell+1}+\text{H.c.}\Big)\nonumber\\
    &+\frac{m}{2}\sum_{\ell=1}^{L}\hat{\sigma}^z_\ell+\frac{g^2a}{2}\sum_{\ell=1}^{L-1}\big(\hat{S}^z_{\ell,\ell+1}\big)^2
    \nonumber\\
    \label{eq:U1QLMPH}
    &
    -a\chi \sum_{\ell=1}^{L-1}(-1)^{\ell+1}\hat{S}^z_{\ell,\ell+1},
\end{align}
and Gauss's law now takes the form 
\begin{align}\label{eq:GjPH}
\hat{G}_\ell=(-1)^{\ell+1}\bigg(\hat{S}^z_{\ell,\ell+1}+\hat{S}^z_{\ell-1,\ell}+\frac{\hat{\sigma}^z_\ell+\mathds{1}}{2}\bigg).
\end{align}
Again restricting the physics to the physical sector, yields $\hat{\sigma}^z_\ell/2=-(\hat{S}^z_{\ell,\ell+1}+\hat{S}^z_{\ell-1,\ell}+\mathds{1}/2)$. Plugging the latter into the Hamiltonian in Eq.~\eqref{eq:U1QLMPH} gives an alternate QLM Hamiltonian \cite{Desaules2022prominent}
\begin{align}\nonumber
    \hat{\tilde{H}}^{(1+1)\text{D}}_{\text{QLM}}=&-2\tilde{\kappa}\hat{\mathcal{P}}\Big(\sum_{j}\hat{S}^x_j\Big)\hat{\mathcal{P}}+\frac{g^2a}{2}\sum_j\big(\hat{S}^z_j\big)^2\\\label{eq:U1QLM_MatterIntegratedOut}
    &-\sum_j\left[2m+a\chi(-1)^j\right]\hat{S}^z_j,
\end{align}
up to an irrelevant constant. Note that we have dropped the link notation and adopted a site-only notation in its place.\footnote{Technically, this form applies to periodic boundary conditions where there is a one-to-one correspondence between the number of sites and the number of links emanating from the sites. When open boundary conditions are concerned, one can properly modify the boundary terms; see, e.g., Sec.~\ref{sec:Rydberg}.} Here, $\hat{\mathcal{P}} \coloneq \prod_j\hat{P}_{j,j+1}$ is a global projector onto the allowed states in the physical sector, where the local two-site projector is defined as
\begin{align}\label{eq:ProjectorQLM}\nonumber
&\hat{P}_{j,j+1} \coloneq \sum_{m_z=-S}^S
(\ket{m_z}\bra{m_z})_j \otimes (\ket{-m_z}\bra{-m_z})_{j+1}\\
&+\sum_{m_z=-S}^{S-1}
(\ket{m_z}\bra{m_z})_j \otimes (\ket{-m_z-1}\bra{-m_z-1})_{j+1}.
\end{align}
It is straightforward to understand the terms of $\hat{P}_{j,j+1}$ by considering the physical sector of the generator in Eq.~\eqref{eq:GjPH}. The first (second) term denotes the allowed electric-field configurations on two neighboring links in the absence (presence) of matter on the site in between, where their eigenvalues must sum to $0$ ($-1$) in order to satisfy Gauss's law.

%%%
%%%
\subsubsection{$\mathbb{Z}_2$ lattice gauge theory and its Ising-model duals
\label{sec:Z2-Ising}}
The $\mathbb{Z}_2$ LGT is another paradigmatic model at the center of many theoretical and experimental works in recent years. Here, we introduce a few of its variants that have been studied in the context of nonequilibrium gauge-theory quantum simulations.

\vspace{0.125 cm}
\noindent
\emph{A typical $(1+1)$D $\mathbb{Z}_2$ LGT:}
For a theory of one-component fermions (mapped to spins via a Jordan--Wigner transformation) and hardcore gauge bosons in $(1+1)$D, the Hamiltonian is given by \cite{schweizer2019floquet}
\begin{align}\label{eq:Z2LGT}
    \hat{H}_{\mathbb{Z}_2}^{(1+1)\text{D}}=-\sum_{\ell=1}^{L-1}\big[J\big(\hat{\sigma}^+_\ell\hat{\tau}^z_{\ell,\ell+1}\hat{\sigma}^-_{\ell+1}+\text{H.c.}\big)+h\hat{\tau}^x_{\ell,\ell+1}\Big].
\end{align}
The Pauli operators $\hat{\sigma}^\pm_\ell = (\hat{\sigma}_\ell^x \pm i\hat{\sigma}_\ell^y)/2$ are the matter creation and annihilation operators on site $\ell$, and the Pauli matrix $\hat{\tau}^{x(z)}_{\ell,\ell+1}$ represents the electric (gauge) field at the link between sites $\ell$ and $\ell+1$.\footnote{The alternative choice $\hat{\sigma}^{x(z)}_{\ell,\ell+1}$ for the gauge (electric) field is also commonly used.} The first term in the Hamiltonian in Eq.~\eqref{eq:Z2LGT} is the minimal coupling with strength $J$, which involves the gauge-invariant tunneling of matter between two neighboring sites, while the second term is the electric field with strength $h$. The generator of the $\mathbb{Z}_2$ gauge symmetry, i.e., the Gauss's law operator, is
\begin{align}\label{eq:Z2LGT_GaussLaw}
    \hat{G}_\ell=-\hat{\tau}^x_{\ell-1,\ell}\hat{\sigma}^z_\ell\hat{\tau}^x_{\ell,\ell+1}.
\end{align}
This operator has two eigenvalues, $\pm1$, and the physical sector is often, by convention, the sector with the $+1$ eigenvalue at all sites. The electric-field term acts as a confining potential such that, for any value of $h\neq0$, particle-antiparticle pairs become bound, leading to confinement. At $h=0$, the system is deconfined. In higher spatial dimensions, which we will discuss later, the confinement-deconfinement transition occurs at a finite nonzero value of $h$.
Other variants of the $\mathbb{Z}_2$ LGT Hamiltonian in $(1+1)$D include, for example, a staggered mass term or a gauge-invariant pairing term.

\vspace{0.125 cm}
\noindent
\emph{A $(1+1)$D $\mathbb{Z}_2$ LGT with a mixed-field Ising-model dual:} There exists a $(1+1)$D $\mathbb{Z}_2$ LGT whose dynamics in the gauge-invariant sector are equivalent to those of the quantum Ising chain~\cite{de2024observation,lerose2020quasilocalized,surace2021scattering}.  
The Hamiltonian of this LGT reads
\begin{align}
\label{eq_Hgauged}
\hat{\tilde{H}}_{\mathbb{Z}_2}^{(1+1)\text{D}}&=
 -g \sum_\ell \Big[  \hat c^\dagger_\ell (\hat b_{\ell,\ell+1}+\hat b^\dagger_{\ell,\ell+1}) \hat c_{\ell+1} 
 \nonumber\\
&+\hat c^\dagger_\ell (\hat b_{\ell,\ell+1} + \hat b^\dagger_{\ell,\ell+1}) \hat c^\dagger_{\ell+1} + \mathrm{H.c.} \Big]+m \sum_\ell \hat c^\dagger_\ell \hat c_\ell 
\nonumber\\
&+ \zeta \sum_\ell  \hat n_{\ell,\ell+1} - \sum_\ell
 \sum_{ \ell'>1}v_{\ell'} \hat n_{\ell,\ell+1} \hat n_{\ell+\ell'}.
\end{align}
Here, $\hat c^\dagger_\ell$ and $\hat c_\ell$ are fermionic creation and annihilation operators on lattice site $\ell$. $\hat b^\dagger_{\ell,\ell+1}$ and $\hat b_{\ell,\ell+1}$ are, respectively, hardcore-boson creation and annihilation operators, or gauge-links operators, residing on the link connecting sites $\ell$ and $\ell+1$, with $\hat n_{\ell,\ell+1}=\hat b^\dagger_{\ell,\ell+1} \hat b_{\ell,\ell+1}$ being the electric field. $g$, $m$, $\zeta$, and $v_r$ represent, respectively, the coupling between matter and gauge fields, fermion mass, energy cost associated with an electric-field flip, and  electric-field's variable-range self-interaction strength. 

The Hamiltonian in Eq.~\eqref{eq_Hgauged} is invariant under $\mathbb{Z}_2$ gauge transformations generated by the local Gauss-law operators $\hat G_\ell= (-1)^{\hat n_{\ell-1,\ell}+\hat n_{\ell,\ell+1}+\hat c^\dagger_\ell \hat c_\ell}$. Consider the gauge-invariant sector where the eigenvalue of $\hat G_\ell$ is $+1$ for all $\ell$ by convention. Due to these local constraints, one can eliminate the fermionic degrees of freedom to obtain an exact description in terms of the gauge fields only. To formally obtain such a representation, first one applies the Jordan--Wigner transformation to turn fermions into spin-$\frac{1}{2}$ operators $\hat \tau_\ell^\alpha$, for $\alpha=+,-,z$, as discussed previously. One can also transform hardcore bosons into spin-$\frac{1}{2}$ operators $\hat \sigma_{\ell,\ell+1}^\alpha$, for $\alpha=+,-,z$ via $ \hat \sigma_{\ell,\ell+1}^+ =  \hat b_{\ell,\ell+1},~ \hat \sigma^-_{\ell,\ell+1} = \hat b_{\ell,\ell+1}^\dagger$, and $\hat \sigma_{\ell,\ell+1}^z = 1-2 \hat n_{\ell,\ell+1}$.\footnote{Note that we have assigned $\hat \sigma$ and $\hat \tau$ operators to fermions and boson operators, which is the opposite of our convention in the previous $\mathbb{Z}_2$ LGT example.} Now a unitary transformation $\hat{\mathcal{U}}$ is introduced such that the transformed Gauss's law operator $\hat{\bar{G}}_\ell=\hat{\mathcal{U}} \hat G_\ell \hat{\mathcal{U}}^\dagger$ only depends on the matter degrees of freedom, whereas the transformed Hamiltonian $\hat{\bar{H}}_{\mathbb{Z}_2}'=\hat{\mathcal{U}}\hat {\tilde{H}}_{\mathbb{Z}_2}'\hat{\mathcal{U}}^\dagger$ only depends on the gauge-field degrees of freedom. This is accomplished by choosing
\begin{equation}
    \hat{\mathcal{U}}=\prod_\ell \exp\left[\frac{i\pi}{2}(\hat \tau_\ell^x-\mathds{1})\frac{\mathds{1}-\hat\sigma_{\ell-1,\ell}^z\hat\sigma_{\ell,\ell+1}^z}{2}\right] \, .
\end{equation}
The transformed constraint $\hat{\bar{G}}_\ell\ket{\Psi}_\text{phys}=\ket{\Psi}_\text{phys}$ with $\hat{\bar{G}}_\ell= -\hat\tau_\ell^z$ decouples the $\tau$ spins, while the transformed Hamiltonian depends only on the $\sigma$ spins: 
\begin{align}
\label{eq_transfH}
    \hat{\bar{H}}_{\mathbb{Z}_2}^{(1+1)\text{D}}= & -g \sum_\ell 
 \hat\sigma^x_{\ell,\ell+1} + m \sum_\ell \frac{1-\hat\sigma_{\ell-1,\ell}^z  \hat\sigma_{\ell,\ell+1}^z}{2} \,
     \nonumber\\
     &-\frac 1 2 \bigg(
 \zeta-\sum_{\ell'>1} v_{\ell'}\bigg)  \sum_\ell  \hat\sigma^z_{\ell,\ell+1} 
 \nonumber\\
 &- \frac 1 4  \sum_\ell
     \sum_{ \ell'>1}v_{\ell'} \hat\sigma^z_{\ell,\ell+1}\hat\sigma^z_{\ell+\ell',\ell+\ell'+1}.
\end{align}
This LGT Hamiltonian is that of a dual-field long-range Ising Hamiltonian (up to irrelevant additive constants):
\begin{equation}
\label{eq:H-Ising}
    \hat H_{\text{Ising}} = - \sum_{i<j}J_{i,j} {\hat\sigma}_i^{z} {\hat\sigma}_j^{z} - \sum_{i}h{\hat\sigma}_i^{z} - \sum_{i}g {\hat\sigma}_i^{x},
\end{equation}
with the identification of the bonds of the fermionic chain connecting sites $\ell$ and $\ell+1$ with sites $i$ of the dual spin chain, and the parameters as
\begin{equation}
\label{eq:parameters}
    m \equiv 2J_1, \quad 
    \zeta \equiv 2h+ \sum_{r=2}^\infty J_r, \quad v_r \equiv  4 J_{r},
\end{equation}
with $J_r \coloneqq J_{i,i+r}$. This model, or slight variations of, is widely used in quantum-simulation studies of confinement, string breaking, and scattering, as will be covered in this Review.

\vspace{0.125 cm}
\noindent
\emph{A $(2+1)$D pure $\mathbb{Z}_2$ LGT:} The $\mathbb{Z}_2$ LGT in $(2+1)$D in the absence of matter has served as another paradigmatic model of quantum-simulation explorations. This model, moreover, has connections to quantum error correction~\cite{kitaev2003fault,kitaev2006anyons} and fermion-to-qubit encodings~\cite{chen2018exact,chen2020exact,chen2023equivalence}. The Hamiltonian most relevant to LGT phenomenology is
\begin{align}\label{eq:Z2Hamiltonian}
    \hat H_{\mathbb{Z}_2}^{(2+1)\text{D}} = \sum_\square \hat W_\square +g \sum_l \hat \sigma_l^z,
\end{align}
where $\hat W_\square \coloneq \prod_{l \in \square} \hat \sigma_l^x$ and $\square$ are the elementary (square) plaquettes of the lattice. Here, $\sigma^x_l$ and $\hat \sigma^z_l$ are Pauli operators on edge $l$. Gauss-law operator is 
\begin{align}\label{eq:GLZ22p1}
    \hat G_\ell = \prod_{l \in \ell} \hat \sigma^z_l\,,
\end{align}
where $\ell$ is a site that connects four links indexed by $l$, and the physical sector of the theory corresponds to the eigenvalue of $\hat G_\ell$ being unity $\forall \ell$. With infinite boundaries, this model maps to an Ising model in $(2+1)$D, widely known as the Wegner duality~\cite{wegner1971duality,wegner2017duality,horn1979hamiltonian}. Such a duality results from a maximal gauge fixing and taking advantage of Gauss's law. For finite systems, close attention must be paid to the effect of  the boundaries on the duality transformation such that the correct symmetry subsectors of the subsystems are identified~\cite{mueller2022thermalization,mueller2025quantum}. The ground state exhibits two phases, topologically order (deconfined) and trivial (confined) phases, separated by a quantum phase transition~\cite{fradkin2013field}.

%%%
%%%
\subsubsection{SU$(N_c)$ lattice gauge theories in $(d+1)$D:
\label{eq:SU(Nc)}}
Quantum-simulation experiments of non-Abelian and higher-dimensional LGTs have started to become reality in recent years, primarily on digital quantum-computing platforms, as will be highlighted in this Review. Most of this studies, as reviewed in Sec.~\ref{sec:non-Abelian}, concern \emph{pure} SU$(N_c)$ LGTs in $(d+1)$D in the absence of matter.\footnote{Fermions, in form of SU$(N_c)$ multiplets, can be added in the standard way, according to, e.g., the Kogut--Susskind prescription; see Refs.~\cite{zohar2015formulation,davoudi2025tasi} for a pedagogical discussion.} The Hamiltonian of a pure SU$(N_c)$ LGT reads:\footnote{$a$ as a superscript refers to a group color index, and must be distinguished from the lattice spacing $a$.}
\begin{align}
\hat H_\text{KS}^{(d+1)\text{D}}&=\frac{1}{2a^{d-2}}\sum_{l}\sum_{a=1}^{N_c^2-1} \left(\hat E^a_l\right)^2
\nonumber\\
&+\frac{a^d}{2a^4g^2}\sum_\square \text{Tr}\left[2-\hat{\mathcal{P}}_\square -\hat{\mathcal{P}}_\square^\dagger\right].
\label{eq:H-E-B-dp1}
\end{align}
where $l$ and $\square$ denote the link and square plaquettes on the lattice. 
There are two types of electric field on each link $l$, $E_{L,l}$ and $E_{R,l}$, with the commutation algebra:
\begin{align}
&[\hat E^a_{L,l},\hat E_{L,l'}^b]=-if^{a,b,c}\hat E_{L,l}^c\delta_{l,l'},
\\
&[\hat E^a_{R,l},\hat E_{R,l'}^b]=if^{a,b,c}\hat E_{R,l}^c\delta_{l,l'},\\
&[\hat E^a_{L,l},\hat E_{R,l'}^b]=0.
\label{eq:xx}
\end{align}
These are conjugate to the gauge-link variable at site $l$, $U_l$, which is an $N_c \times N_c$ matrix in the fundamental representation of the group:
\begin{align}
&[\hat E^a_{L,l},\hat U_{l'}]=gT^a\hat U_l\delta_{l,l'},\\
&[\hat E^a_{R,l},\hat U_{l'}]=g\hat U_lT^a\delta_{l,l'},
\label{eq:xx}
\end{align}
Here, $f^{a,b,c}$ are the group's structure constants and $T^a$ are the group's generators. The electric fields on each link satisfy: $\sum_{a=1}^{N_c^2-1}(\hat E^a_l)^2 \coloneqq \sum_{a=1}^{N_c^2-1}(\hat E_{L,l}^a)^2 \equiv \sum_{a=1}^{N_c^2-1}(\hat E^a_{R,l})^2$.

The plaquette operator $\mathcal{P}_\square$ in Eq.~\eqref{eq:H-E-B-dp1}is defined as: 
$\hat{P}_\square=\prod_{l \in \square} \hat U_l$, where $l$ denotes the links on a square plaquette $\square$ traversed counterclockwise. The trace is taken over the gauge-color space. To connect this lattice Hamiltonian to the continuum one, one realizes the relation $\hat{P}_\square=e^{ia^{(3-d)/2}g\hat B_\square}$ where $\hat B_\square$ is the curl of the vector gauge field $\hat A$ around the plaquette $\square$. Expanding the plaquette term in small $a$ and taking the limit $a \to 0$ recovers the continuum magnetic Hamiltonian proportional to $\hat B_\square^2$.

The physical states in the absence of any color charge are those annihilated by all the $N_c^2-1$ noncommuting Gauss's law operators $\hat{G}^a_\ell=\sum_{l \in v_\ell} \hat E^a_{L,l}-\sum_{l' \in v_\ell'}\hat E^a_{R,l'}$ for all $\ell$, where $v_\ell$ ($v_\ell'$) denotes the set of links emanating from site $\ell$ along the positive (negative) Cartesian directions.

%%%%%%%%%%%%
%%%%%%%%%%%%
\subsection{Quantum-simulation platforms
\label{sec:platforms}}

Quantum simulation is the use of a quantum system to model the physics of another quantum system, via emulation or computation. In other words, quantum simulation can be performed in two main ways: analog simulation and digital computation. It turned out that a hybrid of these modes is also possible, and sometimes desired.

\vspace{0.125 cm}
\noindent
\emph{Analog simulation:} Analog quantum simulation involves mapping a target quantum system to another \textit{experimental} quantum system (i.e., the quantum simulator). The mapping is most useful when degrees of freedom are similar, like mapping quantum spins to quantum spins, fermions to fermions, or bosons to bosons. Nonetheless, more indirect mappings are also possible but may lead to inexact representations, like using the finite-dimensional Hilbert spaces of finite spins to encode the infinite-dimensional Hilbert space of bosons. Once the mapping is established, quantum-simulator's control knobs should be tuned to engineer interactions among the degrees of freedom that mimic those in the target model. This procedure utilizes simulator's intrinsic interactions (such as atom-laser coupling, electric-circuit coupling, coupling to a microwave cavity, etc.) among simulator's degrees of freedom (such as neutral atoms, ions, photons, LC circuits, etc.). Once the mapping and Hamiltonian engineering are (often approximately) accomplished, unitary dynamics in the simulator tracks unitary dynamics in the target model (up to those approximation errors).

\vspace{0.125 cm}
\noindent
\emph{Digital computation:}
Digital quantum computation often employs qubits, i.e., spin-$\frac{1}{2}$ degrees of of freedom, and employs quantum gates, to decompose the time evolution of a Hamiltonian into a sequence of discrete operations. Quantum gates are implemented using the hardware's intrinsic interactions. They often involve the same physical principles and methods that are used in Hamiltonian engineering in the analog mode. Each gate acts for a short period of time, and is optimized to be robust to noise and other imperfections. In a stricter sense, digital computation is synonymous with fault-tolerant computation. Pre–fault-tolerant digital computation, discussed in this Review, still relies on finely tuned (analog) Hamiltonian interactions to implement the desired unitaries, including variable-angle operations. In contrast, fault-tolerant computation is truly digital: it uses a fixed set of discrete unitaries (with digitized errors); variable-angle unitaries must be synthesized down to such sets.

\vspace{0.125 cm}
\noindent
\emph{Hybrid analog-digital simulation:} A hybrid analog-digital quantum simulation takes advantage of simulator's intrinsic degrees of freedom like in the analog mode, but relies on gates to implement time evolution and other tasks as in digital quantum computers. These gates, nonetheless, implement more versatile and natural operations than the standard quantum-computing operations. For example, they can leverage multi-dimensional Hilbert spaces and multi-body interactions. 

All these simulation modes have been explored in recent years for gauge-theory studies, as will become clear throughout this Review.
With this introduction, let us enumerate three qualities that quantum-hardware developers strive to achieve:

\vspace{0.125 cm}
\noindent
\emph{Flexibility:} An analog quantum simulator is, by design, tailored for a specific Hamiltonian, and cannot be easily \textit{programmed} to simulate a different Hamiltonian \cite{georgescu2014quantum,Bloch2008}. This makes analog quantum simulators a lot less flexible than their digital counterparts. Digital quantum computers are \textit{programmable} and can, in principle, simulate any Hamiltonian dynamics with a sufficient number of gates and qubits. Modification of the Hamiltonian often involves updating the quantum circuit rather than the hardware and its control suit.

\begin{figure*}
    \includegraphics[width=0.995\textwidth]{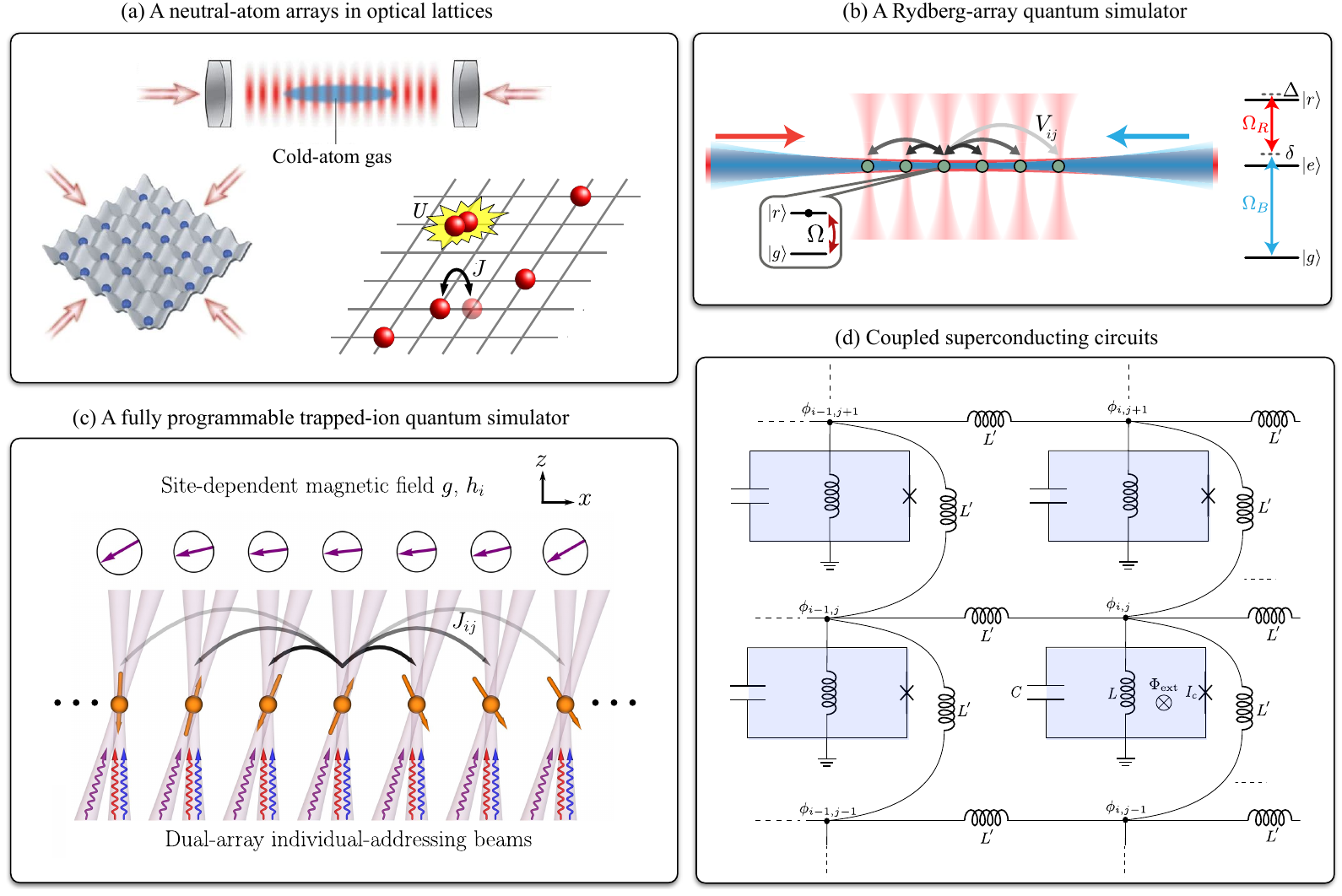}
    \caption{(a) An optical lattice
is formed by interfering continuous-wave lasers, creating a $1$d lattice potential, as depicted in the top graphic. By superimposing $1$d lattices in orthogonal directions, higher-dimensional optical lattice can be created, as depicted in the bottom-left graphic. Ultracold atoms trapped in a sufficiently deep lattice potential are described by a Hubbard model, described by the Hamiltonian in Eq.~\eqref{eq:bhm}, and shown in the bottom-right cartoon. Figure elements are reproduced from Refs.~\cite{bloch2008quantum,schafer2020tools}. (b) Individual atoms are trapped using optical tweezers (vertical red beams) and arranged into arrays. Exciting the atoms $i$ and $j$ (using horizontal blue and red beams) to a Rydberg state, with Rabi frequency $\Omega$ and detuning $\Delta$ (inset), induces coherent interactions $V_{i,j}$ between them (arrows). A two-photon process couples the ground state $|g \rangle$ to the Rydberg state $|r \rangle$ via an intermediate state $|ei\rangle$ with detuning $\delta$, using lasers with single-photon Rabi frequencies of $\Omega_B$ and $\Omega_R$. The resulting Hamiltonian is that in Eq.~\eqref{eq:H_Ryd}. The Figure is reproduced from Ref.~\cite{bernien2017probing}. (c) A trapped-ion quantum simulator in its most programmable form can implement site-(and time-)dependent transverse and longitudinal magnetic fields with strength $g_i$ and $h_i$, respectively, at ion $i$, and spin-spin Ising interactions with variable-range coupling strength $J_{i,j}$ among ion $i$ and ion $j$. These interactions further serve as the basis of a universal single- and two-qubit gate sets. The Figure is reproduced from Ref.~\cite{luo2025quantum}. (d) An example of $LC$ circuits with a Josephson junction element in a $2$d geometry via nearest-neighbor inductive couplings. The dynamics of the flux $ \phi_i$ at each circuit $i$ and its conjugate field is governed by the $(2+1)$D generalization of the Hamiltonian in Eq.~\eqref{eq:sc-hamiltonian-full}, which is the basis of a universal gate set in certain superconducting-qubit platforms. The Figure is reproduced from Ref.~\cite{belyansky2024high}.)
    }\label{fig:platforms}
\end{figure*}

\vspace{0.125 cm}
\noindent
\emph{Control and reliability:} The often approximate Hamiltonian engineering involved in analog quantum simulation leads to unavoidable errors, the control and suppression of which require innovative schemes that are usually platform dependent. Sometimes, when specific types of errors are in play and only local observables are concerned, those errors do not propagate and do not accumulate boundlessly~\cite{trivedi2024quantum,kashyap2025accuracy}. In digital quantum computers, gate errors are a major issue in the current era of noisy intermediate-scale quantum (NISQ)~\cite{preskill2018quantum,bharti2022noisy} devices. Nevertheless, there are various schemes for noise mitigation in the near term~\cite{endo2021hybrid} and error correction in the far term~\cite{devitt2013quantum} that can render computations more reliable. Recent quantum-error-correction experiments have indeed shown promise in approaching fault tolerance~\cite{egan2021fault,ryan2021realization,krinner2022realizing,ryan2022implementing,google2023suppressing,acharya2024quantum,eickbusch2024demonstrating,bluvstein2024logical,campbell2024series,lacroix2024scaling}.

\vspace{0.125 cm}
\noindent
\emph{Scalability:} Current system sizes, realized on both analog quantum simulators and digital quantum computers, are relatively small due to the underlying errors and engineering overhead required. Nonetheless, both types are, in principle, scalable.
On the other hand, scalability is rather architecture dependent. For example, semiconductor fabrication technology has made possible the production of large-scale superconducting quantum processors, but superconducting qubit's coupling is limited to qubit's neighbors~\cite{devoret2013superconducting,gambetta2017building,kjaergaard2020superconducting,huang2020superconducting,siddiqi2021engineering}. Trapped-ion systems, on the other hand, enjoy favorable all-to-all connectivity and qubit stability, but are more challenging to scale in a linear array~\cite{schneider2012experimental,wineland1998experimental,porras2004effective,monroe2021programmable,pagano2025fundamentals}.
Large-scale simulations of gauge theories of relevance to nature would require large-scale fault-tolerant quantum computers, as current estimates indicate~\cite{Shaw:2020udc, ciavarella2021trailhead, kan2021lattice,Lamm:2019bik,haase2021resource,Davoudi:2022xmb,Murairi:2022zdg,rhodes2024exponential,Lamm:2024jnl,Balaji:2025afl}.

A broader overview of analog quantum simulators of gauge theories, beyond studies of nonequilibrium dynamics, can be found in \cite{Halimeh2025cold-atom,bauer2024quantum}, while digital quantum computing of gauge theories is reviewed in more depth in Ref.~\cite{bauer2023quantum,di2024quantum}. In the following, we will briefly introduce a few platforms, summarized in Fig.~\ref{fig:platforms}, that have enabled groundbreaking quantum-simulation experiments of gauge theories in recent years.

\begin{figure*}[t!]
    \includegraphics[width=0.995\textwidth]{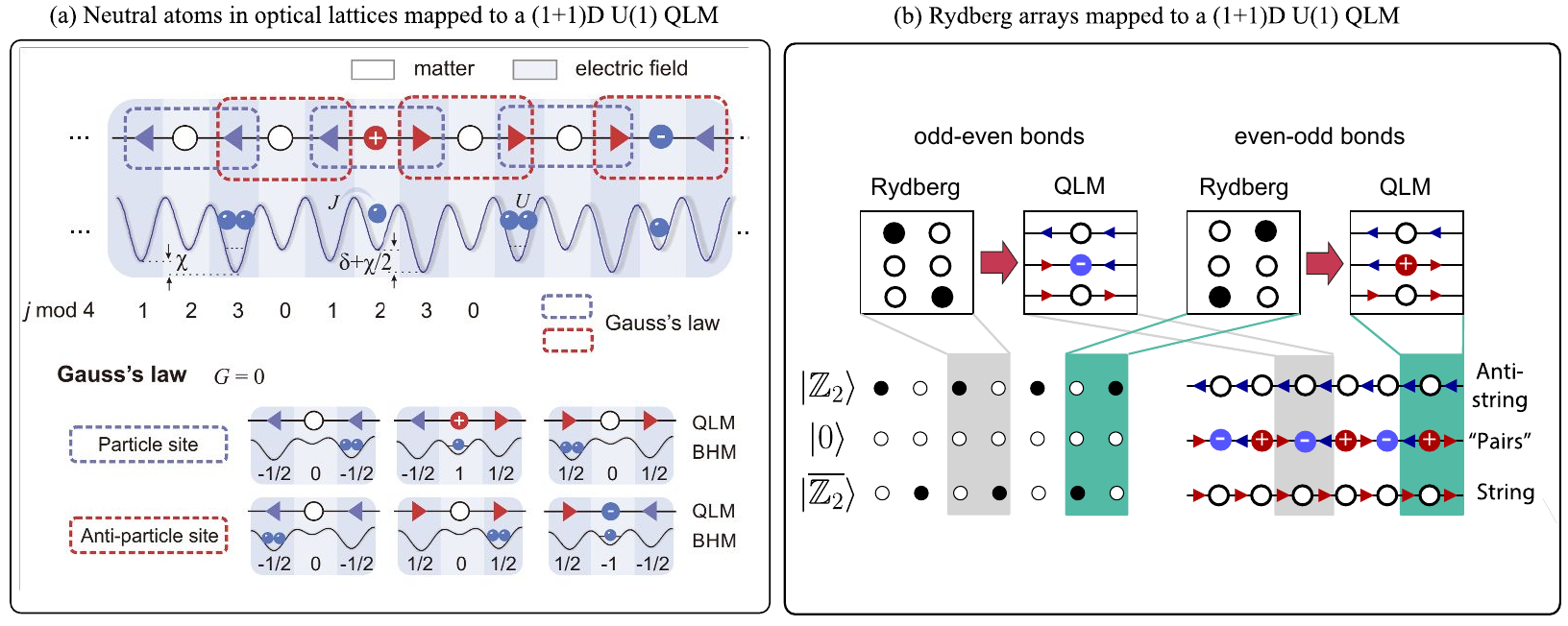}
    \caption{(a) Mapping of a $(1+1)$D spin-$\frac{1}{2}$ U$(1)$ QLM onto an analog optical superlattice quantum simulator. In the regime of strong on-site repulsion strength and staggering potential, the Bose--Hubbard model describing the optical superlattice captures the U$(1)$ QLM in second-order degenerate perturbation theory. In that regime, a bosonic doublon ladder operator on deep wells of the superlattice represents the gauge field on the link of the QLM, while bosonic singlon ladder operators on shallow wells of the superlattice represent the matter creation and annihilation operators on the sites of the QLM. (b) Employing Gauss's law to integrate out the matter degrees of freedom in the $(1+1)$D spin-$\frac{1}{2}$ U$(1)$ QLM allows to map it onto the PXP model of Rydberg-atom arrays, Eq.~(\ref{eq:PXP}). The three allowed configurations of two neighboring sites in the PXP model map directly onto the same number of allowed configurations of matter and gauge fields in the QLM on the even and odd sites, as shown in the boxes. Examples of three special states, i.e., the two infinite strings and the charge-proliferated state, map onto the $\ket{\mathbb{Z}_2}$, $\ket{\overline{\mathbb{Z}_2}}$, $\ket{0}$ states of the Rydberg model, respectively. This equivalence facilitates the implementation of the $(1+1)$D spin-$\frac{1}{2}$ U$(1)$ QLM using Rydberg atoms. Figure is reproduced from Ref.~\cite{Surace2020}.
    }\label{fig:platforms_mapping}
\end{figure*}

%%%
%%%
\subsubsection{Optical superlattices}\label{sec:opticalsuperlattice}
           
Optical superlattices are among prominent platforms for quantum simulation. They have been used to probe the physics of a myriad of condensed-matter systems such as the Fermi--Hubbard model \cite{Esslinger2010FermiHubbard,lewenstein2012ultracold}, Bose--Hubbard model \cite{Greiner2002,Bloch2008,Bloch2012} and extensions thereof, artificial gauge fields \cite{Gerbier2010gaugefields,Aidelsburger2015diss} and topological bands \cite{Baur2014dynamic}. Optical superlattices have also been demonstrated to be a powerful tool in simulating LGTs \cite{schweizer2019floquet,gorg2019realization,aidelsburger2021cold,Halimeh2025cold-atom}.

To put the discussion on a concrete footing, let us consider the $(1+1)$D spin-$\frac{1}{2}$ $\mathrm{U}(1)$ QLM introduced in Sec.~\ref{sec:U1LGT_formulations}.
This model has been experimentally realized in a large-scale Bose--Hubbard quantum simulator~\cite{yang2020observation,zhou2022thermalization,su2023observation,WangQMBSCriticality,zhang2023observation}. The Hamiltonian of a Bose--Hubbard quantum simulator can be written as~\cite{yang2020observation}
\begin{align}
\nonumber
\hat{H}_\text{BHM} =&-J\sum_{j=1}^{N-1} \left( \hat{b}^\dagger_{j}\hat{b}_{j+1}+ \hat{b}^\dagger_{j+1}\hat{b}_{j}\right)  \\\label{eq:bhm}
&+\frac{U}{2}\sum_{j=1}^N \hat{n}_j\left(\hat{n}_j-\mathds{1}\right)+\sum_{j=1}^N \mu_j \hat{n}_j.
\end{align}
Here, $\hat{b}_{j}$ and $\hat{b}^\dagger_{j}$ are the bosonic annihilation and creation operators, respectively, and $\hat{n}_j=\hat{b}_j^\dagger \hat{b}_j$ is the boson density operator. $J$ denotes the tunneling strength between neighboring sites, $U$ is the on-site interaction, and $N=2L$ is the total number of sites in the quantum simulator. Figure~\ref{fig:platforms}(a) presents a schematic of the hardware system and the lattice Hamiltonian. In order to generate the gauge-matter interactions in the U$(1)$ QLM in Eq.~\eqref{eq:U1QLMPH}, the chemical potential term needs to take the form $\mu_j = j  \Delta + (-1)^j \delta/2 + \text{sin}(j \pi/2)\chi/2$, where $\Delta$ is a linear tilt used to suppress long-range single-atom tunneling~\cite{Halimeh2021gauge-symmetry,halimeh2020robustness}, $\delta$ is a staggering potential generated by a period-$2$ optical superlattice separating the system into two sublattices, and $\chi$ is the confinement parameter in the QLM defined in Eq.~\eqref{eq:chi}. The sites with even $j$ (first superlattice) are assigned as the ``matter'' sites while sites with odd $j$ (second superlattice) are assigned as the ``gauge'' sites. For $\delta \sim U/2 \gg J$, a resonant second-order correlated hopping process $\hat{b}_{j-1}(\hat{b}^\dagger_{j})^2\hat{b}_{j+1}+ \text{H.c.}$ (with $j$ odd) is induced, where single bosons on neighboring matter sites are annihilated (created) to form a doublon (hole) on the gauge site in between; see Fig.~\ref{fig:platforms_mapping}(a). This process maps to an effective Hamiltonian of the form in Eq.~\eqref{eq:U1QLM} at $S=1/2$, with the identifications $\tilde\kappa = 4\sqrt{2}J^2/U$ with $\tilde\kappa$ defined in Eq.~\eqref{eq:tildekappa}, and $m=\delta/2-U$~\cite{yang2020observation,Halimeh2022tuning,Cheng2022tunable}. The confining term proportional to $ \chi$ breaks the degeneracy between the two vacuums $\ket{\ldots \triangleleft,\varnothing,\triangleleft,\varnothing,\triangleleft,\varnothing,\triangleleft \ldots} \leftrightarrow \ket{\ldots 2,0,0,0,2,0,0 \ldots}$ and $\ket{\ldots \triangleright,\varnothing,\triangleright,\varnothing,\triangleright,\varnothing,\triangleright \ldots} \leftrightarrow \ket{\ldots 0,0,2,0,0,0,2 \ldots}$, where integers denote the bosonic occupation of each site of the superlattice. Theoretical proposals to extend this mapping to two spatial dimensions \cite{Osborne2025largescale} and larger-$S$ representations of the gauge field \cite{osborne2023spinsmathrmu1quantumlink} have also been put forth.

An alternative realization of the spin-$\frac{1}{2}$ $\mathrm{U}(1)$ QLM on an optical superlattice entails integrating out the matter degrees of freedom through Gauss's law, which gives rise to the Hamiltonian in Eq.~\eqref{eq:U1QLM_MatterIntegratedOut} with $S=1/2$, as explained in Sec.~\ref{sec:formulations}. The Bose--Hubbard Hamiltonian still takes the form of Eq.~\eqref{eq:bhm}, but now with $\mu_j=j\Delta+(-1)^j\delta/2$ \cite{su2023observation}. The mapping becomes valid in the regime $U\approx\Delta\gg J$. Consider a preparation scheme where each well of the superlattice starts off with a single boson. In this regime, the only accessible configurations through hopping processes belong to the set $\{\ket{20},\ket{11},\ket{12},\ket{02},\ket{01}\}$. The spin-up state $\ket{\uparrow}$ is mapped to $\ket{20}$, while the spin-down state $\ket{\downarrow}$ is mapped to $\ket{11}$, $\ket{12}$, $\ket{02}$, or $\ket{01}$. In this regime, the parameters of both models map as $\tilde\kappa \approx \sqrt{2} J$, $m=(\Delta-U)/2$, and $a\chi=\delta$~\cite{su2023observation}. This mapping will be relevant in Sec.~\ref{sec:confinement} and Sec.~\ref{sec:QMBS} when we discuss confinement and quantum many-body scarring, respectively, in this model.

%%%
%%%
\subsubsection{Rydberg-atom arrays}\label{sec:Rydberg}

Rydberg atoms in optical tweezer arrays are another powerful platform for quantum simulation, particularly in relation to strongly correlated and quantum many-body systems~\cite{saffman2010quantum,adams2019rydberg,browaeys2020many}. Their distinctive features include (i) the ability to realize programmable geometries in one, two, and even three spatial dimensions, (ii) the broad interaction range varying from fast-decaying van der Waals potentials ($\,{\propto}1/r^6$ with $r$ being the inter-atom distance) to longer-range dipolar forces (${\propto}\,1/r^3$), and (iii) the accessible fast coherent control they provide over single atoms and the ability to entangle them. Rydberg setups 
are valuable quantum simulators for probing the physics of systems such as Ising models with long-range interactions~\cite{Zeiher2017coherent,Borish2020transverse}, systems with frustrated magnetism~\cite{Glaetzle2015designing}, quantum dimer~\cite{Samajdar2021quantum} and quantum spin liquids~\cite{Semeghini2021probing}, density waves~\cite{Samajdar2020complex}, and supersolid phases~\cite{Homeier2025supersolidity}.

The local spin Hilbert space $\ket{\downarrow}$ and $\ket{\uparrow}$ can be encoded in the electronic ground state, usually denoted by either $\ket{g}$ or $\ket{\circ}$, and the excited Rydberg state, $\ket{r}$ or $\ket{\bullet}$,  respectively, which form a pseudospin-$\frac{1}{2}$. These Rydberg excitations are created by laser light in a tweezer array, as illustrated in Fig.~\ref{fig:platforms}(b).
The dynamics of $N$ Rydberg atoms arranged in a $d$-dimensional array is governed by the Hamiltonian
\begin{align}\label{eq:H_Ryd}   \hat{H}_\text{Rb}= \Omega \sum_{j=1}^N \hat X_j + \sum_{j=1}^N \delta_j  \hat n_j + \frac{1}{2} \sum_{j,j^\prime}^N V_{j,j'}\hat{n}_j\hat{n}_{j'},
\end{align}
where $\hat X \equiv \ketbra{\circ}{\bullet}+\ketbra{\bullet}{\circ}$ is a Pauli-$x$ matrix describing the Rabi oscillations of each atom $j$ with frequency $\Omega$\footnote{Sometimes, it is conventional to define the Rabi frequency as $\Omega/2$ instead.}, $\hat n \equiv \ketbra{\bullet}{\bullet}$ is a density operator corresponding to the Rydberg excitation on a given site, with $\delta_j$ representing the detuning of the $j$th atom away from resonance [Fig.~\ref{fig:platforms}(b)], and $V_{j,j^\prime}=C_6/(\lvert\mathbf{x}_j-\mathbf{x}_{j'}\rvert/a)^6$ is the van der Waals interaction potential, where $\mathbf{x}_j$ is the vector representing the position of site $j$ on a $d$-dimensional grid, $C_6$ is the coupling strength, and $a$ is the lattice spacing (typically set to $a=1$). Note that our definition of $\hat n_j$ expresses the fact that the atoms only interact if they are simultaneously excited into the Rydberg states. The detuning $\delta_j$ can be the same for all atoms, in which case it has the meaning of an overall chemical potential (and, occasionally, the notations $\Delta$ or $\mu$ are used). However, it is also possible to engineer a space-dependent detuning on the transition
between ground and Rydberg states~\cite{Omran2019}, typically with a two-site periodicity of $\delta_j$. 

Due to the fast decay of van der Waals interactions with the distance between the atoms, $V_{j,j^\prime}$ can often be approximated with the nearest-neighbor interaction. For simplicity, we now assume a $1$d array and denote the nearest-neighbor interaction by $V\equiv V_{j,j+1}$. The regime where  
$V \gg \Omega, \delta_j$ is known as the Rydberg blockade: two neighboring atoms are prohibited from being simultaneously in excited (Rydberg) states~\cite{browaeys2020many}. Performing a Schrieffer-Wolff transformation at $V\gg \Omega,\delta_j$, the effective model describing the blockade regime is the Fendley--Sengupta--Sachdev or the ``PXP'' Hamiltonian~\cite{FendleySachdev}
\begin{align}\label{eq:PXP}
\hat{H}_\text{PXP}= \Omega \sum_{j} \hat{P}_{j-1} \hat X_j \hat{P}_{j+1}+ \sum_{j} \delta_j \hat n_j.
\end{align}
Here, the projector $\hat{P}_j \equiv \ketbra{\circ}{\circ}_j$ enforces the Rydberg blockade: the $j$th atom can undergo a Rabi flip from the ground state to the excited state only if both of its neighbors ($j\pm 1$) are in their ground states. This ensures that quantum dynamics generates no neighboring pairs of excitations, $\ket{\ldots \bullet\bullet\ldots}$, anywhere in the system, which would result in a large energy penalty in the Rydberg blockade regime. Note that, if Eq.~(\ref{eq:PXP}) is defined with open boundary conditions, the first and the last atom only have neighbors to one side and the respective boundary terms in $\hat H_\text{PXP}$ are $\hat X_1\hat P_2$ and $\hat P_{N-1}\hat X_N$.

The meaning of the Rabi term $\hat X_j$ is very different in Eq.~(\ref{eq:H_Ryd}) compared to Eq.~(\ref{eq:PXP}): in the former, it describes independent flipping of each atom $j$, while in the latter the flipping is dependent on the state of the atom's neighbors. The latter is a form of a kinetic constraint, which renders the Hamiltonian in Eq.~(\ref{eq:PXP}) intrinsically interacting and impossible to solve via analytic means. Nevertheless, the kinetic constraint provides a very natural framework to encode constraints resulting from Gauss's law, such as in the $(1+1)$D spin-$\frac{1}{2}$ $\mathrm{U}(1)$ QLM. Enforcing Gauss's law on even and odd sites, Eq.~(\ref{eq:GaussLawKS}), it is possible to map the configurations of matter on a given site and gauge fields on neighboring links to the states of a Rydberg-atom dimer~\cite{Surace2020}, as illustrated in Fig.~\ref{fig:platforms_mapping}(b). In this way, the states with infinitely long strings map to the so-called $\ket{\mathbb{Z}_2}\equiv \ket{\ldots{\circ}{\bullet}{\circ}{\bullet}\ldots}$ and $\ket{\overline{\mathbb{Z}_2}}\equiv \ket{\ldots{\bullet}{\circ}{\bullet}{\circ}\ldots}$ states in the Rydberg language, while the charge-proliferated state maps to the so-called polarized state, $\ket{0}\equiv \ket{\ldots{\circ}{\circ}{\circ}{\circ}\ldots}$, as demonstrated in Fig.~\ref{fig:platforms_mapping}(b). The Rydberg states obtained in this mapping are precisely the ones obtained by applying the PXP terms in Eq.~(\ref{eq:PXP}) to the $\ket{0}$ state. Equivalently, upon employing Gauss's law to integrate out the matter degrees of freedom, as done in Sec.~\ref{sec:U1LGT_formulations}, one arrives at Eq.~\eqref{eq:U1QLM_MatterIntegratedOut} which, for $S=1/2$, is equivalent to the PXP model in Eq.~(\ref{eq:PXP}) upon setting 
\begin{equation}\label{eq:QLM_Rydberg_mapping}
\Omega=-2\tilde\kappa \quad \text{and} \quad \delta_j =-2m -  a \chi (-1)^j.    
\end{equation}
Here, the overall constant term has been absorbed in the reference value for the energy. The equivalence between the PXP model and the $(1+1)$D spin-$\frac{1}{2}$ $\mathrm{U}(1)$ QLM (after enforcing Gauss's law) can also be seen from Eq.~(\ref{eq:U1QLM_MatterIntegratedOut}) and the projector in Eq.~(\ref{eq:ProjectorQLM}), which (for $S=1/2$) reduces to $\hat P_{j,j+1}=\mathbb{1}-(\ketbra{\bullet})_j(\ketbra{\bullet})_{j+1}$, which is identical to the Rydberg blockade constraint. 

Thus, the $(1+1)$D spin-$\frac{1}{2}$ $\mathrm{U}(1)$ QLM is natively encoded in a Rydberg-atom array via Eqs.~(\ref{eq:PXP})-(\ref{eq:QLM_Rydberg_mapping}), provided one tunes to the strong blockade regime. We will discuss the rich out-of-equilibrium physics of the PXP model in Sec.~\ref{sec:QMBS}. We note that the programmable nature of Rydberg-atom arrays and the flexible choice of lattice geometry have been proposed as a way of simulating other LGTs models, such as the Schwinger model in Eq.~(\ref{eq:SchwingerKS})~\cite{lerose2024simulatingschwingermodeldynamics}.

%%%
%%%
\subsubsection{Trapped ions}\label{sec:TrappedIons}

Trapped-ion quantum simulators are another example of powerful platforms~\cite{monroe2021programmable} that have been used to experimentally study Ising models~\cite{islam2011onset,pagano2020quantum,zhang2017observation,jurcevic2017direct}, long-range XY~\cite{lewis2023ion} and Heisenberg models~\cite{bermudez2017long}, topological order and edge modes~\cite{nevado2017topological,iqbal2024topological}, Luttinger liquids~\cite{michelsen2019ion}, many-body localization~\cite{smith2016many,morong2021observation}, spin-boson and Holstein-type models~\cite{knorzer2022spin,sun2025quantum}, frustrated magnetism~\cite{kim2010quantum,islam2013emergence,qiao2022observing}, and time crystals~\cite{li2012space,zhang2017observation}. 

Trapped-ion platforms operate in both analog and digital modes. They often feature a linear array of charged ions (Ytterbium, Beryllium, Calcium, etc.) trapped in a Paul trap. The spin states are often encoded in the two lowest internal hyperfine levels, which are separated in energy by an angular frequency $\omega_0$. Besides these quasi-spins (qubits), these systems host trap's (local or normal) motional-mode excitations (phonons) with the $\mathsf{k}$-th mode frequency $\omega_{\mathsf{k}}$. To manipulate the state of the ion, they are addressed by sets of counter-propagating Raman laser beams. For more details on these architectures, see Refs.~\cite{wineland1998experimental,porras2004effective,schneider2012experimental,monroe2021programmable,pagano2025fundamentals}. In the Lamb--Dicke regime~\cite{wineland1998experimental}, where laser-induced ion-phonon coupling is sufficiently small, transitions in the space of coupled spin-phonon system take simple forms, and can be realized through quantum gates.

Explicitly, the carrier transition is obtained by setting $\omega_j^L=\omega_0$, with $\omega_j^L$ being the laser's frequency at ion $j$, and the Hamiltonian corresponding to ion $j$ becomes
\begin{equation}
\hat{H}^{\sigma}_{j}=\frac{\Omega_j}{4} \bigg[(e^{i\phi_j}+e^{-i\phi_j})\hat{\sigma}_j^x+(e^{i\phi_j}-e^{-i\phi_j})i\hat{\sigma}_j^y \bigg].
\label{eq:Hsigma}
\end{equation}
Here, $\Omega_j$ and $\phi_j$ are laser's Rabi frequency and phase at ion $j$, respectively. $\hat{\sigma}_j^{x/y}$ are Pauli operators acting on the space of the qubit. The single-spin rotations of arbitrary angles along the $x/y$ axis of the Bloch sphere can, therefore, be obtained by applying $H^{\sigma}_{j}$ for proper times. Rotations along the $z$ axis, on the other hand, can be implemented using a classical phase shift on the beam controller. 

The blue and red sideband transitions are obtained by setting $\omega_j^L=\omega_0+\omega_\mathsf{k}$ and $-\omega_\mathsf{k}$, respectively. This setting leads to a coupled spin-phonon Hamiltonian. In order to achieve a Hamiltonian proportional to, e.g., $\sigma^y_j$, beatnotes associated to blue and red sidebands can be applied simultaneously with phases that add up to zero. Then the Hamiltonian corresponding to ion $j$ becomes
\begin{align}
&\hat H^{\sigma a}_{\mathsf{k},j}(\phi_{\mathsf{k},j})=-\frac{\eta_{\mathsf{k},j}\Omega_j}{2} (e^{i\phi_{\mathsf{k},j}}\hat a_\mathsf{k}+e^{-i\phi_{\mathsf{k},j}}\hat a_\mathsf{k}^\dagger) \, \hat\sigma_j^y,
\label{eq:Hsigmaa-single}
\end{align}
where $\eta_{\mathsf{k},j}$ is the Lamb--Dicke parameter and $\phi_{\mathsf{k},j}=\frac{1}{2}(\phi^r_{\mathsf{k},j}-\phi^b_{\mathsf{k},j})=\phi^r_{\mathsf{k},j}$, with $\phi^{r(b)}_{\mathsf{k},j}$ denoting the red (blue) sideband laser phase. The spin-phonon rotations of arbitrary angles can be obtained by applying $H^{\sigma a}_{\mathsf{k},j}$ for proper times~\cite{davoudi2021toward}.

Applying simultaneous blue and red sideband transitions detuned from the normal-mode frequencies can effectively induce a spin-spin interaction, which leads to the well-known M\o lmer--S\o rensen (MS) gate~\cite{molmer1999multiparticle, milburn2000ion, solano1999deterministic,cirac1995quantum}. The corresponding Hamiltonian between ions $j$ and $j'$ is
\begin{align}
&\hat H^{\sigma \sigma}_{j,j'}=\Omega_j \Omega_{j'} \sum_\mathsf{m} \eta_{\mathsf{m},j} \eta_{\mathsf{m},j'}\frac{\omega_\mathsf{m}}{(\omega_j^L-\omega_0)^2-\omega_\mathsf{m}^2} \hat \sigma_j^x \hat \sigma_{j'}^x,
\label{eq:Hsigmasigma}
\end{align}
Spin-spin rotations of arbitrary angles can be obtained by applying $H^{\sigma \sigma}_{j,j'}$ for proper times.

The discussions above demonstrate how single- and two-qubit gates, and even qubit-phonon gates, can be realized in trapped ions. This elementary single- and two-qubit gate set allowed the first quantum simulation of time dynamics in a gauge theory in Ref.~\cite{martinez2016real}. Here, the Schwinger model in its fully gauge-fixed form in Eq.~\eqref{eq:SchwingerGaugeFieldsIntegratedOut} was time-evolved using a first-order Trotterization scheme, as discussed in Sec.~\ref{sec:particle-production}. The simulation of an $N$-site model involved only an $O(N)$ gate complexity since the $O(N^2)$ long-range $ZZ$ interactions were implemented in a parallel fashion, leveraging the all-to-all long-range interactions induced by global Raman beams.

The discussions above further indicate the type of Hamiltonians trapped-ion platforms can implement in an analog mode, which is of an Ising type, given in Eq.~\eqref{eq:H-Ising}. Here, $J_{i,j}=Je^{-\beta(r-1)}r^{-\alpha}$, where $r = |i-j|$. By varying the side-band detuning, the parameters $\alpha$ and $\beta$ can be varied within the range $0 \lesssim  \alpha \lesssim 3 $~\cite{monroe2021programmable,porras2004effective} and $\beta > 0$~\cite{kim2009entanglement,nevado2016hidden,feng2023continuous,schuckert2025observation}. The additional transverse and longitudinal fields can be implemented simultaneously using the same beam arrays as those generating the spin-spin coupling, while the  longitudinal field can be applied by employing a second array of tightly focused beams that drive carrier transitions~\cite{de2024observation,luo2025quantum}. Trapped-ion quantum simulators, therefore, have been widely used to study Ising model's dynamics, which is mapped to a confining $\mathbb{Z}_2$ LGT, as discussed in Sec.~\ref{sec:formulations}. Additionally, as proposed in Ref.~\cite{davoudi2020towards}, by introducing extra sets of independent counter-propagating Raman beams, one can further engineer Heisenberg-type Hamiltonians [see also Refs.~\cite{kotibhaskar2024programmable,kranzl2023observation} for other approaches]. Such a Hamiltonian is that of the Schwinger model in its fully gauge-fixed form in Eq.~\eqref{eq:SchwingerGaugeFieldsIntegratedOut}. Here, tuning lasers' frequency and amplitudes can ensure both nearest-neighbor and long-range spin-spin Hamiltonians are accurately engineered, and simultaneously applied.

%%%
%%%
\subsubsection{Superconducting qubits}\label{sec:SCQ}

Superconducting (SC) quantum circuits have emerged as a major  platform for digital quantum computation~\cite{devoret2004implementing,schoelkopf2008wiring,wendin2017quantum,krantz2019quantum,kjaergaard2020superconducting}, and have been used in studies of spin chains, Fermi--Hubbard model~\cite{stanisic2022observing}, topological phases and Chern insulators~\cite{xiang2023simulating}, Anderson localization and disordered systems~\cite{rosen2025flat}, quantum many-body scarring and constrained dynamics~\cite{zhang2023many}, as well as discrete time crystals~\cite{frey2022realization}. Unlike atomic or photonic systems, which rely on the manipulation of individual quantum particles, SC devices are built from nanofabricated macroscopic elements. This fundamentally different design approach gives rise to distinct advantages and limitations in terms of control, noise resilience, error correction, connectivity, and scalability.

At their core, SC circuits are composed of elements such as inductors (L) and capacitors (C), 
which together form quantum harmonic oscillators with evenly spaced energy levels~\cite{krantz2019quantum},
\begin{align}\label{eq:cQED1}
    \hat H_{\text{SC}}^{(0)}= \frac{\hat Q^2}{2C} + \frac{\hat \Phi^2}{2L} =4E_C\hat\pi^2 + \frac{E_L}{2}\hat \phi^2,
\end{align}
where $\hat \Phi$ and $\hat Q$ are flux and charge operators, respectively, with commutation relations $[\hat\Phi,\hat Q]=i$; further $\hat \pi=\hat Q/(2e)$, $\hat \phi=2\pi\hat \Phi/\Phi_0$,  $E_C=e^2/(2C)$, $E_L=(\Phi_0/2\pi)^2/L$ and $\Phi_0=\pi/e$. To introduce anharmonicity, these circuits incorporate nonlinear components such as Josephson junctions (JJ), resulting, in the simplest form, in
\begin{align}\label{eq:cQED2}
   \hat H_\text{SC}^\text{(JJ)}=-E_J\cos\hat \phi\,,
\end{align}
where $E_J=I_c \Phi_0/2\pi$, and $I_c$ is the critical current of the junction. The resulting energy spectra often resemble those of atoms, but with the added benefit of tunability.

A prominent example is the transmon qubit: one operates with sufficient anharmonicity to isolate two energy levels so as to enable coherent control using resonant microwave pulses (typically in the few-GHz range), while suppress transitions to higher levels. More complex designs include the flux qubit, which uses multiple JJ elements arranged in a loop, or the fluxonium qubit, which incorporates an even larger number of JJ elements. SC qubits can be coupled capacitively, i.e., $\hat H_\text{SC}^{\rm (int)}\sim \hat \pi_1 \hat \pi_2$ and inductively, i.e., $\hat H_\text{SC}^{\rm (int)}\sim \hat \phi_1 \hat \phi_2$, leading to interaction terms in the multi-qubit Hamiltonian that, together with single-qubit control, enable universal operations. For a comprehensive overview of SC-qubit architectures and their operational regimes, readout and control, see, e.g., Ref.~\cite{krantz2019quantum}.

As with all pre-fault-tolerant devices, SC circuits can operate in an analog mode too~\cite{puertas2019tunable,ozguler2021excitation,blais2021circuit,roy2021quantum,roy2023quantum,belyansky2024high}, by directly implementing Hamiltonians of the form in \Eqs{eq:cQED1}{eq:cQED2} and their inter-mode couplings, rather than synthesizing gate operations. In this regime, these systems realize strongly coupled bosonic systems with connectivity patterns depending on the hardware. For example, the coupled circuit units can implement the massive sine-Gordon Hamiltonian:
\begin{align}
\label{eq:sc-hamiltonian-full}
    \hat H = \sum_i \bigg[&4E_C\hat \pi_i^2+\frac{E_{L'}(\hat \phi_{i}-\hat \phi_{i-1})^2}{2} 
    \nonumber\\
    &+\frac{E_L \hat \phi_i^2}{2}+E_J\cos(\hat \phi_i-\Phi_{\mathrm{ext}})\bigg],
\end{align}
where $\Phi_{\mathrm{ext}}$ is the external flux threading each loop and $L'$ is the inductance of the inductors coupling the loops. The Hamiltonian in Eq.~\eqref{eq:sc-hamiltonian-full} also describes the bosonozed from of the Schwinger model, as discussed in Sec.~\ref{sec:scat-in-schwinger-model}. A similar Hamiltonian can be realized in $(2+1)$D by coupling the SC circuits in a 2d array, as depicted in Fig.~\ref{fig:platforms}(d).

In contrast to atomic systems, SC qubits are not perfectly uniform as their properties depend on fabrication tolerances.  Like all engineered quantum systems, they are subject to noise, much of which is characteristic of solid-state platforms. However, many noise sources can be systematically mitigated using a high degree of control and tunability in SC circuits; see Ref.~\cite{krantz2019quantum} for an overview. ``Software-based'' noise-mitigation techniques have been extensively tested in the context of quantum-simulation applications; see, e.g., Refs.~\cite{a2022self,urbanek2021mitigating,farrell2024quantum,bennewitz2022neural}. 

SC qubits offer a major advantage in speed: single-qubit gates typically operate on timescales of tens of nanoseconds, while two-qubit entangling gates are performed within a few hundred nanoseconds. Their qubit connectivity is fixed by hardware layout. Common architectures include the heavy-hex topology used by IBM, or rectangular $2$d grids employed by Google---chosen both for fabrication and with surface-code-type quantum error correction in mind. For (pre-fault-tolerant) quantum-simulation applications, however, limited connectivity can pose a constraint: unless the simulated model naturally maps onto a $1$d or $2$d geometry compatible with the hardware layout, additional SWAP operations are needed to route information between distant qubits, increasing the overall circuit depth and error budget. 

Quantum Simulation has been a prime application for SC-based quantum computers,  
both in analog and digital modes~\cite{houck2012chip,marcos2013superconducting}. LGT simulations 
have been at the forefront of this development, including simulation of the Schwinger model~\cite{klco2018quantum,farrell2024quantum,farrell2025digital}, SU$(2)$ and SU$(3)$ LGTs~\cite{a2022self,klco20202,ciavarella2021trailhead,atas2023simulating,ciavarella2022preparation,ciavarella2024quantum,ciavarella2025string,turro2024classical}, or $\mathbb{Z}_2$ LGTs~\cite{charles2024simulating,hayata2024floquet}, and models in higher dimension~\cite{cochran2024visualizingdynamicschargesstrings}. Recent work involving simulation of a $(1+1)$D $\mathbb{Z}_2$ LGT with matter~\cite{Mildenberger2025}, for example, addressed key hardware connectivity constraints in realizing three-body interactions through the use of SWAP operations.

%%%%%%%%%%%%%%%%%%%%%%%%%%%%%%%%%%%%%%%%%%%%%%%%%%%%%%%%%%%%%%%%%%%%%%%%%%%%%%%%%%%%%%%%%%%%%%%%%%%%%%%%%%%%%%%%%%%%%%%%%%%%%%%%%%%%%%%%%%%%%%%%%%%%%%%%%%%%%%%
\section{
Particle production and string breaking in gauge theories
\label{sec:particle-string}}

The theory of the strong interaction, QCD, exhibits the striking feature that its fundamental degrees of freedom---quarks and gluons---are not directly observable at low energies. This property, known as confinement~\cite{wilson1974confinement,greensite2011introduction}, implies that quarks and gluons, which carry non-Abelian SU$(3)$ color charge of QCD, are never found in isolation in nature. Instead, they form color-neutral bound states---mesons and baryons---which constitute the bulk of visible matter in the universe. When a high-energy, high-resolution probe is used, with energies on the order of a few GeV or more or distances on the order of $\sim 0.1$ fm, QCD becomes asymptotically free: the effective interaction between quarks and gluons weakens, and they behave approximately as free particles~\cite{gross1973ultraviolet}---a behavior that has been firmly established in deep inelastic scattering (DIS) experiments~\cite{breidenbach1969observed,bjorken1969asymptotic,bethke2007experimental,blumlein2013theory} and which underpins high-energy QCD phenomenology in modern particle colliders~\cite{ellis2003qcd}.

However, a central open question is: what happens to the quarks and gluons that are liberated at high energies as they evolve into the hadrons observed in detectors? Due to SU$(3)$ gauge invariance, quarks and gluons are connected by extended structures known as color flux tubes or color strings, which are themselves manifestations of gluonic gauge fields. Confinement can be understood as the phenomenon whereby separating a quark-antiquark pair, connected by such a string or flux tube, requires an ever-increasing amount of energy, diverging in the limit of infinite separation. Before this occurs, however, the energy stored in the color string becomes sufficient to create a new quark-antiquark pair from the vacuum---a process known as string breaking. This mechanism restores color neutrality by forming new mesons, and thus ensures that color-charged objects remain confined. A mechanism for this phenomenon was proposed in a seminal work of K. Wilson~\cite{wilson1974confinement}, in which a lattice-gauge-theory method was developed to study the confinement problem.

String breaking---or quark fragmentation---is a key ingredient in QCD collider phenomenology. However, it remains classically difficult to compute from first principles, as it is a time-dependent phenomena and involves real-time dynamics at strong couplings. As a result, string breaking is typically modeled using phenomenological, often stochastic, frameworks. Prominent examples include the Lund string model~\cite{andersson1998lund}, which underlies event generators like \textsc{Pythia}~\cite{sjostrand2020pythia}, and cluster fragmentation models used, e.g., in \textsc{Herwig}~\cite{gieseke2004herwig}. These models treat fragmentation as a local, sequential probabilistic process, and do not capture essential quantum features such as coherence, phase information, or entanglement between particles. While they have been remarkably successful in describing experimental data, they are not derived from the fundamental dynamics of QCD and can suffer model uncertainties.
Fragmentation functions~\cite{metz2016parton}---analogous to parton distribution functions~\cite{soper1997parton,ellis2003qcd}---are experimentally extracted probability distributions that encode the contribution of fragmentation to scattering cross sections. However, their use relies on specific kinematic approximations and the assumption of factorization~\cite{collins2011foundations}, meaning that the fragmentation process is treated as an incoherent process, and independent of the rest of the scattering event. 

A long-term goal is to compute string breaking---or quark fragmentation---\emph{ab initio} using lattice-gauge-theory methods applied to QCD. While simulating QCD itself remains a formidable challenge, significant progress can be made by studying prototype models that capture key features. One such example is the $(1+1)$D Schwinger model~\cite{coleman1975charge,coleman1976more}, which exhibits confinement and string breaking via a linear potential, similar to QCD. However, important differences must be noted: the Schwinger model is Abelian, and its confinement mechanism is geometric, stemming from the spacetime dimensionality, whereas confinement arises from the non-Abelian structure of the gauge group of QCD in $(3+1)$D.
Nevertheless, quantum simulations of such lower-dimensional models provide valuable testing grounds to study nonperturbative dynamics, benchmark emerging quantum hardware, and develop techniques that may ultimately be essential for tackling the complexity of QCD itself. As we will see below, quantum-simulation experiments in this direction 
are already underway. Besides the string-breaking dynamics, we review in this Section experimental progress in realizing closely related phenomena in gauge theories, such as particle production, confinement dynamics, and metastability and false-vacuum decay. 

%%%%%%%%%%%%
%%%%%%%%%%%%
\subsection{Schwinger effect and pair production
\label{sec:particle-production}}
\begin{figure*}[t!]
    \includegraphics[width=0.895\textwidth]{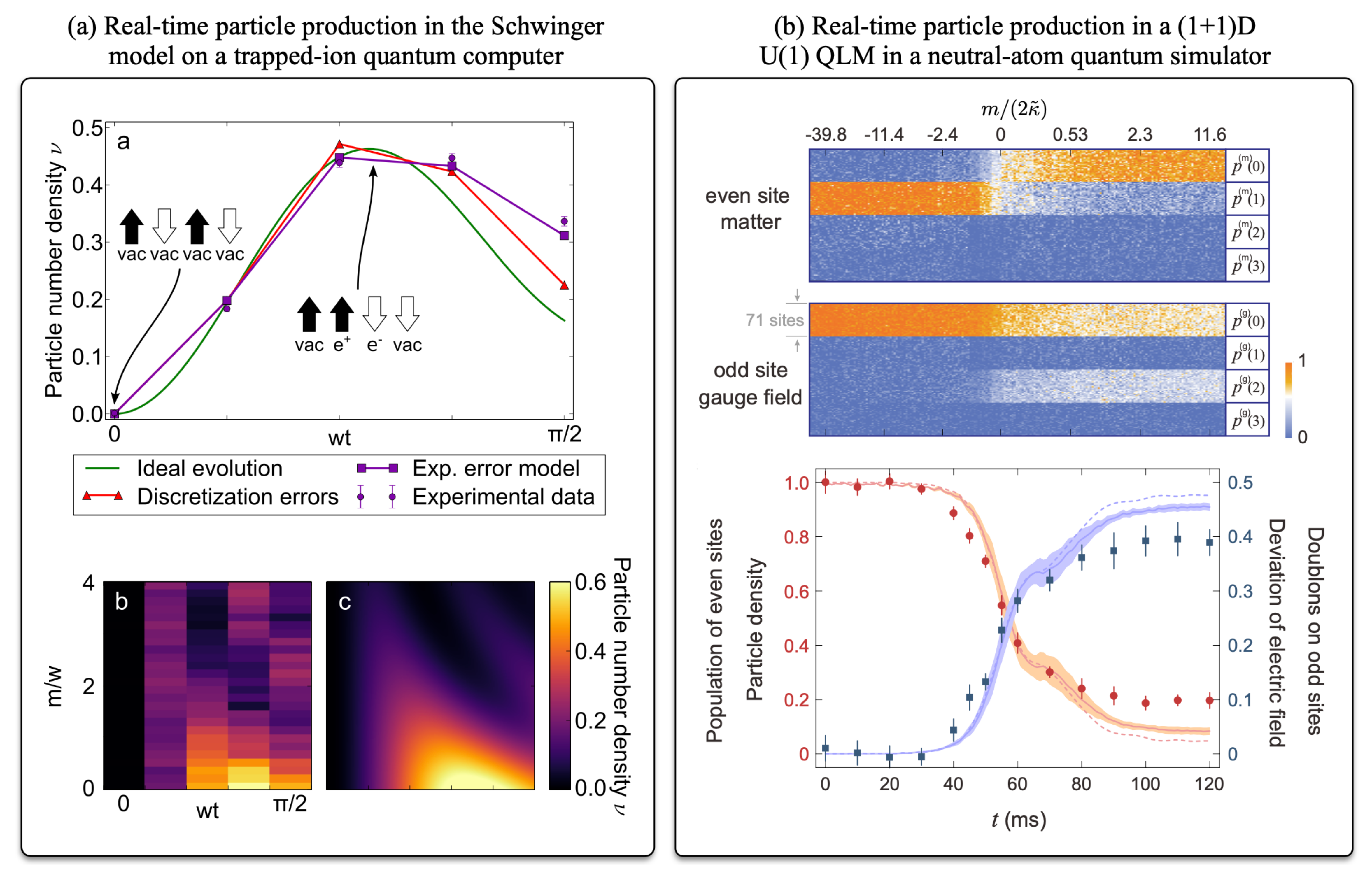}
    \caption{(a) Time evolution of the Schwinger model in its fully fermionic form starting from the `empty' state ($\ket{\Psi_0}=\ket{\uparrow\downarrow\uparrow\downarrow}$) in a trapped-ion quantum computer using a first-order Trotterization scheme. Plotted in the top panel is the resulting particle-number density $\nu=\sum_{N=1}^\ell [(-1)^\ell \hat\sigma_\ell^z+1]/(2N)$ for a four-site system. The lower panels plot the same quantity obtained from experiment (left) and from theory (right) for different values of the fermion mass and for $J=w$ [with $w \equiv -\kappa / (2a)$ and $J \equiv g^2a/2$ in Eq.~\eqref{eq:SchwingerGaugeFieldsIntegratedOut}]. For details, see Ref.~\cite{martinez2016real} from which the Figure is reproduced. (b) Time-resolved observation of the spontaneous breaking of charge conjugation and parity symmetry in a $(1+1)$D U$(1)$ QLM during a mass ramp on an analog optical superlattice quantum simulator. Starting in an initial state where each shallow well of the optical lattice hosts a single boson, equivalent to a charge-proliferated state in the gauge-theory picture, the singlons on the shallow wells bind into doublons in the deep wells in between, which corresponds to the annihilation of a neighboring electron-positron pair and a flip in the electric field on the link between them. The total electric flux density goes from zero at the beginning of the ramp to close to its maximal value of $0.5$ at the end of the experiment. Throughout all accessible times, there is good agreement between experimental results and numerical benchmarks. Figure is reproduced from Ref.~\cite{yang2020observation}.
    }\label{fig:pair-production}
\end{figure*}
Spontaneous creation of particle-antiparticle pairs out of the quantum-field-theory vacuum is a hallmark of relativistic dynamics. A much-sought-after phenomenon is the Schwinger effect~\cite{schwinger1951gauge}, in which particle-antiparticle pairs are created in a strong electric field at a constant rate per unit volume. One can start to study the pair creation in simple nonequilibrium settings such as in a quantum quench, in which the trivial vacuum (empty state) is evolved under the gauge-theory Hamiltonian. The Hamiltonian dynamics allow for the creation of all states consistent with the symmetries, while the initial state's finite energy provides the energetics for pair creation. Such conditions were created in the first experimental quantum-simulation studies of pair production in gauge theories, as will be reviewed in the following.

In a first gauge-theory quantum-simulation experiment~\cite{martinez2016real}, the pair-creation phenomena was experimentally observed in the quench dynamics of the Schwinger model. A four-site system was prepared in the initial state $\ket{\Psi_0}=\ket{\uparrow\downarrow\uparrow\downarrow}$, corresponding to a bare-vacuum state where all sites are unoccupied by matter particles. The state was evolved via a Trotterization scheme under the Hamiltonian in Eq.~\eqref{eq:SchwingerGaugeFieldsIntegratedOut} with the gauge fields integrated out, and with $\theta=\pi$. These dynamics were realized on a quantum computer composed of four trapped ions, using the implementation scheme outlined in Sec.~\ref{sec:TrappedIons}. Electron-positron pairs were created out of the vacuum throughout time dynamics, as shown in Fig.~\ref{fig:pair-production}(a). Lower values of the fermionic mass $m$ lead to greater matter production, while larger mass values suppress this production.  Subsequent Schwinger-model quantum simulations explored the same physics on other hardware~\cite{klco2018quantum,nguyen2022digital}.
\begin{figure*}
    \includegraphics[width=0.995\textwidth]{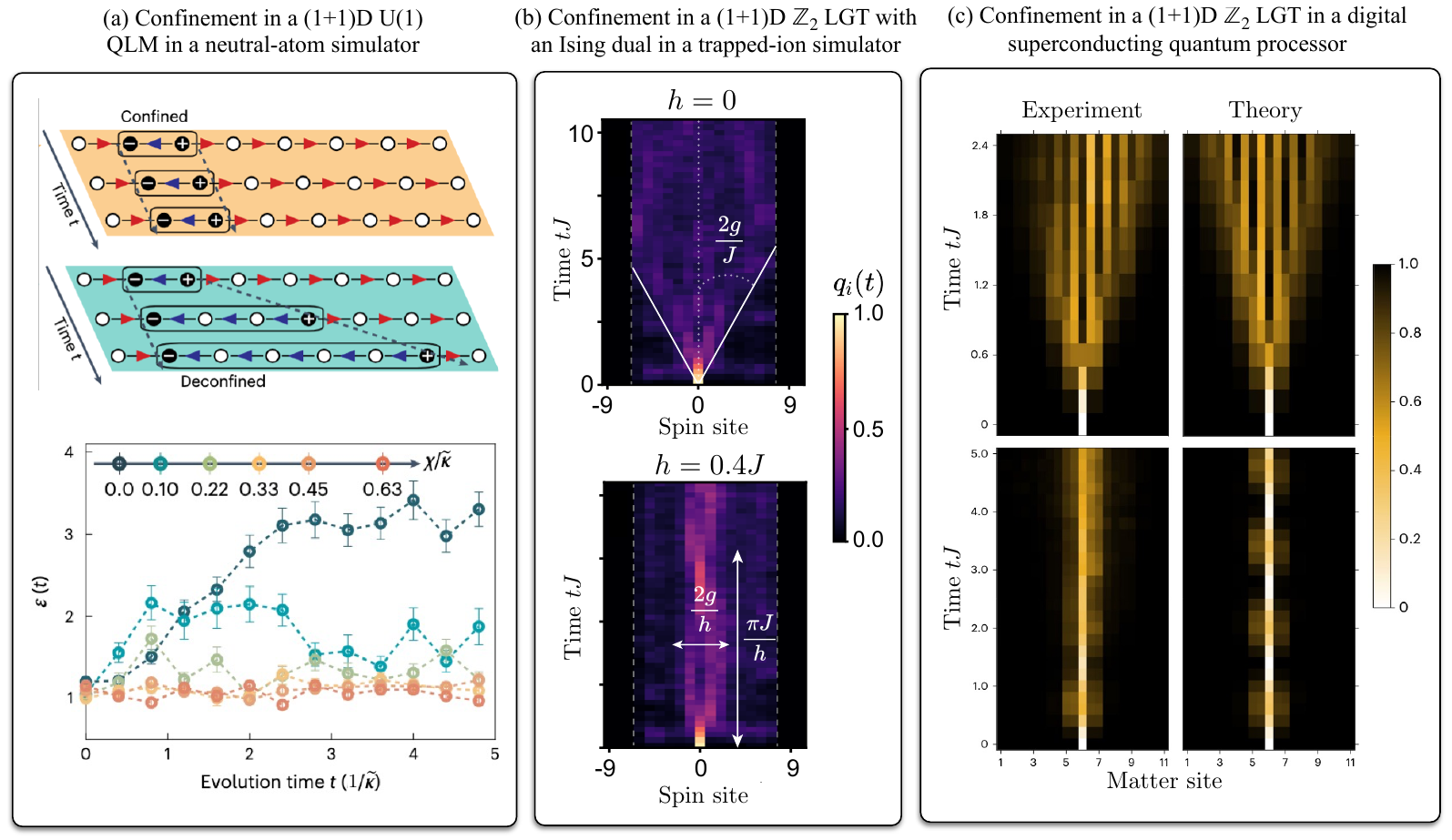}
    \caption{(a) Shown in the top is the schematic depiction of the (de)confined dynamics of an electron-positron pair in the $(1+1)$D U$(1)$ QLM on an analog optical-superlattice quantum simulator. When $\chi=0$, the pair spread ballistically away from each other as there is no penalty on the stretching of the electric string between them. When 
    $\chi>0$, the string stretching is energetically penalized, leading to confinement. Shown in the bottom is a quantum-gas-microscope measurement of the length of this string $\mathcal{E}(t)$ in the wake of quenches at different values of 
    $\chi$, corroborating the schematic depiction. 
    Figure is reproduced from Ref.~\cite{zhang2023observation}. 
    (b) Evolution of a localized charge after a quench of the $g$ coupling (top) and both $g$ and $h$ couplings (bottom) in a mixed-field Ising model with the Hamiltonian in Eq.~\eqref{eq:H-Ising} using a programmable trapped-ion analog quantum simulator. Depicted in both panels is the local charge density $q_i=\langle1-\sigma^z_{i-1}\sigma^z_{i}\rangle/2$. The superimposed lines and arrows in the top panel correspond to a single-charge approximation of dynamics: in the top panel, lines $i=\pm v_{\rm{max}} t$ with $v_{\rm{max}}=2g$ highlight the maximum speed of propagation of the charge, and in the bottom, arrows highlight amplitude ($2g/h$) and period ($\pi/h$) of the coherent oscillations. Figure is reproduced from Ref.~\cite{de2024observation}. 
    (c) Experimental results with theoretical benchmarks for (de)confinement dynamics of a single matter particle at the central site of a $(1+1)$D $\mathbb{Z}_2$ LGT on a digital superconducting-qubit quantum computer. Shown in the top is the limit of a weak electric field, for which the dynamics is deconfined. The bottom corresponds to large values of the electric field, for which the matter particle is confined up to all experimentally accessible times. Figure is reproduced from Ref.~\cite{Mildenberger2025}.
    }\label{fig:1p1D_confinement}
\end{figure*}

As another example, an analog quantum simulator was used in Ref.~\cite{yang2020observation} to probe the Coleman phase transition~\cite{coleman1976more} in the spin-$\frac{1}{2}$ $\mathrm{U}(1)$ quantum link model, Eq.~\eqref{eq:U1QLM}, at $\theta=\pi$. This setup was reviewed in Sec.~\ref{sec:opticalsuperlattice}. Using a $71$-site Bose--Hubbard optical superlattice, the system was prepared in the charge-proliferated state $\ket{\ldots\triangleright,e^-,\triangleleft,e^+,\triangleright,e^-,\triangleleft,e^+,\triangleright,\ldots}\leftrightarrow\ket{\ldots,0,1,0,1,0,1,0,1,0,\ldots}$, which is the ground state of the $\mathrm{U}(1)$ quantum link model at $m/\tilde\kappa\to-\infty$. The tunneling $J$, the on-site interaction strength $U$, and the staggering potential $\delta$ in Eq.~\eqref{eq:bhm} are then tuned such that the fermionic mass $m$ is adiabatically ramped to $m/\kappa\to+\infty$, where the $\mathrm{U}(1)$ quantum link model has the two degenerate ground states $\ket{\ldots \triangleleft,\varnothing,\triangleleft,\varnothing,\triangleleft,\varnothing,\triangleleft \ldots} \leftrightarrow \ket{\ldots 2,0,0,0,2,0,0 \ldots}$ and $\ket{\ldots \triangleright,\varnothing,\triangleright,\varnothing,\triangleright,\varnothing,\triangleright \ldots} \leftrightarrow \ket{\ldots 0,0,2,0,0,0,2 \ldots}$. By monitoring the occupation of single bosons in the shallow wells and the occupation of doublons in the deep wells of the optical superlattice, it was possible to measure both the matter density as well as the electric flux density during this adiabatic ramp, as shown in Fig.~\ref{fig:pair-production}(b). Whereas the former started at unity and the latter at zero, toward the end of the ramp, the former reached a value close to zero and the latter reached $0.5$, indicating that the wave function had a large overlap with the matter-free vacuums. Recall that the Coleman phase transition is related to the spontaneous breaking of a global $\mathbb{Z}_2$ symmetry due to charge conjugation and parity symmetry conservation in Eq.~\eqref{eq:U1QLM} at $\theta=\pi$. The experiment starts in the $\mathbb{Z}_2$-symmetric phase where the order parameter (the electric flux) is zero, but during the ramp, there is a spontaneous breaking of this global $\mathbb{Z}_2$ symmetry where the flux becomes nonzero.

%%%%%%%%%%%%
%%%%%%%%%%%%
\subsection{Confinement in real-time dynamics}
\label{sec:confinement}
Once the particle pairs are produced in the dynamics, one may ask what happens to the microscopic dynamics of such pairs. In a confining theory, the pairs are connected by an electric-field flux, i.e., a string, whose energy grows with the string length as particles fly away from each other. Such phenomena can be probed in quantum-simulation experiments of real-time dynamics of simplified models. Here, both the pair dynamics and the isolated charge dynamics can be probed, in contrast to what is observed in QCD, in which quarks cannot be separated and studied in isolation.

Among the first confinement studies in a quantum simulation experiment is that reported in Ref.~\cite{zhang2023observation}. This work uses an optical superlattice quantum simulator of the spin-$\frac{1}{2}$ $\mathrm{U}(1)$ QLM with matter integrated out; see Eq.~\eqref{eq:U1QLM_MatterIntegratedOut}. Using pattern-programmable addressing beams, the product state $\ket{\cdots,2,0,2,0,1,1,2,0,2,0,\cdots}\to\ket{\ldots,\triangleright,\varnothing,\triangleright,e^-,\triangleleft,e^+,\triangleright,\varnothing,\triangleright,\varnothing,\ldots}$ was prepared such that an electron-positron pair (a bare meson) was created in an empty background. This state was then quenched at a large mass value (in order to suppress matter fluctuations) and at a given value of the $\theta$ angle. Using a quantum gas microscope, the length $\mathcal{E}(t)$ of the electric-flux string between the electron and positron was measured during the time evolution. As shown in Fig.~\ref{fig:1p1D_confinement}(a), at $\chi = 0$, the electron and positron spread ballistically away from each other, and the length of the string grows linearly in time. However, as $\chi$ was tuned away from $0$, the length of the string is restricted to just one site, which is its initial value, indicating confinement. Confinement arises as a direct result of the nonvanishing $\theta$-term, which imposes an energy penalty on the growth of the electric string between the electron and positron, bounding them into a stable meson throughout the time evolution.

Confinement dynamics have also been probed in trapped-ion analog quantum simulators. Consider the mixed-field Ising model in Eq.~\eqref{eq:H-Ising}, which is dual to a $(1+1)$D Ising model, as described in Sec.~\ref{sec:formulations}. A charge in the LGT formulation is encoded as a kink in the dual formulation, where a kink is a pair of anti-aligned spins $\uparrow\downarrow$ or $\downarrow\uparrow$. The charge confinement in this model was studied in Ref.~\cite{de2024observation} using a programmable trapped-ion quantum simulator. To make the best use of quantum resources, a simpler system was simulated which comprised of infinitely long arrays of static spins to the left and right of a finite domain of dynamical spins. While only the dynamical spins are encoded in the quantum simulator, an additional site-dependent longitudinal field, $\Delta h_i$, was applied to the dynamical spins to emulate interactions with the static spins, and alleviate the undesired effects of hard boundaries. The experiment constituted of initializing the system with a static charge: the region to the left of the charge is the classical vacuum state with the electric field $\langle\sigma^z_i\rangle=1$, and the region to the right of the charge is a (semi-infinite) classical string state with $\langle\sigma^z_i\rangle=-1$. When $h=0$ (no confinement), the charge was observed to ballistically propagate from the center until it hits the static boundary and bounces back. However, when $h \neq 0$ (confinement), the charge's propagation was observed to be limited in space, indicating a Wannier-Stark localization phenomenon~\cite{wannier1960wave}: The string tension imparts a constant acceleration to the charge. However, on a lattice, the momentum cannot grow unbounded and a constant acceleration generates a periodic variation of momentum in time. The charge, therefore, coherently oscillates around its initial position, akin to Bloch oscillations~\cite{bloch1929quantenmechanik}. This effect was observed in the experiment, as shown in Fig.~\ref{fig:1p1D_confinement}(b).

The confining dynamics of a single charge was also probed in Ref.~\cite{Mildenberger2025}, in which a $(1+1)$D $\mathbb{Z}_2$ LGT dynamics were implemented on a superconducting quantum processor. In this model, described by the Hamiltonian in Eq.~\eqref{eq:Z2LGT}, the electric field, once again, acts as a confining potential. In the experiment, a single matter particle in an otherwise empty lattice was prepared. This initial state was subsequently quenched with the $\mathbb{Z}_2$ LGT Hamiltonian. It was observed that at vanishing values of $h$, the matter particle spreads ballistically on the lattice, while at $h>0$, it remains localized during the time evolution; see Fig.~\ref{fig:1p1D_confinement}(c). 

\begin{figure*}
    \includegraphics[width=0.995\textwidth]{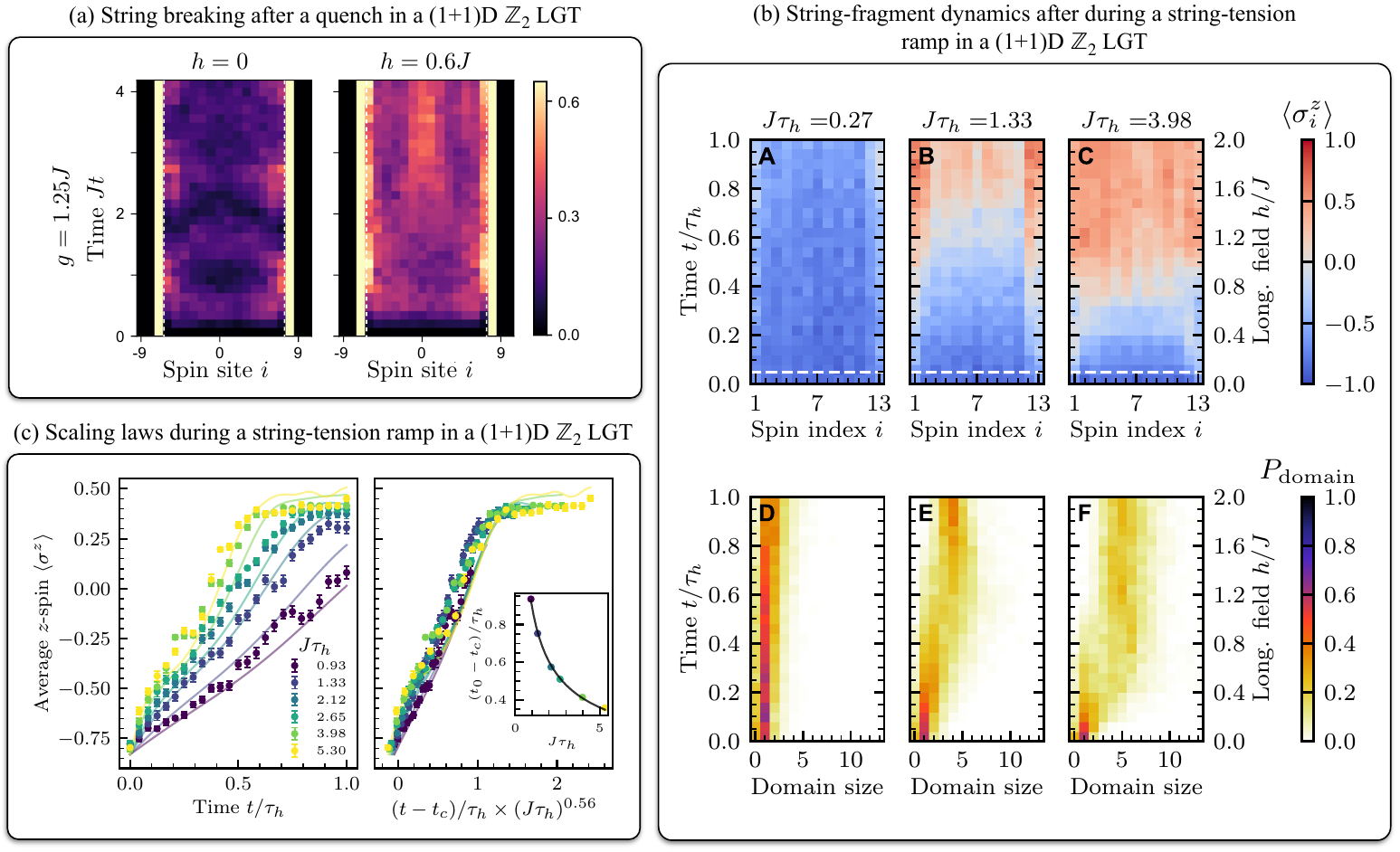}
    \caption{
    (a) Evolution of a string after a quench of the $g$ coupling (left) and both $g$ and $h$ couplings (right) in in a mixed-field Ising model with the Hamiltonian in Eq.~\eqref{eq:H-Ising} using a programmable trapped-ion analog quantum simulators. Depicted in both plots is the local charge density $q_i$.
    (b) String-breaking dynamics for 13 spins with an initial state $\ket{\Psi_\downarrow}$ at the transverse-field value $g=1.2J$ and during a linear ramp the longitudinal-field $h$ from $0$ to $2.0J$. Time evolution of individual-spins magnetization, $\langle \sigma_i^z \rangle$, is depicted in the top plots, while the probability, $P_\text{domain}$, that the largest connected domain of $\uparrow$-spins has size $n$ is plotted in the bottom plots, for $J\tau_h=0.27$, $1.33$, and $3.98$, respectively from left to right, where $2J/\tau_h$ is the speed of the linear ramp. The dashed white lines in the top plots indicate the moment when $h=h_c$.
    (c) Dynamics of $\langle\sigma^z\rangle$ during the $h$-ramp for a range of $\tau_h$, using the same linear-ramp protocol as in part (b). Rescaled time traces from the left panel collapse to a single curve in the right panel using a power-law scaling with exponent $0.56$. The time difference $|t_0 - t_c|$ (where $t_0$ is when $\langle\sigma^z\rangle$ changes sign and $t_c$ is when $h(t) = h_c$) versus $\tau_h$ fits a power law with exponent $0.56$ (black line), as shown in the inset. Solid lines indicate numerical simulations without decoherence. Points in the inset in the right panel correspond to numerical simulations.
    }\label{fig:string-breaking-ion}
\end{figure*}
%

%%%%%%%%%%%%
%%%%%%%%%%%%
\subsection{String dynamics and breaking}
\label{sec:string-breaking}
String breaking can be probed indirectly via the electric potential among static charges as a function of the distance between the charges, as has been studied in many classic lattice-QCD work over the years; see, e.g., Refs.~\cite{bali1998glueballs,aoki1999static,pennanen2000string,bali2000static,duncan2001string,bernard2001zero,kratochvila2003observing,bali2005observation}. The linearly rising confining potential eventually saturate to a constant value, signaling the breaking of the string. Here, we focus on direct probing of string breaking in quantum-simulation experiments, with a particular emphasis on the real-time dynamics of the string. We highlight recent $(1+1)$D experiments on an analog trapped-ion quantum simulator, as well as $(2+1)$D experiments on superconducting-circuit and Rydberg-atom quantum simulators. Other relevant work will be mentioned only briefly.

%%%
%%%
\subsubsection{String dynamics in $(1+1)$D LGTs
\label{sec:string-breaking-1D}}
String-breaking dynamics can be probed in trapped-ion analog quantum simulators. A suitable testing ground is the mixed-field Ising model [Eq.~\eqref{eq:H-Ising}] mentioned in the previous Section, where the confining potential among domains of flipped spin induces nontrivial dynamics. The longitudinal field $h$ generates a \emph{string tension}. The transverse field introduces quantum fluctuations that couple charge and string dynamics, with coupling strength $g$. The string breaks and forms new charge pairs as both $h$ and $g$ are increased. Through a series of experiments, Refs.~\cite{de2024observation,luo2025quantum} study string breaking in such a model. The same boundary conditions, as discussed in Sec.~\ref{sec:confinement} for this setting, were applied using frozen exterior spins to alleviate the boundary effects. 

The first set of experiments~\cite{de2024observation} concerned string breaking in quench processes, and away from a resonant regime (where couplings are tuned to induce a large overlap with the most favorable broken-string configuration). Instead, both $h$ and $g$ where turned on instantaneously starting from an initial state involving two static charges at the two ends of the ion chain, making up a string. As the $h$ and $g$ values are quenched, the nonequilibrium conditions breaks the string by creating new charges. The observed string evolution, nonetheless, proceeds via a mechanism distinct from the conventional Schwinger mechanism (that is a uniform, spontaneous creation of charge pairs in the bulk of the system). Charge-pair formation in the experiment systematically occurred at the string edges, i.e., near the static charges, as schematically shown in Fig.~\ref{fig:string-breaking-ion}(a) for select examples. For vanishing or weak string tension, dynamical charge pairs perform coherent oscillations confined to the edges, but for larger string tension, these charge pairs propagate and spread from the edges toward the bulk. This phenomenon occurs over a time scale that does not sensitively depend on Hamiltonian parameters. This is in stark contrast with the exponential enhancement, as a function of string tension and coupling strength, of the rate of spatially uniform charge-pair formation in the bulk associated with the Schwinger mechanism. A perturbative approach in small $g/J$ and $h/J$ was developed in Ref.~\cite{de2024observation} to capture the origin of the effect: the main contribution to nonequilibrium string dynamics arises from the quantum diffusion of a single pair of charges generated by the dynamical protocol near the edges; see Ref.~\cite{de2024observation} for more discussions.

The second set of experiments~\cite{luo2025quantum} concerned a more controlled dynamical protocol compared to a quench, in which the Hamiltonian parameters were ramped with various speeds. The infinitely slow ramp starting from the string state is an adiabatic evolution that passes through an avoided-level crossing (with a gap exponentially small in system size). It  converts the string to a broken string with two charge pairs forming at the edges (hence a complete screening of the bulk electric field). The infinitely fast limit is the quench, which misses all avoided-level crossing and leads to a more complex final state as in the first set of experiments above. The intermediate speeds, therefore, could reveal rich string-breaking dynamics. To probe these dynamics, a linear ramp $h(t)/J=2t/\tau_h$ was implemented in an $\ell=13$ ion experiment starting from the interacting ground state, $\ket{\Psi_\downarrow}$ (a state with the largest overlap to a bare string state), $\ket{\downarrow \cdots \downarrow}$. This initial state was created using an initial slow ramp of $g$ at $h=0$. The local magnetization $\langle \sigma_i^z \rangle$ was measured, as the longitudinal field $h$ was ramped from $0$ to $2J$, well beyond the transition point $h_c(\ell=13)=0.1J$. The results are shown in the top panels of Fig.~\ref{fig:string-breaking-ion}(b). Again, string breaking initiates preferentially near static boundary quarks. For short ramp durations, the magnetization remains nearly constant while for slower ramps, it changes sign at $h>h_c(\ell)$. Such a behavior indicates a transition via bubble formation, or localized domains of positive magnetization, rather than a collective spin flip as in the adiabatic case. The growth of the largest domain size as a function of ramp time was further analyzed in this work. The bottom panels of Fig.~\ref{fig:string-breaking-ion}(b) display $P_\text{domain}(n)$, the probability that the largest connected domain of $\uparrow$-spins has size $n$, as a function of domain size $n$ and time elapsed during the $h$-ramp. For a very short ramp duration, bubble formation is minimal, and the largest domain size is typically $n=1$. For longer ramp times, domain sizes grow during the ramp, with $P(n)$ peaked around $n=4,5$ at the end of the ramp. Last but not least, it was observed in Ref.~\cite{luo2025quantum} that scaling the time by a fixed power of the ramp speed collapses all magnetization values for different ramps to a single curve, as shown in Fig.~\ref{fig:string-breaking-ion}(c). This behavior points to conjectured Generalized Kibble--Zurek scaling laws~\cite{zurek1985cosmological,kibble1980some} near the transition point of certain discontinuous phase transitions~\cite{qiu2020observation,surace2024string}. We come back to this point in Sec.~\ref{sec:FV} where we discuss the closely related phenomena of false-vacuum decay and metastability.

%%%
%%%
\subsubsection{String dynamics in $(2+1)$D LGTs}
\begin{figure*}
    \includegraphics[width=0.995\textwidth]{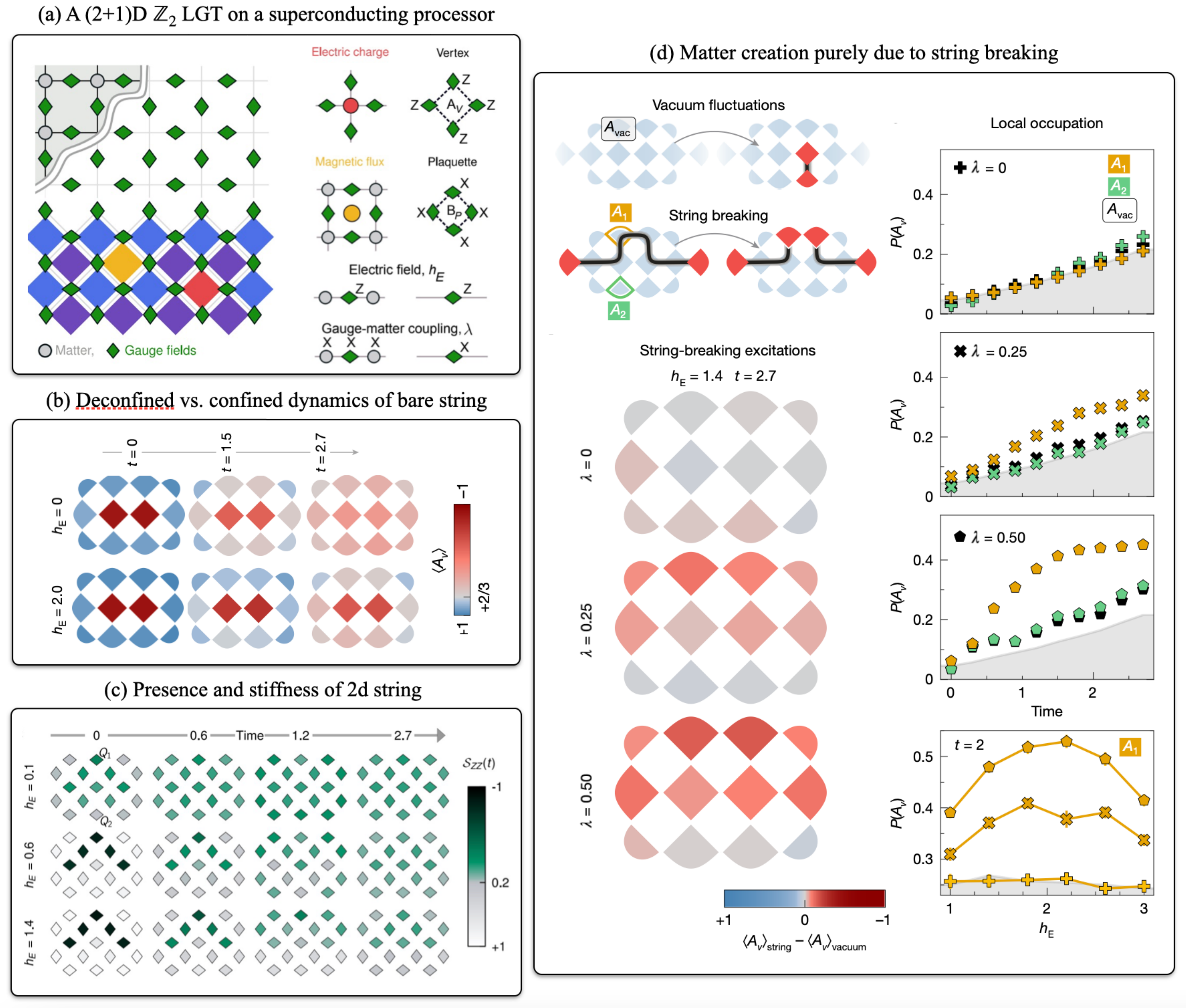}
    \caption{(a) The $(2+1)$D $\mathbb{Z}_2$ LGT implemented by on a superconducting-qubit quantum computer (Google's \texttt{Sycamore} quantum processor) to probe string dynamics and breaking. Matter (gauge) degrees of freedom are denoted as circles (diamonds). Through the local $\mathbb{Z}_2$ gauge symmetry, the matter degrees of freedom are integrated out, and the Hamiltonian in Eq.~\eqref{eq:Z2LGT2d} is obtained. 
    In the experiments, the mass and plaquette-term strength are set to $J_E=J_M=1$. The electric-field strength $h_E$ can be thought of as a confining potential between two charges. 
    (b) Dynamics of two electric excitations on adjacent vertices (red tiles) on top of the interacting vacuum. The coupling that induces dynamics to the excitations is set to $\lambda = 0.25$. The quantity shown is the average density of electric excitations as measured by $\langle \hat{A}_v\rangle$ for $h_E = 0$ (top) and $h_E = 2.0$ (bottom), corresponding to deconfined and confined phases of the model, respectively.
    (c) An out-of-equilibrium string is prepared on top of the vacuum at various values of $h_E$ and allowed to evolve in time. A special two-time qubit correlator is measured to track the string dynamics, revealing three distinct regimes: one where the string is not well-defined, a second where a string is formed but is floppy, and a third where the string is rather stiff and exhibits dynamics in only the upper half of the grid.
    (d) String breaking dynamics is achieved by tuning to the resonance $h_E=2J_E$, whereby a link-length segment of the string is broken and its energy is used to create a neighboring pair of particles; see the text for details. Figure is reproduced from Ref.~\cite{cochran2024visualizingdynamicschargesstrings}.}\label{fig:Z2LGT_2p1D_Google}
\end{figure*}

In this Section, we turn to $(2+1)$D experiments that probe string-breaking dynamics. The first experiment reported in Ref.~\cite{cochran2024visualizingdynamicschargesstrings} realizes such dynamics using a superconducting quantum processor (Google's \texttt{Sycamore}). The $(2+1)$D $\mathbb{Z}_2$ LGT in the presence of matter on a square lattice is described by the combination of the two Hamiltonians in Eqs.~\eqref{eq:Z2LGT} (upon a trivial basis transformation) and \eqref{eq:Z2Hamiltonian}. It reads:
\begin{align}\label{eq:Z2LGT2d}
    \hat{H}=-\lambda\sum_l\hat{\tau}^x_l-h_E\sum_l \hat{\tau}^z_l-J_E\sum_v\hat{A}_v-J_M\sum_p\hat{B}_p.
\end{align}
Here, $\hat{\tau}^x_l$ and $\hat{\tau}^z_l$ are the Pauli operators representing, respectively, the gauge and electric fields at link $l$, $\hat{A}_v=\prod_{l\in v}\hat{\tau}^z_l$ is the star operator, which is
the product of the electric fields on the four links connecting to vertex $v$, and $\hat{B}_p=\prod_{l\in p}\hat{\tau}^x_l$ is the plaquette operator, which is the product of the gauge fields on the four links of plaquette $p$; see Fig.~\ref{fig:Z2LGT_2p1D_Google}(a). Moreover, $\lambda$ is the minimal coupling, $J_E$ is the mass, $h_E$ is the electric-field strength, and $J_M$ is the magnetic-field strength. The Hamiltonian in Eq.~\eqref{eq:Z2LGT2d} arises from integrating out the hardcore-bosonic matter fields using Gauss's law. The Gauss's law operator in this case is $\hat{G}_v=\hat{A}_v\hat{\sigma}^z_v$, where $\hat{\sigma}^z_v$ denotes matter occupation at vertex $v$. By working in the gauge sector where the eigenvalue of $\hat{G}_v$ is $+1$ everywhere, one obtains Eq.~\eqref{eq:Z2LGT2d}. At $\lambda=h_E=0$, one recovers the famous Toric code \cite{kitaev2003fault}, which is at the heart of several quantum-error-correction codes. At small values of $\lambda/J_E$ and $h_E/J_M$, the ground state of the Hamiltonian in Eq.~\eqref{eq:Z2LGT2d} is in a deconfined topologically ordered phase. At large $\lambda/J_E$ and small $h_E/J_M$, the ground state is in the Higgs phase. At small $\lambda/J_E$ and large $h_E/J_M$, the ground state is in the confined phase, which is where excitations are bound via strings \cite{Fradkin1979phase,Trebst2007breakdown,Vidal2009low-energy,Wu2012phase}. This is also the phase where string dynamics and breaking can be reliably probed. As an example, in the thermodynamic limit, the critical point separating the deconfined and confined phases corresponds to $h_E\approx 0.33$ at $J_M=J_E=1$ and $\lambda=0.25$~\cite{Vidal2009low-energy,Wu2012phase,Bloete2002cluster,Xu2025critical}.

In the quantum simulation, realized on a $72$-qubit \texttt{Google} \texttt{Sycamore} quantum processor, the star and plaquette terms are set to unity strength, i.e., $J_M=J_E=1$. The ground state is approximately obtained using the variational weight-adjustable loop ansatz (WALA) \cite{Dusuel2015mean-field,Sun2023parametrized} on a two-dimensional grid of $35$ qubits. Of those, $17$ are link qubits (diamonds) and $18$ are ancilla qubits (circles), which is equivalent to a system of $3\times4$ matter sites; see Fig.~\ref{fig:Z2LGT_2p1D_Google}(a). The ansatz parameters are optimized on a classical machine such that the energy of the approximate ground state $\ket{\Psi_0}$ is minimized as a function of $h_E$. After preparing this low energy-density initial state, which approximates well the ground state of the Hamiltonian in Eq.~\eqref{eq:Z2LGT2d}, a series of experiments are carried out to probe the dynamics of strings, several of which we review in the following. Time evolution is at the core of these experiments, which was performed using an efficient Suzuki--Trotter expansion of the time-evolution operator under the Hamiltonian in Eq.~\eqref{eq:Z2LGT2d}. This is achieved by using ancilla qubits at each vertex (where the integrated-out matter had lived) and the center of each plaquette.

First, a pair of matter particles, or excitations, is prepared on neighboring sites by applying a single Pauli-X to $\ket{\Psi_0}$ at the link between them. This out-of-equilibrium initial state will now exhibit dynamics that can probe (de)confinement in two spatial dimensions. Setting $\lambda=0.25$, the dynamics is probed for $h_E=0$ (deconfined phase) and $h_E=2$ (confined phase). Markedly distinct behavior is observed in both cases on the quantum processor, as depicted in Fig.~\ref{fig:Z2LGT_2p1D_Google}(b).
In the case of $h_E=0$, there is no energy penalty on the growth of the string between the two excitations, and so they spread very quickly throughout the two-dimensional grid. However, their spread is significantly suppressed for $h_E=2$, where the large electric field penalizes any spread in the excitations and thus keeps the string close to its initial length. 

To create a string on this grid, a set of Pauli-X gates are applied to $\ket{\Psi_0}$ along a line of link qubits that connect two ancilla qubits at vertices on either end of the grid. This is equivalent to creating a pair of excitations along the grid with a string connecting them. In order to observe the fluctuations of the string over time, the two-time correlator $S_{ZZ}(t)=\Re[\langle \hat{\tau}^z(t)\hat{\tau}^z(0)\rangle]\times\langle\hat{\tau}^z(0)\rangle$ is measured for each qubit using a Hadamard test with an auxiliary register. This correlator is specifically designed as a product of two contributions, $\Re[\langle \hat{\tau}^z(t)\hat{\tau}^z(0)\rangle]$ and $\langle\hat{\tau}^z(0)\rangle$, such as to probe the \textit{presence} (second term) and the \textit{stiffness} (first term) 
of the string. The first term is sensitive to changes in the string over time, while the second term checks whether a string has been created on top of $\ket{\Psi_0}$, which is expected to occur only in the confined phase. Measurements of $S_{ZZ}(t)$ reveal three distinct regimes; see Fig.~\ref{fig:Z2LGT_2p1D_Google}(c). When $h_E=0.1$, the system is in the deconfined phase, and applying the aforementioned Pauli-X sequence does not create a \textit{bona fide} string at $t=0$. Over the evolution time, the $S_{ZZ}(t)$ spreads all over the grid and takes on roughly the same value at each qubit. Repeating the experiment with $h_E=0.6$, which is in the confined phase, a well-defined string is observed at $t=0$. However, because of the relatively small value of the electric field, the string tension is weak, and $S_{ZZ}(t)$ indicates that the string is equally likely to be found in the upper or lower halves of the system at later evolution times. The string can be described as floppy in this regime. Setting to a larger value of $h_E=1.4$, which is deep in the confined regime, leads to a two-time correlator $S_{ZZ}(t)$ exhibiting dynamics mostly in the upper half of the grid, with very low probability that the string will oscillate in the lower half. This shows that deep in the confined regime the string is rather stiff, despite freely moving in the upper half of the grid (where the original string was created).

Due to energy conservation, the above prepared string cannot dynamically change length unless the system is at or near a resonance that allows matter creation \cite{xu2025stringbreakingdynamicsglueball}. The smallest piece of a string that can break has the length of a single link. This results in an energy change of $2h_E$. This change can be compensated by dynamically creating matter excitations at the two neighboring matter sites connected by the link, resulting in an energy change of $4J_E$. As such, this lowest-order resonance condition for breaking the string under Hamiltonian in Eq.~\eqref{eq:Z2LGT2d} is $h_E=2J_E$. To investigate string breaking deep in the confined phase, the electric-field strength is set to $h_E=1.4$ and the difference in matter occupation at each qubit between the string state and the vacuum, $\Delta A_v=\langle \hat{A}_v\rangle_\text{string}-\langle \hat{A}_v\rangle_\text{vacuum}$, is measured at an evolution time $t=2.7$. The vacuum here is $\ket{\Psi_0}$, i.e., the WALA state before the application of the gate sequence creating the string. The reason for considering this difference is that at finite mass $J_E=1$, there will always be matter production unrelated to the string breaking. As such, any difference between the matter occupation in the string state and that in the vacuum state at late times can be attributed to string breaking. For $\lambda=0$, $\Delta A_v\approx0$, indicating that matter occupation is roughly the same in both states. This makes sense because matter creation can only occur through the minimal-coupling term $\propto\lambda$, which is equivalent to gauge-invariant tunneling and pairing in the matter fields. When the minimal coupling is set to $\lambda=0.25$ or $0.5$, the picture drastically changes and $\Delta A_v$ becomes nonzero on qubits along which the string was initialized, signaling string breaking.

To investigate this behavior more thoroughly, the probability $P(A_1)$ of matter creation can be measured at a qubit around the midpoint of the initial string configuration, along with the probability $P(A_2)$ of creating matter at its mirror-symmetric qubit with respect to the horizontal middle of the grid. In addition to these quantities, the corresponding probability of matter creation in the vacuum state at either qubit is also measured (in the absence of the string both qubits are equivalent). As shown in Fig.~\ref{fig:Z2LGT_2p1D_Google}(d), all three quantities follow the same trend for $\lambda=0$. However, at finite $\lambda$, $P(A_1)$ grows distinctly from the other two, which remain very similar in their dynamics. The larger $\lambda$ is, the greater the growth of $P(A_1)$ over time. The string breaking seen here is not exactly at the resonance $h_E=2J_E$. To see if this resonance leads to a greater probability of matter creation, $P(A_1)$ is measured as a function of $h_E$ at $t=2$ for each considered value of $\lambda\in\{0,0.25,0.5\}$. Since no string breaking occurs for $\lambda=0$, $P(A_1)\approx0$ for all values of $h_E$. But when $\lambda=0.25$ or $0.5$, $P(A_1)$ has a maximum at $h_E=2=2J_E$.
\begin{figure*}
    \includegraphics[width=0.995\textwidth]{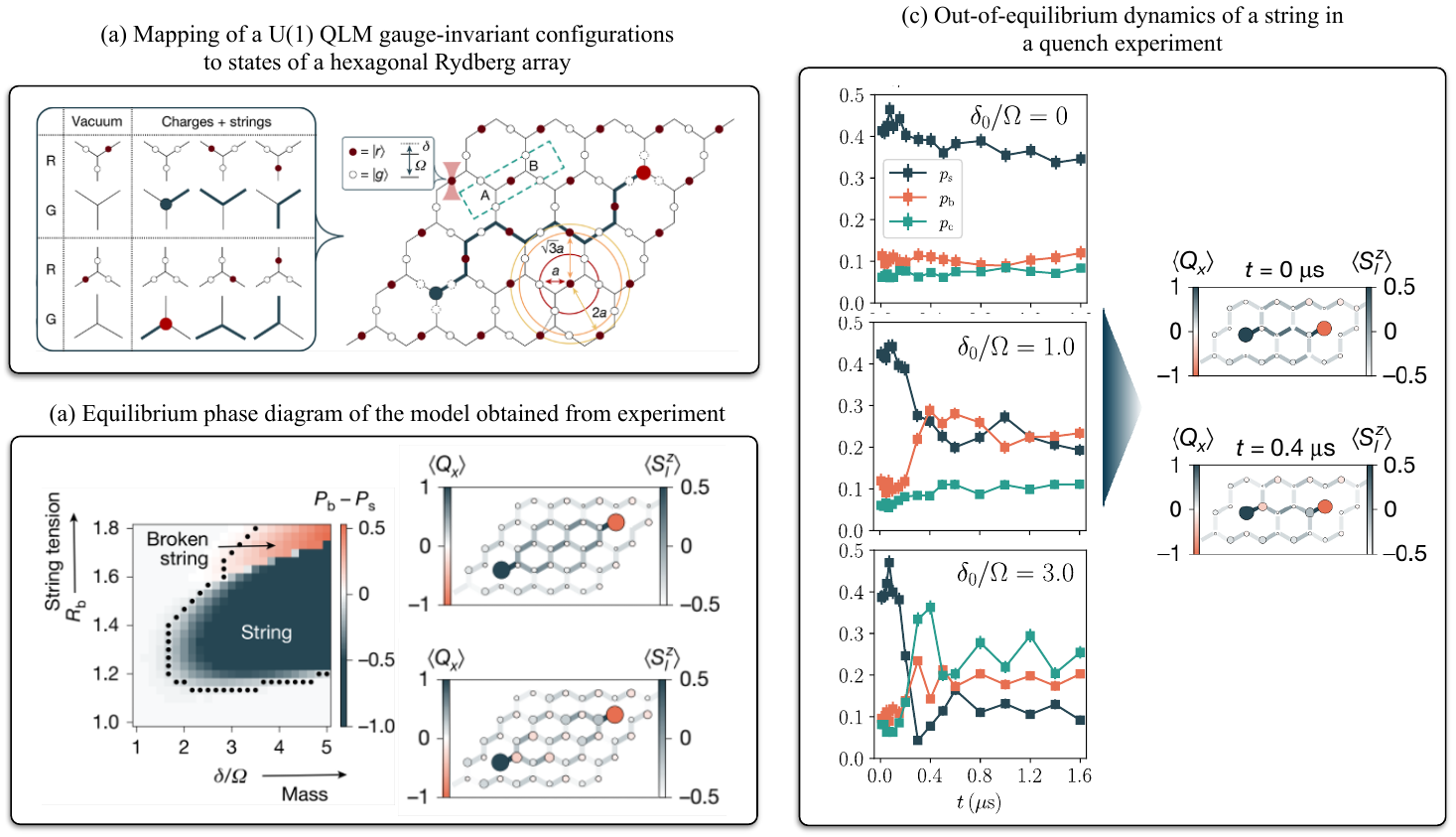}
    \caption{(a) Mapping of a U$(1)$ QLM on a 2d lattice onto an analog hexagonal Rydberg-atom array quantum simulator. The mapping involves integrating out the matter degrees of freedom through Gauss's law, leaving only the gauge degrees of freedom to be represented by the Rydberg atoms. Removing the three atoms around a given vertex is equivalent to placing a static charge at that vertex. Two such static charges give rise to a string between them. In the table in the left, equivalence between atom configurations satisfying the Rydberg blockade constraint (R) and the corresponding gauge-invariant states (G) is shown for the sites on the A (upper) and B (lower) sublattices in each unit cell in the right. (b) Mapping the equilibrium phase diagram of the model at different values of the Rydberg blockade radius $R_\text{b}$ and $\delta/\Omega$, which reveals three regimes: one where strings are not well-defined, a second where they are, and a third where they are broken. Plotted in the left are probability of broken string $P_\text{b}$ subtracted from that of the unbroken string $P_\text{s}$. Plotted in the right is the charge expectation value at each site $x$, $\langle Q_x \rangle$, and the spin expectation value at each link $l$, $\langle S^z_l \rangle$. (c) Out-of-equilibrium string-breaking dynamics upon a quench of $\delta_0/\Omega$ by tuning near a resonance condition allowing the creation of particle-antiparticle pairs. Probabilities in the left are those of broken $P_\text{b}$, unbroken $P_\text{s}$ strings and of fully charged configuration $P_\text{c}$. Quantities in the right are those in the right panels of part (b) at two time slices of the evolution for $\delta_0/\Omega = 1.0$. The figure is reproduced from Ref.~\cite{gonzalez2024observation}.
    }\label{fig:U1QLM_2p1D_QuEra}
\end{figure*}

String breaking on a 2d lattice has been recently studied in a Rydberg-atom quantum simulator as well~\cite{gonzalez2024observation}. This experiment uses a programmable Rydberg tweezer array by QuEra Inc., where ${}^{87}$Rb atoms are placed on the sites of a Kagome lattice; see Fig.~\ref{fig:U1QLM_2p1D_QuEra}. In such a setup, the atoms are initially in their ground state $\ket{g}$, after which a laser couples them to the excited (Rydberg) state $\ket{r}$ with Rabi frequency $\Omega$ and detuning $\delta$. The atoms interact according to the Hamiltonian in Eq.~\eqref{eq:H_Ryd}, up to a global factor of $\frac{1}{2}$ as well as a sign flip for the detuning term. The goal is to simulate dynamics of a $\mathrm{U}(1)$ QLM on a 2d lattice where one can utilize the Rydberg blockade inherent to this system, along with the geometry of the lattice, to enforce Gauss's law locally. The generator of Gauss's law in this model is 
\begin{align}
    \hat{G}_x=(-1)^{s_x}\sum_{l\in x}(-1)^{s_l}\hat{S}^z_l-\hat{Q}_x,
\end{align}
where $\hat{S}^z_l$ is a spin-$\frac{1}{2}$ operator representing the electric field at a link $l$ adjacent to site $x$, the dynamical charge is $\hat{Q}_x=\hat{a}_x^\dagger\hat{a}_x-[1-(-1)^{s_x}]/2$, and $\hat{a}_x^{(\dagger)}$ are hardcore-bosonic ladder operators. Here, $s_x=0$ ($1$) for sites $x$ in the A (B) sublattice, while $s_l=0$ ($1$) for links $l$ connecting to a site in the same (a different) unit cell; see Fig.~\ref{fig:U1QLM_2p1D_QuEra}(a). The sector of Gauss's law that the experiment seeks to realize is that associated with a staggered
configuration of static charges on the lattice, i.e., $\hat{G}_x\ket{\Psi}=q_x\ket{\Psi}$, where $q_x=(-1)^{s_x}/2$.

The experiment is performed in the confined phase at large $\delta/\Omega$ at a Rydberg blockade radius $R_\text{b}=[C_6/(2\Omega)]^{1/6}$ such that $1.2\lesssim R_\text{b}\lesssim1.8$, which is associated with the nematic phase \cite{Samajdar2021quantum}. In this phase, the confined charges and strings emerge as low-energy excitations. In the blockade regime $R_\text{b}^6\gg\delta/\Omega$, only one atom can be in the Rydberg state within the blockade volume. Based on the geometry of the lattice that presents only three atoms in a unit cell, only four possible configurations are permitted: one with all three atoms in the ground state, and three with a single atom in the Rydberg state. These configurations map exactly to the gauge-invariant configurations allowed by Gauss's law. The effective Hamiltonian realized on the Rydberg quantum simulator takes the form
\begin{align}\nonumber
    \hat{H}_\text{eff}&=\frac{\Omega}{2}\sum_{x,y}\Big(\hat{a}_x^\dagger\hat{S}^+_{x,y}\hat{a}_y+\text{H.c.}\Big)+\frac{\delta}{2}\sum_x(-1)^{s_x}\hat{a}_x^\dagger\hat{a}_x\\\label{eq:QuEraLGT}
    &+\frac{1}{2}\sum_{(l,l')\neq\langle l,l'\rangle}V_{l,l'}\bigg[(-1)^{s_l}\hat{S}^z_l+\frac{1}{2}\bigg]\bigg[(-1)^{s_{l'}}\hat{S}^z_{l'}+\frac{1}{2}\bigg],
\end{align}
where $\langle \cdots \rangle$ denotes nearest neighbor pairs. Strings are identified as $S^z_l=+1/2$ on a link $l$. $\hat{S}^+_{x,y}$ represents the gauge field at the link connecting sites $x$ and $y$. Static charges can be created at a vertex by removing the three corresponding neighboring atoms. Note that magnetic (plaquette) interactions are absent from this effective Hamiltonian, and evolution occurs under gauge-invariant matter hopping, staggered mass, and an electric term only.

First, the experiment probes string breaking in equilibrium. The quantum simulator used (QuEra's \texttt{Aquila}) has $59$ atoms, tailored into a system of $L_0=L_1=5$ unit cells, with $\{0,1\}$ denoting the two axes of the hexagonal lattice. The static charges are separated by $d_0=d_1=2$ unit cells. To prepare the ground state of the Hamiltonian in Eq.~\eqref{eq:QuEraLGT} with the two static charges, one starts in the disordered phase at a very large negative $\delta$ with a fixed value of the Rydberg blockade radius $R_\text{b}$. Upon applying $\Omega$ globally and then adiabatically sweeping $\delta$, the system effectively evolves close\footnote{Since the sweep goes through a phase transition, the final state will contain some defects and is, therefore, never exactly the true ground state.} to the desired ground state at some positive value of $\delta/\Omega$. The ground state is mapped as a function of $R_\text{b}$, which is related to the string tension, and $\delta/\Omega$, which is proportional to the mass. Intuitively, when the string tension is weak, strings cannot form regardless of the mass. On the other hand, when the mass is small, matter will proliferate in the system, and strings cannot form regardless of the strength of the string tension. Therefore, strings can be expected to occur at relatively high values of $\delta/\Omega$ and $R_\text{b}$, which is indeed what the experiment finds; see Fig.~\ref{fig:U1QLM_2p1D_QuEra}(b). Furthermore, in the large-mass regime, the strings break at sufficiently large values of $R_\text{b}$. This phase diagram is mapped by measuring the probability of the six possible minimal string configurations between the two static charges, which the ground state is expected to be a superposition thereof; and the probability of broken strings manifest in a matter particle next to each static charge that screen it. 

Second, the experiment probes the string-breaking dynamics via a quench. The simulation now involves a system of $L_0=5$ and $L_1=3$ unit cells, with the static charges separated by a distance of $d_0=2$ along the horizontal direction. This setup corresponds to $31$ atoms. Using the same adiabatic protocol for the equilibrium case, an initial state is prepared at $R_\text{b}=1.2$ and $\delta/\Omega=2.3$, which is in the regime $2m>\sigma d$, where $m$ is the mass, $d$ is the string length, and $\sigma \propto R_b^6$ is the string tension. In this regime, creating matter particles to screen the static charges is energetically unfavorable, and the string should remain intact. The initial probabilities for unbroken and broken strings are measured to be $P_\text{s}\approx0.4$ and $P_\text{b}\approx0.1$, respectively. To break the string, the initial state should be quenched to a regime where $2m\approx\sigma d$, which is a resonance condition for the string to break through the creation of a particle-antiparticle pair of mass $2m$. To test this, the initial string state is quenched at various values of $\delta_0/\Omega=0,1,3$, where $\delta_0$ is a local detuning $\delta\to\delta-\delta_0 s_l$ for $l$ when the Rydberg atoms are in the ground state. Due to this detuning, an extra linear contribution is added to the bare string tension $\sigma\to\sigma(R_\text{b})+\delta_0$. For a given value $\delta_0^\ast/\Omega$ for which the effective resonance condition $2m\approx[\sigma(R_\text{b})+\delta_0^\ast]d$ is reached, one can expect that $P_\text{b}$ will reach a maximal value after some time $t^\ast$. For $\delta_0=0$, which means there is no quench, the values of the probabilities remain roughly the same over time, though $P_\text{s}$ exhibits a drop in its value attributed to decoherence. Quenching at $\delta_0/\Omega=1$ shows $P_\text{s}$ and $P_\text{b}$ evolving over time to roughly the same value, indicating an equal probability of having broken and unbroken strings in the late-time wave function, which is also evident from the real-space configuration snapshot shown in the right panel of Fig.~\ref{fig:U1QLM_2p1D_QuEra}(c). For the quench at $\delta_0/\Omega=3$, a strong decay in $P_\text{s}$ is observed, indicating strong string breaking, while $P_\text{b}$ increases in value. Furthermore, $P_\text{c}$, the probability of fully charged configuration shows a marked increase over time, indicating the proliferation of particle-antiparticle pairs along the grid distance between the two static charges.

Quantum simulation of string-breaking statics and dynamics remains an active frontier of research. Here, we focused the discussion to a few highlights to enable a detailed presentation of the simulation strategies and the physics observed in experiments. For other relevant progress on both theory and experiment fronts, see Refs.~\cite{crippa2024analysis,borla2025stringbreaking21dmathbbz2,alexandrou2025realizingstringbreakingdynamics,Xu2025tensor,xu2025stringbreakingdynamicsglueball,cobos2025real}. These studies remain limited to simpler lower-dimensional models and small system sizes. Nonetheless, they pave the way toward future experiments with increased complexity and scale.

%%%%%%%%%%%%
%%%%%%%%%%%%
\subsection{Metastability and false-vacuum decay}\label{sec:FV}

\begin{figure*}
    \centering
    \includegraphics[width=0.995\linewidth]{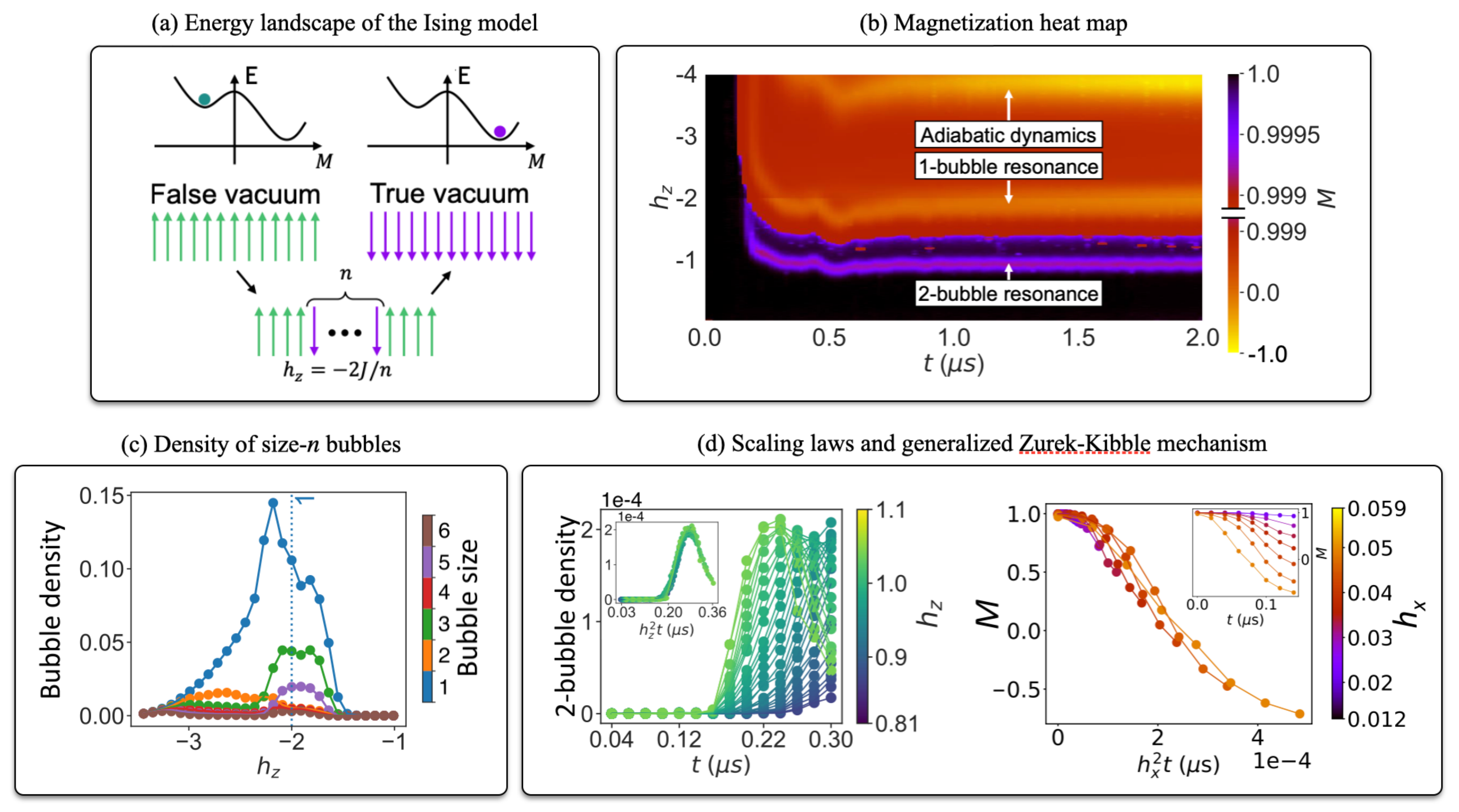}
    \caption{(a) A semiclassical energy landscape as a function of the order parameter $M$ can exhibit a local metastable minimum, dubbed the false vacuum (green dot), whose energy is higher than the true-vacuum energy (purple dot). In the Ising model in Eq.~(\ref{eq:H-Ising}) with the ferromagnetic coupling $J>0$ restricted to nearest-neighbor spins, the false vacuum is represented by the polarized $|\uparrow \uparrow \ldots \uparrow\rangle$ state, while the true vacuum is the other polarized state $|\downarrow \downarrow \ldots \downarrow\rangle$. The decay unfolds via the creation of quantized true-vacuum bubbles, which are the simple spin flips in the limit of large longitudinal magnetic field $h$. Bubbles of size $n$ occur at the resonance condition $h=-2J/n$ (see text for details).
    (b) The heat map of the magnetization $M$ versus time $t$ and the longitudinal-field magnitude $h \equiv h_z$ at the transverse-field strength $g \equiv h_x = 0.002$. The adiabatic dynamics and the $n = 1$ bubble resonance are observed on the larger scale, while the $n = 2$ bubble resonance can only be resolved in the 4th decimal of $M$, as the rate of dynamics decreases by an order of magnitude. 
    (c) Bubble-density measurements at $J = 1$ and different $h_z$ magnitudes, with $h_x = 0.002$ and $t = 2µs$. The bubble size $n = 1$ is dominant around its resonance value $h_z = -2J$, indicated by vertical dotted lines. (d) The left panel shows the 2-bubble density at $h_x = 0.002$ as a function of time at various $h_z$ values. Inset displays the collapse of different curves when time is rescaled by $h^2_z$ in accordance with Landau-Zener theory~\cite{sinha2021nonadiabatic}. The right panel shows magnetization $M$ at the $h_z = -J$ resonance as a function of rescaled time $h^2_xt$, for different values of $h_x$. The measured magnetization curves follow the same $h^2_x$ scaling law, suggesting that the effective Hamiltonian governing the dynamics is proportional to $h^2_x$. The inset shows the raw data obtained on the quantum annealer without rescaling. The Figure is reproduced from Ref.~\cite{Vodeb2025}.
}
    \label{fig:FV}
\end{figure*}

The phenomena of particle production and string breaking have a direct connection to the phenomenon of metastability, whereby the system gets trapped in a state---also called the false vacuum---which does not correspond to the global minimum; see Fig.~\ref{fig:FV}(a). This behavior is typically associated with first-order phase transitions, and occurs in a variety of physical and chemical systems, including possibly the Universe as a whole~\cite{Coleman1977,callan1977fate,kobsarev1974bubbles,guth1981inflationary,turner1982our,kibble1980some}. 

In the example of a $(1+1)$D U$(1)$ QLM in Sec.~\ref{sec:particle-production}, the false and the true vacuums correspond to states
$|\ldots \triangleleft, \varnothing, \triangleleft, \varnothing, \triangleleft, \varnothing, \triangleleft \ldots\rangle$ and 
$|\ldots \triangleright, \varnothing, \triangleright, \varnothing, \triangleright, \varnothing, \triangleright\ldots\rangle$, respectively. Their energy splitting can be controlled by the confinement parameter $\theta$, inducing a first-order phase transition. Since the false vacuum does not correspond to the true energy minimum, thermal or quantum fluctuations are expected to allow the system to tunnel to the true vacuum---a process known as false-vacuum decay. The dynamics of false-vacuum decay consist of `bubbles' of true vacuum forming in the background of false vacuum, e.g., $|\ldots \triangleleft, \varnothing, \triangleleft, e^-, \triangleright, \varnothing, \triangleright, \varnothing, \triangleright, e^+, \triangleleft, \varnothing, \ldots\rangle$. By analogy with the familiar example of supercooled water, the size of a bubble is determined by balancing the energy reduction proportional to the bubble volume and energy increase proportional to the bubble surface. Bubbles are typically assumed to undergo isolated quantum-tunneling events, but the process is difficult to study in general due to the nonperturbative nature of the tunneling dynamics. 

Aspects of the false-vacuum decay have recently been probed in tabletop experiments using various quantum-simulation platforms~\cite{Zenesini2024, Vodeb2025, zhu2024probingfalsevacuumdecay, luo2025quantum}.
Utilizing the mapping of a $(1+1)$D $\mathbb{Z}_2$ LGT onto the Ising model in Eq.~(\ref{eq:H-Ising}), Refs.~\cite{Vodeb2025,luo2025quantum} probed bubble formation in the presence of confinement; see also Ref.~\cite{zhu2024probingfalsevacuumdecay} for a related study in the $(1+1)$D U$(1)$ QLM. By tuning the parameter that drives a first-order phase transition, e.g., the longitudinal field $h$ in the Ising model, these simulations initialize the system in the metastable false vacuum state (i.e., an infinitely long string) and observe its decay into the true vacuum (i.e., the broken string). 
The case of false-vacuum decay and bubble formation, along with the associated scaling laws in a quantum-simulation experiment of the mixed-field Ising model using programmable trapped-ion quantum simulators~\cite{luo2025quantum} was presented in Sec.~\ref{sec:string-breaking-1D} in the context of string-breaking dynamics. Here, we discuss, in further detail, a large-scale quantum simulation of false vacuum decay using on a D-wave quantum annealer~\cite{Vodeb2025}. 

A quantum annealer is a purpose-built quantum processor  designed to solve optimization problems. The building blocks of a quantum annealer are the same superconducting qubits introduced in Sec.~\ref{sec:SCQ}, but the principle of operation is different from digital processors discussed so far. An optimization problem is mapped into a physical Hamiltonian of qubits in an annealer---typically an Ising-type spin model on an arbitrary graph. The solution corresponds to the system’s lowest energy state, which is typically reached by adiabatic evolution, i.e., by slowly modulating the Hamiltonian. However, beyond optimization problems, the same device can be used to directly perform analog quantum simulation of the physical Hamiltonian, such as the Ising model in Eq.~(\ref{eq:H-Ising}). This approach has the advantage of large system size: D-wave's annealer \texttt{Advantage system5.4} with 5564 qubits used in Ref.~\cite{Vodeb2025} allowed the observation of the formation of large true-vacuum bubbles comprising up to ${\sim}300$ flipped spins. 

Furthermore, the discrete nature of qubit systems gives a direct window into quantized bubble creation, in which a cascade of bubble sizes is seen to emerge by tuning the longitudinal field, as seen in Fig.~\ref{fig:FV}(b)-(c). Due to the limited coherence of the annealer, Ref.~\cite{Vodeb2025} focused on a semiclassical regime where the transverse field $g$ is much weaker than the longitudinal field $h$ and the Ising coupling $J$. In that regime, the bubble size ($n$) is determined by balancing the gain in surface energy, $4J$, and the reduction in volume energy, $2h n$. Hence, the formation of bubbles of size $n$ comes at no extra cost at the resonance $h=-2J/n$. Thus, if  $g$ is suddenly ramped from zero, the dynamics remain frozen unless $h$ takes one of the discrete values $-2J/n$, at which the magnetization $M$ can exhibit nontrivial modulation, as seen in Fig.~\ref{fig:FV}(b). Note that the dynamics seen around $h\approx -4$ in Fig.~\ref{fig:FV}(b) are not related to bubbles, but simply due to $|h|$ being very large and forcing the system to follow its modulation during the annealer protocol. Moreover,  Fig.~\ref{fig:FV}(c) confirms that the dominant bubbles at $h=-2J$ resonance are indeed single-spin flips ($n=1$), consistent with theoretical expectation. 

In a two-level approximation, tunneling events to different $n$-bubbles can be thought of as Landau-Zener transitions, where the metastable (false vacuum) state  and an $n$-bubble state at the appropriate resonance are the two states involved in the anticrossing~\cite{sinha2021nonadiabatic}. This model can be used to derive scaling laws for the bubble density and magnetization in the scenario probed in Fig.~\ref{fig:FV}(b), which are found to be in good agreement with the annnealer data, Fig.~\ref{fig:FV}(d) [see also Refs.~\cite{zhu2024probingfalsevacuumdecay,luo2025quantum} for similar scaling collapses in other false-vacuum decay realizations]. These scaling laws may be viewed as a generalization of the Kibble--Zurek effect~\cite{zurek1985cosmological,kibble1980some,qiu2020observation,surace2024string}, which describes nucleation of domains of an ordered phase as the system is driven through a symmetry-breaking phase transition at a finite rate. While the conventional Kibble--Zurek picture applies to continuous phase transitions, its applicability in the context of first-order transitions associated with false-vacuum decay requires further investigation. While these $(1+1)$D systems are still tractable classically and even analytically~\cite{Rutkevich1999,surace2024string}, larger-scale quantum simulations, and those of $(2+1)$D and higher dimensional theories on complex lattice topologies, may reveal other rich phenomenology of the bubble dynamics.
\begin{figure*}[t!]
    \includegraphics[width=1\textwidth]{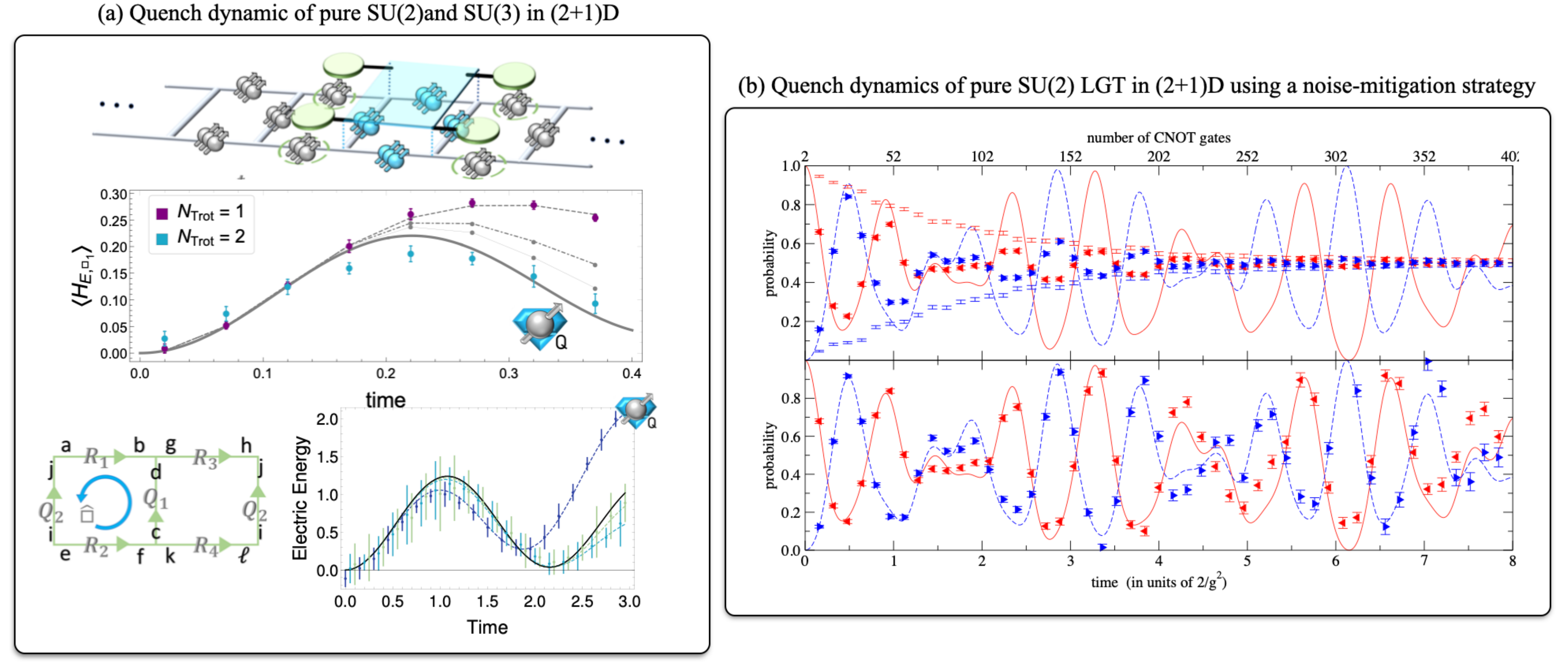}
    \caption{(a) The top panel shows the lattice distribution of the qubit registers associated with each gauge link in a $1$d array of plaquettes. The action of the plaquette operator is denoted by the light-blue square, and is controlled by the four neighboring qubit registers (denoted by the light-green circles) to enforce the Gauss’s laws. Figure is reproduced from Ref.~\cite{klco20202}. The middle panel plots the expectation value of the electric-energy contribution of the first plaquette in the two-plaquette lattice with PBCs in a truncated pure SU$(2)$ LGT computed on IBM’s \texttt{Tokyo} quantum processor. The dashed, dot-dashed, and thin gray lines are the $N_\text{Trot} = 1, 2, 3$ values, while the thick gray line corresponds to the exact time evolution. Figure is reproduced from Ref.~\cite{klco20202}. The lower panel plots the electric energy of the two-plaquette system in a truncated pure SU$(3)$ LGT in the irrep basis using IBM's \texttt{Athens} quantum processor, reproduced from Ref.~\cite{ciavarella2021trailhead}. Evolution in both plots is under a first-order Trotterization of the Hamiltonian in Eq.~\eqref{eq:H-E-B-dp1} (with the fields appropriately adopted for SU$(2)$ and SU$(3)$ theories). (b) Time evolution by on a two-plaquette lattice in a truncated pure SU$(2)$ LGT. In both panels, the red solid (blue dashed) curve is the exact probability of the left (right) plaquette being measured to have $j =1/2$. The red left-pointing (blue right-pointing) triangles in the upper panel are the data computed using IBM's \texttt{lagos} quantum processor. The red (blue) error bars without symbols are the mitigation data computed on the same processor from the same circuit but with half the steps forward in time and half backward in time. The triangles in the lower panel are the results obtained by deducing the noise from the forward-backward circuit and using it to normalize the all-forward time-evolution circuit, as explained in Ref.~\cite{a2022self}. Figure is reproduced from Ref.~\cite{a2022self}.
    }\label{fig:nonabelian}
\end{figure*}

For example, beyond the initial bubble formation,  Ref.~\cite{Vodeb2025} found that subsequent dynamics of 1d bubbles follow an intricate scenario where large bubbles cannot spread in isolation, but only through an interaction whereby one bubble enlarges itself by reducing the size of the other. Once reduced to the smallest size of one lattice site, the bubble can then move freely along the system. These results imply that false-vacuum dynamics is a complex many-body process: it can be viewed as a heterogeneous gas of bubbles, where the smallest `light' bubbles bounce around in the background of larger `heavy' bubbles that directly interact with each other. Direct observation of bubble interactions and its impact on far-from-equilibrium dynamics of the false vacuum remains another interesting direction for future quantum simulations. Such simulations likely need higher coherence levels and the ability to probe higher-order correlation functions and quantum entanglement. 

%%%%%%%%%%%%
%%%%%%%%%%%%
\subsection{Toward quantum simulating non-Abelian gauge-theory dynamics
\label{sec:non-Abelian}}

Simulating non-Abelian LGTs necessarily involves encoding more degrees of freedom, and imposition of more Gauss's laws, as described in Sec.~\ref{eq:SU(Nc)} for the case of pure SU$(N_c)$ LGTs within the Kogut--Susskind formulation. Engineering gauge-invariant interactions in analog quantum simulators, therefore, presents more challenges. While theoretical proposals have been put forward~\cite{zohar2013cold,zohar2013quantum,tagliacozzo2013simulation,stannigel2014constrained,mezzacapo2015non,kasper2020jaynes,luo2020framework,kasper2023non,bender2018digital,ott2021scalable}, e.g., to use built-in gauge protection is ultracold polar molecules in optical tweezer arrays and alkaline-Earth-like atoms in optical superlattices \cite{zohar2013quantum,Halimeh2024spinexchange,depaciani2025quantumsimulationfermionicnonabelian}, they are yet to be experimentally demonstrated. Digital quantum simulation constitutes a more feasible path toward non-Abelian LGT studies at present times.

In fact, several works in recent years have demonstrated small-scale digital quantum simulation and SU$(2)$ and SU$(3)$ LGTs on quantum hardware~\cite{klco20202,ciavarella2021trailhead,a2022self,atas20212,atas2023simulating,ciavarella2022preparation,farrell2023preparationsI,farrell2023preparationsII,than2024phase,chernyshev2025pathfinding,ciavarella2024quantum,ciavarella2025string,turro2024classical}. These studies are often limited to system sizes where non-Abelian constraints can be maximally solved such that the evolution is mapped to a much smaller subspace, or impose severe truncations on the gauge degrees of freedom such that the qubit requirement and interaction complexity are considerably reduced. We highlight in Fig.~\ref{fig:nonabelian} a selection of such results using IBM's quantum processors for Trotterized evolution after a quantum quench. These results include short-time dynamics of electric-field excitations in a two-plaquette system in pure SU$(2)$ and SU$(3)$ LGTs obtained in Refs.~\cite{klco20202,ciavarella2021trailhead} [Fig.~\ref{fig:nonabelian}(a)], and long-time evolution of a system of two plaquettes upon employing an effective noise-mitigation technique obtained in Ref.~\cite{a2022self} [Fig.~\ref{fig:nonabelian}(b)]. 

Upon scaling such simulations to larger system sizes, one can start to study the nontrivial interplay between chromo-electric and chromo-magnetic interactions in nonequilbrium conditions, relevant for high-energy particle colliders. Furthermore, upon introducing matter fields, particle-production and string-breaking mechanisms can be explored in such models; see, e.g., Ref.~\cite{ciavarella2025string} for one recent example in a $(1+1)$D SU$(2)$ LGT.

%%%%%%%%%%%%
%%%%%%%%%%%%
\subsection{Open questions}

Recent progress in quantum-simulation experiments of string dynamics and breaking, including the phenomenon of bubble formation in related quantum phase transitions, are commendable achievements. Nonetheless, several directions remain to be explored.

On the one hand, it would be interesting to discern genuinely $(2+1)$D effects in string-breaking dynamics. Even when the string is produced on a $2$d lattice, its dynamics in the confined phase can be effectively those in $(1+1)$D when the plaquette term is absent in the dynamics, as is the case in Ref.~\cite{gonzalez2024observation}. On the other hand, it would be interesting to initialize strings in far-from-equilibrium configurations that can give rise to other rich dynamics including glueball formation~\cite{xu2025stringbreakingdynamicsglueball}. From a collider-physics standpoint, the more relevant processes are high-energy collisions of confined excitations, which lead to stretching and breaking of strings present in the initial states, and the formation of new particles in the aftermath of the collision. Aspects of such physics will be reviewed in the next Section. The phenomenon of metastability, bubble formation, and many-bubble dynamics will likely also bear rich phenomenology in higher dimensions that merit detailed studies in the coming years.

Importantly, since string breaking and hadron production has its origins in QCD, it would be pertinent to dive deeper into this phenomenon in non-Abelian LGTs, particularly in $(2+1)$D and ultimately $(3+1)$D, and investigate its feasibility on modern quantum hardware. Experiments observing non-Abelian string breaking would bear great relevance to collider physics and begin to shape these setups into truly complementary venues for studying QCD phenomena.

%%%%%%%%%%%%%%%%%%%%%%%%%%%%%%%%%%%%%%%%%%%%%%%%%%%%%%%%%%%%%%%%%%%%%%%%%%%%%%%%%%%%%%%%%%%%%%%%%%%%%%%%%%%%%%%%%%%%%%%%%%%%%%%%%%%%%%%%%%%%%%%%%%%%%%%%%%%%%%%
\section{Scattering processes in gauge theories
\label{sec:scattering}}
Scattering is a critical tool in many areas of physics. A vast amount of knowledge about nature over the past century is gained by employing particle colliders of ever-increasing power and sophistication. In a successful theory-experiment interplay, a complex theory of nature, the Standard Model of Particle Physics, was born and confirmed~\cite{weinberg2004making,oerter2006theory}. Particle colliders with even higher collision energies and luminosities are planned into the current century~\cite{fcc2019fcc1,fcc2019fcc2,fcc2019fcc3,anderle2021electron,aihara2019international,shiltsev2021modern,accardi2016electron,khalek2022science,khalek2022snowmass}, in quest for unlocking mysteries such as matter-antimatter asymmetry, quantum gravity, and the nature of dark matter. Analytical and computational techniques that enable describing the fate of particle collisions have also become increasingly sophisticated~\cite{andersson1983parton,buckley2011general,buckley2010systematic,anderson2013snowmass}. They, nonetheless, remain limited when it comes to end-to-end simulations of high-energy collisions, including incorporating complexities of initial and final states, and the various phases of matter formed shortly after the collisions. 

Colliding particles can either be elementary particles of the SM, such as electrons, muons, and neutrinos, or be composite ones such as protons, neutrons, and atomic nuclei. The final state, among other factors, depends on the total initial center-of-mass energy of colliding particles. Sometimes this energy is unknown, such as in neutrino-nucleus scattering in long-baseline experiments~\cite{adams2013long,abi2020volume}, and the goal is to reconstruct it using the final-state composition, which constitutes a highly complex problem~\cite{alvarez2018nustec,ruso2022theoretical}. Other times, the initial energy is known but the distribution of energy and momentum among elementary constituents of composite colliding particles may be uncertain, hence limiting the final-state predictions~\cite{ball2013parton,dulat2016new,bozzi2011impact,gavin2012w,gao2018structure}.

Even with known initial-state energy and structure, the final-state outcome may be complicated to predict and analyze. The reason is, detectors only probe the asymptotic particles flying out of the collision point, and register classical snapshots of their energy and charge deposits. Hence, access to events near the collision point is challenging and requires indirect probes. An example of such probes is the fate of heavy quarks, which can be traced to their movement in the hot and dense matter created shortly after the collision, i.e., quark-gluon plasma~\cite{mclerran1986physics,harris1996search,rischke2004quark,braun2007quest}. Nonetheless, how exactly the plasma cools down and turns into composite hadrons and other particles is not clear from a first-principles perspective. The celebrated string breaking and fragmentation discussed in Sec.~\ref{sec:string-breaking} contributes to this \emph{hadronization} process but in ways that has so far only been modeled based on simpler theories and simplifying assumptions~\cite{andersson1983parton,buckley2011general,buckley2010systematic}. Jets~\cite{sterman1977jets}, or sprays of energetic particles, originating from quark and gluon remnants in the post-collision stage, are also hard to analyze~\cite{almeida2009top,gallicchio2011quark,altheimer2012jet,larkoski2020jet}. Finally, the multitude of partonic showers, such as gluon emissions and recombinations~\cite{nagy2018parton,forshaw2020building,sterman1996partons,catani2002qcd,hoche2009qcd}, exhibit complex quantum correlations. These correlations are hard to track, hence semi-classical assumptions are built in the descriptions~\cite{sjostrand2006pythia,bahr2008herwig,gleisberg2009event}, which could lead to inaccuracies~\cite{nachman2021quantum}.

These complexities can only be tackled theoretically with full real-time simulations of the collision processes. Such simulations are believed to be realistic with far-term large-scale reliable quantum simulators/computers~\cite{bauer2023quantum,bauer2024quantum,beck2023quantum,di2024quantum}. In the meantime, near-term quantum simulators/computers can provide a powerful playground for designing, conducting, and analyzing end-to-end quantum simulations of scattering in simpler gauge theories or their finite-dimensional partners. Since accessing scattering observables in real-time simulations is a new frontier, efforts with simpler models are still highly nontrivial and valuable. Importantly, these studies merit pursuing on their own right, as they illuminate our understanding of post-scattering phenomenology, which can be fed into, and enhance, models used in analyzing experimental outcome of particle colliders.

In this Section, we review the rapidly evolving developments in studying scattering problems of relevance to nuclear and high-energy physics using particle simulators, covering both analog and digital approaches, and various models of interest. An important aspect of this program is preparing moving particle wave packets. Another frontier is strategies for learning properties of the complex final state and attempts to construct the scattering $S$ matrix. Tensor-network studies have also been advanced in parallel to illuminate the problem in $(1+1)$D. We will only mention such tensor-network studies (without elaboration) when relevant, prioritizing works that focus on quantum-simulation proposals and/or hardware implementation.

%%%%%%%%%%%%
%%%%%%%%%%%%
\subsection{Analog quantum simulation of scattering processes
\label{se:scatt-analog}}
Quantum simulation of nonequilibrium dynamics in analog quantum simulators has often been limited to quench processes, and to simple initial states. More complex state preparations and long evolution times before and after collision, as those needed for particle-scattering experiments, often require a level of control and coherence that is only available in digital quantum computers. Nonetheless, recent work has included proposals for analog quantum simulation of scattering in simpler, low-dimensional LGTs, that may become experimental reality in upcoming years. We review relevant developments in this Section, with a focus on concrete scattering-simulation proposals for specific hardware.

%%%
%%%
\subsubsection{Scattering in confining spin models with a gauge-theory dual}
A natural starting point for exploring rich phenomena produced in particle collisions is studying spin models with a gauge-theory dual, as discussed in Sec.~\ref{sec:Z2-Ising}. Certain Ising spin chains with long-range interactions~\cite{liu2019confined,vovrosh2022dynamical}, or with short-range interactions but subject to a longitudinal magnetic field~\cite{mccoy1978two, kormos2017real}, host composite (bound) particle excitations~\cite{vovrosh2021confinement,vovrosh2021simple,tan2021domain,vovrosh2025meson}. Such models can generally be mapped to a gauge theory in $(1+1)$D, as shown in Sec.~\ref{sec:Z2-Ising}, and further have a natural mapping to a variety of spin quantum simulators. They, therefore, offer a first playground for setting up a scattering problem in simulation experiments. We review in this Section an analog quantum-simulation proposal to study mesonic scattering in such models~\cite{bennewitz2024simulating}. Other studies include numerical exploration of scattering processes in Ising and other spin models~\cite{milsted2022collisions,jha2025real,karpov2022spatiotemporal,vovrosh2022dynamical}.

Consider the Hamiltonian in Eq.~\eqref{eq:H-Ising}, with the Ising spin-spin coupling taking either an exponentially decaying form, $J_{i,j}=J_0e^{-\beta|i-j|}$, or a power-law decaying form, $J_{i,j}=J_0|i-j|^{-\alpha}$, with $J_0>0$, $\beta>0$, and $\alpha>1$. The bare excitations, i.e., eigenstates of $\hat{H}_{\text{Ising}}$ in Eq.~\eqref{eq:H-Ising} when $g=0$, are associated with spin flips. A single spin flip at position $i$, from a fully spin-up bare vacuum state, corresponds to two adjacent spin domain walls, or kinks. When this excitation forms a bound state, it will be denoted as $\ket{1,i}$, and is called a $1$-meson state in the following. The dressed eigenstates are those with a nonvanishing $g$. Two-kink excitations are bound if the interactions decay sufficiently slowly. For short-range interactions, both confined and deconfined excitations are possible in the low-energy spectrum. It would, therefore, be interesting to investigate the scattering outcome in both the long- and short-range scenarios. 

A scattering protocol in an analog spin quantum simulator goes as follows. First, a bare 1-length meson, i.e., a bound two-kink state is created on the left side of the spin chain. Then, applying one of the two schemes described below for time $\tau_\text{prep}$ generates a Gaussian meson wave packet of the form:
\begin{align} 
    \ket{\Psi_g(x_0, k_0)} 
    & = \sum_{i=1}^N \psi^g_i(x_0, k_0 ) \ket{1,i},
\end{align}
where
\begin{align}
    \psi^g_i(x_0, k_0 ) & = \frac{1}{\mathcal{N}} e^{-(x_i-x_0)^2/(2 \Delta_x^2) + i k_0 x_i }.
\end{align}
Here, $N$ is the number of spins in the chain, the $1$-meson wave packet is centered at $x_0$ and $k_0$, $\Delta_x$ is the width of the (position-space) wave packet, and $\mathcal{N}$ is a normalization factor. The bare wave packet does not move, as excitations are dispersionless when $g=0$. Such a stationary property ensures that the wave packet does not spread out as it is  prepared. A similar procedure is applied to the right side of the chain, either simultaneously or subsequently. Once the two bare meson wave packets are created, the transverse field proportional to $g$ is turned up adiabatically. This process produces dressed mesonic wave packets. It further encodes a nonzero velocity in the wave packets, since excitations in the interacting theory have a nontrivial dispersion. Once the wave packets scatter, the transverse field is adiabatically turned off via a second ramp, so that excitations become undressed, making their measurements and interpretation straightforward in terms of the bare excitations.

The wave-packet state preparation can be achieved in spin quantum simulators in at least two ways. First, when a spin blockade mechanism is possible, such as in Rydberg arrays or in trapped-ion systems with long-range interactions, a Gaussian wave packet is prepared by driving each spin with a site-dependent transverse field, that is, by adding to the system's bare Hamiltonian the term $-\sum_i g_i \cos( \omega_i t + \phi_i) \hat{\sigma}_i^x$. The driving frequency is $\omega_i = E_i - E_0$, where $E_i$ is the energy of the eigenstate $\ket{1, i}$ of the bare Hamiltonian and $E_0$ is the energy of the all-spin-up bare vacuum state. As shown in Ref.~\cite{bennewitz2024simulating}, the Gaussian wave packet is produced resonantly upon setting the transverse-field amplitude to $g_i = \Gamma  J_0 \psi^g_i(x_0, 0)$ with a tunable parameter $\Gamma$, and the site-dependent phase shift to $\phi_i = - k_0 x_i$. If driven slowly enough, the state prepared at $ J_0 \tau_\text{prep} = \pi/\Gamma$ is the desired wave packet.

An alternative scheme to prepare the wave packet is to use a quantum bus, such as phonons in trapped-ion system~\cite{wineland1998experimental,porras2004effective} or photons in cavities~\cite{koch2010time,blais2021circuit}. Consider a time-dependent Hamiltonian $\hat{H}_{\text{JC}}(t) = \sum_{i,k} \big( A_{ik}  e^{i \delta_k t} \hat \sigma^-_i \hat a_k + \text{h.c.} \big)$ describing the coupling between a (bosonic) quantum bus and spins~\cite{jaynes1963comparison}. Here, $\hat{\sigma}^+$ ($\hat{\sigma}^-$) are the spin raising (lowering) operator at site $i$ in the $z$ basis, $\hat a_k^\dagger $ ($\hat a_k$) is the boson creation (annihilation) operator for mode $k$, $A_{ik}$ are the site- and mode-dependent amplitudes, and $\delta_k = \omega_k - \nu$, with boson mode frequency $\omega_k$ and drive frequency $\nu$. The idea of this  scheme is to first start from a state initialized with all spins up and no occupied bosonic modes, then create an excitation in the target boson mode and, finally, use the evolution under the Hamiltonian above to transfer the bosonic excitation to the chain in such a way that the spin excitation is distributed according to a wave-packet profile. This is achieved by setting the amplitude $A_{ik} = \Omega_{0} B_{ik} \psi^g_i(x_0, k_0)$ for target boson mode $k$, such that $B_{i k_t} = 1$ with a tunable Rabi frequency $\Omega_0$. By driving the system at the target-mode frequency and evolving for $\tau_\text{prep} = \pi/(2 \Omega_0)$, the all-spin-up state with one quantum in the target boson mode turns into the desired wave packet state while annihilating the boson.

To benchmark the quality of the wave-packet preparation, and to investigate parameter ranges that lead to interesting scattering outcome, Ref.~\cite{bennewitz2024simulating} presents a numerical study based on exact diagonalization of evolution in a chain of 24 spins. For scattering in the power-law model with $\alpha=1.5$ and a range of $g$ values, only elastic scattering is detected. For the exponentially decaying model with $\beta=1$, both elastic and inelastic scattering occurs as $g$ increases. The region of flipped magnetization between the outgoing particles signals an inelastic scattering channel composed of a pair of unbound kinks. For values of $g>1$, up to $25\%$ probability is observed for such an unbound-kinks scattering channel. This sizable probability, as well as the distinct signature of the associated measurement, means that inelastic processes may be be detectable in near-term spin quantum simulators.

The protocols proposed in Ref.~\cite{bennewitz2024simulating}, while technically feasible for current hardware, require evolution times that are up to an order of magnitude longer than the coherence limit of current e.g., trapped-ion analog quantum simulators~\cite{feng2023continuous,joshi2022observing}. Future research, therefore, may need to improve upon these schemes to enable the first accurate hadron-simulation scatterings in analog quantum simulators.

%%%
%%%
\subsubsection{Scattering in a quantum link model} 
As mentioned in Sec.~\ref{sec:U1LGT_formulations}, QLMs are anticipated to recover the continuum limit of the corresponding gauge theory in the limit of an infinite-link Hilbert space~\cite{buyens2017finite, zache2022toward,halimeh2022achieving}. Their finite-dimensional formulations are ideal for realization of gauge-theory-like dynamics in quantum simulators, as these models respect local symmetries and can exhibit confining dynamics. A first analog-simulation scattering experiment is proposed for a QLM in Ref.~\cite{su2024cold}, suitable for a cold-atom platform. We review basic ingredients of this proposal, along with the scattering phenomenology expected to arise from an Abelian QLM, with and without a CP-violating term.

\begin{figure*}[t!]
\centering
\includegraphics[width=0.995\linewidth]{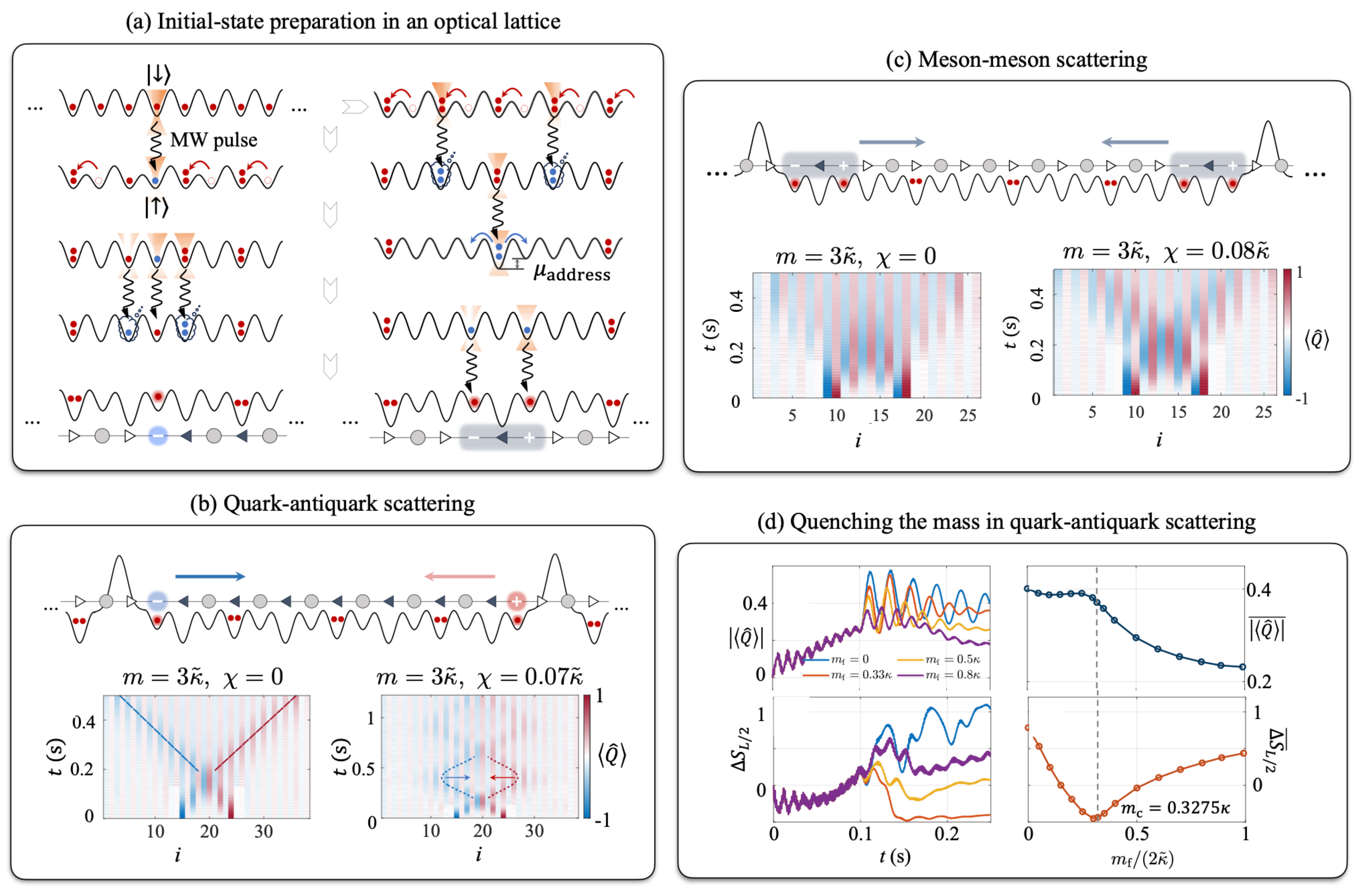}
\caption{(a) The initial-state preparation for a quark and a quark-antiquark pair starting from uniformly occupied single or alternately occupied double-boson states in a uniform and nonuniform lattice potential, respectively. Red (blue) dots denote atoms in hyperfine state $\ket{\downarrow}=\ket{F=1,m_F=-1}$ ($\ket{\uparrow}=\ket{F=2,m_F=-2}$). Addressing tweezer beams induce local AC-stark shift on the $\ket{\uparrow}$ state, such that the addressed atoms can be individually transferred between $\ket{\downarrow}$ and $\ket{\uparrow}$ states with a microwave (MW) pulse. Successive operations of the pulses as shown prepares the desired initial states. 
(b) A schematic of particle-antiparticle collision in a Bose--Hubbard optical lattice quantum simulators, Eq.~\eqref{eq:bhm}, is shown in the top panel. A single (anti)particle excitation in the gauge theory is represented by a single boson, or red dot, on the (even) odd matter site. Gray dots are unoccupied matter sites. The corresponding electric fluxes are denoted by arrows. The bottom panels show expectation value of charge density on matter sites $\langle \hat{Q} \rangle$ illustrating the collision of moving particle (blue) and antiparticle (red) wave packets in the deconfined phase ($\chi=0$) and confined phase ($\chi=0.035$).
The particle and antiparticle undergo an elastic collision in the confined phase while in the confined phase, due to the creation of a string potential, the particle and antiparticles undergo acceleration and deceleration, leading the multiple collisions. (c) The same quantities as in the previous part but for the meson-meson collision. The initial particle-antiparticle pairs are delocalized after the collision in the deconfined phase while they remain rather stable when there is confinement. 
(d) The left panels show the magnitude of the average charge density around the center ($i \in [16,23]$) as well as the half-chain van-Neumann entropy difference relative to the equivalent case of a matter-free vacuum initial state, as a function of time, when a mass quench at $t=0.1$s is introduced from $m=0$ to $m=m_f$. The right panels are the time-averaged quantities as a function $m_f/(2\tilde\kappa)$. These quantities signal Coleman's critical point $m_c$, at which point the system undergoes a $\mathbb{Z}_2$ symmetry-braking phase transition. Above $m_\text{c}$, pair production is exponentially suppressed.
Figure is reproduced from Ref.~\cite{su2024cold}.
}
\label{fig:qlm-scattering}
\end{figure*}

The model Hamiltonian is that in Eq.~\eqref{eq:U1QLM} with $a=1$ and $S=\frac{1}{2}$, and both deconfined ($\chi = 0$) and confined ($\chi \neq 0$) phases are studied. The single (anti)particle, and a particle-antiparticle pair, are prepared in experiment by a series of laser  pulses on an easily prepared product state of bosons, as described in Fig.~\ref{fig:qlm-scattering}(a). To prepare moving wave packets, Ref.~\cite{su2024cold} proposes applying a potential barrier via single-site addressing~\cite{weitenberg2011single,islam2015measuring, zhang2023observation}. A barrier placed left (right) to the original wave packet reflects the left-(right-)moving momentum components of localized particles to the right (left), thus shifting the center of the momentum distribution of each particle to a finite value. Once the wave packet has moved away, the barrier is removed. While this method does not create wave packets with sharp momentum values, it can be used as a first approximation to moving wave packets and ease the state-preparation step. For a meson state, the barrier forbids the antiparticle from hopping to the right, meanwhile the particle hops to the left, and the antiparticle follows afterward. In the presence of a confining potential, it is easy to show that it is energetically favorable for the particle-antiparticle pair to move together as a composite particle. 
  
To illuminate scattering dynamics, Ref.~\cite{su2024cold} provides a numerical simulation of the outcome of quark-antiquark [Fig.~\ref{fig:qlm-scattering}(b)] and meson-meson [Fig.~\ref{fig:qlm-scattering}(c)] scatterings using experimentally feasible parameters. Spontaneous pair creation and annihilation are suppressed in the limit $m \gtrsim 2\tilde\kappa$.\footnote{$\kappa$ in this reference is related to $\tilde{\kappa}$ in the Hamiltonian in Eq.~\eqref{eq:U1QLM} via $\kappa \to 2\tilde\kappa$. We adhere to the notation introduced in this Review for consistency.} Thus, the particle-antiparticle scattering is elastic in this limit, i.e., particle and antiparticle cannot annihilate each other out, or will not combine with newly created pairs out of vacuum, as verified in the bottom-left panel of Fig.~\ref{fig:qlm-scattering}(b). In the confined case, the particle and antiparticle exchange momentum, and recoil away from each other after the collision. As they move apart, the string energy increases with the inter-particle distance. This leads to the conversion of the kinetic energy to potential energy, which decelerate the particle and antiparticle. The particle and antiparticle reach zero velocity, at which point they start accelerating toward each other, causing the second collision. This cycle of deceleration, acceleration, and collisions continues throughout the evolution, as depicted in the bottom-right panel Fig.~\ref{fig:qlm-scattering}(b). Such oscillatory string dynamics form a particle-antiparticle bound state, i.e., a meson.

For two particle-antiparticle pairs, scattering dynamics in the deconfined case is trivial. The elementary particles and antiparticles are not bound into a meson, and the original wave packets delocalize after the collision; see the bottom-left panel of  Fig.~\ref{fig:qlm-scattering}(c), producing significant entropy. In the confined regime, the particle and antiparticle wave packets remain localized after the collision. Their relative position remains unchanged since the particle and antiparticle can not tunnel through each other in the large-mass limit. The original mesons, therefore, are rather stable under the collision; see the bottom-right panel of  Fig.~\ref{fig:qlm-scattering}(c). The dominance of an elastic scattering channel is a consequence of the band structure of the mesons. The energetics of the initial state forbids the creation of mesons in the higher energy bands, as analyzed in Ref.~\cite{su2024cold}.

In order to create more inelasticity in particle-antiparticle scattering, Ref.~\cite{su2024cold} proposes a mass quench at the collision point, from $m=0$ to $m=m_f$. Around the Coleman critical point, the vacuum background quickly thermalizes, while the wave packets exhibit slower thermalization. Figure~\ref{fig:qlm-scattering}(d) plots in the top right the magnitude of the average charge density near the center where particles collide. The colliding particle-antiparticle pair has a lower entanglement entropy than the vacuum background, as is indicated in the bottom-left plot displaying the difference of half-chain entanglement entropy $\Delta S_{L/2}$ between the time-evolved state and the vacuum.\footnote{The half-chain entanglement entropy is defined as $S_{L/2} \coloneq -\tr \hat \rho_{L/2} \ln \hat \rho_{L/2}$, with $\hat \rho_{L/2}$ being the reduced density matrix for one half of the chain. We will discuss in more detail the reduced density matrix and the associated Schmidt decomposition in Sec.~\ref{sec:thermalization}.} This results shows that the presence of the wave packets delay the onset of thermalization after the quench. Furthermore, different $m_\text{f}$ values exhibit distinct dynamics. The late-time average of quantities with respect to the final mass is shown in the right panels of Fig.~\ref{fig:qlm-scattering}(d). Quantities dip at the critical point $m_\text{c}$, providing a signature of criticality in the dynamics.
  
%%%
%%%
\subsubsection{Scattering in the lattice Schwinger model
\label{sec:scat-in-schwinger-model}}
Scattering in the U$(1)$ LGT can also be studied in different analog quantum simulation platforms. For example, a recent proposal in Ref.~\cite{belyansky2024high} uses a bosonized formulation of the Schwinger model, introduced in Sec.~\ref{sec:U1LGT_formulations}, to map it to a circuit-QED platform~\cite{frisk2019ultrastrong,forn2019ultrastrong}. As discussed in Sec.~\ref{sec:SCQ}, these platforms host native bosonic degrees of freedom and offer a high degree of control, hence are ideal simulators of bosonic dynamics. In the following, we review the proposal of Ref.~\cite{belyansky2024high}, along with the scattering phenomenology in this model obtained with tensor-network methods. Scattering in the Schwinger model using tensor networks has further been studied in Refs.~\cite{rigobello2021entanglement,papaefstathiou2024real}.
\begin{figure*}[t!]
\centering
\includegraphics[width=0.995\linewidth]{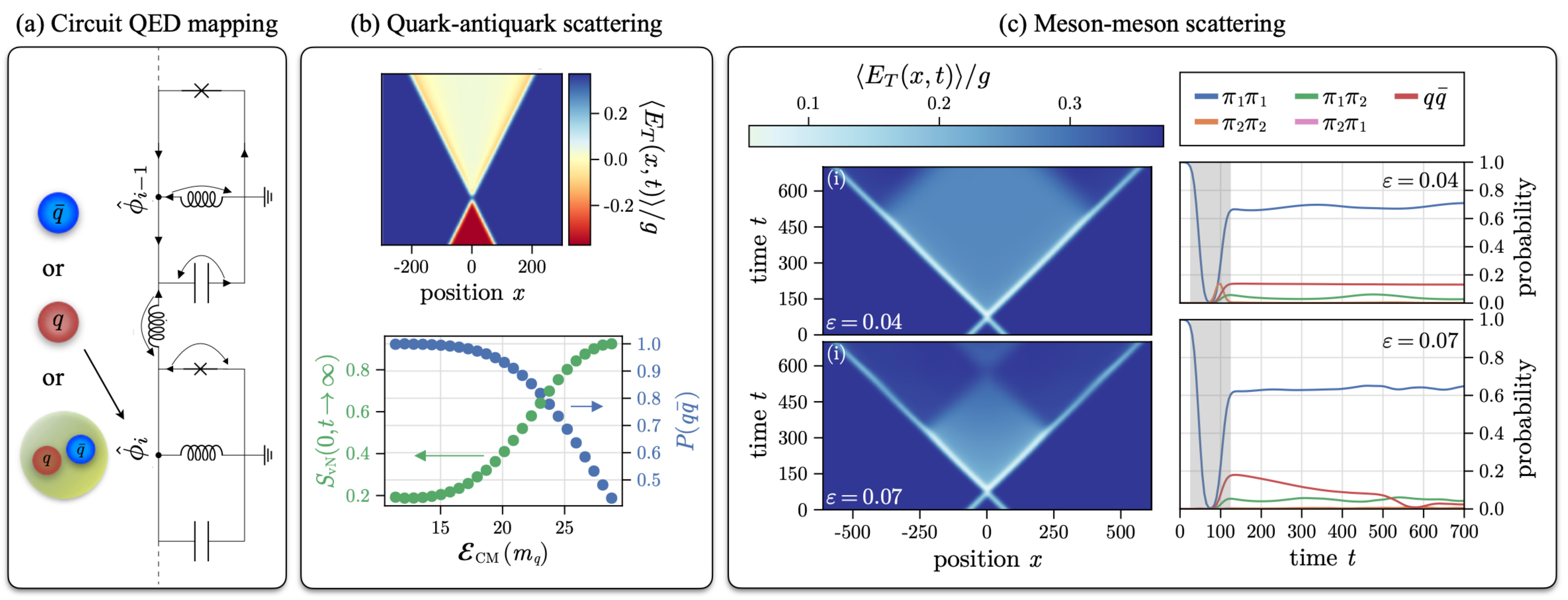}
\caption{
(a) The degrees of freedom in the lattice Schwinger model (quarks or gluons in the deconfined phase and mesons in the confined phase) can be mapped to a bosonic scalar field on a lattice as in Eq.~\eqref{eq:Schwinger-bose-lattice}, and be encoded as fluxonium degrees of freedom in a circuit-QED platform. (b) Spatiotemporal distribution of the electric field in quark-antiquark scattering in the deconfined phase of the lattice Schwinger model is shown in the top panel. The wave packets are centered at $p_0=\pm0.6$, corresponding to $\mathcal{E}_{\text{CM}}/m_q= 28.8$. The bottom panel displays elastic-scattering probability (right, blue) and asymptotic von Neumann entanglement entropy for the $x=0$ cut (left, green) as a function of the center-of-mass energy for the quark-antiquark scattering in the deconfined regime. (c) Spatiotemporal distributions of the electric field in meson-meson scattering in the confined phase with $\varepsilon=\pi-\theta=0.04$ and $0.07$ are shown in the left panels. The wave packets are centered at $p_0=\pm0.6$, corresponding to $\mathcal{E}_{\text{CM}}/{m_{\pi_{1}}}= 5.95$. The right panel displays probabilities of two-particle states $f,f'$ ($f,f' \in [\pi_{1},\pi_{2},q,\bar{q}]$) for the same meson-meson scattering as in the left. The state cannot be fully captured by a basis of asymptotic particles near the initial collision (shaded region), and the secondary collision at $t\approx550$ for $\varepsilon=0.07$. The remaining parameters are set to $\mu^2=0.1$ and $\lambda=0.5$. Figure is reproduced from Ref.~\cite{belyansky2024high}.
}
\label{fig:schwinger-scattering}
\end{figure*}

The bosonized form of the lattice Schwinger model on a spatial lattice, described by the Hamiltonian in Eq.~\eqref{eq:Schwinger-bose-lattice}, can be realized in a simple superconducting circuit consisting of a chain of inductively coupled fluxoniums, as described in Eq.~\eqref{eq:sc-hamiltonian-full} and Fig.~\ref{fig:platforms}(d); see also Fig.~\ref{fig:schwinger-scattering}(a). Concretely, the following relations ensure an exact mapping of the massive Schwinger model to the circuit-QED Hamiltonain:
\begin{align}
\frac{E_{L'}\beta^4}{8E_C}=1,~\chi = \frac{8E_C}{\beta^2},~\mu^2=\frac{E_L\beta^4}{8E_C}, \lambda = \frac{E_J\beta^2}{8E_C},
\end{align}
along with $\theta=\Phi_{\mathrm{ext}}-\pi$. A complete protocol for preparation and evolution of mesonic wave packets for a scattering experiment is described in Ref.~\cite{belyansky2024high}, including a method for preparing initial wave packets of bosonic particles using two ancillary qubits. Explicitly, the system is cooled down to its ground state in the confined phase. Two ancillary qubits~\cite{jordan2012quantum}, far away from each other, are then added to the system, and are initialized in the excited state. Subsequently, they are coupled to the original circuit, resulting in the decay of each qubit's excitation into the system, producing two wave packets of quasiparticles. Exciting multi-particle states can be suppressed in the weak-coupling limit. Note that since the quarks are topological excitations in the model, they do not couple to local operators, and need to be prepared by other means. Local density~\cite{zhang2023superconducting} or the output field at the edges \cite{forn2017ultrastrong,vrajitoarea2022ultrastrong} are among quantities that can be measured in a circuit-QED platform using standard techniques, allowing partial characterization of the final state of scattering.

To gain insight into the scattering phenomenology in the deconfined ($\theta=\pi$) and confined ($\theta \neq \pi$) regimes of the Schwinger model, Ref.~\cite{belyansky2024high} conducts a numerical study based on uniform matrix-product-state (MPS) ground- and quasiparticle-state ansatzes~\cite{zauner2018variational,haegeman2012variational,haegeman2013elementary}, at fixed values of model parameters. The spatiotemporal distribution of the electric field for a collisions with a large center-of-mass (CM) energy $\mathcal{E}_{\text{CM}}/m_q= 28.8$ is plotted in the top panel of Fig.~\ref{fig:schwinger-scattering}(b). 
Initially, the quark and antiquark are separated, and the electric field between them is equal in magnitude but opposite in sign to the field outside, representing the two degenerate ground states. After the collision, an increase of the post-collision electric field is observed, signaling additional charge production. The lowest-order inelastic channel is the four-quark production ($q\bar{q}\rightarrow q\bar{q}q\bar{q}$), which corresponds to quark fragmentation. The two inner particles screen the electric field produced by the outer two. An inelastic processes can also be characterized by the generation of significant von Neumann entanglement entropy, $S_\mathrm{vN}(x,t)=-\tr(\rho_{>x}(t) \ln \rho_{>x}(t))$ (with $\rho_{>x}(t)$ being the reduced density matrix for sites $y > x$), across the collision point $x=0$. As is shown in the bottom panel of Fig.~\ref{fig:schwinger-scattering}(b), the asymptotic ($t \to \infty$) entanglement increases as a function of the collision energy. Furthermore, by projecting the post-collision state onto a basis of asymptotic two-particle states, the momentum-resolved elastic scattering $S$-matrix can be obtained. The elastic scattering probability $P(q\bar{q})$ is displayed in blue in the same plot as a function of the collision energy. It decreases monotonically with energy, falling below $0.5$ around $\mathcal{E}_{\text{CM}}/m_q\gtrsim 28$.

To study mesons-meson scattering, Ref.~\cite{belyansky2024high} considers the weakly confined regime, i.e., $\theta = \pi-\varepsilon$ with $\varepsilon\ll 1$. At higher energies, the scattering phenomenology turned out to be richer than the deconfined case. There are multiple stable scalar meson excitations, which are labeled by $\pi_{j}\, (j=1,2,...)$, with increasing masses $m_{\pi_{j}}$. For $\pi_{1}\pi_{1}$ collisions, the meson wave packets are prepared as before, centered at $p_0=\pm0.6$ with $\mathcal{E}_{\text{CM}}/{m_{\pi_{1}}}=5.95$ for $\varepsilon=0.07$. The spatiotemporal distributions of the electric-field are displayed in the left panel of Fig.~\ref{fig:schwinger-scattering}(c). Before the collision, the background electric field is only locally disturbed by the charge-neutral mesons. After the collision, the mesons partially fragment into a quark-antiquark pair. The quarks are joined by an electric-field string. This field screens the background electric field (light-blue regions) inside the collision cone. As the quarks travel away from each other, their kinetic energy converts to the potential energy of the string. Eventually, they turn and propagate toward each other, causing a second collision. The right panel of Fig.~\ref{fig:schwinger-scattering}(c) displays the probabilities of the lightest two-meson state, $\ket{\pi_1,\pi_2}$, and the quark-antiquark state $\ket{q\bar{q}}$. The dominant scattering channel is the flavor-conserving elastic scattering, $\pi_{1}\pi_{1}\rightarrow \pi_{1}\pi_{1}$, and only a smaller probability is observed for exciting one of the outgoing mesons to $\pi_{2}$. There is also a substantial $q\bar{q}$ component, which decreases in time, indicating the occurrence of string breaking.

Future experimental realization of this proposal in strongly coupled SC circuits can unravel rich scattering phenomenology in regimes of even higher energy and entanglement generations, which may be beyond the reach of such classical computations.

%%%%%%%%%%%%
%%%%%%%%%%%%
\subsection{Digital quantum computation of scattering processes}
%%%%%%%%%%%%%%%%%%%%%%%%%%%%%
%%%%%%%%%%%%%%%%%%%%%%%%%%%%%
The scattering proposals described so far correspond to mapping the physical system to an analog quantum simulator. Unfortunately, analog quantum simulators have limited ability to simulate arbitrary complex dynamics accurately. In particular, no analog quantum simulator of practical use presently exists for non-Abelian gauge theories or for three-dimensional models. Digital quantum computers, therefore, are the primary choice when it comes to simulating gauge field theories of relevance to Nature, assuming large-scale and error-corrected machines will become available. Here, we review the leading algorithms and implementations of the scattering problem in quantum field theories in general, and in gauge-field theories, in particular.

%%%
%%%
\subsubsection{Scattering in scalar field theories} 
In a set of pioneering works~\cite{jordan2011quantum,jordan2012quantum}, an algorithm was put forward for simulating scattering in a quantum field theory consisting of interacting real scalar fields, with an interaction of the form $\lambda \phi^4$. This algorithm, which nowadays is known as the Jordan-Lee-Preskill (JLP) algorithm after the authors, is polynomial in the number of particles, their energy, and the desired precision. Here, we briefly review the basic features of the algorithm, and subsequent extensions and improvements, before focusing our attention on examples of gauge-theory scattering on quantum hardware.

The JLP algorithm assigns, at each lattice site, a register of $\mathcal{O}(\phi_\text{max}/\delta_\phi)$ qubits. This register stores values of a digitized scalar field $\phi$ whose magnitude is at most $\phi_\text{max}$ and whose resolution is $\delta_\phi$. The lattice holds $V/a^d$ lattice points, where $V$ is the lattice volume, $a$ is the lattice spacing, and $d$ is spatial dimensionality. As shown in Refs.~\cite{jordan2011quantum,jordan2012quantum}, a quantum state at energy scale $E$ can be represented with fidelity $1-\epsilon$ if each register consists of $O\left( \log \left( E \mathcal{V}/\epsilon \right)\right)$ qubits. With this, the quantum algorithm for simulating scattering of two particle wave packets proceeds as follows: i) prepare the ground state of the free theory using appropriate algorithms [such as that by Kitaev-Webb~\cite{kitaev2008wavefunction}], ii) create wave packets of the free theory, iii) create wave packets of the interacting theory by evolving the wave packets of the free theory under a Hamiltonian whose interaction term is turned on adiabatically during time $\tau$, iv) evolve the wave packets of the interacting theory for time $t$ during which scattering occurs, v) perform measurements of the outcome: either evolve the final state to its noninteracting counterpart by adiabatically turning off the interactions, and measure the number operators of the momentum modes of the free theory, or as in particle detectors, choose small regions and measure the total energy operator in each region via a quantum phase-estimation algorithm~\cite{nielsen2010quantum}.
  
How does the gate count scale in this algorithm as a function of system size, system parameters, and the accuracy? Theoretical inaccuracies arise from the use of truncated and digitized fields, and a discretized finite-size lattice. Algorithmic inaccuracies stem from approximate digitized time evolution or deviation from nonadiabaticity during state preparation. An attempt is made in Refs.~\cite{jordan2011quantum,jordan2012quantum} to quantify these inaccuracies, and eventually deduce the algorithmic cost. Step i) involves the use of Kitaev-Webb algorithm that constructs multivariate Gaussian superpositions. It can be performed with a classical processing step that goes as $\widetilde{O}(V^{2.376})$ (where the $\widetilde{O}$ notation hides additional logarithmic factors). Step ii) involves turning a nonunitary particle-creation operator into a unitary one using an ancilla qubit, and requires using a Trotterized approximation of the unitary. In order to create $n_i$ incoming wave packets, they all need to be apart by distance $r$, such that the error $\epsilon_\text{overlap}$ associated with the wave-packet overlaps is suppressed as $e^{-r/m}$ for a particle of mass $m$. This means the volume $V$ must be taken to scale as $n_i \log(1/\epsilon_\text{overlap})$. Step iii) involves an adiabatic time evolution for time $\tau$, which can be simulated using a $k^\text{th}$-order product formula by a quantum circuit of $O((\tau V)^{1+\frac{1}{2k}})$ gates. Similarly, the time evolution in step iv) can be implemented with $O((t V)^{1+\frac{1}{2k}})$ gates via a $k^\text{th}$-order product formula. Measurements involving quantum phase estimation require $O(V^{2+\frac{1}{2k}})$, where again a $k^\text{th}$-order product formula could be used. To turn these volume scalings to error-threshold scalings, Refs.~\cite{jordan2011quantum,jordan2012quantum} invokes weak-coupling and strong-coupling assumptions to relate the desired accuracy to the lattice spacing, hence the total volume of the simulation (note that for example, adiabatic state preparation scales polynomially with the energy gap to the next excited state, and this gap closes toward the continuum limit). It also considers the question of renormalization of the mass and coupling and finite-volume effects. Overall, the conclusions remain the same: the JPL algorithm requires resources that are polynomial in system size, energy, and accuracy, hence outperforming classical algorithms that exhibit exponential scaling for the same problem in general.

The JLP algorithm has been extended or improved in a variety of ways~\cite{brennen2015multiscale,bagherimehrab2022nearly,klco2019digitization,barata2021single,turco2024quantum,vary2023simulating,liu2022towards}. Moreover, other ideas for extracting scattering information have also been proposed. For example, if interested in only exclusive processes (i.e., those in which all final-state particles are specified and measured), scattering cross section can be accessed via the $S$-matrix formalism. Here, one may create initial and final states of relevance to the process (with any efficient method available), and use quantum algorithms based on, e.g., a Hadamard test~\cite{nielsen2010quantum}, to evaluate overlap between the time-evolved initial state and of the final state. This idea is explored in Refs.~\cite{li2024scattering,briceno2023toward} for scalar field theories. Inclusive decay rates (i.e., the probability of decaying to all allowed final states) can be accessed via an application of the optical theorem, which relates the decay rate to the imaginary part of the transition-operator matrix, as proposed in Ref.~\cite{ciavarella2020algorithm}. Finally, as with lattice-QCD calculations of scattering amplitude using classical computing, L\"uscher's formalism~\cite{luscher1986volume,luscher1991two} and generalizations of~\cite{briceno2018scattering,hansen2019lattice,davoudi2021nuclear} can be used to access scattering amplitudes in the low-energy regime (where up a few final-state particles can be created). The idea is that scattering amplitudes in infinite space can be related to the low-lying spectrum of the interacting theory in a finite space. These energies can either be accessed via Monte Carlo simulations of Euclidean correlation functions, or by quantum computing the spectrum using a variety of algorithms such as variational quantum eigensolvers and quantum phase-estimation algorithms. Such a route is explored in a simple effective field theory in Ref.~\cite{sharma2024scattering}; see also Refs.~\cite{gustafson2019benchmarking,gustafson2021real}.

A hardware implementation of scattering in $(1+1)$D scalar field theory was presented in Ref.~\cite{zemlevskiy2024scalable}. The simulation was performed on a spatial lattice of $L=60$ sites, and each scalar field was truncated in the field basis such that only two qubits was used to encode its Hilbert space. To enable a near-term simulation, both the interacting vacuum state, two-wave-packet state, and time evolution were performed using variational quantum circuits~\cite{peruzzo2014variational,mcclean2016theory,cerezo2021variational,tilly2022variational}. The variational algorithms used is based on scalable algorithms developed earlier in Refs.~\cite{farrell2024scalable,farrell2024quantum,gustafson2024surrogate} for preparation of vacuum and hadronic wave packets in the Schwinger model [which are improved versions of the ADAPT-Variational Quantum Eigensolver (VQE) algorithm of Refs.~\cite{grimsley2019adaptive,feniou2023overlap,van2024scaling}]. The distance between the fidelity of the parametrized state with respect to the desired density matrix was maximized in small system sizes where the fidelity can be computed exactly. In theories with a finite correlation length, local observables are exponentially close to their infinite-volume values. This implies that the structure of the circuits that create these states also converges exponentially. By tracking the variational parameters' dependence on $L$, the converged value at small values of $L$ can be used to generate the wave function in much larger system sizes, without the need for direct optimization. Similar ideas are applied to the time-evolved state, although the applicability of the methods is now limited to simple final states (i.e., low-energy, low-entanglement output). This 120-qubit simulation was executed on the IBM's \texttt{Heron} quantum processor. The results, upon noise mitigation, indicated the effect of interactions and were found to be in agreement with classical MPS simulations. 
\begin{figure*}[t!]
\centering
\includegraphics[width=0.995\linewidth]{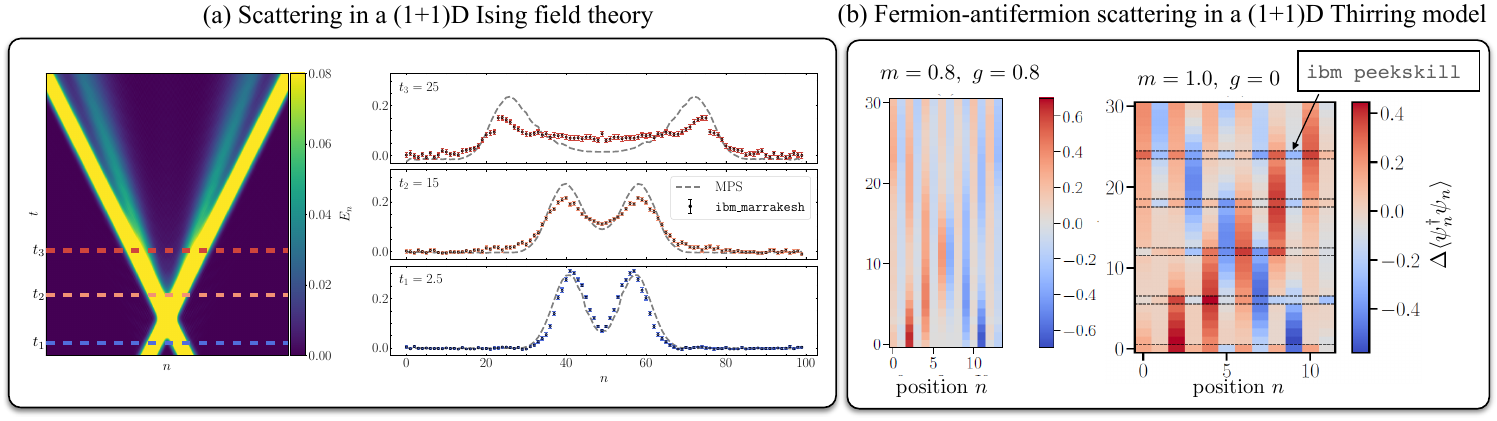}
\caption{(a) The energy density throughout the scattering process as a function of lattice position $n$ and time $t$ is plotted in the left side of the panel in an Ising field theory, obtained from a MPS circuit simulator. In the right side of the panel, results from simulations of scattering using $L = 100$ qubits of IBM's \texttt{marrakesh} quantum processor are shown for a selection of times marked in the left plot. For details, see Ref.~\cite{farrell2025digital}.
(b) The difference in the expectation value of the particle-density operator $\langle \hat{\psi}_n^\dagger \hat{\psi}_n \rangle$ at site $n$ in the evolved state compared to the vacuum as a function of time. The plot in the left corresponds to an interacting case, and that in the right to the noninteracting case, both using exact numerics. Nonetheless, in the noninteracting plot, the exact numerical results at shown time slices are replaced by the results obtained from the IBM's \texttt{peekskill} quantum processor, as detailed in Ref.~\cite{farrell2025digital}.
Figures are reproduced from Refs.~\cite{farrell2025digital,chai2023entanglement}.
}
\label{fig:scattering-digital}
\end{figure*}
%

%%%
%%%
\subsubsection{Scattering in fermionic field theories}
Scattering observables can also be obtained in fermionic field theories using quantum computing. In fact, Jordan, Lee, and Preskill presented an extension of their scalar-field-theory algorithm for a fermionic theory~\cite{jordan2014quantum}, focusing on the massive Gross--Neveu model with the continuum Lagrangian density:
\begin{align}
\mathcal{L} = \sum_{j=1}^{N_f} \bar{\psi}_j (i\gamma^\mu \partial_\mu - m) \psi_j 
+ \frac{g^2}{2} \bigg( \sum_{j=1}^N \bar{\psi}_j\psi_j \bigg)^2 \,,
\label{eq:GN-model}
\end{align}
for $N_f$ fermionic fields $\psi_j$ in $(1+1)$D, and for mass $m$ and coupling $g$. Here, $\overline{\psi}_j = \psi_j^\dagger \gamma^0 $ and $\gamma_\mu$ are Gamma matrices in $(1+1)$D. Once the model is discretized on a spatial lattice, the algorithm goes as follows. One first prepares the ground state of the Hamiltonian with both the bare interaction term ($g_0^2$) and the nearest-neighbor lattice-site interactions turned off. Then via Trotterization, the nearest-neighbor interactions are turned on adiabatically, hence obtaining the ground state of the noninteracting theory. Then one turns on the interaction term, while adjusting the bare parameter $m_0$ to compensate for the renormalization of the physical mass. Next, particle wave packets are excited by introducing a  source term in the Hamiltonian, i.e., a term sinusoidally varying in time and space so as to resonantly select the desired  mass and momentum of particle excitations. Now evolving in time via the full massive Gross--Neveu Hamiltonian, scattering will occur. Finally, the measurement step goes as the original JLP algorithm. The improvement in this algorithm is due to its different state-preparation step: the adiabatic step is used to prepare the interacting vacuum, and a faster resonant method is used to excite particles out of this interacting vacuum. One advantage is in reducing the time during which the wave packets are generated, hence ameliorating the wave-packet broadening during preparation. The algorithmic analysis of Ref.~\cite{jordan2014quantum} arrives at a gate complexity of $O(\epsilon^{-8-o(1)})$ in the high-accuracy regime for the state preparation step in a one-flavor theory (notation $o$ mean the order of magnitude in the asymptotically large regime of a parameter, in this case the order of product formula used in the adiabatic state preparation). The dominant cost is associated with the adiabatic state preparation as in the scalar-field-theory case. To eliminate the adiabatic state-preparation cost in the same theory, Ref.~\cite{hamed2018faster} proposes obtaining a MPS representation of the interacting vacuum using classical computing, and translating it directly into a quantum circuit. Making this replacement in the above algorithm results in a gate complexity $O(p^{-3.23-o(1)})$, which is a significant improvement.

Small-scale demonstrations of fermionic scattering algorithms on quantum hardware have become reality recently. For example, Ref.~\cite{chai2023entanglement} develops and implements an algorithm for wave-packet generation and their scattering in the staggered lattice formulation of the Thirring model. The Lagrangian density of the model in the continuum has the same free form as that in Eq.~\eqref{eq:GN-model}, with the interaction term replaced by $\frac{g}{2}(\overline{\psi} \gamma_\mu \psi)(\overline{\psi} \gamma^\mu \psi)$. The idea is to construct (an)a (anti)fermion creation operator for momentum mode $k$ assuming the free-fermion limit. These creation operators can be combined with a Gaussian wave-packet profile to obtain the (anti)fermion wave packet. In the weak-coupling limit, this should from a reasonable initial state, but for arbitrary interactions (including for composite initial states), it needs to be optimized; see Sec.~\ref{sec:scatt-gauge-theory}. One can then proceed to prepare a fermion and an antifermion wave packet at given energy and perform the Hamiltonian dynamics, as demonstrated in Ref.~\cite{chai2023entanglement}. The results are shown in Fig.~\ref{fig:scattering-digital}(a). The quantity plotted here is $\Delta \langle \hat{\psi}_n^\dagger \hat{\psi}_n \rangle$, the difference in the expectation value of the particle density operator at site $n$ in the evolved state compared to that in the vacuum, as a function of time. The plot in the left corresponds to an interacting case, and that in the right to the noninteracting case. The noninteracting plot incorporates data obtained from the IBM's \texttt{peekskill} quantum processor, replacing the exact numerical results at shown time slices. Hardware simulations for the interacting theory were beyond reach, but will likely be feasible in current hardware using a range of noise-mitigation strategies; see Sec.~\ref{sec:scatt-gauge-theory} for an example.

Simulations of scattering in a $(1+1)$D Ising field theory have been also performed using MPS methods in Ref.~\cite{jha2025real} and on a quantum processor in Ref.~\cite{farrell2025digital}. An Ising field theory arises upon tuning the parameters of the Hamiltonian in Eq.~\eqref{eq:H-Ising} with nearest-neighbor Ising couplings, such that the scale-invariant ratio $\frac{g-1}{h^{8/15}}$ is kept fixed~\cite{Zamolodchikov:1989hfa,Zamolodchikov:1989fp,jha2025real}. By changing the energy of the initial state, Ref.~\cite{jha2025real} demonstrates clear distinctions between elastic scattering, scattering near a resonance, and particle production via inelastic scattering. The subsequent quantum-hardware implementation was enabled by an efficient, constant-depth particle wave-packet preparation algorithm based on $W$-state preparation, and further incorporates the scalable-ADAPT-VQE algorithm of Refs.~\cite{farrell2024scalable,farrell2024quantum}. The scattering simulation was performed on 100 qubits of IBM's \texttt{marrakesh} quantum processor. The outcome of this simulation is depicted in Fig.~\ref{fig:scattering-digital}(a). While the MPS simulations of the scattering clearly indicate four particle tracks in the late-time final state, the large depth of the circuits required limits the evolution on the hardware to early times, where signatures of additional particle productions are hard to detect. These simulations, nonetheless, demonstrates advances in digital quantum simulation of real-time scattering on quantum computers.

Scattering in fermionic theories can also be studied using a hybrid analog-digital approach. As an example, Ref.~\cite{garcia2015fermion} proposes quantum simulation of fermion-antifermion scattering in $(1+1)$D, mediated by a continuum of bosonic modes, using a circuit-QED platform. The Hamiltonian of the model in continuum is:
\begin{align}
\label{eq:H-ffA}
H = & \int dp \ \omega_p (b^{\dag}_{p}b_{p} + d^{\dag}_{p}d_{p}) + \int dk \ \omega_k a^{\dag}_ka_k  \nonumber \\ 
& \hspace{3 cm} +  \int dx \ \psi^{\dag}(x)\psi(x)A(x).
\end{align}
Here, $A(x)=i\int dk \ \lambda_k \sqrt{\omega_{k}}( a^{\dag}_k e^{-i k x} - a_k  e^{i k x} )/\sqrt{4\pi}$ is a bosonic field, with coupling constants $\lambda_k$, and $\psi(x) = \int dp\left( b_p  e^{i p x} +  d_p^{\dag} e^{-i p x} \right)/\sqrt{4\pi\omega_p}$ is the fermionic field. $b^{\dagger}_p$ ($b_p$) and $d^{\dagger}_p$ ($d_p$) are the corresponding fermionic and antifermionic creation (annihilation) operators for mode frequency $\omega_p$. $a^{\dagger}_k$ ($a_k$) is the creation (annihilation) bosonic operator associated with frequency $\omega_k$. To simplify the problem for mapping to a circuit-QED device, one may consider one fermionic and one antifermionic field comoving modes interacting via a continuum of bosons. The comoving-mode creation operators are constructed such that they create normalizable propagating wave packets when applied to the vacuum. The circuit-QED analog simulator has the following components. An open transmission line  supports the continuum of bosonic modes. It interacts with two superconducting qubits simulating the fermions (when mapped to qubits via a Jordan--Wigner transformation) and one ancilla qubit. A second one-dimensional waveguide supports a single mode of the microwave field and interacts with the two superconducting qubits. The ancilla qubit is used to implement boson-only Hamiltonian terms in the evolution. A sequence of single- and two-qubit gates, and qubit-boson gates are then implemented to simulate the evolution of the particle and antiparticle excitations interacting via the continuum bosonic mode.

A proposal for simulating the Yukawa model, a model of fermions interacting with a bosonic quantum field, has also been proposed in Ref.~\cite{davoudi2021toward}, using the hybrid analog-digital approach in trapped-ion quantum computers with the aid of phonon modes; see also Refs.~\cite{casanova2011quantum, casanova2012quantum, lamata2014efficient, mezzacapo2012digital} for earlier work in other models. Single fermion-antifermion dynamics coupled to a single bosonic mode are demonstrated experimentally in trapped-ion systems in Ref.~\cite{zhang2018experimental} but larger simulations, simulations in other platforms, and simulations of quantum field theories, remain to be realized.

%%%
%%%
\subsubsection{Scattering in gauge field theories
\label{sec:scatt-gauge-theory}}
Scalable simulations of scattering processes in gauge theories have not yet become practical, as qubit and gate requirements are high. The reasons are, initial-state preparation in confining gauge theories can be nontrivial, and the interactions are complex. Algorithmic progress in recent years, nonetheless, has established a path to digital quantum simulation of time dynamics in these theories in both the near- and far-term eras of quantum computing~\cite{byrnes2006simulating,Shaw:2020udc,haase2021resource,kan2021lattice,ciavarella2021trailhead,kane2022efficient,Davoudi:2022xmb,sakamoto2024end,gustafson2024primitive,rhodes2024exponential}. Furthermore, there has been significant progress in vacuum and hadronic-state preparation~\cite{klco2018quantum,kokail2019self,atas20212,farrell2024scalable,crippa2024analysis,fromm2024simulating,than2024phase,xie2022variational}, including for hadronic wave packets for scattering processes~\cite{farrell2024quantum,davoudi2024scattering,davoudi2025quantum,chai2025towards,farrell2025digital}. In the following, we first briefly overview recent works in preparation of hadronic wave packets in $(1+1)$D gauge theories, then provide examples of the first hardware implementations of scattering in these theories.

\begin{figure*}[t!]
\centering
\includegraphics[scale=0.685]{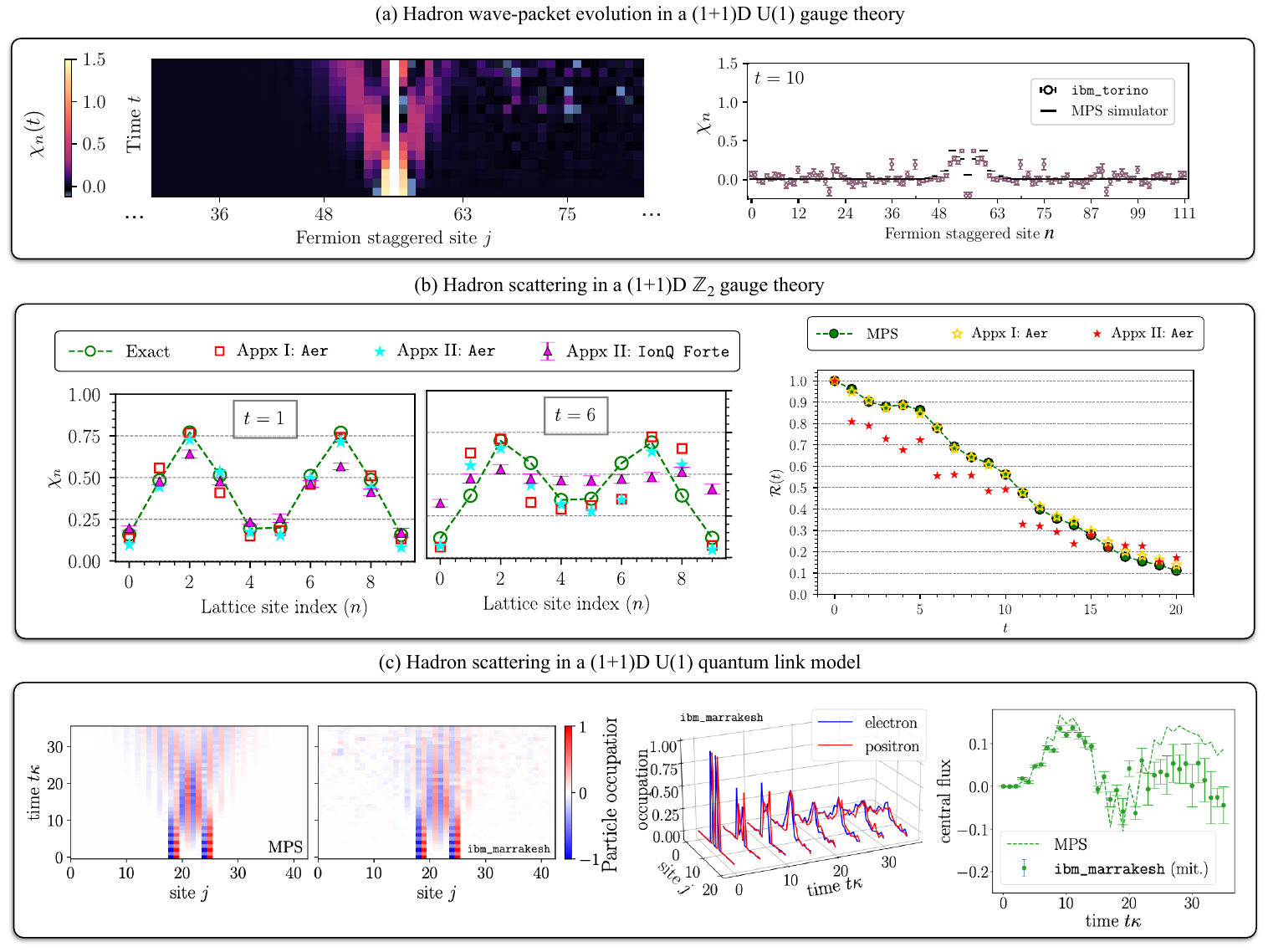}
\caption{(a) Time evolution of a hadronic wave packet in a 112-site lattice Schwinger model from Ref.~\cite{farrell2024quantum}. The left plot is particle-density expectation value, $\chi_n$, as a function of time past since a localized wave-packet is created in the middle of the lattice, while the right plot is the same quantity at a fixed time. The left side of the plot in the left is computed using an MPS method, while the right side is generated by IBM's \texttt{torino} quantum processor. See Ref.~\cite{farrell2024quantum} for details. (b) The left and middle plots are particle density, $\chi_n$, as a function of lattice site $n$, for two-hadron-wave-packet scattering in a $\mathbb{Z}_2$ LGT for a system of $L=10$ fermion sites, at two distinct times during Trotterized time evolution. \texttt{Aer} refers to circuit-emulator results and IonQ's \texttt{Forte} refers to hardware results. The right plot is the initial-state survival probability, $\mathcal{R}(t)$, as a function of Trotterized time for a system size $L=26$, using only the \texttt{Aer} circuit emulator compared with the MPS result, obtained for two approximations to the initial wave-packet states as descried in the text. For further detail, see Ref.~\cite{davoudi2025quantum}. 
(c) The particle occupation number (left three plots) and the central electric flux (the right plot) for two bare-meson collision with given mass and confinement parameters; see Ref.~\cite{schuhmacher2025observation} for details. The system involves 43 fermion sites and the evolution involves 35 Trotter steps. The hardware simulations are performed on IBM's \texttt{marrakesh} quantum processor while the classical results are obtained via an MPS method. Figures are reproduced from Refs.~\cite{farrell2024quantum,davoudi2025quantum,schuhmacher2025observation}.
}
\label{fig:hadron-scattering-digital}
\end{figure*}

It is possible to prepare interacting hadronic wave packets within confined LGTs on a quantum computer without the need for adiabatic evolution. As detailed in Ref.~\cite{davoudi2024scattering} for $(1+1)$D theories, the full state preparation involves generating the interacting vacuum using a variational quantum eigensolver, then optimizing a reasonable mesonic excitation ansatz in each momentum sector in the interacting theory, and finally building and circuitizing the full Gaussian wave-packet creation operator  using efficient algorithms. The ansatz is similar to that proposed in Ref.~\cite{rigobello2021entanglement} in the context of tensor-network simulations of scattering in the Schwinger model, but is further optimized to perform well for arbitrary coupling values, including toward the continuum limit. The meson excitations are constructed out of bare mesonic operators of arbitrary length. The momentum of the quark and antiquark are distributed such that large momentum differences are suppressed by an exponentially decaying form, containing a free parameter to be optimized. The examples of a $\mathbb{Z}_2$ LGT and a U$(1)$ LGT in $(1+1)$D with one flavor of staggered fermions [i.e., Eq.~\eqref{eq:Z2LGT} when a staggered mass is included and Eq.~\eqref{eq:SchwingerKS}, respectively] were tested in Ref.~\cite{davoudi2024scattering}, showing high fidelities for systems of up to 10 staggered sites. The interacting mesonic creation operator involves $O(L^3)$ bare mesonic operators, where $L$ is the number of fermionic sites. Upon prioritizing mesons with smaller sizes, the wave-packet circuits were executed on Quantinuum's \texttt{H1-1} quantum processor for a 6-site theory, requiring 13 qubits (a single ancilla was used to implement the preparation algorithm). The hardware results, for both physical basis-state probabilities and for local particle density were found to be in agreement with the exact results.

Another approach to simulating meson scattering in the $(1+1)$D $\mathbb{Z}_2$ LGT coupled to staggered fermions was presented recently in Ref.~\cite{chai2025towards}. Meson creation operators are constructed using a quantum subspace expansion (QSE) method~\cite{mcclean2017hybrid,mcclean2020decoding,takeshita2020increasing,yoshioka2022variational}. The method obtains meson states without the need for variational quantum eigensolvers, using MPS ansatzes to determine the states classically. More explicitly, the meson-state ansatz is a linear combination of bare-meson operators with unknown coefficients. These coefficients are optimized by solving a generalized eigenvalue problem for the energy and charge-conjugation eigenvalues, obtained from MPS, along with a normalization equation~\cite{chai2025towards}. The quantum-circuit decomposition for preparing meson wave packets uses Givens rotations~\cite{jiang2018quantum,kivlichan2018quantum}, and involves a CNOT gate complexity $O(L^2)$ and a circuit depth $O(L)$, where $L$ is the number of lattice sites.

A scalable algorithm for preparing hadron wave packets in the lattice Schwinger model was presented in Ref.~\cite{farrell2024quantum}. The Hamiltonian is that in Eq.~(\ref{eq:SchwingerGaugeFieldsIntegratedOut}), in which the gauge degrees of freedom are integrated out. However, this procedure is done in such a way to retain a CP-symmetric form. Explicitly. the electric Hamiltonian in the total-charge zero sector (with zero background electric field) takes the CP-symmetric form~\cite{farrell2024quantum}
\begin{align}
\hat{H}_{E}^{(Q=0)} = & \frac{g^2}{2} \bigg \{\sum_{\ell=0}^{\frac{L}{2}-2}\bigg ( \sum_{\ell'=0}^n \hat{Q}_{\ell'}\bigg )^2 + \sum_{\ell=\frac{L}{2}+1}^{L-1}\bigg ( \sum_{\ell'=\ell}^{L-1} \hat{Q}_{\ell'}\bigg )^2 
\nonumber\\
& +\frac{1}{2}\bigg [\bigg (\sum_{\ell=0}^{\frac{L}{2}-1} \hat{Q}_\ell \bigg )^2 + \bigg (\sum_{\ell=\frac{L}{2}}^{L-1} \hat{Q}_\ell \bigg )^2 \bigg ] \bigg \}.
\label{eq:HelCP}
\end{align}
Here, $\hat{Q}_\ell=-\hat\psi^\dagger_\ell\hat\psi_\ell+\frac{1}{2}[1-(-1)^\ell]$ is the electric-charge operator at site $\ell$. This Hamiltonian only depends on the fermionic fields, but consists of $O(L^2)$ interacting terms. This can be costly for near-term implementations. Nonetheless, since interactions among charges in a confining theory with a mass gap are screened beyond the Compton wavelength of the lightest excitation, an effective Hamiltonian can be formed to only consist of $O(N)$ interacting terms, as developed in Ref.~\cite{farrell2024quantum}. Note that open boundaries break translational symmetry, hence momentum eigenstates can only be defined approximately in the large-system limit. The preparation of the hadron wave packet is performed in two steps. First, the interacting vacuum  is prepared across the whole lattice using an improved VQE algorithm and workflow~\cite{farrell2024scalable,grimsley2019adaptive,feniou2023overlap,van2024scaling}. This VQE is then extended to the preparation of localized states, by adaptively constructing low-depth circuits that maximize the overlap with an adiabatically prepared hadron wave packet. Since the wave packets are localized, these circuits can be determined on a sequence of small lattices using classical computers. They are then scaled to prepare wave packets on large lattices for subsequent quantum-computer simulations. Using multiple error-mitigation strategies, up to 14 Trotter steps of time evolution are performed for generated wave packets in Ref.~\cite{farrell2024quantum}, using 112 qubits of IBM's \texttt{torino} quantum processor. The propagation of hadrons is clearly identified, with results that compare favorably with MPS simulations, as plotted in Fig.~\ref{fig:hadron-scattering-digital}(a).

Quantum simulation of hadron scattering has also become reality recently using digital quantum hardware, albeit in simple low-dimensional gauge theories. For example, two-hadron scattering in a $\mathbb{Z}_2$ LGT in $(1+1)$D was studied in Ref.~\cite{davoudi2025quantum}. The model Hamiltonian is that presented in Eq.~\eqref{eq:Z2LGT} with a staggered mass term and with periodic boundary condition.\footnote{With the alternative but equivalent convention: $\hat \tau_n^x \leftrightarrow \hat \tau_n^z$.} Nonetheless, to reduce the qubit-resource cost, an equivalent formulation was used in which, via a gauge transformation and imposing the Gauss's laws, only one gauge degree of freedom had to be retained at the cost of an all-to-all interaction among the charges. This degree of freedom then only require one qubit to be encoded in the quantum circuit. The Hamiltonian reads:
\begin{align}
    \hat H^{(1+1}_{\mathbb{Z}_2,\text{PBC}} &= \frac{1}{4} \sum_{n=0}^{N-2}( \hat\sigma^{x}_{n}\hat\sigma^{x}_{n+1} + \hat\sigma^{y}_{n}\hat\sigma^{y}_{n+1})
        +\frac{\alpha_L}{4}(\hat\sigma^{x}_{L-1}\hat\tau^{x}_{L-1}\hat\sigma^{x}_{0}
        \nonumber\\
        & +\hat\sigma^{y}_{L-1}\hat\tau^{x}_{L-1}\hat\sigma^{y}_{0})+
        \frac{m}{2}\sum_{n =0}^{L-1}{(-1)^{n+1}  \hat\sigma^{z}_{n}}+\epsilon \hat\tau^{z}_{L-1} 
        \nonumber\\
       & +\epsilon \sum_{n=0}^{L-2}\gamma_n
        \hat\tau^{z}_{N-1}\left(\prod_{j=0}^n \hat\sigma^{z}_{j}\right).
    \label{eq:Z2-H-MGF}
\end{align}
Here, $\alpha_L = (-1)^{L/2+1}$ for the subspace $Q=L/2$ (with $Q$ being the total fermion number) and $\gamma_n = e^{-i\pi\sum_{m=0}^{n} (1-(-1)^m)/2}=i^{n}$ for even $n$ and $i^{n+1}$ for odd $n$. $\hat \sigma$ and $\hat \tau$ operators act on the fermion and gauge-boson qubits, respectively. The term in the last line of Eq.~\eqref{eq:Z2-H-MGF} involves interactions between all fermion qubits (but one) and the gauge-boson qubit. This term, nonetheless, can be simulated with a linear-depth circuit in Trotterized time evolution, as shown in Ref.~\cite{davoudi2025quantum}. The hadronic wave-packet ansatz is an improved version of that developed in Ref.~\cite{davoudi2024scattering}: bare mesons of increasing length are added to the ansatz systematically, which facilitates the optimization process, and increases wave-packet fidelities order by order, given the finite correlation length in the system. This ansatz, in particular, achives much higher fidelities in both larger system sizes and larger momentum states than that in Ref.~\cite{davoudi2024scattering}. The two isolated wave packets are generated using a VQE procedure that optimizes the ansatz parameters. The wave packets are then evolved in a Trotterized scheme and collide. The left and middle plots in Fig.~\ref{fig:hadron-scattering-digital}(b) depict the scattering results obtained from exact simulation, circuit emulator, and quantum-processor results obtained from IonQ's \texttt{Forte} quantum processor for a system of $N=10$ fermion sites. Unfortunately, decoherence limits evolving the wave packets to longer times. The left panel depicts, for a larger system size of $N=26$, the survival probability of the initial state as a function of time, which is a diagonal entry of the $S$-matrix. In all plots, two approximations for the initial-state preparation is used, Appx I with higher fidelity with the exact initial state (obtained using an MPS method) and Appx II with lower fidelity. While local quantities were shown to be rather insensitive to the difference in the initial-state fidelity, the survival probability is seen to deviate significantly for the less accurate Appx II. This observation implies the importance of high-fidelity initial-state preparation for hadron scattering observables such as $S$-matrices.

Simultaneously, a quantum-simulation experiment of meson scattering in a spin-$\frac{1}{2}$ U$(1)$ quantum-link model [i.e., Eq.~\eqref{eq:QLMmapping} with $S=1/2$] was reported in Ref.~\cite{schuhmacher2025observation} using IBM's \texttt{marrakesh} quantum processor. Here, a tunable CP-violating $\theta$ parameter controls confinement in the model, and allows for both electron-positron and meson-meson scattering, as described in Sec.~\ref{sec:U1LGT_formulations}. To simplify the state-preparation step,  Ref.~\cite{schuhmacher2025observation} starts from the bare meson excitations (instead of dressed-meson wave packets) which delocalize through time evolution. In order to restrict the motion of the left (right) excitations, their coupling to right (left) side of the chain is turned off originally, so that the excitations move toward each other and collide. While this initial state does not have well-defined interacting-particle momentum, its scattering outcome still represent nontrivial dynamics, and allows quantum resources to be primarily allocated to the time-evolution stage of the scattering. Fig.~\ref{fig:hadron-scattering-digital}(c) presents the outcome of a 43-qubit meson-meson simulation (involving 43 fermionic sites after the gauge degrees of freedom being integrated out) with 35 Trotter steps of digitized time evolution. While the initial state involves two well-isolated bare electron-positron pairs, their early-time collision dynamics smears the initial bare-meson states, in agreement with the MPS simulations. Nonetheless, long-time features of the scattering state cannot yet be reproduced by the quantum processor due to the effect of decoherence.

Last but not least, first attempts at toy models of interesting electroweak and beyond-the-Standard-Model scattering processes have been reported, albeit with simplified initial states. For example, a time-dependent probability amplitude was computed for a prototype model of single-$\beta$ decay of a proton to a neutron in Ref.~\cite{farrell2023preparationsII}, and of neutrinoless double-$\beta$ decay of two neutrons to two protons in Ref.~\cite{chernyshev2025pathfinding}, in a $(1+1)$D QCD coupled to flavor-changing weak interactions. The former uses Quantinuum's \texttt{H1} processor for a lattice of a single physical site (requiring 16 qubits) and the latter employs IonQ's \texttt{Forte} processor for a lattice of two physical sites (requiring 32 qubits). Complete simulations of asymptotic $S$-matrices, involving large lattices and large-time evolutions, require qubit and gate counts that are beyond today's hardware capacity and capability. Nonetheless, these examples, and other examples presented in this Section, demonstrate the rapid progress in methodologies and implementations of scattering problems on quantum hardware.

%%%%%%%%%%%%
%%%%%%%%%%%%
\subsection{Open questions
\label{sec:scattering-open-Qs}}
The progress to date has only started to scratch the surface of the rich field of high-energy scattering dynamics. As mentioned in the Section opening, the ultimate goal of this program is to provide insights into, and eventually detailed understanding of, scattering in collider experiments, and unravel the role of Standard-Model interactions in scattering phenomenology in real time. None of the quantum-simulation proposals and experiments to date have, nonetheless, gone beyond Abelian LGTs and $(1+1)$D models. Even then, the scattering simulations have been limited to short times and simple scattering outcomes (often elastic scattering). How to move beyond these simple models and scenarios, clearly, constitutes a main open question for the field.

To observe inelastic processes, even in the simple scenarios considered to date, requires either fine tuning of the model parameters to ensure abundant kinematically allowed nearby particle channels in the spectrum for the initial state to turn to, or to increase the energy of the incoming particles to allow for a range of energetic excitations in the final state. The former strategy may be challenging if the model spectrum is hard to access, such as in higher-dimensional theories, and the latter strategy requires creating high-energy initial wave packets that appear more challenging to prepare with the current algorithms.

Even if one achieves rich inelastic final states, characterizing the content of the state in terms of single- and multi-particle channels is challenging and requires dedicated strategies. For example, to compute the scattering cross section for two particles to end asymptotically in an $k$-particle state requires creating the $k$-particle state and finding its overlap with the asymptotic final state using known algorithms---a process that can be costly. Alternatively, one can compute the multi-field correlation functions and access the asymptotic scattering amplitudes via a Lehmann–Symanzik–Zimmermann (LSZ) reduction formula~\cite{lehmann1955formulierung}, or closely related schemes~\cite{
haag1958quantum,haag1958asymptotic}, but these are also involved algorithms for near-term hardware~\cite{li2024scattering,briceno2023toward,ciavarella2020algorithm,turco2024quantum,vary2023simulating,turco2025creation}.

Perhaps more interesting is probing nonasymptotic states, i.e., those phases of matter that are rather short-lived after the scattering event, such as quark-gluon plasma. It would be interesting to develop tools to extract information about such phases via local and nonlocal measurements, or efficient state-tomography tools, including entanglement-Hamiltonian tomography. Such tools can potentially allow for understanding the mechanism and pace of thermalization in high-energy scattering processes; see Sec.~\ref{sec:thermalization}.

Finally, one needs to develop a framework for turning the real-time scattering-simulation results to the input needed in particle-collider experiments, including the design and optimization of event generators used in analyzing experimental data~\cite{sjostrand2020pythia,gieseke2004herwig}. Filling this gap requires understanding the various elements of these generators and their deficiencies due to lack of first-principles dynamical, quantum input. For example, the phenomenon of string breaking and quark fragmentation discussed in the previous Section can be studied in scattering aftermath, mimicking realistic conditions in collider experiment. Only then can the quantum simulations begin to prove their worth in the realm of nonequilibrium dynamics of particle collisions.

%%%%%%%%%%%%%%%%%%%%%%%%%%%%%%%%%%%%%%%%%%%%%%%%%%%%%%%%%%%%%%%%%%%%%%%%%%%%%%%%%%%%%%%%%%%%%%%%%%%%%%%%%%%%%%%%%%%%%%%%%%%%%%%%%%%%%%%%%%%%%%%%%%%%%%%%%%%%%%%
\section{Thermalization dynamics in gauge theories
\label{sec:thermalization}}

Thermalization in isolated quantum many-body systems and quantum field theory is a complex and multifaceted topic across various scientific fields such as atomic, molecular, and optical physics, condensed-matter physics and material science, high-energy and nuclear physics, and cosmology. While many systems eventually exhibit thermodynamic behavior, understanding the precise dynamics that drive thermalization remains a challenging and open problem. 

Gauge theories play a pivotal role in this area. For example, when large ions, smaller ions, or even single protons are accelerated to nearly the speed of light and collided at facilities like the Relativistic Heavy Ion Collider (RHIC) or the Large Hadron Collider (LHC), they may create a state of matter known as the quark-gluon plasma~\cite{mclerran1986physics,harris1996search,rischke2004quark,braun2007quest}. This plasma is so hot that it no longer consists of hadronic matter, such as protons and neutrons, rather constituent quarks and gluons, as governed by the laws of QCD. Ultra-relativistic heavy ion collisions generate temperatures of the order of $5\times10^{12}$~K~\cite{adam2016direct}. Similar temperatures existed fractions of milliseconds after the big bang, thus they also provide a view into the early universe. Experiments at RHIC and LHC have provided evidence that the quark-gluon plasma is a near-perfect fluid with a shear viscosity to entropy density ratio $\eta /s~\le 0.2$, remarkably close to a bound $\eta /s=1/4\pi$ obtained from the AdS/CFT duality~\cite{policastro2001shear}. Understanding how this extreme form of matter behaves far from equilibrium, and whether it eventually thermalizes, has intrigued researchers for decades, fueling a large and ongoing scientific program~\cite{vogt2007ultrarelativistic,florkowski2010phenomenology}. 

Despite many experimental and theoretical studies~\cite{chatterjee2010electromagnetic,van2006heavy,foka2016overview,guangyou2020soft,luo2017search,metag1993near,lovato2022long}, no first-principle approach can quantitatively predict the dynamics of matter throughout the stages of an ultra-relativistic heavy ion collision. Instead, the understanding is based on a composition of various effective descriptions, ranging from the color-glass condensate effective field theory~\cite{gelis2010color}, to classical-statistical approximations~\cite{berges2012overpopulated,kurkela2012ultraviolet}, QCD kinetic theory~\cite{arnold2003effective,schlichting2019first}, holography~\cite{maldacena1999large},  or hydrodynamics~\cite{gale2013hydrodynamic,florkowski2018new}.

In recent years, it has been recognized that certain features of nonequilibrium systems are universal, i.e., are independent of their microscopic details~\cite{hohenberg1977theory}. One manifestation is similarity between ultracold Bose gases at $\mu$K temperatures and overoccupied quark-gluon plasmas~\cite{berges2015universality}. As a result, large-scale experiments with synthetic quantum systems, like ultracold gases, have been proposed to study fundamental systems such as QCD~\cite{berges2024ultracold}. Digital universal quantum computation, on the other hand, is more difficult to scale up, and time evolution must be implemented through circuit-based methods like Trotterization, rather than implemented naturally. As a result, progress in the digital quantum simulation of gauge-theory thermalization has lagged behind that of simpler models, which are more easily mapped onto atomic, molecular, or optical systems.
\begin{figure*}
    \centering
    \includegraphics[scale=0.25]{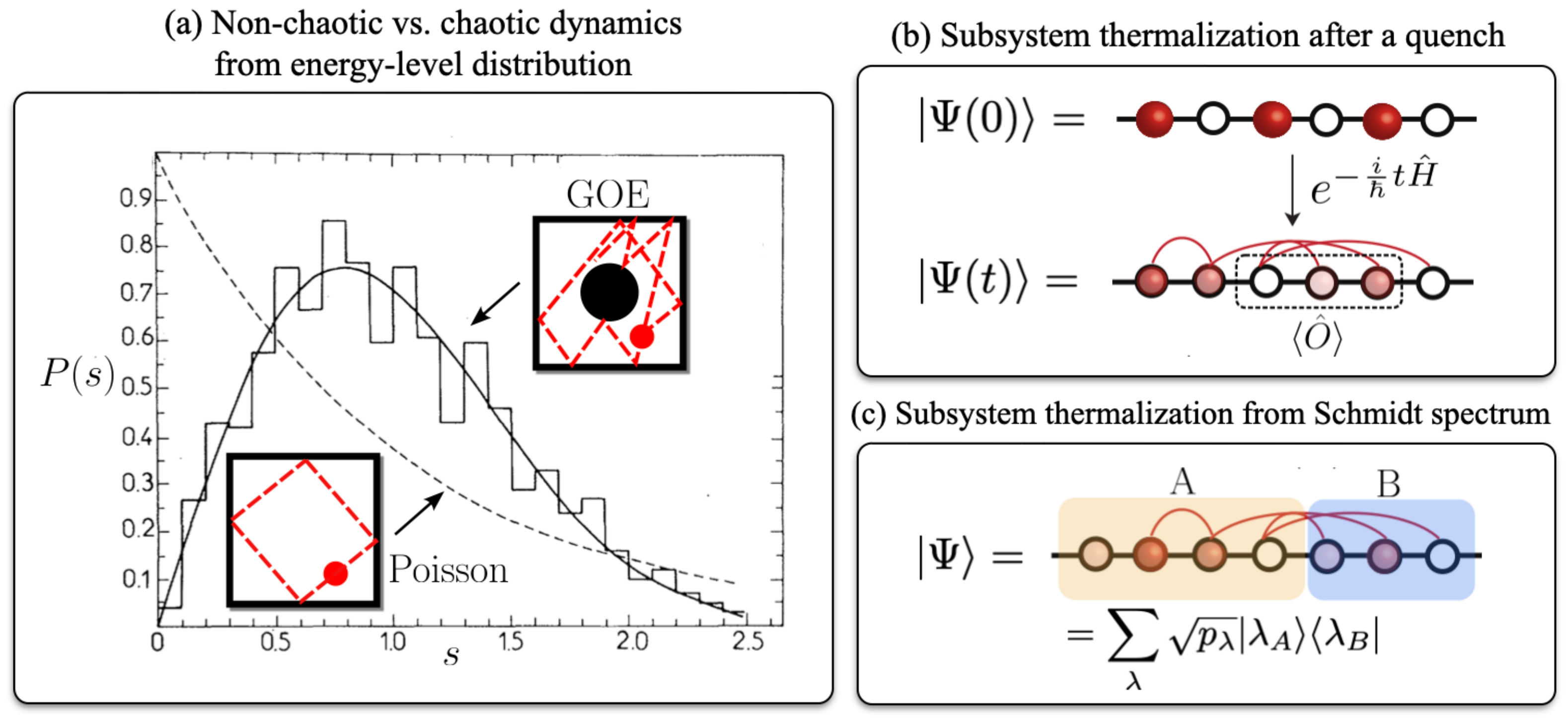}
    \caption{
    (a) A standard method for probing thermalization in few-body quantum systems is the statistics $P(s)$ of energy-level spacings, $s=E_{n+1}-E_n$. In an integrable system, e.g., a particle inside a square-shaped region, the levels follow the Poisson distribution. By contrast, for the Sinai billiard, i.e., the square with a circular region cut out, the particle's motion becomes chaotic and the levels follow a GOE distribution. Panel is reproduced  from Ref.~\cite{bohigas1984characterization}.
    (b) In an isolated many-body quantum system, thermalization can be conveniently probed using a quantum quench: the system is prepared in an out-of-equilibrium initial state $|\Psi(0)\rangle$ and evolves under unitary dynamics. While the system remains in a pure state $|\Psi(t)\rangle$ at all times, thermalization occurs at the level of subsystems, e.g., a local observable $\hat O$ defined for three spins in the middle of the chain (dashed line) approaches the thermal value at late times. This process is governed by the ETH~\cite{DeutschETH,SrednickiETH}. (c) Beyond local observables, thermalization can also be characterized using bipartite entanglement of the state. The distribution of the Schmidt coefficients $\lambda_k$
    mimics the behavior of the energy spectrum, providing a complementary window into the process of thermalization.
    }
    \label{fig:therm}
\end{figure*}

In quantum many-body systems, the notion of thermalization has historically been subtle, since several concepts that apply naturally in classical systems do not transfer straightforwardly. Early studies approached thermalization through the lens of chaos and random-matrix theory~\cite{berry1977level,bohigas1984characterization}, often by analogy with the semiclassical limit. A standard diagnostic from random-matrix theory is illustrated in \Fig{fig:therm}(a): the distribution $P(s)$ of nearest-neighbor energy-level spacings, defined as $s \coloneq E_{n+1}-E_n$. For integrable models which do not thermalize, $P(s)$ follows a Poisson distribution (as exemplified by a particle confined to a square region, with typical classical trajectories shown), whereas for chaotic systems, it follows the Wigner–Dyson distribution of the Gaussian Orthogonal Ensemble (GOE) (exemplified by the Sinai billiard, i.e., a circular obstacle in a square region); see Refs.~\cite{kudo2003unexpected,muller2004semiclassical,OganesyanHuse,gomez2011many,gubin2012quantum,atas2013distribution,torres2017dynamical} for exemplary studies.

These studies ultimately culminated in the Eigenstate Thermalization Hypothesis (ETH)~\cite{DeutschETH,SrednickiETH} which now also applies to many-body systems. The ETH posits that thermal behavior of local observables emerges from specific properties of energy eigenstates. The ETH has been since extended to thermal behavior in subsystems~\cite{garrison2018does,dAlessio2016}, thereby establishing a direct connection with entanglement~\cite{nandkishore2015many,polkovnikov2016thermalization}. Subsystem and local-observable thermalization, as well as the role of entanglement, have already been explored in quench experiments~\cite{Kaufman2016}, where a Hamiltonian parameter is abruptly changed and the subsequent dynamics are studied via local observables; see~\Fig{fig:therm}(b) for an illustration. Such quenches are often experimentally straightforward to implement, but it remains to be seen how these insights extend to physically relevant scenarios in nature. 

Moreover, concepts originating from random-matrix theory are reinterpreted in new settings. For example, the bipartite entanglement of a state can be analyzed through the statistical distribution of its Schmidt coefficients $ \sqrt{p_\lambda}$, 
\begin{align}\label{eq:schmidtcoeff}
    | \Psi \rangle = \sum_\lambda \sqrt{p_\lambda} | \lambda_A \rangle \otimes | \lambda_B\rangle
\end{align}
defined in \Fig{fig:therm}(c). This distribution may mirror the behavior of energy-level statistics~\cite{Geraedts2016,rakovszky2019signatures,chang2019evolution}. LGTs are expected to exhibit especially rich and unconventional behavior in this regard. For example, while it has long been recognized~\cite{casini2014remarks,aoki2015definition,ghosh2015entanglement} that the entanglement structure inherent to gauge theories is unique, 
the topic of thermalization within this context 
has just started to gain attention~\cite{mueller2022thermalization}. In the remainder of this Section, we will first discuss theoretical studies of LGT thermalization, including the ETH and the role of entanglement, followed by discussion of the first quantum-simulation results.

%%%%%%%%%%%%
%%%%%%%%%%%%
\subsection{Lattice-gauge-theory thermalization}\label{sec:thermalizationGT}
Given the challenges of engineering gauge theories of the Standard Model on quantum simulators, and the limited circuit depths of digital quantum computers, thermalization has so far been studied only in simple models and small systems. We first review exploratory theoretical works before turning to the quantum-simulation experiments.

%%%
%%%
\subsubsection{Gauge-theory thermalization and ETH}

The ETH concerns the conditions under which local observables show thermal behavior in closed quantum systems~\cite{DeutschETH,SrednickiETH}.  The ETH relates the late-time average of observables with the properties of these observables when measured in a `typical', i.e., mid-spectrum, eigenstate. Concretely, the ETH states that matrix elements $A_{n,n^\prime} \coloneq \langle n | \hat A | n^\prime \rangle$ in the energy eigenbasis $ | n \rangle $ of an observable $\hat{A}$ have the form
 \begin{align}\label{eq:ETHmastereq}
     A_{n,n'} = \langle \hat A \rangle_{\rm MC}^{(\mathcal{E})}\delta_{n,n'} + e^{-S(\mathcal{E})/2} f(\mathcal{E},\epsilon)R_{n,n'},
 \end{align}
where $\mathcal{E} \coloneq (E_n+E_{n'})/2$, $\epsilon \coloneq E_n-E_{n'}$, $\langle A \rangle_{\rm MC }^{(\mathcal{E})}$ is the microcanonical average of $\hat{A}$ at energy $ \mathcal{E}$ and is a smooth function, $S$ is the thermodynamic entropy which is extensive in the number of degrees of freedom, $f$ is a smooth positive and real function,  and $R_{n,n'}$ is a (model dependent) random variable with zero mean and unit variance. The conditions for the ETH are typically met in nonintegrable systems~\cite{rigol2008thermalization}---though exceptions do exist (see Sec.~\ref{sec:ergodicitybreaking}). However, whether (and how) gauge theories obey the ETH remains an open question, given that they are systems with an extensive number of local symmetries, and a complex Hilbert-space structure. As we describe below, quantum simulators have recently started to probe this question of applicability and implications of the ETH in gauge theories.

While the ETH assumes nondegeneracy of eigenstates, it is likely broadly applicable when they are accounted for---for instance in systems with non-Abelian symmetries~\cite{murthy2023non}. Aspects of ETH were studied theoretically in Ref.~\cite{yao20232} in a Kogut--Susskind formulation of a $(2+1)$D SU$(2)$ LGT with the Hamiltonian in \Eq{eq:H-E-B-dp1}. The study presents numerical evidence that the model obeys the ETH, in the form of \Eq{eq:ETHmastereq}, for a few selected observables. The electric flux, $\sum_{a=1}^3(\hat E_l^a)^2| j,m_L,m_R \rangle_l=j(j+1)| j,m_L,m_R \rangle_\ell$ in this study is cut off, with $j\le j_{\rm max}=1/2$. Here, $j$ is the total angular momentum on the link $l$, and $m_L$ and $m_R$ are the third component of the angular momentum on the left and right of the link, respectively, i.e., $|m_{L/R}| \leq j$. In addition, this work presents an analysis of the distribution of the eigenenergies of \Eq{eq:H-E-B-dp1}, analogously to \Fig{fig:therm}(a), which is found to be consistent with GOE. Follow-up work~\cite{ebner2024eigenstate} extends this analysis to a larger electric-field cutoff, while Refs.~\cite{muller2023simple,ebner2024eigenstate} consider a a chain of plaquettes on a two-dimensional honeycomb lattice with periodic boundary conditions (still with $j_{\rm max}=1/2$). While these works suggest the applicability of ETH in non-Abelian LGTs, directions for further exploration include understanding the dependence on the truncation scheme, the choice of observables (as ETH should hold for all local observables), and the role of finite-volume effects.

%%%
%%%
\subsubsection{Gauge-theory thermalization from entanglement spectra}

Entanglement is an important probe of thermalization~\cite{kaufman2016quantum,neill2016ergodic,polkovnikov2016thermalization,garrison2018does}. Various measures are studied in this context: von-Neumann  and R\'enyi entanglement entropies, defined, respectively, by the relations:
\begin{align}\label{eq:vonNeumann_Renyi}
    &S_{\rm vN} \coloneq - \text{Tr}_A\left\{\hat{\rho}_A \log(\hat{\rho}_A)\right\}\,,\\
    &S_{\rm R}^{(k)} \coloneq \frac{1}{1-k} \log\left\{\text{Tr}_A(\hat{\rho}_A^k)\right\}\,,
\end{align}
where $ \hat{\rho}_A \coloneq \text{Tr}_{B} \ketbra{\psi}{\psi}$ is the reduced density matrix of a subsystem $A$ formed by bipartitioning a pure quantum state $\ket{\Psi}$ to $A \cup B$, as in Fig.~\ref{fig:therm}(c). The corresponding entanglement Hamiltonian~\cite{li2008entanglement}, 
\begin{align}\label{eq:EntanglementHamiltonian}
    \hat{H}_{\rm E} \coloneq -\log(\hat{\rho}_A)\,,
\end{align}
has recently become a target for quantum-simulation experiments~\cite{islam2015measuring,pichler2016measurement,dalmonte2018quantum,lukin2019probing,elben2023randomized,brydges2019probing,elben2020mixed,vermersch2018unitary}.

The entanglement Hamiltonian of a gauge theory has been studied theoretically in   Ref.~\cite{mueller2022thermalization}. The model is a $\mathbb{Z}_2$ LGT in $(2+1)$D described by the Hamiltonian in \Eq{eq:Z2Hamiltonian}, and the corresponding Gauss's law operator in \Eq{eq:GLZ22p1}. The focus is on the entanglement Hamiltonian, \Eq{eq:EntanglementHamiltonian}, and its spectral structure, which serve as state-dependent indicators for chaotic dynamics. Because of Gauss's law, the entanglement Hamiltonian has a rich symmetry structure, $H_{\rm E} = \bigoplus_a H_{\rm E}^{(a)}$, where $a$ label superselection sectors. By accounting for all these symmetries, Ref.~\cite{mueller2022thermalization} observed signatures of quantum chaos in the distribution of time-dependent entanglement-Hamiltonian spectra. 

More specifically, starting from an initial unentangled product state, a rapid growth in entanglement was observed under time evolution. Simultaneously, the distribution of entanglement gap ratios $P(r)$ evolved from an initial state of no level repulsion, eventually saturating to a regime of level repulsion within dimensionless time scales typically on the order of one. Here, the entanglement gap ratio is defined as~\cite{OganesyanHuse}
\begin{align}\label{eq:gapratio}
r_\lambda \coloneq \frac{\min(\delta_\lambda, \delta_{\lambda-1})}{\max(\delta_\lambda, \delta_{\lambda-1})},
\end{align}
where $\delta_\lambda \coloneq \xi_{\lambda} - \xi_{\lambda-1}$, and $p_\lambda \coloneq e^{-\xi_\lambda}$ and $|\lambda_A \rangle$ are the Schmidt coefficients and vectors of $\hat{\rho}_S$ from \Eq{eq:schmidtcoeff}, respectively:
\begin{align}\label{eq:gapratio}
\hat{\rho}_A = \sum_{\lambda} e^{-\xi_\lambda} | \lambda_A \rangle \langle \lambda_A|\,.
\end{align}
At later times, the approach to thermalization is marked by a self-similar, universal form of the Schmidt spectrum, $p_\lambda(t)$, given by $p_\lambda (t) = \tau^{-\alpha} P(\tau^\beta \lambda)$, where $\tau = gt$ is dimensionless and
 $P(x)$ is a universal function, and the scaling exponents are $\alpha = 0.8 \pm 0.2$ and $\beta = 0.0 \pm 0.1$. This qualitatively mimics Kolmogorov wave turbulence~\cite{nazarenko2011wave}, a classical example of universality in which, for instance, classical plasmas exhibit similar momentum-space scaling of one-body densities, enabling rapid transport of momentum and energy across scales and thus driving thermalization~\cite{berges2014jhep}.  Finally, at later times, the system reaches thermal equilibrium, as indicated by the saturation of the von Neumann entanglement entropy to a value consistent with a Gibbs ensemble. This occurs, nonetheless, at scaled times much larger than one.
\begin{figure*}[t!]
    \centering
    \includegraphics[width=0.995\linewidth]{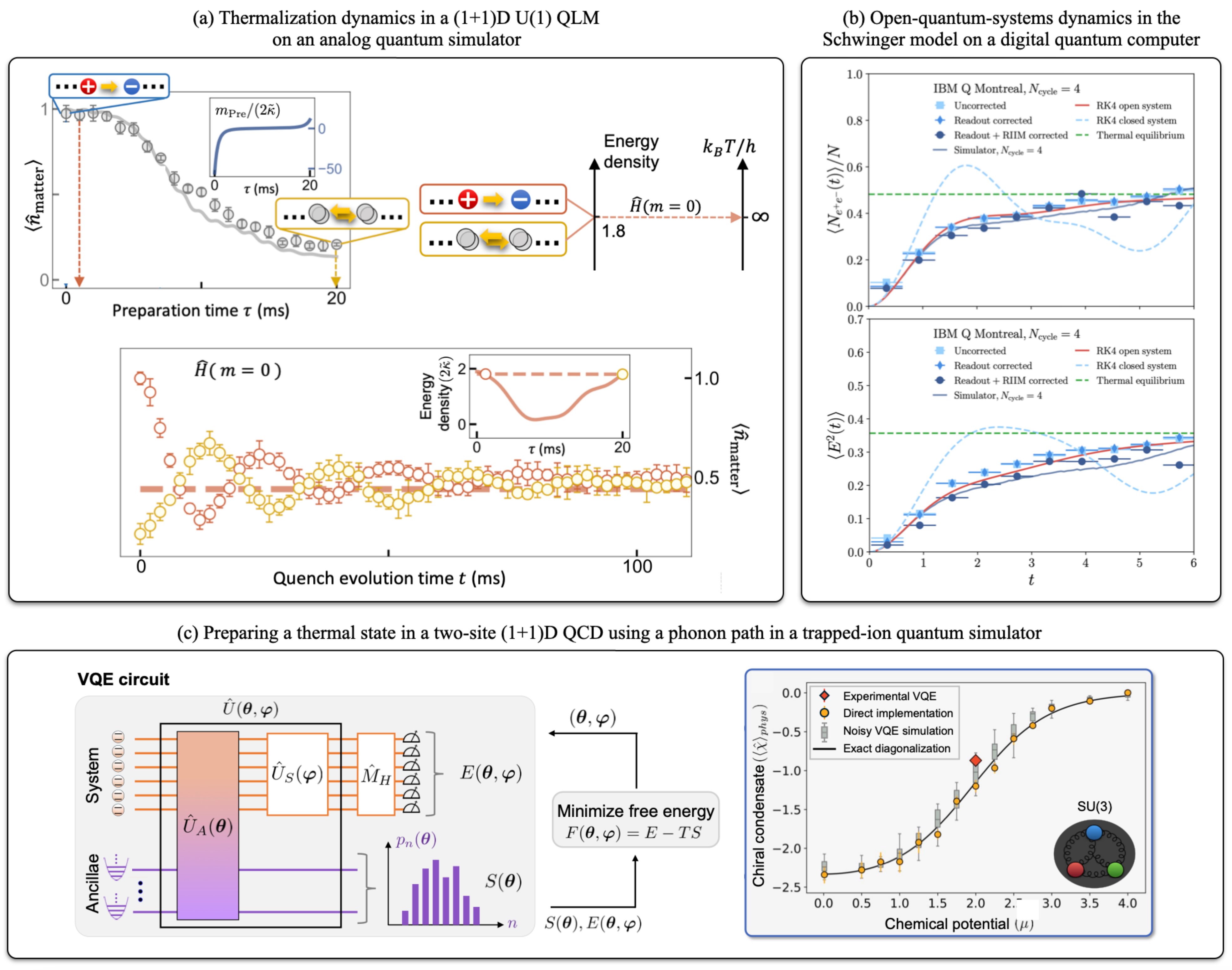}
    \caption{(a) Analog quantum simulation of gauge-theory thermalization using a 71-site Bose--Hubbard quantum simulator that emulates a $(1+1)$D spin-$\frac{1}{2}$ U$(1)$ QLM. First, an adiabatic ramp of the mass is performed in the top-left panel, taking a ``fully matter-filled'' state to almost ``matter-empty'' state with preparation time $\tau$ and corresponding mass parameter $m_\text{Pre}/(2\tilde\kappa)$, as shown in the inset. Two initial states with equal energy density are then chosen, according to the top-right panel. All energy densities are plotted with respect to the ground state of the evolution Hamiltonian. The evolution is then quenched and the time dependence of the volume-averaged local charge density is monitored in the lower panel. The resulting observable values in the steady state are then compared to a canonical thermal ensemble (dashed lines) whose temperature is determined from the energy density. The insets show the energy density evolution during state preparation, and the circles mark the chosen initial states. Figure is reproduced from Ref.~\cite{zhou2022thermalization}.
    (b) Open-quantum-system dynamics of the Schwinger model weakly coupled to a bath of scalar fields via a Yukawa coupling. Shown are the particle-antiparticle pair density $\langle \hat N_{e^+e^-}\rangle$ normalized by the system size $N=2$ (top) and the electric-energy density $\langle \hat E^2 \rangle$ (bottom) as a function of simulation time $t$, obtained using IBM's \texttt{montreal} quantum processor. Various noise-mitigation techniques are applied, as explained in Ref.~\cite{de2022quantum}, from which the Figure is reproduced. (c) The Figure illustrates a scheme that leverages phonon degrees of freedom in trapped-ion devices as an ancillary register to variationally prepare thermal states of system qubits. Shown on the right is a quantum computation of the chiral condensate, an order parameter of chiral symmetry breaking, as a function of temperature for QCD in $(1+1)$D. Figure is reproduced from Ref.~\cite{than2024phase}.} 
    \label{fig:fig_zhou_thermalization}
\end{figure*}
%

%%%%%%%%%%%%
%%%%%%%%%%%%
\subsection{Quantum simulating gauge-theory thermalization}

Despite substantial progress in quantum simulations of LGTs---both digital and analog, only a few experiments have directly addressed their thermalization dynamics. While many $(1+1)$D  models can be mapped onto analog platforms that allow continuous-time evolution, models that cannot, particularly in higher dimensions, require digital simulation. In these cases, time evolution is typically implemented via Trotterization, which limits simulations to short timescales due to circuit-depth constraints. Early-time signatures of thermalization are, therefore, are more suitable targets in the near term.

%%%
%%%
\subsubsection{Thermalization dynamics in a $(1+1)$D U$(1)$ QLM}
An early example of an analog quantum simulation of thermalization in a gauge theory is that presented in Ref.~\cite{zhou2022thermalization} using a  71-site optical superlattice. The dynamics studied are governed by the Hamiltonian of a $(1+1)$D U$(1)$ QLM; see \Eq{eq:U1QLM} with $\chi=0$. The model, as described in Sec.~\ref{sec:opticalsuperlattice}, is
a tilted Bose--Hubbard Hamiltonian with a staggered potential, where bosonic degrees of freedom are mapped to fermionic ones using a Jordan--Wigner transformation. The gauge symmetry can be controlled by tuning the parameters of the Bose--Hubbard model and selecting the initial state. By performing quantum quenches over a range of initial conditions and model parameters, local-observables' evolution toward stationary values can be studied.

A selection of the results in this study are displayed in \Fig{fig:fig_zhou_thermalization}(a). 
The top-left panel shows the initial-state preparation step, where an adiabatic change of the mass term is performed from a ``matter-filled'' to an almost ``matter-empty'' state with  mass parameter $m_{\rm Pre}/(2\tilde\kappa)$ depending on the preparation time $\tau$. 
Here, two distinct gauge-invariant initial states are considered such that they have the same energy, as shown in the top-right panel. To prepare such states, the simulation starts in the fully matter-filled initial state, which is the ground state in the limit of infinite mass. The mass is then adiabatically ramped. At select points along the ramp, one may stop, obtaining the initial state for the subsequent quench experiment. The quench is performed with a Hamiltonian at $m=0$. The matter density is then measured after the quench. Results for this quantity averaged over 71 sites are shown in the bottom panel as a function of time elapsed after the quench, and are compared to a canonical ensemble (dashed lines) to which they converge. These results demonstrate the approach of the chosen local observable toward its stationary thermal value. This value is insensitive to the initial-states' microscopic details, and is only determined by the finite energy of the state. In a follow-up experiment in Ref.~\cite{WangQMBSCriticality}, the focus was put on an Ising quantum phase transition in the aforementioned model, via studying thermalization dynamics, from a $\mathbb{Z}_2$-gauge-invariant initial state, across a quantum critical point. 

These studies highlight the unique advantage of analog quantum simulators in thermalization studies: they enable continuous time evolution, and can access asymptotically late times. They, nonetheless, are still limited to simpler gauge theories with Abelian dynamics and lower dimensionality.

%%%
%%%
\subsubsection{Thermalization, thermodynamics, and open-quantum dynamics} 

An alternative approach to considering the thermalization of an \textit{isolated} quantum system is studying \textit{open quantum systems}---both from a dynamics point of view, i.e., how open quantum systems are driven to equilibrium, but also from a practical point of view, i.e., how ancillary degrees of freedom can be used as a reservoir to prepare thermal states. We will discuss first steps in both directions in the context of LGTs

An open-quantum-systems approach is pursued in Ref.~\cite{de2022quantum} where, rather than considering unitary evolution, a Lindblad evolution in the U$(1)$ lattice Schwinger model was simulated. To simulate a thermal bath, the Schwinger model was
weakly  coupled to a scalar field $\hat\phi$ 
using a Yukawa-type interaction:
\begin{align}
    \hat H_{\rm Yukawa}^{(1+1)\text{D}}= \lambda_{\rm Yukawa} \sum_\ell \hat\phi_\ell \hat\psi^\dagger_\ell \hat\psi_\ell
\end{align}
in the quantum Brownian-motion limit~\cite{schieve2009quantum}, yielding a Schr\"odinger-picture Lindblad equation for the system's density matrix. The nonunitary evolution is mapped onto a quantum algorithm using the Stinespring dilation theorem~\cite{stinespring1955positive}, utilizing ancilla qubits~\cite{nielsen2010quantum,cleve2016efficient}. The algorithm is then run on IBM's \texttt{montreal} quantum processor, using $N=2$ system qubits and one ancillary qubit. Results are shown in \Fig{fig:fig_zhou_thermalization}(b). Plots display the evolution of the number density of fermion/antifermion pairs in the top panel and the average electric-field squared in the bottom panel. These local observables asymptote to the expected thermal values at late times.

In another direction, Ref.~\cite{than2024phase}, demonstrates the use of ancillary degrees of freedom---native to a quantum computing platform---to prepare thermal states of system qubits. The study investigates the phase diagram of (two-color SU$(2)$ and three-color SU$(3)$) QCD in 1+1D, leveraging the controllability of phonon modes in a trapped-ion quantum computer alongside qubit degrees of freedom encoded in the ions’ internal states. This is feasible because a trapped-ion device realizes phonon excitations in three dimensions, of which only one is required for gate operations, leaving the others available as motional ancillae~\cite{leibfried2003quantum}.

Following the Kogut--Susskind Hamiltonian for SU$(N_c)$ LGTs in $(1+1)$D [akin to \Eq{eq:SchwingerKS} with $\theta=0$ and upon integrating out the gauge degrees of freedom akin to \Eq{eq:SchwingerGaugeFieldsIntegratedOut}], the study focuses on a single hypercell comprising two staggered fermion sites and their connecting gauge-link Hilbert space. Each ion encodes a qubit in its hyperfine-split electronic ground state. Phonon states along the radial $x$ direction are used for gate operations, while phonons in the radial $y$ direction serve as an ancilla register. This ancilla register is used to prepare Gibbs states of the system qubits according to the Kogut--Susskind Hamiltonian Hamiltonian. The Gibbs-state preparation employs a VQE protocol, as illustrated in the left panel of \Fig{fig:fig_zhou_thermalization}(c): A unitary circuit $\hat{U}_A(\boldsymbol{\theta})$, parametrized by a set of variational parameters $\boldsymbol{\theta}$, entangles the system and ancilla registers, thereby tuning a probability distribution $\tilde{p}_j(\boldsymbol{\theta})$ over bitstrings $| j \rangle = | j_1 \dots j_N \rangle$ of the $N$ system qubits. A second unitary, $\hat{U}_S(\boldsymbol{\varphi})$, acts on the qubit register to create approximate energy eigenstates. In a feedback loop, the parameters $(\boldsymbol{\theta}, \boldsymbol{\varphi})$ are iteratively updated via classical optimization to minimize the free-energy-like functional
\begin{align}
\langle \hat H \rangle - T \langle \hat{S} \rangle, 
\end{align}
where $\langle \dots \rangle$ denotes expectation values with respect to $\hat{\rho}(\boldsymbol{\theta},\boldsymbol{\varphi})$, and $\hat{S}$ is the entropy inferred from measurements of the ancilla register, and $T$ the temperature. At the conclusion of the variational search, the Gibbs state is prepared, after which gauge-invariant measurements are performed using a projection procedure.

The quantum-computed phase diagram, i.e., the chiral-condensate order parameter as a function of temperature, is displayed in the right panel of \Fig{fig:fig_zhou_thermalization}(c) for QCD in $(1+1)$D, showing clearly a transition from a chirally broken phase at low temperatures to the chirally symmetric phase at large temperatures. While this computation, for two lattice sites, clearly can be classically emulated, it nevertheless demonstrates the basic ingredients necessary to quantum compute phase diagrams of non-Abelian gauge theories.

The use of motional ancillae represents a significant advancement, as it effectively extends the accessible Hilbert space of trapped-ion quantum simulators---an important consideration given the inherent size limitations of ion chains. The ability to control motional modes constitutes a crucial step toward employing this platform for studies of thermalization and quantum thermodynamics, functioning as a tailored heat bath, even if fully universal control of those modes has not yet been achieved.

Together these references show that thermalization and thermodynamics can be studied in ways other than following the nonequilibrium evolution of isolated quantum systems. Similar ideas, although mostly theoretical, of preparing thermal states have been explored in recent years; see, e.g. Refs.~\cite{brandao2019finite,lu2021algorithms,motta2020determining,temme2011quantum,bilgin2010preparing,zhu2020generation,verdon2019quantum,chowdhury2020variational,schuckert2023probing} and in particular Refs.~\cite{davoudi2023towards,ikeda2024real,chen2024minimally,ballini2024quantum,zhang2025quantum,cheng2025variational} for LGT applications.

%%%
%%%
\subsubsection{Universal thermalization dynamics in a $(2+1)$D $\mathbb{Z}_2$ LGT.}
\begin{figure*}[t!]
    \centering
    \includegraphics[width=0.985\linewidth]{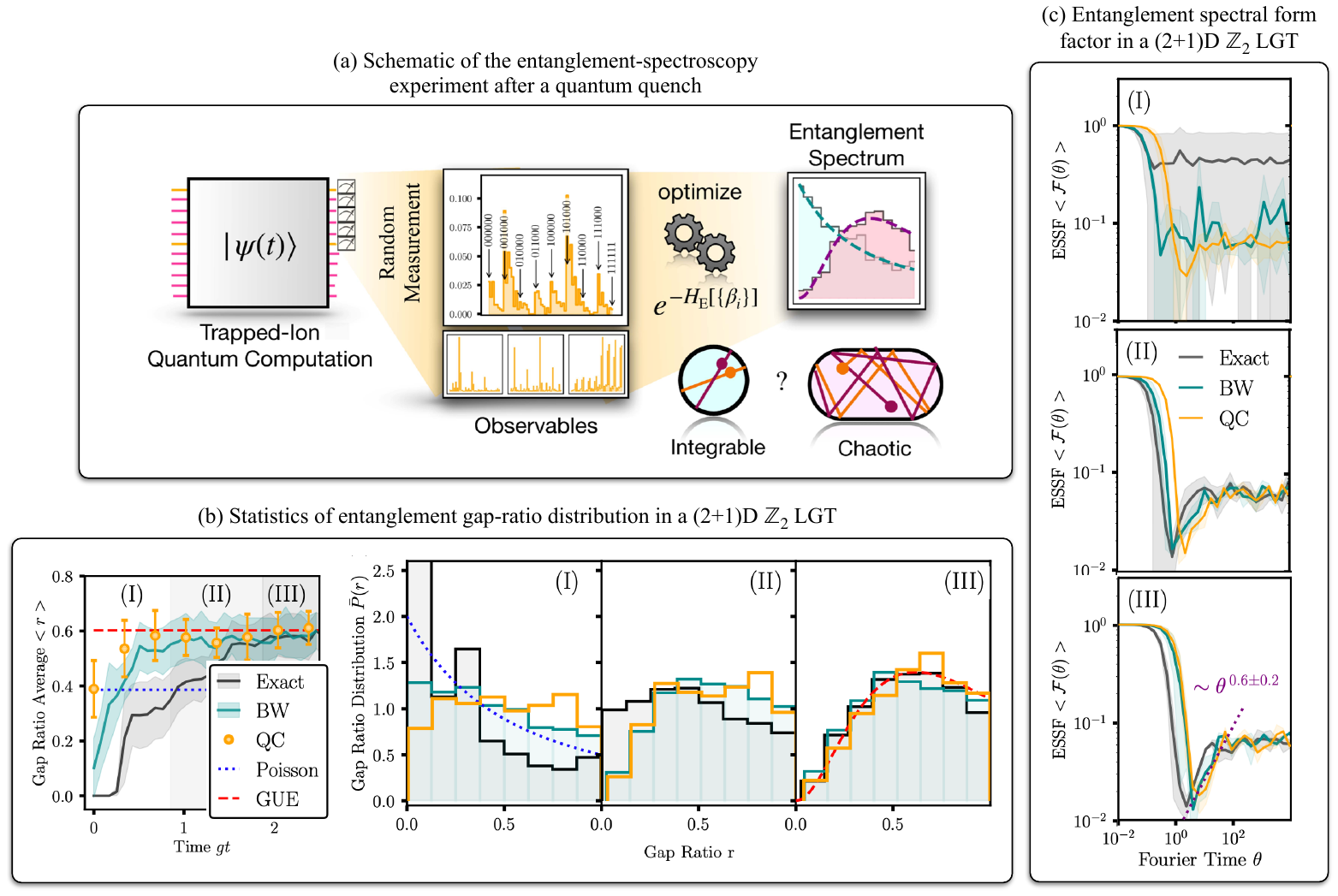}
    \caption{
    (a) A trapped-ion system consisting of 15 optically controlled ions is used to digitally quantum compute time evolution of a $(2+1)$D $\mathbb{Z}_2$ LGT, starting from randomly drawn gauge-invariant product states. A random-measurement-based tomography protocol consists of measuring the probabilities of bitstrings in a random basis. From this, the protocol extracts a classical approximation of a reduced density matrix, parameterized in terms of an entanglement Hamiltonian, $H_{\rm E}$. The statistical properties of the entanglement Hamiltonian are then studied. (b) The average gap ratio, i.e., the averaged distribution of \Eq{eq:gapratio} across the spectrum of $H_{\rm E}$ and averaged over initial states and symmetry sectors of the reduced density matrix, shows a transition from initial nonrepulsive (nonchaotic) level statistics to later level repulsion indicating quantum chaos. The distributions of entanglement spectrum gap ratios reflect this behavior, showing initially a Poisson (blue dotted), transitioning through a transient phase, and ultimately converging to a distribution consistent with a Gaussian Unitary Ensemble (GUE). (c) The averaged entanglement spectral form factor, \Eq{eq:defESFF}, shows the build-up of a ramp-plateau feature indicating the emergence of quantum chaos. Figure is reproduced from Ref.~\cite{mueller2025quantum}.
    } 
    \label{fig:fig_thermalization_EHT}
\end{figure*}
Thermalization dynamics of a $(2+1)$D $\mathbb{Z}_2$ LGT [with the Hamiltonian in Eq.~\eqref{eq:Z2Hamiltonian}] was quantum-simulated in Ref.~\cite{mueller2025quantum} using a digital trapped-ion quantum computer. This work utilized random-measurement based~\cite{brydges2019probing,huang2020predicting,elben2023randomized} entanglement-Hamiltonian tomography~\cite{dalmonte2018quantum,kokail2021entanglement} to look for early signatures of quantum chaos, hence thermalization, in the subsystem's entanglement spectrum in a nonequilibrium state. Such a state was created by quenching the evolution from an initial gauge-invariant product state. As illustrated in \Fig{fig:fig_thermalization_EHT}(a), a randomized measurement protocol can be implemented to approximate the reduced density matrix of a subsystem within a thermalizing quantum state. This protocol parameterizes the reduced state in terms of an entanglement Hamiltonian $H_{\rm E}$, defined in \Eq{eq:EntanglementHamiltonian}, which is optimized to reproduce a set of random measurements obtained from experiments. A crucial ingredient is inspired by the Bisognano-Wichmann (BW) theorem~\cite{bisognano1975duality,bisognano1976duality}: The entanglement Hamiltonian is composed of finitely many local terms, allowing for an efficient parameterization, i.e., using only polynomially many parameters as function of subsystem size, which can be constrained by the available data.

 Specifically, the experiment determined the distribution of entanglement gap ratio, defined in \Eq{eq:gapratio}, average gap ratio over given times, as well as the entanglement spectral form factor (ESSF)~\cite{chang2019evolution}, defined as 
\begin{align}\label{eq:defESFF}
    \mathcal{F}(\phi) \coloneq \Big\langle \frac{1}{\mathcal{R}^2} \sum_{\lambda,\lambda'} e^{i\phi[\xi_\lambda -\xi_{\lambda'} ]} \Big\rangle\,,
\end{align}
Here, $\mathcal{R} \coloneq \lim_{\alpha \rightarrow 0} \exp\{ \frac{1}{1-\alpha} \log(\sum_\lambda e^{- \alpha\xi_\lambda})\}$ is the rank of $H_{\rm E}$, and $\langle \cdot \rangle$ is the average over statistically similar states. A ramp-plateau feature in this latter quantity indicates quantum chaos.
The results for these quantities are shown in \Fig{fig:fig_thermalization_EHT}(b,c). 
Figure~\ref{fig:fig_thermalization_EHT}(b) points to a transition from initial nonrepulsion of the levels, consistent with a Poisson distribution, to level repulsion at later times, consistent with a Gaussian unitary ensemble (GUE). This behavior in a gauge theory is found to be in agreement with a general scenario as outlined in Ref.~\cite{chang2019evolution}, and theoretically for the same LGT with Ref.~\cite{mueller2022thermalization}.
Figure~\ref{fig:fig_thermalization_EHT}(c) shows the ESSF, obtained from the same data. These data demonstrate a transition toward a ramp-plateau structure at late times that indicates quantum chaos. 

The above results, which follow the first experimental measurement of an entanglement Hamiltonian in Ref.~\cite{kokail2021entanglement}, constitute a first step toward experimentally probing quantum thermalization via constraining the entanglement Hamiltonian. They demonstrate that universal features of the entanglement spectrum can be reliably extracted in quantum-simulation experiments. Such features, importantly, emerge at early times accessible in current digital quantum computers.

%%%%%%%%%%%%
%%%%%%%%%%%%
\subsection{Open questions}

As reviewed in this Section, quantum-simulation experiments have begun to explore the thermalization dynamics of LGTs, progressing from analog platforms to digital quantum computers. 
Nonetheless, simulating thermalization in gauge theories of the Standard Model, for now, remains somewhat distant, requiring scalable fault-tolerant quantum computers. In the meantime, simpler gauge theories can offer valuable insights into thermalization processes, with potential input for QCD and early-universe phenomenology. 

As simulations scale up, they can begin to resolve the underlying mechanisms and time scales of thermalization in gauge theories. Importantly, while quantum simulation has so far primarily addressed Abelian LGTs, the next step are non-Abelian LGTs. Concepts from quantum-information theory, such as entanglement, have opened new avenues to tackle these challenges, while new techniques and interdisciplinary connections between high-energy and nuclear physics, condensed-matter theory, and atomic, molecular, and optical physics have sparked novel and pressing questions.

One central question is the extent to which the ETH governs thermalization in gauge theories, especially for non-Abelian LGTs and in high-energy-physics experiments. Closely related is the role of quantum chaos, with many diagnostic tools to probe chaos, ergodicity, and information scrambling, being largely unexplored in the context of LGTs.
Another promising direction concerns the computational complexity of quantum many-body states. While it is strongly suspected—though not proven—that thermalization is classically hard to simulate, the precise observables and regimes for which this holds remain unclear. While entanglement generated during thermalization ultimately presents significant challenges to, e.g., classical tensor networks, alternative techniques~\cite{angrisani2024classically,schuster2024polynomial} may offer insights for specific observables and models. To address this, measures such as nonstabilizerness~\cite{gottesman1997stabilizer,bravyi2005universal} or non-Gaussianity~\cite{zhuang2018resource} not only quantify the quantum resources needed for simulating LGTs on quantum devices, but may also provide insights into thermalization itself~\cite{liu2022many,leone2022stabilizer,turkeshi2025pauli}.

More broadly, under what conditions a system can evade thermalization under time scales polynomial in system size, and whether gauge-theory dynamics exhibit distinct behaviors, remain intriguing questions, which we discuss in more detail the following Section. 

%%%%%%%%%%%%%%%%%%%%%%%%%%%%%%%%%%%%%%%%%%%%%%%%%%%%%%%%%%%%%%%%%%%%%%%%%%%%%%%%%%%%%%%%%%%%%%%%%%%%%%%%%%%%%%%%%%%%%%%%%%%%%%%%%%%%%%%%%%%%%%%%%%%%%%%%%%%%%%%
\section{Ergodicity breaking in gauge theories}\label{sec:ergodicitybreaking}

\begin{figure*}[htb]
    \centering
    \includegraphics[width=0.995\linewidth]{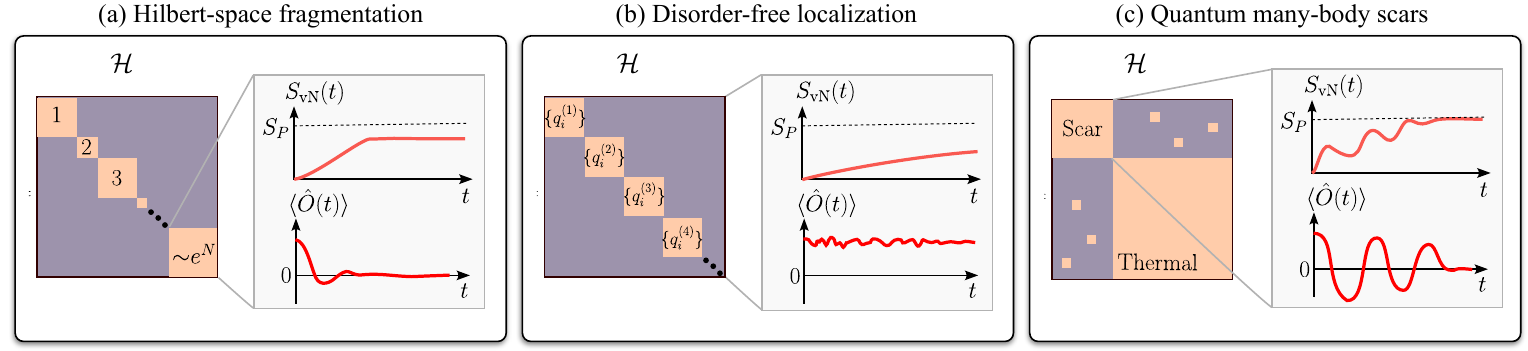}
    \caption{
    (a) Hilbert-space fragmentation leads to the Hilbert space $\mathcal{H}$ breaking up into dynamically disconnected sectors, whose number can scale exponentially with the number of degrees of freedom $N$. The sectors differ in their size, ranging from one-dimensional to exponentially large in $N$. The latter can exhibit thermalizing dynamics, e.g., a local observable $\hat O$ quickly relaxes to its equilibrium value and entanglement entropy $S_\mathrm{vN}$, Eq.~(\ref{eq:vonNeumann_Renyi}), grows linearly with time. However, due to fragmentation, $S_\mathrm{vN}$ may not reach the Page bipartite entropy $S_P{=}(N/2)\ln 2{-}1/2$ of a random state of $N$ qubits~\cite{Page1993}. (b) In disorder-free localization, the sectors are of equal size and depend on the configuration of background charges $\{q_i\}$---the eigenvalues of the gauge-symmetry generators.  For initial states that are equally distributed among the sectors, the system exhibits a breakdown of thermalization akin to many-body localization: entanglement spreads slowly and local observables never relax to zero. (c) In systems hosting quantum many-body scars, most initial states undergo thermalizing dynamics, similar to panel (a). However, if the system is prepared in special initial states that belong to the ``scar'' subspace, the dynamics of entanglement and local observables exhibit regular oscillations, while thermalization is significantly delayed. }
    \label{fig:ergbreak}
\end{figure*}

Generic relaxation behavior of isolated quantum many-body systems at a nonzero energy density has been established---via a plethora of numerical and experimental examples---to follow the ETH paradigm, as discussed in Sec.~\ref{sec:thermalization}. The presence of gauge symmetries, however, can fundamentally alter this behavior in some cases, such that thermalization is significantly delayed or potentially avoided altogether---a type of behavior broadly termed \emph{ergodicity breaking}. This Section focuses on three paradigms of ergodicity breaking in LGTs, which are intimately tied to the structure of the physical Hilbert space constrained by gauge symmetries---see Fig.~\ref{fig:ergbreak} for a summary of their main phenomenology.

Gauge invariance introduces nontrivial dynamical constraints that can ``shatter'' the Hilbert space into disconnected sectors, resulting in Hilbert-space fragmentation (HSF), Fig.~\ref{fig:ergbreak}(a). The fragmentation, typically inherited from Gauss’s law or conserved local charges, prevents the system from exploring the full set of allowed configurations, which can impact thermalization and leave some nonergodic signatures in the dynamics, even in the absence of integrability or disorder. 

Another paradigm of ergodicity breakdown is disorder-free localization (DFL), in which translationally invariant states of matter and gauge fields correspond to superpositions of effectively disordered charge sectors. The averaging over different charge sectors can significantly inhibit the spreading of correlations and entanglement, similar to Anderson localization of a single particle in a disordered potential
and its interacting cousin---the many-body localization (MBL); see Fig.~\ref{fig:ergbreak}(b). 
Unlike the latter, which rely on spatial randomness, DFL originates in the correlated background of conserved charges or electric-flux configurations, which can occur in spatially uniform LGTs. 

A third paradigm of ergodicity breaking are quantum many-body scars (QMBSs), where a small subspace of nonthermal eigenstates coexists with an otherwise thermal spectrum, Fig.~\ref{fig:ergbreak}(c). QMBSs have been explored prominently in models like the PXP model, which can be mapped onto a spin-$\frac{1}{2}$ U$(1)$ QLM after integrating out matter fields. A common feature of QMBS systems is their extreme sensitivity to the initial condition: while for generic initial states, QMBS systems undergo chaotic dynamics, for special initial conditions they can exhibit coherent dynamics in the form of persistent quantum revivals. Thus, QMBSs provide an example of ``weak'' ergodicity breaking, bridging between full thermalization and its complete breakdown as in DFL.  

In this Section, we discuss in more detail the above three paradigms of quantum nonergodic behavior, which lead to the breakdown of the ETH and the assumptions of quantum statistical mechanics in gauge-theory settings. We note that quantum many-body dynamics and ergodicity breaking are actively developing fields and we do not aim to provide their exhaustive review; rather, we seek to highlight the role played by LGTs as prime settings for studying different forms of ergodicity breaking that stem from local conservation laws. As will become apparent below, different forms of ergodicity breaking are not mutually exclusive and can indeed coexist in many models of LGTs, producing a rich spectrum of behaviors far from equilibrium. We will return to the implications and open questions related to all the different forms of ergodicity breaking in LGTs at the end of this Section.

%%%%%%%%%%%%
%%%%%%%%%%%%
\subsection{Hilbert-space fragmentation}

To formally define HSF,  illustrated in Fig.~\ref{fig:ergbreak}(a), one needs to introduce the notion of a Krylov subspace $\mathcal{S}$: the set of all vectors obtained by the repeated action of the Hamiltonian $\hat H$ on some ``root'' vector $\left|\Psi_0\right\rangle$:
\begin{equation}\label{eq:krylovsubspace}
\mathcal{S}(\hat H,\left|\Psi_0\right\rangle) \coloneq \operatorname{span}\left\{\left|\Psi_0\right\rangle, \hat H\left|\Psi_0\right\rangle, \hat H^2\left|\Psi_0\right\rangle, \cdots\right\}.    
\end{equation}
By definition, $\mathcal{S}$ is closed under the action of $\hat H$, hence the dynamics initialized in $|\Psi_0\rangle$ remain confined to $\mathcal{S}$ for all subsequent times.

In many systems, including various LGTs, the subspace $\mathcal{S}(\hat H,\left|\Psi_0\right\rangle)$ can further decompose into dynamically disconnected sectors beyond conventional symmetry-invariant blocks. In other words, even after resolving all symmetries,  $\mathcal{S}\left(H,\left|\Psi_0\right\rangle\right)$ may not span the entirety of states with the same quantum numbers as $\left|\Psi_0\right\rangle$. Thus, HSF can be formally stated as
\begin{equation}\label{eq:HSFdef}
\mathcal{H}=\bigoplus_{\mathbf{s}} \bigoplus_{i} \mathcal{S}(\hat H,|\Psi_i^{(\mathbf{s})}\rangle),    
\end{equation}
where $\mathbf{s}$ labels the symmetry quantum numbers and $i$ labels the Krylov subspaces generated from root states $|\Psi_i^{(\mathbf{s})}\rangle$ belonging to the symmetry sector $\mathbf{s}$. 

For a review of HSF, we direct the reader to Ref.~\cite{MoudgalyaReview}; here, we only recall a few key properties. Generally, one can distinguish between ``strong" and ``weak" HSF, depending on whether or not the ratio of the largest Krylov subspace to the Hilbert space within a given global symmetry sector vanishes in the thermodynamic limit. 
Most saliently, different fragments can exhibit vastly different dynamical properties. For example, some fragments (even though exponentially large in system size) may be integrable, while others may be nonintegrable. Furthermore, it is important to distinguish exact from approximate HSF. In the exact case,
the fragments are exactly decoupled from one another. This can occur due to an `intertwined' action of symmetries, e.g., charge and dipole-moment conservation~\cite{Sala2019,Pretko2019HSF,Khemani2019_2,Moudgalya2019,MoudgalyaCommutant}, such that even within a single sector, invariant under the action of all symmetry generators, there is an additional emergent block-diagonal structure of the Hamiltonian.  
In the approximate HSF, the fragments are simply the energy minibands and their coupling is controlled by the ratio of certain terms in the Hamiltonian. A classic example is the Hubbard model, Eq.~(\ref{eq:bhm}), in the regime where the interaction $U$ is much larger than the hopping $J$. In this case, the subspaces are typically reconnected in second (or higher) order in $J/U$ in perturbation theory, hence the dynamics may eventually escape the initial fragment after a sufficiently long time. 

Examples of HSF abound in LGTs. In a $(1+1)$D U$(1)$ LGT coupled to dynamical fermions, where Gauss's law enforces a local constraint that forbids certain hopping processes, HSF can give rise to ``bubble'' states that are completely inert and others with configurations with restricted mobility~\cite{Mukherjee2021HSF}. A minimal U$(1)$ LGT is constructed in Ref.~\cite{Hu2024HSF} with local conserved charges directly identified with Gauss's law generators; these charges exactly fragment the Hilbert space into invariant subspaces, some of which reduce to the U$(1)$ QLM. In Ref.~\cite{Khudorozhkov2022HSF}, a nonintegrable $(2+1)$D quantum spin-$\frac{1}{2}$ model with subsystem U$(1)$ symmetries is shown to exhibit HSF due to conservation of magnetization along every row and column. Finally, the Kogut--Susskind formulations of LGTs in $(d+1)$ dimensions naturally give rise to HSF~\cite{ciavarella2025generic}. Working in the electric-field (i.e., group-representation) basis, one can perform a perturbative expansion in terms of the irreducible representations of the gauge group SU$(2)$, systematically including contributions up to an error of order $1/(g^6\varepsilon)$, where $g$ is the gauge coupling and $\varepsilon$ is a cutoff defined in terms of the group's Casimir. Thus, in the framework of Ref.~\cite{ciavarella2025generic}, the Hilbert space decomposes into dynamically disconnected subspaces, each associated with a fixed set of group representations. It has recently been shown that HSF can be leveraged in Kogut--Susskind LGTs to estimate the size of truncation errors in the group-representation basis \cite{ciavarella2025truncation}, highlighting the usefulness of HSF in making predictions related to the quantum-field-theory limit of LGTs.

In contrast to the abundance of theoretical evidence for HSF, direct experimental probes of HSFs in LGTs are currently lacking. Recent experiments on general Fermi- and Bose--Hubbard models, i.e., not in the regime where they map to LGTs,  have probed signatures of HSF by sampling the dynamics for various initial states belonging to different fragments~\cite{Scherg2020,Adler2024HSF,Honda2025HSF,Loh2025}. While the observed dynamics clearly reflected the different nature of the fragments, in practice one cannot exhaustively sample all the fragments as their number may be exponential in system size. This highlights the need for sharper experimental probes of HSF that extend beyond sampling the fragments and checking the dynamics of initial states residing in them. A particular challenge in this context is the possibility of ``quantum'' HSF, whereby the root state in Eq.~(\ref{eq:HSFdef}) is an entangled state in the computational basis. While there has been some progress for certain classes of Hamiltonians~\cite{regnault2022integercharacteristicpolynomialfactorization}, an efficient method for diagnosing quantum HSF is an important open problem, both in theory and experiment.   

%%%%%%%%%%%%
%%%%%%%%%%%%
\subsection{Disorder-free localization}

A defining feature of LGTs is the presence of local symmetries, which result in an extensive number of conserved quantities. The latter are the eigenvalues, or `charges,' of the local generators of the gauge symmetry. In standard LGT formulations in the context of elementary particle physics, Gauss's law requires a projection onto a particular sector, e.g., the zero charge sector, fixing a subspace of gauge-invariant physical states~\cite{kogut1975hamiltonian}. Here, instead, we take a different approach, focusing on {\it unconstrained} gauge theories~\cite{prosko2017simple}, in which the physics is not constrained to a single gauge or charge sector, and states can also be in superposition over many charge sectors. This entails that gauge-field link operators become proper physical degrees of freedom (i.e., to not be integrated out as is the case in $(1+1)$D LGTs). Unconstrained LGTs have been employed in numerical quantum Monte Carlo (QMC) studies where the Gauss's law emerges spontaneously at low temperatures~\cite{assaad2016simple}. Most importantly, this fresh perspective leads to new physics and the nonequilibrium phenomenon of DFL, i.e., Fig.~\ref{fig:ergbreak}(b).  

%%%
%%%
\subsubsection{Basic idea of DFL in a $\mathbb{Z}_2$ LGT}\label{sec:DFL_1}
 
The exactly soluble $\mathbb{Z}_2$ LGT in $(1+1)$D provided the first and simplest setting for DLF~\cite{Smith2017a}. To explain the main phenomenology of DFL, we consider a $1$d chain of spinless fermions $\hat f_i$ coupled with spin-$\frac{1}{2}$ operators $\hat \sigma^{\alpha}_{i,i+1}$ on the bonds, governed by the Hamiltonian 
\begin{align}
\label{eq:DFL_Z2}
	\hat{H}_{\mathbb{Z}_2}^{'(1+1)\text{D}}=&- J \sum_{i} \left(\hat f_i^{\dagger}  \hat \sigma^z_{i,i+1} \hat f_{i+1} + \text{H.c.} \right) 
    \nonumber\\
    & + h \sum_i \hat \sigma^x_{i-1,i} \hat \sigma^x_{i,i+1}.
\end{align}
Similar to Eq.~(\ref{eq:Z2LGT}), this model has a local $\mathbb{Z}_2$ symmetry, $\hat f_i \to \eta_i \hat f_i$, as well as $\sigma^z_{i,i+1} \to \eta_i \sigma^z_{i,i+1} \eta_{i+1}$ with $\eta_i=\pm 1$, but note the different form of the last term giving dynamics to the gauge link variables. There are still an extensive number of conserved charges $\hat q_i=(-1)^{\hat f_i^{\dagger} \hat f_i} \hat \sigma^x_{i-1,i} \hat \sigma^x_{i,i+1}$ with eigenvalues $\pm 1$; these operators commute with the Hamiltonian in Eq.~\eqref{eq:DFL_Z2}.

\begin{figure*}[htb]
	\centering
	\includegraphics[width=0.9\linewidth]{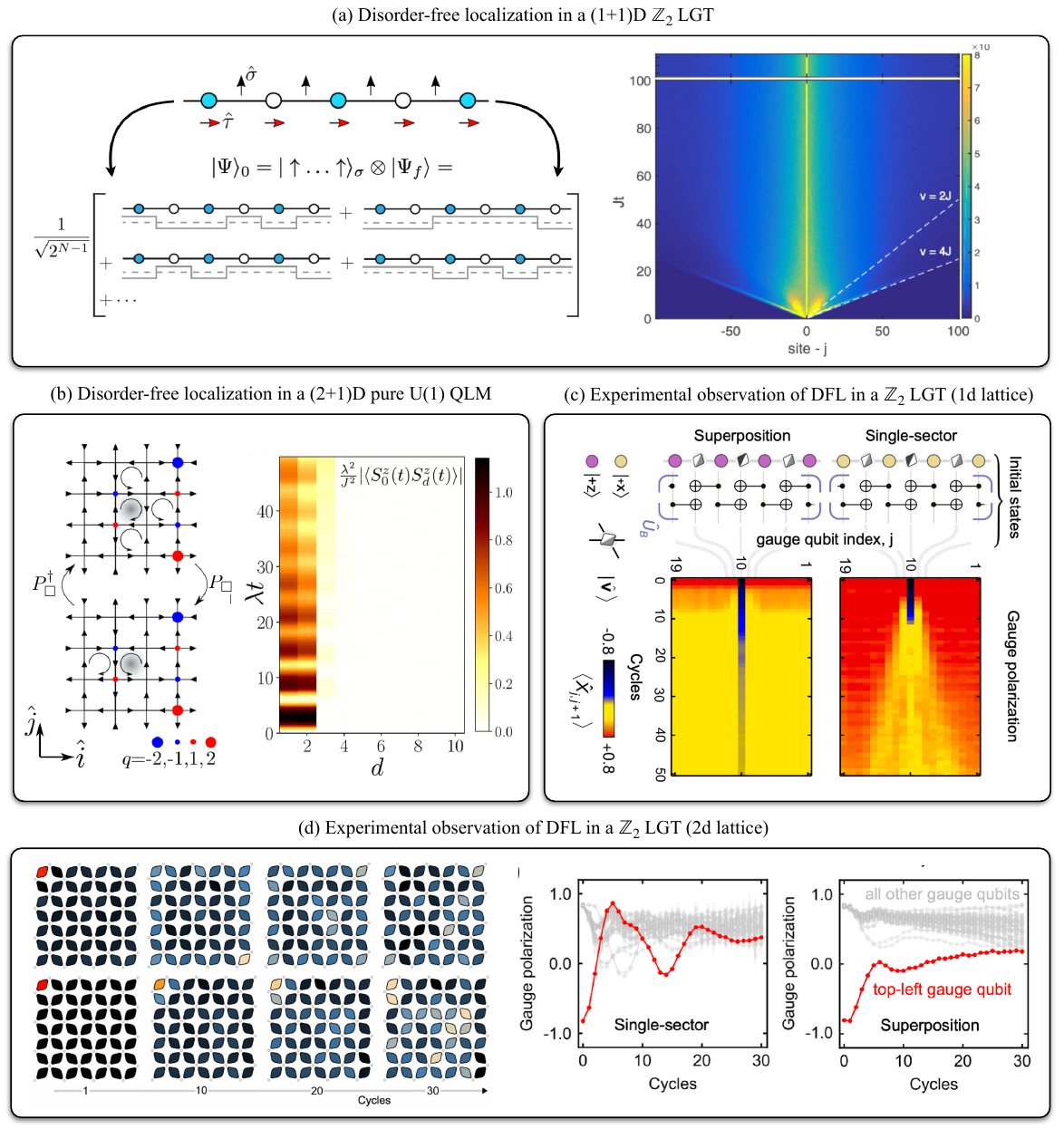}
	\caption{
    (a) The left panel displays the mapping of the translationally invariant initial state of the $(1+1)$D $\mathbb{Z}_2$ LGT in Eq.~(\ref{eq:DFL_Z2}) to a superposition over an extensive number of charge sectors. Note that each typical charge sector has a disordered static charge configuration. Figure is reproduced from Ref.~\cite{smith2018dynamical}. 
    The right panel shows the spreading of the fermionic-matter density-density correlation function following a quench from the initial state shown in the left, i.e., only even sites are occupied by fermions. Correlations quickly reach a steady state with an exponential decay away from the central site set by an emergent localization length, corroborating the DFL phenomenon. Figure is reproduced from Ref.~\cite{Smith2017a}. (b) Illustration of a $(2+1)$D U$(1)$ QLM showing DFL in a strongly interacting $(2+1)$D pure LGT described by the Hamiltonian in Eq.~\eqref{eq:DFL_U1link} and reproduced from Ref.~\cite{karpov2021disorder}. The right panel shows the halting of correlation spreading, with an emergent localization length similar to the $(1+1)$D case. (c) Experimental implementation of DFL in a modified $\mathbb{Z}_2$ LGT described by the Hamiltonian in Eq.~\eqref{eq:DFL_Z2_Google} on the \texttt{Google Willow} device. 
    In am implementation on a $1$d lattice, the correlations of the gauge field in a single sector appear `delocalized', while the superposition initial state remains `localized', revealing a signature of DFL. Figure is reproduced from Ref.~\cite{Gyawali2024}.
    (d) Similar type of dynamics as in the previous panel but on a $2$d lattice with 81 superconducting qubits. The excitation initialized at the upper left corner remains localized in the superposition initial state (lower panels) but quickly spreads in the homogeneous initial state (upper panels). Figures are reproduced from Ref.~\cite{Gyawali2024}.
	}
	\label{fig:DFL}
\end{figure*}

The key idea is that treating the bond spins as physical degrees of freedom, one can write down simple states which are superpositions of an {\it extensive} number of charge sectors. For example, a fully $z$-polarized state can be expressed as
\begin{align}
\label{eq:DFL_Z2_Istate}
|\Psi\rangle_0	&= | \uparrow \dots \uparrow \rangle_{\sigma} \otimes | \Psi_f\rangle \nonumber\\
&= \frac{1}{\sqrt{2^{N-1}}} \! \! \sum_{\{ q_i\} = \pm 1} | q_1 \dots q_N \rangle_{\sigma} \otimes | \Psi_\text{f}\rangle\,, 
\end{align}
which is illustrated in Fig.~\ref{fig:DFL}(a). Here, $\ket{\Psi_\text{f}}$ is an arbitrary state of the fermions, and subscript $\sigma$ denotes the state of the bond spins. The quench dynamics starting from this initial state is localized. This becomes transparent when block-diagonalizing the Hamiltonian in charge sectors, which is achieved with the help of the Ising-chain duality $\hat \tau^z_i = \hat \sigma^x_{i-1,i} \hat \sigma^x_{i,i+1}$, $\hat \sigma^z_{i,i+1} = \hat \tau^x_i \hat\tau^x_{i+1}$. Using the composite fermions, $\hat c_i=\hat \tau^x_i \hat f_i$, the Hamiltonian can be rewritten as
\begin{align}
\label{eq:DFL_Z2_Hcharge}
	\hat{H}_{\{ q_i\}}=- J \sum_{i} \left(\hat c_i^{\dagger} \hat c_{i+1} + \text{H.c.}\right) + 2h \sum_i q_i \hat c^{\dagger}_i \hat c_i,
\end{align}
up to a constant term. Crucially, the dynamics of each sector $\{ q_i\}$ is independent of other sectors and governed by a simple free-fermion tight-binding model in a discrete-potential background set by the conserved charge configuration. Since a typical configuration of charges in the sum of Eq.~(\ref{eq:DFL_Z2_Istate}) is disordered, the fermions are Anderson localized~\cite{Anderson58}, and the Hamiltonian in Eq.~\eqref{eq:DFL_Z2_Hcharge} describes an Anderson insulator. As a result, ergodicity is broken in a standard quench setup,
\begin{align}
\label{eq:DFL_Z2_TimeEv}
	e^{-it \hat{H}_{\mathbb{Z}_2}^{'(1+1)\text{D}}} |\Psi\rangle_0 \propto \sum_{\{ q_i\} = \pm 1}  e^{-it \hat{H}_{\{ q_i\}}} |\Psi_\text{f}\rangle\,.
\end{align}
For example, in contrast to ETH-obeying states, the fermion-density inhomogeneities do not decay and the associated correlation spreading is absent beyond a short localization length (tuneable via $h$), as seen in Fig.~\ref{fig:DFL}(b).

Remarkably, despite the localization of matter degrees of freedom, DFL leads to nontrivial dynamics of nonlocal quantities, as diagnosed by out-of-time-order correlators~\cite{smith2019logarithmic} and entanglement spreading~\cite{smith2017absence}. The origin of the latter is the dephasing of different charge sectors with respect to each other, which results in a volume-law scaling of the half-system entanglement [after a tuneable timescale set by $\tau \propto (h/J)^2$ for the $\mathbb{Z}_2$ LGT in  Eq.~(\ref{eq:DFL_Z2})]. Depending on the type of gauge theory, the approach to the long-time value of the entropy can be anomalously slow, reminiscent of strongly disordered interacting systems~\cite{Moore12,Serbyn13-2}. Similarly, the dynamics of the link variables need not be localized. Nonetheless, the explicit construction of a basis of static degrees of freedom and dynamically localized matter shows that DFL is an example of a so-called `quantum disentangled liquid'~\cite{grover2014quantum, smith2017absence}. Explicitly, the construction resembles  the conjuncture than in fluids consisting of two (or more) species of indistinguishable quantum particles with a large mass ratio, the light particles might localize on the heavy particles~\cite{grover2014quantum}. 

The example model in Eq.~(\ref{eq:DFL_Z2}) above is the simplest of its kind with an exact solution. However, we stress that DFL only relies on the local symmetry; neither 
solvability nor the mapping to Anderson insulators are required. For example, DFL has been confirmed by adding density interactions to the matter fermions of Eq.~(\ref{eq:DFL_Z2}) or more general gauge-field dynamics~\cite{smith2017absence}, like a standard transverse field for the spin variables, which spoils the locality of the duality transformation and its explicit mapping to a disordered model. 
%

%%%
%%%
\subsubsection{DFL in U$(1)$ LGT and quantum link models}\label{sec:DFL_2}

DFL also appears in more complex LGTs, for example those corresponding to continuous gauge groups such as U$(1)$~\cite{brenes2017many,Giudici2020}.
Similarly to the $\mathbb{Z}_2$ case discussed above, one can still choose simple translationally invariant initial states as product states of matter fermions and field link operators:
\begin{equation}\label{initialst}
    \begin{split}
        |\Psi\rangle_0= \big(\otimes_{n}|\tilde{E}_{n}  \rangle \big) \otimes |\Psi_\text{f} \rangle.
    \end{split}
\end{equation}
For example, in a basic spin-$1$ QLM, one can chose an equal-weight superposition of local link eigenvalues, $|\tilde{E}_{n}  \rangle  =\frac{1}{\sqrt{3}}\Big(\ket{-1}_{n}+\ket{0}_{n}+\ket{1}_{n}\Big)$. The state is a superposition of an extensive number of charges, i.e., Eq.~(\ref{eq:DFL_Z2_TimeEv}) holds.

As was shown in Ref.~\cite{brenes2017many}, matter fermions are localized, realizing DFL. The qualitatively new feature is that there is no 
solvable limit and the theory is always strongly interacting. In $(1+1)$D, one can still integrate out the gauge field but the final Hamiltonian has charge-sector-dependent long-range interactions; see Eq.~(\ref{eq:SchwingerGaugeFieldsIntegratedOut-terms}). Interestingly, the confinement substantially increases localization and leads to a very slow growth of the entanglement.

It is also possible to find an exactly solvable U$(1)$ LGT by replacing the standard term for the gauge-field dynamics $\hat{E}_{n}^2 \to (\hat{E}_{n}-\hat{E}_{n-1})^2$, where again a mapping to a free-fermion Anderson insulator can be found~\cite{papaefstathiou2020disorder}. Interestingly, the system is still fully localized even though in the U$(1)$ case, the disorder configuration is spatially correlated. Whether other exactly solvable LGTs exist, in which the emergent disorder is even more strongly correlated such that it can give rise to delocalization or mobility edges\footnote{Mobility edges refer to the existence of a critical energy that separates localized from extended states, e.g., in the 1d Anderson model with correlated disorder~\cite{izrailev2012anomalous}.} remains an open question.

%%%
%%%
\subsubsection{DFL beyond one dimension}\label{sec:DFL_3}

Both localization and LGT phenomena strongly depend on spatial dimensionality. For example, in $(1+1)$D, all eigenstates of free-fermion models with uncorrelated disorder are localized~\cite{kramer1993localization}, but in higher dimensions, mobility edges and full delocalization can appear depending on the presence of symmetries like time reversal~\cite{altland1997nonstandard}. For LGTs, 
the gauge field does not have its own independent dynamics in $(1+1)$D and can be integrated out.
Therefore, an important question is how to generalize DFL beyond $(1+1)$D and whether new features appear.

Several works have studied DFL in two-dimensional 
LGTs~\cite{smith2018dynamical,karpov2021disorder,chakraborty2022disorder,chakraborty2023spectral,Homeier2023realistic,osborne2023disorderfreelocalization21dlattice}. For example, a basic extension of the $(1+1)$D $\mathbb{Z}_2$ LGT in Eq.~(\ref{eq:DFL_Z2}) replaces the transverse spin interaction by the `star operator' on a square lattice. This preserves the 
solvability of the model and connection to Anderson localization~\cite{smith2018dynamical} because in a typical configuration of background charges, matter fermions hop in a disordered binary potential. In the strong potential limit, the binary disorder potential can be understood to create a percolating background of accessible hopping paths. Nonetheless, whether the localization-delocalization transition coincides with the classical percolation threshold, i.e., threshold for the formation of long-range connectivity in random systems and networks~\cite{shante1971introduction}, has remained inconclusive.

A similar connection between DFL and classical percolation appears in a spin-$\frac{1}{2}$ U$(1)$ QLM~\cite{karpov2021disorder}. The Hamiltonian is:
\begin{equation}\label{eq:DFL_U1link}
    H_{\text{QLM}}^{(2+1)\text{D}} = \sum_{\square} \left[ \lambda \left( P_{\square} +P_{\square}^{\dagger}\right) - J \left( P_{\square} +P_{\square}^{\dagger}\right)^2  \right] .
\end{equation}
The plaquette operators $P_{\square}= S^{\dagger}_{\mathbf{r},\mathbf{i}} S^{\dagger}_{\mathbf{r+i},\mathbf{j}}S^{-}_{\mathbf{r+j},\mathbf{i}}S^{-}_{\mathbf{r},\mathbf{j}}$ with the usual raising and lowering operators spin $S^{\pm}_{\mathbf{r},\mathbf{\mu}}$ on link $\mathbf{\mu}$ emanating from site $\mathbf{i}$. The second term in Eq.~\eqref{eq:DFL_U1link} induces dynamics; see Fig.~\ref{fig:DFL}(c). However, whether the configuration of a given plaquette can change strongly depends on its surroundings. Using a variational network ansatz, the halting of correlation spreading was confirmed numerically in this model~\cite{karpov2021disorder}. Remarkably, the highly constrained dynamics of typical charge sectors of the quench setup can be (approximately) mapped to a classical percolation problem, providing a bound on localization distinct from the mechanism for Anderson localization (namely disorder-induced quantum interference)~\cite{karpov2021disorder}.  
So far, extensions to three-dimensional models or LGTs with dynamical matter beyond $\mathbb{Z}_2$ have not been explored and provide challenging open problems. 

%%%
%%%
\subsubsection{Stability of DFL}\label{sec:DFL_4}

DFL touches upon a number of fundamental questions in localization and many-body quantum physics. In the context of Anderson localization~\cite{Anderson58}, and later in its interacting many-body localization (MBL) extension~\cite{Basko06,Mirlin05,OganesyanHuse}, it had been a long-standing question whether a \emph{clean} quantum many-body system can be localized. Note that this requires both the Hamiltonian and the initial state of a quantum quench setup to be translationally invariant, hence it excludes systems in the presence of external electromagnetic-field gradients, which can lead to the so-called Stark localization~\cite{Wannier1962,vanNieuwenburg2019,schulz2019stark}.  Interaction-induced localization without disorder was first proposed in the context of solid helium by Kagan and Maksimov~\cite{kagan1984localization} and later generalized to light-heavy particle mixtures for quasi-MBL~\cite{schiulaz2015dynamics,yao2016quasi}. However, these proposals did not show infinite time memory, rather a long-time transient behavior akin to localization. The {\it local} symmetry of LGTs, as explained in the examples above, allow for the DFL phenomenon to exist, thus, providing conditions for localization in a clean systems~\cite{Smith2017a}.
 
An obvious feature of DFL is that the reliance on local conserved charges makes them fine-tuned. One way out is to postulate the local gauge symmetry to be a fundamental property of nature but then superposition of many charge sectors is unphysical. Moreover, in any quantum simulator, local symmetry will need to be engineered, which will practically entail gauge-symmetry-breaking errors. It has been shown that violating the exact conservation of charges turns DFL into a transient phenomenon, and DFL becomes a prethermal phenomenon if the gauge-symmetry violation is small~\cite{smith2018dynamical}. With an eye toward experimental realizations of LGTs, different schemes have been proposed to mitigate the effect of unwanted interactions not compatible with local gauge symmetry~\cite{lang2022disorder}. In Ref.~\cite{halimeh2022enhancing}, it was shown how the addition of translationally invariant local terms not only stabilizes but even enhances DFL up to tuneable long-time scales. The idea is to add local pseudogenerators of the gauge group that act identically to the full generator in a single superselection sector, but not outside of it. Unwanted terms with nonzero matrix elements between different charge sectors are then dynamically suppressed via a quantum Zeno-like effect and DFL is restored. 

%%%
%%%
\subsubsection{Quantum parallelism for DFL}\label{sec:DFL_5}

We note that the idea of coupling local ancilla spins for efficient disorder averaging~\cite{paredes2005exploiting,Enss2017mbl,Andraschko2014purification}---an example of quantum parallelism---also falls in the DFL category. The two share the idea that a superposition of an extensive number of disordered configurations (labeled by charges or ancillas), whose dynamics evolve independently, can be translationally invariant. However, the difference is that in the LGT formulation of DFL disorder emerges dynamically from the local gauge symmetry. As a result, the link gauge operators are interacting with nontrivial dynamics themselves while the local ancillas are noninteracting and only used to prepare the initial state. In addition, the LGT connection has proven very useful for constructing new models showing DFL with an ever-increasing number of complex phenomena discussed in the following. 

The DFL construction allows to choose LGT theories particularly suited for implementation on available quantum simulators. It has enabled the realization of DFL in a recent experiment on a \texttt{Google} QPU~\cite{Gyawali2024}. The quantum quench dynamics is simulated from a Trotterized time evolution under two minimal LGT Hamiltonians in $(1+1)$ and $(2+1)$D, described by a slight modification of the Hamiltonian in Eq.~\eqref{eq:Z2LGT}:
\begin{align}
\label{eq:DFL_Z2_Google}
	\hat{H}_{\mathbb{Z}_2}=- \sum_{\langle \ell,\ell' \rangle} \left( J \hat \sigma^z_\ell \hat \tau^z_{\ell,\ell'} \sigma^z_{\ell'}   - h \hat \tau^x_{\ell,\ell'} \right) +\sum_\ell \mu \hat \sigma^x_\ell.
\end{align}
Note the unusual form or the gauge-matter coupling compared to the standard $\mathbb{Z}_2$ LGT.
Using a superconducting quantum processor, translationally invariant initial states 
can be efficiently prepared; see Fig.~\ref{fig:DFL}(d). These states can be in a single charge sector or in an extensive superposition of random charge configurations. In the former `delocalized' state, energy spreads ballistically with a clear light cone, but in the latter `localized' state, energy excitations remain localized, realizing DFL. Remarkably, even on a 2d lattice including 81 qubits, the localization of an initial perturbation is clearly visible; see Fig.~\ref{fig:DFL}(e). However, proving true localization is difficult for the time scales and system sizes that can be reached currently. Confirmation of DFL in $(2+1)$D LGTs makes an important target of improved quantum simulator architectures in the future, and in turn can be used for benchmarking their capabilities~\cite{halimeh2024disorder}.

%%%%%%%%%%%%
%%%%%%%%%%%%
\subsection{Quantum many-body scarring}\label{sec:QMBS}

Quantum many-body scars (QMBSs) are a form of weak ergodicity breaking that is a quintessential aspect of far-from-equilibrium behavior of many LGTs. The key manifestations of QMBSs include revivals from special initial conditions, illustrated in Fig.~\ref{fig:ergbreak}(c), accompanied by the presence of rare ETH-violating eigenstates. In general, QMBSs can occur due to a variety of mechanisms~\cite{ShiraishiMori, Dea2020, Pakrouski2020, MoudgalyaCommutant, Buca2023, LerosePappalardi, Pizzi2025} and we direct the reader to several recent reviews for their in-depth survey~\cite{Serbyn2021, MoudgalyaReview, ChandranReview, PapicReview}. Here, we focus solely on a subset of QMBS phenomena with direct relevance to LGTs. 

In the following, we first discuss in detail the dynamical phase diagram of a U$(1)$ QLM as a function of mass and confinement, as an archetype of  QMBSs. This discussion, will be phrased in the language of the $(1+1)$D PXP model, which is relevant for the existing experimental realizations of QMBSs. Following this discussion, we mention several extensions of QMBSs to other types of LGTs and beyond $(1+1)$D. 

\begin{figure*}[htb]
    \centering
    \includegraphics[width=0.995\linewidth]{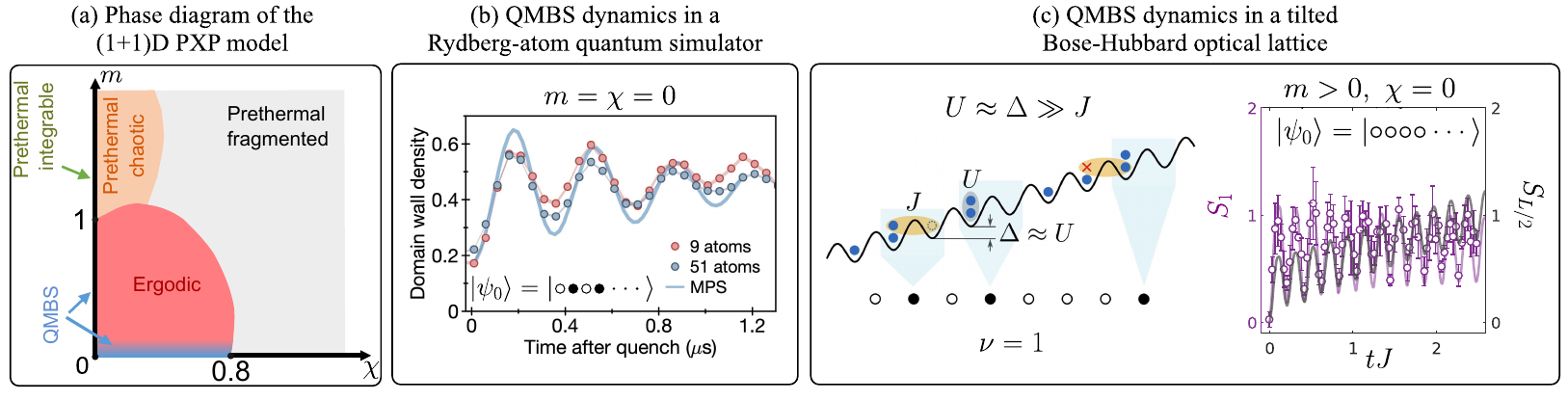}
    \caption{
    (a) The numerical phase diagram of the PXP model in Eq.~(\ref{eq:PXP}) as a function of mass $m$ and confinement parameter $\chi$. As explained in Sec.~\ref{sec:Rydberg}, this is an equivalent representation of the $(1+1)$D U$(1)$ QLM with $S=\frac{1}{2}$. Both axes are in units of $\Omega$ [or, equivalently, can be expressed in terms of $\kappa$ according to Eq.~(\ref{eq:QLM_Rydberg_mapping})]. Our main interest will be the QMBS regimes, which lie in the vicinity of the two axes, as marked. Figure is reproduced from~\cite{Desaules2023}. 
    (b) Experimental observation of QMBS dynamics in Rydberg-atom arrays, which realized the PXP model in Eq.~(\ref{eq:PXP}) with $m=\chi=0$. Persistent revivals---a hallmark of QMBS dynamics---were observed in the density of domain walls after quenching the atoms from the $|\mathbb{Z}_2\rangle \equiv \ket{{\circ}{\bullet}{\circ}{\bullet}\cdots}$ initial state. Figure is reproduced from Ref.~\cite{bernien2017probing}.
    (c) Experimental observation of QMBS dynamics in a tilted Bose--Hubbard optical lattice, which realized the $(1+1)$D U$(1)$ QLM with mass $m$ but without confinement. In contrast to Sec.~\ref{sec:opticalsuperlattice}, the setup shown here relies on a simpler mapping at boson filling factor $\nu=1$, and tuning the system to a resonance $U\approx\Delta \gg J$. At this resonance, the dominant hopping process is $\ket{\ldots 11 \ldots}\leftrightarrow \ket{\ldots 20 \ldots}$ and the PXP excitations, $\bullet$, live on the bonds between the lattice sites. The doublon configuration $\ket{\ldots 20 \ldots}$ in the Bose--Hubbard model maps to an excitation in the PXP model, while all other configurations are mapped to an empty site, $\circ$.     Plot on the right demonstrates the persistence of QMBS dynamics at nonzero mass values (here $m/\Omega = 0.84$) and from a different initial state---the polarized state $\ket{0}\equiv \ket{{\circ}{\circ}{\circ}{\circ}\cdots}$. Strongly suppressed entanglement growth was found by measuring the R\'enyi entanglement entropy of single-site subsystems $S_1$, which is seen to be in qualitative agreement with the numerical simulation of the entropy for a larger (half-chain) subsystem, $S_{L/2}$. Figures are reproduced from~\cite{su2023observation}.
    }
    \label{fig:QMBS}
\end{figure*}

%%%
%%%
\subsubsection{The PXP model and its phase diagram}\label{sec:QMBSPXP}

We recall that after imposing the Gauss's law, the $\mathrm{U}(1)$ QLM in $(1+1)$D maps onto a spin model, Eq.~(\ref{eq:PXP}), also known as the PXP model~\cite{FendleySachdev,Lesanovsky2012,Surace2020}. In this mapping, detailed in Eq.~(\ref{eq:QLM_Rydberg_mapping}), the prefactor $\kappa/(a\sqrt{S(S+1)})$ sets the overall energy scale and plays the role of the Rabi frequency $\Omega$, 
the mass $m$ can be interpreted as the chemical potential, and the confinement term $\chi$ assumes the role of staggered spin magnetization. Below, we will use both nomenclatures interchangeably. As described in Fig.~\ref{fig:platforms_mapping}(b), we will further primarily focus on the spin-$\frac{1}{2}$ case with the local basis states denoted by $\circ$ (ground state) and $\bullet$ (excited Rydberg state). 

The dynamics of the PXP model in Eq.~(\ref{eq:PXP}) are constrained: a spin can only change its state if both of its neighbors are in the $\circ$ state. The only allowed processes are of the form $\cdots{\circ}{\circ}{\circ}\cdots \leftrightarrow \cdots{\circ}{\bullet}{\circ}\cdots$, while any process generating nearest-neighbor excitations, e.g., $\cdots{\bullet}{\circ}{\circ}\cdots \leftrightarrow \cdots{\bullet}{\bullet}{\circ}\cdots$, is forbidden. This gives rise to HSF, where the fragments are distinguished by the number of nearest-neighbor excitations, with the largest sector containing no such excitations. This sector contains, among exponentially many other states, the fully polarized state $\ket{0}\equiv \ket{{\circ}{\circ}{\circ}\cdots}$ and the $\mathbb{Z}_2$ state,  $\ket{\mathbb{Z}_2}\equiv \ket{{\circ}{\bullet}{\circ}{\bullet}\cdots}$, which will be important below. The distribution of energy-level spacings in this sector converges to the GOE ensemble mentioned in Sec.~\ref{sec:thermalization}, implying that the sector is quantum-chaotic and the PXP model should thermalize~\cite{Turner2018a}.

The HSF in the PXP model described above occurs due to the kinetic constraint and it is present for all values of $m$ and $\chi$. In the sector containing no neighboring $\ldots {\bullet}{\bullet}\ldots$ excitations, the dynamical phase diagram of the PXP model as a function of $m$ and $\chi$ was numerically mapped out in Ref.~\cite{Desaules2023}; see Fig.~\ref{fig:QMBS}(a).\footnote{We note that some of our conventions for $\Omega$, $m$ and $\chi$ in Eq.~(\ref{eq:QLM_Rydberg_mapping}), occasionally differ by a factor of two from the literature, e.g., Ref.~\cite{Desaules2023} and~\cite{su2023observation}. This difference has been accounted for in Fig.~\ref{fig:QMBS} and the text.} The phase diagram contains several regimes, including an ergodic phase and various types of prethermal phases when $m$ or $\chi$ are much larger than $\Omega$, some of which exhibit HSF~\cite{Chen2021}. Note that this is an \emph{additional} HSF which occurs when $m$ or $\chi$ are much larger than $\Omega$. The regimes associated with QMBSs occur either in the deconfined phase with sufficiently small mass or, in the massless case, for sufficiently weak confinement. Below, we discuss these regimes separately, focusing on the deconfined case which has also been probed experimentally.

%%%
%%%
\subsubsection{Scarring at the massless point in the deconfined phase}\label{sec:QMBSdeconfined}

The origin of the phase diagram in Fig.~\ref{fig:QMBS}(a) corresponds to the massless point in the deconfined phase of the U$(1)$ QLM (i.e., $m=\chi=0$). Quantum simulators based on Rydberg-atom arrays (Sec.~\ref{sec:Rydberg}) observed the first signatures of QMBSs in this strongly interacting regime known as the Rydberg blockade~\cite{bernien2017probing}; see Fig.~\ref{fig:QMBS}(b). Specifically, these experiments detected a striking difference in the dynamics between the previously mentioned product states $\ket{0}$ and $\ket{\mathbb{Z}_2}$: while both of them effectively form an ``infinite-temperature'' ensemble, the quench from the $\ket{0}$ state was observed to lead to rapid thermalization, while the $\ket{\mathbb{Z}_2}$ initial state exhibited robust oscillations in the dynamics of local observables such as the density of ``domain walls'', i.e., the number of $\ldots {\circ}{\circ}\ldots$ pairs in the time-evolved state, plotted in Fig.~\ref{fig:QMBS}(b).

The strong sensitivity of the dynamics to the initial condition is reminiscent of the phenomenon of quantum scars of a single particle inside a stadium billiard~\cite{Heller84}. Quantum scarring in a billiard has a semiclassical origin: nonthermalizing dynamics ensues from special initial conditions that are directly linked with the particle's unstable periodic orbits~\cite{HellerLesHouches}.  
By contrast, the periodic `orbit' of the PXP model resides in the many-body Hilbert space, passing through the  $\ket{\mathbb{Z}_2}\equiv \ket{{\circ}{\bullet}{\circ}{\bullet}\cdots}$ state and its partner $\ket{\overline{\mathbb{Z}_2}}\equiv \ket{{\bullet}{\circ}{\bullet}{\circ}\cdots}$, as shown in Ref.~\cite{wenwei18TDVPscar}. When Rydberg atoms are initialized on this orbit---which is conveniently done in experiment by preparing the atoms in $\ket{\mathbb{Z}_2}$ product state---the full quantum dynamics tends to cluster around the orbit even at late times, rather than uniformly dispersing across the Hilbert space. 
Quantum fluctuations can be added to this semiclassical description, without fundamentally changing the picture of the dynamics~\cite{Turner21}. Moreover, the orbit approach can be extended to other classes of initial states~\cite{Michailidis2019,Kerschbaumer2025,ren2025scarfinderdetectoroptimalscar,petrova2025findingperiodicorbitsprojected}, including $|\mathbb{Z}_k\rangle$ states (where an excitation $\bullet$ occurs on every $k$th site) for which QMBSs were observed in recent experiments~\cite{Loh2025}. 

A complementary understanding of QMBSs in the PXP model is based on properties of its ergodicity-breaking eigenstates~\cite{Turner2018a}. These eigenstates form distinct families, which can be identified by their overlap  with $\ket{\mathbb{Z}_2}$ state. Various schemes have been put forward to approximate these QMBS eigenstates~\cite{Turner2018b, Choi2018, Iadecola2019,Omiya22}. While these schemes differ in details, they share an underlying algebraic picture of QMBS eigenstates: these states form an approximate ``restricted su$(2)$-spectrum-generating algebra''~\cite{BernevigEnt,MotrunichTowers}. The spectrum-generating algebra is defined by a local operator $\hat{Q}^{\dagger}$ which obeys the following relation with the Hamiltonian:
\begin{equation}\label{eq:RSGA}
[\hat H, \hat{Q}^{\dagger}]=\omega \hat{Q}^{\dagger},    
\end{equation}
where $\omega$ is some energy scale. Such operators---also known as dynamical symmetries---occur in many models in both condensed-matter and high-energy physics~\cite{arno1988dynamical}. However, in conventional examples, the algebra in Eq.~\eqref{eq:RSGA} is defined over the full Hilbert space, while in the QMBS case it only holds over a smaller subspace spanned by QMBS eigenstates $(\hat{Q}^{\dagger})^n\left|\Psi_0\right\rangle$, where $\left|\Psi_0\right\rangle$ is some eigenstate (e.g., the ground state) of $\hat H$ and $n=0,1,2,\ldots$ is an integer. Such QMBS eigenstates, via Eq.~(\ref{eq:RSGA}), are equally spaced in energy, with the spacing $\omega$. This property leading to the coherent dynamics when the system is prepared in a state which overlaps with QMBS eigenstates. For the PXP model, however, the algebra~in Eq.~\eqref{eq:RSGA} is obeyed only approximately, hence the revivals slowly decay over time, as seen in Fig.~\ref{fig:QMBS}(b). Nevertheless, the picture behind Eq.~(\ref{eq:RSGA}) allows to construct deformations of the PXP model that improve the algebra structure and, thereby, \emph{enhance} the QMBS revivals~\cite{ KhemaniLaumannChandran, Choi2018,Bull2020}. This provides further support to Eq.~(\ref{eq:RSGA}) as the archetypal mechanism of QMBSs in the PXP model.  

%%%
%%%
\subsubsection{Scarring in the presence of mass and confinement}\label{sec:QMBSmass}

The  mass term in the $(1+1)$D U$(1)$ QLM has a nontrivial effect on QMBS dynamics, which became apparent after the PXP model was experimentally realized in a system of ultracold atoms in an optical lattice~\cite{su2023observation}. 
This setup is described by the Hamiltonian in Eq.~(\ref{eq:bhm}); it can directly realize the U$(1)$ QLM with nonzero mass and confinement parameter upon placing bosons at filling factor $\nu=1/2$~\cite{Halimeh2022tuning}. However, this setup requires a period-4 superlattice, which is experimentally challenging. Fortunately, in the absence of confinement ($\chi = 0$), there exists a much simpler mapping to the PXP model when the Hamiltonian in Eq.~(\ref{eq:bhm}) is tuned to the resonance $U\approx \Delta \gg J$~\cite{su2023observation} at filling factor $\nu=1$; see Fig.~\ref{fig:QMBS}(c) for an illustration. The bosonic hopping amplitude then plays the role of the Rabi frequency, $\Omega = \sqrt{2}J$, while detuning from the resonance acts as a chemical potential, $m= (U - \Delta)/2$. The states with the maximal number of doublons, $|2020 \cdots\rangle$ and $|0202 \cdots\rangle$, are the analogs of $|\mathbb{Z}_2\rangle$ states in the PXP model, while the Mott-insulator state $|1111\cdots\rangle$ is the $|0\rangle$ spin state. Crucially, the preparation of these initial states only requires a period-2 superlattice. 

While the dynamics from the $|\mathbb{Z}_2\rangle$-type initial state was found to be consistent with earlier results in Rydberg-atom arrays, the Bose--Hubbard realization of the PXP model in Ref.~\cite{su2023observation} revealed that QMBS dynamics persist in other initial states, such as the $\ket{0}$ state, provided that the mass is sufficiently large, e.g., $m/\Omega = 0.84$ in Fig.~\ref{fig:QMBS}(d). The capability to interfere two copies of the system allowed these experiments to access the R\'enyi entropy of small subsystems, which confirmed the slow spreading of quantum entanglement that accompany the QMBS revivals.  

The existence of a broad QMBS regime, tunable by mass $m$, opened up the possibility to consider more complex setups, e.g., prepare the system in the ground state at some mass $m$ and then suddenly quench the mass term to a different value $m^\prime \neq m$. While this protocol recovers the familiar $\ket{\mathbb{Z}_2}$ and $\ket{0}$ initial states in the limits $m \to \pm \infty$, it allows to access more general QMBS regimes in the $m-m^\prime$ ``dynamical phase diagram''. In particular, this allows to study an interplay of QMBS dynamics with quantum criticality since the $(1+1)$D U$(1)$ QLM undergoes an equilibrium phase transition at $m_c / \Omega=0.655$, associated with the spontaneous breaking of a global $\mathbb{Z}_2$ symmetry~\cite{coleman1976more}. Intriguingly, QMBS dynamics were found to persist even when preparing the system in a highly entangled ground state at the critical point, i.e., choosing $m=m_c$~\cite{Daniel2023}. This is surprising because preparing the system in the $\ket{\mathbb{Z}_2}$ state ($m\to-\infty$) and quenching to the critical point ($m^\prime = m_c$) was previously found to give rise to fast thermalization~\cite{Yao2022Criticality,WangQMBSCriticality}. Nevertheless, the tunability of QMBSs can effectively provide a ``bridge'' for coherent quantum dynamics to extend across an equilibrium phase transition: preparing the system in the ground state at $m=m_c$ and then quenching with $m^\prime \neq m_c$ results in robust revivals; in fact, there exists an entire line of a robust QMBS regime across the $m-m^\prime$ dynamical phase diagram~\cite{Daniel2023}. This suggests that quantum criticality, at least in the $(1+1)$D U$(1)$ QLM, can have a complex interplay with QMBS dynamics, rather than invariably restoring thermalization. 

On the other hand, confinement is expected to have a two-fold effect on QMBS dynamics~\cite{Desaules2023}. From an LGT perspective, confinement implies that any particle-antiparticle pair experiences an energy cost $\propto \chi d$, where $d$ is the distance between the particle and antiparticle~\cite{Surace2020,Halimeh2022tuning,Cheng2022tunable}. In the PXP model, a single particle-antiparticle pair on top of the vacuum takes the form of a single ``defect'' on top of the $|\mathbb{Z}_2\rangle$ state, e.g., $\ket{\cdots {\bullet}{\circ} {\bullet}{\circ}{\circ}{\circ}{\bullet}{\circ}{\bullet}\cdots}$. For such initial states, the spreading of correlations is suppressed at late times at finite values of $\chi$ compared to $\chi=0$~\cite{Desaules2023}, 
implying that confinement enhances the QMBS dynamics. Surprisingly, the effect is nonmonotonic: 
around $\chi^*/\Omega \approx 0.7$, the QMBS revivals are most pronounced. Thus, in addition to limiting the spreading of quasiparticles, confinement 
interplays nontrivially with the su$(2)$ algebra; a feature that would be interesting to explore in future quantum simulations of this and other, more complex LGTs.

%%%
%%%
\subsubsection{Extensions to other types of LGTs and higher dimensions}\label{sec:QMBSextensions}

Beyond the $(1+1)$D U$(1)$ QLM, various examples of $\mathbb{Z}_2$ LGTs have been explored as potential hosts of QMBSs~\cite{Iadecola2019_3,MotrunichTowers, vanVoorden2020,Aramthottil2022, Ge2024}. These models can sustain towers of QMBS eigenstates described by a restricted spectrum-generating algebra, conceptually similar to Eq.~(\ref{eq:RSGA}) discussed above. In the continuum field theory, some of these eigenstates can also be interpreted as meson excitations~\cite{James2019PRL,Robinson2019PRB}. However, while in Eq.~(\ref{eq:RSGA}) $\hat Q^\dagger$ is typically a local operator, there can exist other types of eigenstates that have a more nonlocal algebraic structure. For example, in a $(1+1)$D $\mathbb{Z}_2$ LGT coupled to a dynamical spin-chain as a matter field, Ref.~\cite{Ge2024} found types of ``nonmesonic'' QMBSs generated by nonlocal operators. These nonmesonic QMBSs are somewhat reminiscent of fractionalized excitations such as spinons. These are in contrast to the mesonic ones that can be modeled as condensates of local quasiparticles such as magnons~\cite{Iadecola2019}.

Since QMBSs represent highly excited eigenstates, they are difficult to access beyond $(1+1)$D systems. Consequently, the QMBS landscape in higher dimensions remains largely unexplored. For PXP-type models on bipartite $2$d lattices---those that can be divided into two disjoint sublattices, say $A$ and $B$, such that every nearest neighbor of a site in $A$ lies in 
$B$ and \emph{vice versa}, analogs of $(1+1)$D QMBS phenomena have been theoretically predicted~\cite{Lin2D,Michailidis2D}, and also observed experimentally~\cite{Bluvstein2021}. The large boundary-to-surface ratio in accessible $2$d lattices can have a perturbing effect on QMBS dynamics in the bulk of the system, resulting in weaker revivals compared to $(1+1)$D~\cite{Bluvstein2021}. However, it was shown that this limitation can be greatly mitigated by periodic driving of the chemical potential~\cite{Bluvstein2021}. The driving-induced stabilization of QMBS revivals has been understood within simple toy models where the driving mechanism can be mapped to a discrete time crystal~\cite{Maskara2021}. However, the relevance of such toy models for the experimental driving protocol in Ref.~\cite{Bluvstein2021} is not transparent~\cite{Hudomal2022Driven}. 

When it comes to $(2+1)$D LGTs, various examples of QMBS states have been identified using a combination of analytical constructions supported by small-scale numerical simulations.  In particular, Refs.~\cite{Banerjee2021,biswas2022,Sau2024} explore a class of Hamiltonians $\hat H=\hat{\mathcal{O}}_{\text {kin }}+\lambda \hat{\mathcal{O}}_{\text {pot }}$, where the kinetic term $\hat{\mathcal{O}}_{\text {kin}}$ is purely off-diagonal and the potential term $\hat{\mathcal{O}}_{\text {pot}}$ is diagonal in a chosen basis. Examples of systems include U$(1)$ QLM on arbitrary geometries (ladders and isotropic $2$d lattices) or quantum dimer models. At $\lambda=0$, this model contains exact midspectrum eigenstates with energy $E=0$ whose number grows exponentially with system size. Such ``zero modes'' also appear in the $(1+1)$D PXP model~\cite{Turner2018b,Iadecola2018,lin2018exact,Karle2021,Ivanov2025,ivanov2025exactarealawscareigenstates} and in a more general class of systems under the name of ``Fock-space cages''~\cite{tan2025interferencecagedquantummanybodyscars,benami2025manybodycagesdisorderfreeglassiness,jonay2025localizedfockspacecages,nicolau2025fragmentationzeromodescollective}.
Provided that the zero-mode states violate the ETH, e.g., via a subvolume scaling of the entanglement entropy, they can be viewed as special examples of QMBS states. At any nonzero $\lambda$, the massive degeneracy of the zero-mode subspace is lifted, but some special linear combinations that simultaneously diagonalize $\hat{\mathcal{O}}_{\text{kin}}$ and $\hat{\mathcal{O}}_{\text{pot}}$ can survive as QMBSs---a form of ``order-by-disorder" mechanism in the Hilbert space. 

A variation of the previous construction are so-called lego and sublattice scars $\left|\Psi_s\right\rangle$~\cite{Sau2024}, which satisfy $\hat{\mathcal{O}}_{\text {pot}, \square}\left|\Psi_s\right\rangle=\left|\Psi_s\right\rangle$ for all elementary plaquettes on one sublattice and $\hat{\mathcal{O}}_{\text {pot}, \square}\left|\Psi_s\right\rangle=0$ on the other, while being simultaneous zero modes or nonzero integer-valued eigenstates of $\hat{\mathcal{O}}_{\text{kin}}$. 
Intuitively, such conditions that enforce a form of locality at the sublattice level, induce a strong constraint on the correlations in states $|\Psi_s\rangle$, resulting in a breakdown of ETH, making such states QMBSs. Related constructions exist for a large class of $(2+1)$D pure gauge theories~\cite{Budde2024, miao2025exactquantummanybodyscars}. Intriguingly, when coupling such theories to matter, it has been found numerically that the stability of QMBS revivals can depend sensitively on the exchange statistics, with bosonic matter showing more robust QMBSs compared to fermionic matter~\cite{osborne2024quantummanybodyscarring21d}. However, these simulations of dynamics in $(2+1)$D are currently restricted to cylinders with only a few plaquettes along one of the axes, hence further work is needed to determine the impact of particle statistics on QMBSs in the isotropic limit. 

%%%%%%%%%%%%
%%%%%%%%%%%%
\subsection{Open questions}\label{sec:NonErgopenQs}

In this Section, we have reviewed three paradigms of quantum nonegodicity (HSF, DFL and QMBS) that play a profound role in the physics of LGTs far from equilibrium. The standard probe for all of these phenomena is a quantum quench protocol, where relaxation to equilibrium and memory of the initial state are used as diagnostics of nonthermalizing behavior. Understanding the mechanisms of ergodicity breaking is essential for developing an overarching theory of thermalization in LGTs, in particular for understanding whether the traditional assumption of thermal equilibrium for strongly coupled LGTs may or may not be justified, due to the thermalization bottlenecks arising from local constraints and gauge fields. In the following, we discuss some outstanding open problems in this context. 

As emphasized in the introduction to this Section, different facets of nonergodicity can be intertwined in certain models, complicating their identification in finite-size systems. For example, in the Schwinger model, HSF can be present alongside DFL and give rise to similar phenomenology, such as the appearance of nonoverlapping minibands due to a large separation of energy scales in the model~\cite{Papic2015}. Thus, HSF can mimic the nonergodic behavior otherwise associated with DFL~\cite{Jeyaretnam2025}. In sufficiently large systems, one expects the minibands to reconnect and the distinction between HSF and DFL to become clear, however, this regime may be challenging to reach in classical simulations.

The limitations of finite system size and finite evolution times, shared by both classical simulations and current quantum simulations, must be carefully taken into account when it comes to understanding the nature of DFL and its relation to quantum interference or underlying classical percolation physics~\cite{karpov2021disorder,karpov2022spatiotemporal}. In particular, the relation between DFL and MBL is naturally affected by the ongoing debate about the asymptotic stability of MBL in the thermodynamic limit and especially in higher dimensions~\cite{Suntajs2020,Abanin2021,Sels2021, Morningstar2022,Sierant2025}, which directly translates into the stability of asymptotic localization in DFL sectors. In generic LGTs, the interactions between matter in each charge sector are effectively nonlocal~\cite{brenes2017many}, making the connection with MBL phenomena, which normally assume local interactions, less transparent. Thus, for LGTs, the existence of localization in asymptotic limits of infinite system size and infinite time cannot be inferred from generic MBL studies.

We note that, although counterintuitive at first sight from a fundamental-physics point of view, the DFL idea that wave functions can be in superposition over many charge sectors can be more widely applicable than initially assumed. It could be especially relevant to quantum many-body systems where the gauge-theory description is an  emergent (low-energy) property, for example, in the description of quantum spin liquids~\cite{savary2016quantum,knolle2019field} . Indeed, it has been shown that in the Kitaev honeycomb model, which describes a $\mathbb{Z}_2$ quantum spin liquid, quench dynamics from simple product states of spin $\frac{1}{2}$ leads to slow correlation spreading. The latter can be understood by rewriting the initial state as an extensive superposition of gauge-flux sectors, providing a direct link to DFL~\cite{rademaker2019quenching,zhu2021subdiffusive,yogendra2023emergent}. Whether similar localization phenomena govern the physics of fractionalized phases beyond the example of the solvable Kitaev model is an interesting question for future research. Of direct experimental relevance to solid-state systems is also the idea that electron-phonon coupled systems can show DFL-like transient dynamics, e.g., a coherent state of phonons with (quasi-)conserved phonon occupations can act like DFL, and could possibly explain spectroscopy experiments~\cite{sous2021phonon}.
 
In a broader experimental context, including condensed-matter experiments,  an important open question concerns the impact of gauge constraints on transport of conserved quantities, such as energy or particle number. It has been well established that general chaotic models typically exhibit diffusive transport, while integrability in $(1+1)$D can give rise to faster, i.e., ballistic transport~\cite{BertiniRMP,GopalakrishnanVasseurReview}. On the other end of the spectrum, DFL leads to a complete absence of transport. Hence, one may wonder about the impact of HSF and QMBSs on transport properties of LGTs. Intuitively, constraints and HSF are expected to slow down transport and can indeed lead to subdiffusive behavior~\cite{Singh2021,Richter2022}. Curiously, the study of energy transport in the PXP model found a broad regime of \emph{superdiffusion}, with the dynamical exponent intriguingly close to the Kardar-Parisi-Zhang value $z=3/2$~\cite{Ljubotina2023}. The superdiffusion persists across a broad range of masses $m$, however, the effect of the confinement parameter $\chi$ has not been explored. The observed superdiffusion is at odds with other many-body models, e.g., the Heisenberg ferromagnet where superdiffusion arises due to a combination of integrability and non-Abelian symmetry~\cite{BertiniRMP}, neither of which are present in the PXP model. While this may be a special property of the $(1+1)$D U$(1)$ QLM, a systematic exploration of transport in other LGTs and higher dimensions is much needed. 

Another important question concerns the fate of ergodicity-breaking phenomena in the continuum limit of LGTs. Consider the QLM formulation in Eq.~(\ref{eq:QLMmapping}), and an alternate formulation in which one truncates the gauge field $\hat U_{j,j+1}$ by representing it with an operator $\hat{\tau}_{j, j+1}^{+}$, which has the same matrix structure as $\hat{S}_{j, j+1}^{+}$, but with each of the latter's nonzero matrix elements replaced by 1. While in the $S\to\infty$ limit, both formulations recover the continuum Kogut--Susskind theory, the QMBS signatures exhibit somewhat different scaling with $1/S$.  For example, while the second type of truncation results in robust QMBS signatures that are well-converged in $1/S$~\cite{Desaules2022prominent}, this is less clear for the QLM formulation~\cite{Desaules2022weak}. This difference could be attributed to relatively small system sizes accessible to numerics, which become challenging in the large-$S$ regime.  Nevertheless, the observed sensitivity of QMBSs to the details of the truncation scheme leaves open the question of whether QMBSs are an intrinsic property of continuum LGTs. We note, however, that QMBSs appear as a general feature of the weak-coupling or perturbative regime in general $(1+1)$D quantum field theories~\cite{Delacretaz2023} and holographic models~\cite{Dodelson2022}. These studies suggest that QMBSs can exist in other regimes of continuum gauge theories beyond the large-$S$ limit discussed above, which warrants further exploration. 

In our discussion of DFL, we emphasized that localized dynamics do not require external disorder. However, in experiment, imperfections are invariably present, e.g., the gauge protection might not be perfect, the system may not be fully isolated from the environment, the couplings may be spatially nonuniform, etc. Thus, it is important to understand the impact of such imperfections on ergodicity-breaking phenomena discussed in this Section. Among these, QMBSs as the more fragile ones are expected to be particularly sensitive to disorder, although some QMBS signatures can survive weak disorder in the form of resonances with a finite lifetime~\cite{MondragonShem2020}. Moreover, the \emph{type} of disorder potential can significantly affect the ability to localize the dynamics~\cite{Huang2021,Sierant2021} and it may exert different impact on different types of QMBS eigenstates~\cite{Surace2021disorder}. 

Most of the results in this Section pertain to Abelian LGTs, while the rich symmetry structure of \emph{non-Abelian} LGTs and its impact on thermalization dynamics remain largely unknown. The possibility of DFL in non-Abelian LGTs is just starting be  explored~\cite{cataldi2025disorder}, while evidence of robust QMBSs in a $(1+1)$D matter-coupled hardcore-gluon SU$(2)$ LGT was recently found in Ref.~\cite{Calajo2025} [see also Ref.~\cite{Ebner2024} for a related study without dynamical matter]. The non-Abelian gauge fields, which introduce more intricate dynamical constraints, might yield qualitatively new types of ergodicity-breaking phenomena without direct analogs in Abelian models, warranting a systematic investigation. For example, it would be interesting to understand qualitative differences between HSF in Abelian versus non-Abelian LGTs, and the impact of non-Abelian gauge field on DFL, given its tendency to destabilize MBL~\cite{AbaninRev}.

Finally, the theoretical studies of ergodicity-breaking mechanisms discussed in this Section have traditionally benefited from quantum entanglement, which has proven to be a powerful lens for understanding the process of thermalization and its breakdown. However, entanglement quantifies only one aspect of quantum-information complexity of nonergodic states, and it is important to understand if new insights can be gained from other measures of quantum correlations. In this context, recent studies of nonstabilizerness or ``magic resources''~\cite{knill2004faulttolerantpostselectedquantumcomputation,BravyiKitaev2005} have unearthed a large variety of behaviors in QMBS states. For example, while QMBS states in the $(1+1)$D PXP model can have many-body nonstabilizerness beyond single qubits~\cite{smith2024nonstabilizernesskineticallyconstrainedrydbergatom}, those in a $(2+1)$D $\mathbb{Z}_2$ LGT are stabilizer states with zero magic~\cite{hartse2024stabilizer}. A systematic understanding of nonstabilizerness and other diagnostics of quantum complexity in nonergodic states has, therefore, still to be achieved. 
This understanding could prove useful for potential applications in quantum technology, where nonergodic states like QMBs have already been proposed for quantum sensing and metrology protocols~\cite{Dooley2023,desaules2022extensive}.

%%%%%%%%%%%%%%%%%%%%%%%%%%%%%%%%%%%%%%%%%%%%%%%%%%%%%%%%%%%%%%%%%%%%%%%%%%%%%%%%%%%%%%%%%%%%%%%%%%%%%%%%%%%%%%%%%%%%%%%%%%%%%%%%%%%%%%%%%%%%%%%%%%%%%%%%%%%%%%%
\section{Dynamical quantum phase transitions in gauge theories
\label{sec:DQPT}}

Digital quantum computers and analog quantum simulators provide an opportunity to study dynamical topological effects. 
Topology plays a central role in the physics of gauge theories. For instance, the QCD vacuum is topologically nontrivial, possessing a $\theta$-vacuum structure~\cite{callan1976structure,jackiw1976vacuum}; certain field configurations  tunnel (instantons)~\cite{schafer1998instantons} or dynamically `hop over the barrier' (sphalerons)~\cite{klinkhamer1984saddle} between different vacuums. Importantly, the structure of the QCD vacuum is directly tied to the so-called `strong CP problem':  the parameter $\theta$ is an undetermined angle which can theoretically take any value between $0$ and $2\pi$; yet it is experimentally found to be consistent with $\theta\approx 0$ [approximately $\theta\le 10^{-10}$~\cite{abel2020measurement}]. At $\theta = 0$, QCD does not violate CP symmetry---a unique choice among the infinite possibilities. Numerous explanations have been proposed for this apparent fine tuning; yet none have been supported by experimental evidence to date~\cite{peccei1977cp,nelson1984naturally,barr1984solving}. 

A consequence of the topologically nontrivial structure of the QCD vacuum is a violation of axial-charge conservation of the fermionic content of the theory, via the so-called axial anomaly, which is responsible, e.g., for the large mass of the $\eta`$ meson~\cite{witten1979current,veneziano1979u} or the decay of neutral pions into photons~\cite{adler1969axial,bell1969pcac}. Moreover, transitions between distinct topological sectors of QCD in ultra-hot, deconfined matter are conjectured to drive novel electric transport phenomena in ultra-relativistic heavy-ion collisions, exemplified by the chiral magnetic effect (CME) and related phenomena~\cite{kharzeev2014chiral}. The CME is connected to the topological structure of QCD, which induces an imbalance of axial charge through the chiral anomaly. In the presence of a magnetic field, this imbalance can drive dissipation-free transport of both electric and axial charge. Analogous anomaly-induced transport phenomena have been observed in certain semimetals~\cite{kharzeev2013anomaly}, and the CME is conceptually akin to anomaly-induced topological transport phenomena such as the quantum Hall and quantum spin Hall effects~\cite{kane2005quantum}; see, e.g., Refs.~\cite{kharzeev2020real,chakraborty2022classically,ikeda2024real} for quantum-simulation related work. A nontrivial $\theta$-vacuum structure is not unique to QCD; similar structures arise in supersymmetric Yang--Mills theory~\cite{kac2000vacuum}. Additionally, an axial anomaly is present in QED${}_{1+1}$~\cite{coleman1976more,melnikov2000lattice}, making this theory a valuable prototype model for studying these phenomena. 

In essence, from a cosmological perspective, the strong CP problem requires understanding the nonequilibrium dynamics of the early universe---specifically, how the $\theta$ parameter evolved to its present-day value---and the topological structure of QCD. In this context, quantum-simulation experiments offer a promising avenue for exploration. As we will discuss below, connections to nonequilibrium phenomena in quantum simulators
may yield unexpected insights into these longstanding questions. However, at present, quantum simulation rely on relatively simple models and. Moreover, to generate nonequilibrium dynamics, a common approach is to study a quantum quench. In a quench, as discussed throughout this Review, instead of modeling the detailed out-of-equilibrium initial conditions, the system is prepared in the ground state of a given (simpler) Hamiltonian, after which the Hamiltonian parameters are abruptly changed, leading to out-of-equilibrium evolution. 

In the quench set-up (though strictly not tied to it), so-called \textit{Dynamical Quantum Phase Transitions} (DQPTs)~\cite{heyl2013dynamical,heyl2018dynamical} have been studied in a variety of quantum many-body systems, see, e.g., Refs.~\cite{jurcevic2017direct,budich2016dynamical,flaschner2018observation,tian2019observation,halimeh2020quasiparticle,xu2020measuring,bhaskar2024timescales}. The key insight is that, while systems out of equilibrium generally cannot be described thermodynamically, for instance via (canonical) partition functions, DQPTs are analogs of equilibrium thermal phase transitions. Concretely, the central object is the Loschmidt amplitude,
\begin{align}\label{eq:Loschmidtecho}
    \mathcal{L}(t) \coloneq \langle \Psi_0 | e^{-it\hat H} | \Psi_0 \rangle\,,
\end{align}
where $| \Psi_0 \rangle$ is typically a ground state while $\hat H$ is the `quench' Hamiltonian, i.e., where a parameter is changed relative to the initial Hamiltonian of which $| \Psi_0 \rangle$ is the ground state, leading to nonequilibrium evolution. In analogy with the free energy in statistical physics, one defines a rate function
\begin{align}\label{eq:ratefunction}
    \Gamma(t) \coloneq - \lim_{V\rightarrow \infty} \frac{1}{V} \log|{\mathcal{L}}(t)|\,.
\end{align}
Here, $V$ denotes the system's volume. DQPTs appear as nonanalyticities of \Eq{eq:ratefunction} or zeros of \Eq{eq:Loschmidtecho}. These typically emerge as short-time phenomena, in contrast to late times where the scaling of the (time-averaged) Loschmidt echo serves as a measure of the `effective' Hilbert-space size explored by an initial state~\cite{goussev2012loschmidt}. It is important to distinguish DQPTs from more general dynamical phase transitions that are defined by the long-time behavior of an order parameter, e.g., see Refs.~\cite{Moeckel2008interaction,Sciolla2010quantum,Smacchia2015exploring,Halimeh2017prethermalization,Mori2018,Zhang2017DPT}.

%%%%%%%%%%%%
%%%%%%%%%%%%
\subsection{Dynamical topological quantum phase transitions in gauge-theory dynamics}
In the following, we discuss theoretical studies of DQPTs in LGTs. These studies have been performed in $(1+1)$D and $(2+1)$D Abelian models, with or without matter. These works focus on different aspects of DQPTs, including connections to the underlying quantum phase transitions or dynamics related to the topological structure of the model.

%%%
%%%
\subsubsection{DQPTs in U$(1)$ quantum link models}
Consider the spin-$\frac{1}{2}$ U$(1)$ QLM in Eq.~\eqref{eq:U1QLM} with $\chi=0$. Recall that in this limit, the model exhibits an equilibrium quantum phase transition at $\tilde{\kappa}_c/m=1.526$, which separates a symmetry-broken phase with two degenerate ground states (for $\kappa<\kappa_c$), as enumerated in Eq.~\eqref{eq:vacuua} and to be called $\ket{\Psi_\pm}$ in the following, from a paramagnetic one (for $\kappa>\kappa_c$), with the order parameter $\mathcal{E} \coloneq L^{-1}\sum_{\ell}\langle S^z_{\ell,\ell+1}
\rangle$.

For the case of $\Upsilon$-fold degenerate ground-state manifolds, one can generalize $\mathcal{L}(t)$ to the full return probability $\sum_{\beta=1}^{\Upsilon} \mathcal{L}_\beta(t),
$ where $\mathcal{L}_\beta(t)$ is the Loschmidt echo defined with the vacuum state $\ket{\Psi_\beta}$ in place of $\ket{\Psi_0}$ in Eq.~\eqref{eq:ratefunction}. In the thermodynamic limit $L\to \infty$, there is always a contribution that dominates the total rate function $\Gamma(t)$, which is associated with the vacuum state with the least Loschmidt echo~\cite{heyl2013dynamical}. Whenever the dominant branch of the rate function switches from one to the other
vacuum, one obtains a kink in $\lambda(t) \coloneq \text{min}_\beta (\{L_\beta(t)\})$; hence a DQPT occurs.

To numerically investigate DQPTs in the QLM model, as proposed in Ref.~\cite{huang2019dynamical}, one prepares the system initially at $\tilde{\kappa}=0$ in one of the two ground states, e.g., $\ket{\Psi_-}$. At $t=0$, $\tilde{\kappa}$ is suddenly switched to $\tilde{\kappa}>0$ and $\lambda(t)$ and $\mathcal{E}(t)$ are monitored. Numerically obtained data for these quantities for a quench across the underlying quantum phase transition to $\tilde\kappa=3m>\tilde\kappa_c$ are shown in Fig.~\ref{fig:DQPT1}(a). DQPTs are observed in $\lambda(t)$ at a series of critical times in the form of kinks caused by a crossing of the two rate functions $\lambda_{+}(t)$ and $\lambda_{-}(t)$.
Those points mark when the time-evolved state $|\Psi_{-}(t)\rangle$ switches between being closer to $|\Psi_{+}\rangle$ to being closer to $|\Psi_{-}\rangle$ and vice versa.
These intervals are consistent with when $\mathcal{E}(t)$ changes sign across a DQPT. A closer inspection of the two time intervals as a function of $\tilde{\kappa}/m$ and system size shows that these values tend to approach each other away from the critical point.

A DQPT of similar nature was further observed in the same work in a $(2+1)$D QLM without matter, described by the Hamiltonian in Eq.~\eqref{eq:DFL_U1link}. Once again, two degenerate vacuums of the model are identified in one phase and a quench was performed to drive the system suddenly across the equilibrium quantum phase transition. The cusp singularity in the Loschmidt eco, consistent with the change in the equilibrium order parameter, was observed~\cite{huang2019dynamical}, confirming that the DQPT applies to higher-dimensional gauge theories, and in the absence of matter, as well.

DQPTs were also numerically studied in $(1+1)$D spin-$S$ U$(1)$ QLMs where a more nuanced picture emerged, showing that certain \textit{branch} DQPTs can emerge in the lowest-lying rate function itself, and not as intersections of lowest-lying rate functions \cite{VanDamme2022DQPTspinS}. Zero-mass quenches in these models also established a direct connection between rate-function-intersection DQPTs and QMBSs~\cite{VanDamme2023Anatomy}.

%%%
%%%
\subsubsection{DQPTs in the massive Schwinger model with a $\theta$ term}
\begin{figure*}
    \centering
    \includegraphics[width=0.95\textwidth]{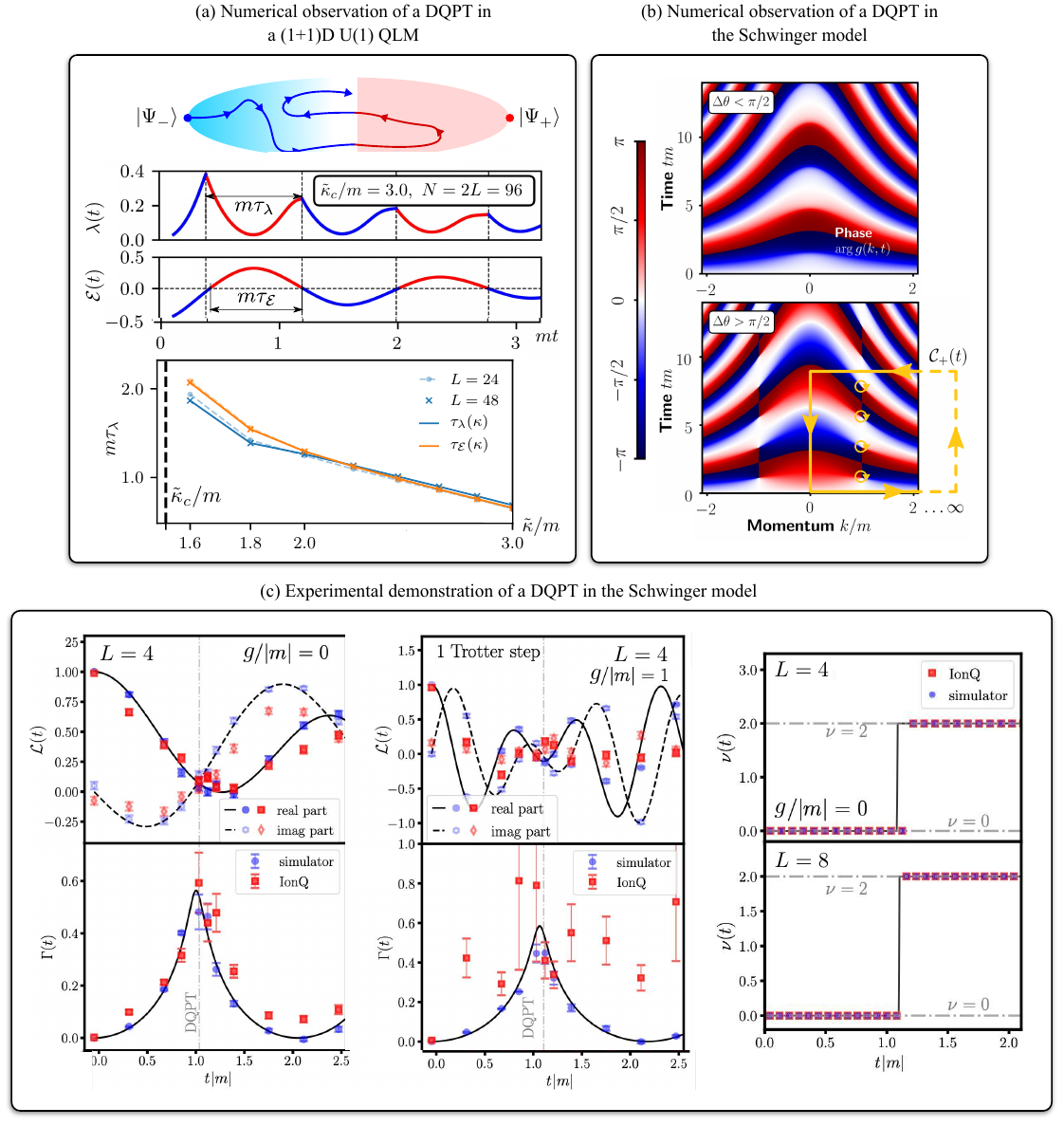}
    \caption{
    (a) DQPTs in a spin-$\frac{1}{2}$ $(1+1)$D U$(1)$ QLM can be studied by initiating the system in one of the two degenerate ground state of the model in one phase ($\ket{\Psi_-}$) and drive the dynamics into the other phase using a quench. During various intervals of the evolution, the state is either closer to one or the other vacuum state, and switching between the two is signified by the cusps of the rate function $\lambda(t)$ as shown. Additionally, an order parameter $\mathcal{E}(t)$, related to the equilibrium phase transition of the model, changes sign between the cusps. The consistency of the time intervals ($\tau_\lambda$ for $\lambda(t)$ and $\tau_{\mathcal{E}}$ for $\mathcal{E}(t)$) is further demonstrated in the lower plot. Figure is reproduced from \cite{huang2019dynamical}. (b) The massive Schwinger model with a $\theta$-term shows DQPTs following a quantum quench of the angle $\theta$. For small quenches with $\Delta \theta < \pi/2$, no significant features are observed in the phase of  a time-ordered correlator in \Eq{eq:unequalcorr} (top panel). In contrast, for $\Delta \theta \ge \pi/2$, vortices appear at critical times $t_c$ and momenta $k_c$ (bottom panel), indicating topological transitions. Plots correspond to the continuum, infinite-volume limit of the Schwinger model at $g = 0$. Figure is reproduced from Ref.~\cite{zache2019dynamical}. (c) DQPTs in the lattice Schwinger model computed on IonQ's \texttt{Aria} quantum processor. The Loschmidt amplitude (left and middle top panels) and the corresponding rate function (left and middle bottom panels) are shown for a system with $N = 4$ qubits and one ancilla, implemented via an interferometric scheme. Shown are results at both zero coupling and  finite coupling ($g/m = 1$); results shown correspond to a single Trotter step for the interacting case. An interferometric protocol enables quantum computation of a topological index defined in \Eq{eq:topo}, and plotted in the right panel for $N=4,8$ system sizes. Figure is reproduced from Ref.~\cite{mueller2023quantum}.
\label{fig:DQPT1}}
\end{figure*}   

Returning to the topic of the QCD vacuum structure, one can mimic the dynamics of topological transitions by rapid quenches of the $\theta$-angle. A controlled arena for this purpose is the Schwinger model; see \Eq{eq:QED_continuum} with a nonvanishing $\theta$ term. As was shown in Ref.~\cite{zache2019dynamical}, nonequilibrium evolution after a quantum quench of the $\theta$-parameter in the Schwinger model exhibits DQPTs of topological nature.

Focusing first on the noninteracting case and in the continuum limit, the system is prepared in the ground state, $| \Omega_{\theta}\rangle$ of $H^{(1+1)\text{D}}_{\text{QED}}(\theta) \equiv H_{\theta}$, followed by abrupt evolution with $H^{(1+1)\text{D}}_{\text{QED}}(\theta') \equiv H_{\theta'}$, where $\Delta\theta \coloneq \theta'- \theta\neq 0$. A key observable is the phase $\varphi_{\Delta \theta}(k,t)$ of the gauge-invariant time-ordered Green's function, 
\begin{align}\label{eq:unequalcorr}
    g_{\Delta \theta}(k,t)\coloneq \int dx \, e^{-ikx} \langle \hat\Psi^\dagger(x,t) e^{-ie \int_0^xdx' \hat A(x',t) } \hat\Psi(0,t) \rangle,
\end{align}
where $\langle \dots \rangle = \langle \Omega_{\theta} | \dots | \Omega_{\theta}\rangle$ and $\hat O(t) = e^{it \hat H_{\theta'}} \hat O e^{-it\hat H_{\theta'}}$. This quantity is plotted in \Fig{fig:DQPT1}(b). The time evolution of the phase exhibits 
qualitative differences when the quench is weaker versus stronger than $\Delta \theta_c=\frac{\pi}{2}$. For $|\Delta\theta|<\Delta \theta_c $ [upper panel of \Fig{fig:DQPT1}(b)],
no special structure is observed in $\varphi_{\Delta \theta}(k,t)$, other than exhibiting the dispersion of the respective momentum mode. However, for $|\Delta\theta|\ge \Delta \theta_c $ [lower panel of \Fig{fig:DQPT1}(b)], vortices form at critical times and momenta, coinciding with DQPTs of the model. These DQPTs are of topological nature~\cite{tian2019observation,xu2020measuring}, as indicated by a topological parameter
\begin{align}\label{eq:topo}
   \nu \coloneq n_+ - n_-,
\end{align}
with
\begin{align}\label{eq:topo}
   n_\pm (t) \coloneq \frac{1}{2\pi}\oint_{\mathcal{C}_{\pm}(t)} d\bm{z} \left[\tilde{g}^*(\bm{z}) \partial_{\bm{z}} \tilde{g}(\bm{z})\right].
\end{align}
Here, $\tilde{g}(\bm{z}) \coloneq {g}(\bm{z})/|{g}(\bm{z})|$, $g(z) \equiv g_{\Delta \theta}(k,t)$, and $\mathcal{C}_{\pm}(t)$ is a rectangular path enclosing the left/right half of the $z \equiv (k,t)$-plane; it runs (counterclockwise) along $(0,0) \leftrightarrow (0,t) \leftrightarrow (\pm \infty,t) \leftrightarrow (\pm \infty,0) \leftrightarrow
(0,0)$, as shown in \Fig{fig:DQPT1}(b).

While the DQPT observed at $\Delta \theta = \frac{\pi}{2}$ can be analyzed analytically in the zero coupling limit, a central question was whether such topological transitions persist in the interacting theory as well. To investigate this question, Ref.~\cite{zache2019dynamical} presents exact numerical simulations of this model at finite $g/m$ on lattices of up to $N=20$ sites, employing staggered fermions with the Hamiltonian in \Eq{eq:SchwingerKS}. The study demonstrated that the DQPT indeed survives at finite coupling, $g/m \approx 1$. This result is significant because it lies in a regime where, for larger systems, classical computations become intractable and quantum computers are required. This observation constitutes a nontrivial physics result: many topological effects rely on the weak-coupling limit even to define topological invariants. From a methodological perspective, observing a robust topological effect at intermediate coupling values suggests that it can serve as a 'standard candle' for benchmarking quantum devices. Its topological nature makes it resilient to noise, thereby enabling the probing of hardware limitations. Once a certain simulation quality is reached, the effect appears robustly.

%%%%%%%%%%%%
%%%%%%%%%%%%
\subsection{Quantum-simulation experiments of dynamical quantum phase transitions in LGTs}

The study in Ref.~\cite{zache2019dynamical} identifies
%DQPTs as 
signatures of DQPTs linked to changes in the topological $\theta$ angle. While QCD is more complex, similar mechanisms may have played a role in the early universe---an area where quantum computers could offer insights. Currently, the phenomenon in the Schwinger model is experimentally accessible, making it a promising near-term quantum-simulation target.  

The first such quantum simulation was performed in Ref.~\cite{mueller2023quantum} using a digital trapped-ion quantum computer (IonQ's \texttt{Aria} processor). The simulation consists of preparing the initial state in the momentum-space eigenbasis of $H_\theta$ (at zero coupling), followed by time evolution with $H_{\theta'}$. The quench is hereby realized by a circuit realizing a Bogoliubov transform from $H_\theta$ to the (noninteracting) $H_{\theta'}$ eigenbasis, followed by time evolution. The time evolution separately occurs in momentum space (where the noninteracting part of the Hamiltonian is diagonal) and position space (where the interaction part of the Hamiltonian is diagonal), involving basis changes between the two bases.

The Loschmidt amplitude in Eq.~\eqref{eq:Loschmidtecho} is nontrivial to compute on a quantum computer. To compute this quantity, an interferometric scheme is employed: an ancilla qubit prepared in the state $(|0\rangle + |1\rangle)/\sqrt{2}$ coherently controls the time-evolution operator, generating a superposition of evolved and unevolved states. Measurement of $\sigma^x$ and $\sigma^y$ on the ancilla yields the real and imaginary parts of $\mathcal{L}(t)$, respectively. Since time evolution preserves particle number, a symmetry-based error-mitigation scheme is applied by discarding measurement outcomes that violate particle-number conservation (analogous results for $N = 8$ sites are not shown). The experiment is further repeated for $N = 4$ and finite coupling $g/m = 1$, where the inclusion of interactions necessitates deeper circuits due to the required basis transformation. These results, shown in the middle panel of \Fig{fig:DQPT1}(c), are significantly noisier, though the approximate location of the DQPT remains discernible.

Finally, the nonequal-time correlator in Eq.~\eqref{eq:unequalcorr} is quantum computed at $g=0$ and for $N=4$ sites, using a generalized interferometry scheme. The corresponding topological order parameter $\nu$ in \Eq{eq:topo} is shown in the right panel of  \Fig{fig:DQPT1}(c). Despite the correlator itself being noisy, $\nu(t)$ tracks the DQPT accurately owing to its topological nature. The entanglement during the DQPT can also be investigated, as done in Ref.~\cite{mueller2023quantum}, using randomized-measurement protocols~\cite{elben2023randomized}, and via quantum computing Renyi entropies and extracting entanglement Hamiltonians~\cite{dalmonte2022entanglement,brydges2019probing,kokail2021entanglement}.

Further quantum-simulation studies for observing DQPTs in the Schwinger model~\cite{pomarico2023dynamical}, 
in QLMs and employing superconducting-circuit quantum processors~\cite{pedersen2021lattice}, are reported in recent years. Additionally, Ref.~\cite{jensen2022dynamical} investigates the impact of quantum-hardware noise on DQPT signatures. Specifically, based on classical emulation using Lindblad dynamics for open quantum systems, the characteristic features of DQPTs in a $(1+1)$D U$(1)$ QLM were shown to remain robust even under noise levels typical of near-term devices, although the overall magnitudes of observables are suppressed.

%%%%%%%%%%%%
%%%%%%%%%%%%
\subsection{Open questions}

DQPTs, recognized only  over the past decade, were initially regarded as a theoretical curiosity, in quest for finding universality in nonequilibrium physics. However, the example above, which connects DQPTs to topological dynamics stemming from the $\theta$-vacuum structure of a gauge theory, suggests that such phase transitions may have deeper implications, including for the strong CP problem in QCD.

Many open questions remain to be investigated: What implications might these insights have for the dynamics of the early universe? Could quantum computers help constrain or even rule out candidate solutions for the strong CP problem, such as axion models, by providing access to the real-time dynamics of strongly coupled QCD? Moreover, if the value of the QCD $\theta$-angle changed during the early universe, might there be observable ``echoes'' of that transition today? And since direct interferometry of the universe’s wavefunction is impossible, what (equal-time) observables can one measure today to infer this early-universe physics? To address these questions, the phenomena reviewed here need to be examined in higher dimensions and in non-Abelian theories. One also needs to find out if these effects persist in strongly coupled regimes, across all relevant length scales, and in the continuum limit.

Moreover, quantum simulators also hold promise for advancing our understanding of anomalous transport phenomena in QCD~\cite{kharzeev2014chiral}. Within the Schwinger model, this physics can be modeled by studying electric and axial currents in the presence of a time-dependent $\theta$ parameter, with a spatially varying $\theta$ providing an effective description of the fluctuating topological transitions in QCD. For example, Ref.~\cite{kharzeev2020real} investigates a quantum quench of a time-dependent (but spatially homogeneous) $\theta$-parameter in the massive Schwinger model, using both numerical simulations and an IBM's circuit emulator. Other questions include: What is the relationship between these DQPTs and dynamical topological transitions, such as sphalerons (classical field configurations), and how would these latter be described in a fully quantum theory~\cite{klinkhamer1984saddle}? 

Another persisting question is whether there are different types of DQPTs. Whereas originally DQPTs were connected to an order parameter changing sign during its dynamics~\cite{heyl2013dynamical}, as demonstrated in the U$(1)$ QLM example in this Section, the picture was shown to be more nuanced~\cite{Karrasch2013,Andraschko2014DQPT,Vajna2014}. In particular, it was demonstrated that \textit{branch} (aka \textit{anomalous}) DQPTs can emerge without the order parameter ever dynamically changing sign~\cite{Halimeh2017dynamicalphasediagram,Halimeh2017probingtheanomalous}. This was explicitly shown in spin chains both at zero~\cite{Homrighausen2017anomalous} and at finite temperature~\cite{Lang2018concurrence,Lang2018dynamical}. It was later shown that these branch DQPTs were connected to confinement~\cite{halimeh2020quasiparticle} in spin chains, with subsequent work extending this observation to LGTs~\cite{osborne2023probingconfinementdynamicalquantum}, and further relating the onset time of branch DQPTs to meson mass~\cite{osborne2024mesonmasssetsonset}. The question thus arises whether branch DQPTs can be used as a ``smoking gun'' for confinement in future quantum simulation experiments of LGTs. Such a probe could be useful as currently ascertaining confinement in LGTs requires investigating several observables, and is proven challenging at finite temperatures, see, e.g., Ref.~\cite{kebrivc2023confinement}.

Beyond their role in studying DQPTs, Loschmidt echoes are broadly relevant across many areas of quantum physics~\cite{goussev2012loschmidt}, and have become attractive targets for experiments. However, measuring the full Loschmidt echo is experimentally challenging, as it typically becomes exponentially small at long times. A promising approach introduced in Ref.~\cite{Halimeh2021localmeasures} and experimentally explored in Ref.~\cite{karch2025probingquantummanybodydynamics} involves subsystem Loschmidt echoes, which remain accessible at both short and long times. Going beyond the short-time DQPT regime, the Loschmidt echo was shown to probe the effective size of the explored Hilbert space at late times. Subsystem Loschmidt echoes, therefore, provide valuable tools for quantifying the effective dimension and structure of the accessible Hilbert space, offering insights into phenomena such as ergodicity breaking and Hilbert-space fragmentation, and establishing a direct connection to the physics discussed in Secs.~\ref{sec:thermalization} and~\ref{sec:ergodicitybreaking}.

Last but not least, from a more general point of view, DQPTs provide a window into out-of-equilibrium quantum many-body universality, which is an open question. Indeed, truly out-of-equilibrium critical exponents can be extracted from DQPTs, as has been shown in spin models \cite{halimeh2019dynamicalquantumphasetransitions,Trapin2021,Halimeh2021localmeasures,Wu2020dqpt}. Extending such investigations to LGTs can shed further light on their out-of-equilibrium universality.

%%%%%%%%%%%%%%%%%%%%%%%%%%%%%%%%%%%%%%%%%%%%%%%%%%%%%%%%%%%%%%%%%%%%%%%%%%%%%%%%%%%%%%%%%%%%%%%%%%%%%%%%%%%%%%%%%%%%%%%%%%%%%%%%%%%%%%%%%%%%%%%%%%%%%%%%%%%%%%%
\section{Outlook
\label{sec:outlook}} 

Quantum simulation offers a promising lens into the intriguing out-of-equilibrium dynamics of gauge theories. It is driving a convergence of theoretical predictions and experimental observations; is expanding intersections of nuclear and high-energy physics with condensed-matter and statistical physics, quantum many-body theory, and quantum information science; and is igniting new questions and perspectives along the way. This Review constitutes a collection of methodologies, results, and discussions in support of this status. With a focus on a multitude of nonequilibrium phenomena of relevance to gauge theories, from particle production and string breaking, to hadronization and thermalization, to nonergodicity and dynamical quantum phase transitions, we highlighted the reach of nonequilibrium quantum-simulation studies, both theoretically and experimentally. As evident from the work covered in this Review, the field has come a long way from the first experimental demonstrations of particle production in small (few-site) simple lattice gauge theories in quench processes, to larger simulations of more nontrivial models and phenomena, with complex initial states, and in more relevant processes such as particle collisions. This vibrant field of study is moving steadily and quickly but is yet to meet many important milestones.

Standard-Model gauge theories are defined in $(3+1)$D, and the strong and weak forces involve non-Abelian gauge groups. Furthermore, many gauge theories of relevance to condensed-matter physics go beyond $(1+1)$D, and those of importance in quantum error correction and fermion-to-qubit encodings involve $2$d geometries. Quantum simulation of higher-dimensional and non-Abelian gauge theories, therefore, are at the cutting-edge of developments, in both theory and experiment. In fact, small-scale dynamical simulations in such models have been reported in recent years, as highlighted in this Review. To scale these simulations toward continuum and thermodynamics limits requires quantum-computational resources that are unfortunately well beyond current capabilities of the hardware. For example, real-time dynamics in QCD for parameter ranges and system sizes relevant to the continuum and thermodynamic limits are estimated to require $\sim 10^{10}$ qubits and $\sim 10^{25}-10^{50}$ non-Clifford gates, depending on the algorithm used and various reasonable precision goals~\cite{kan2021lattice,rhodes2024exponential,davoudi2025tasi}. Problems with such computational complexity require necessarily fault-tolerant \textit{quantum supercomputers}.

It remains to be seen if the quantum-computational cost of the Standard-Model theories can be significantly reduced. This goal underlies an active area of research, with at least two main thrusts. One thrust concerns theoretical developments which aim to achieve more efficient formulations of non-Abelian gauge theories on higher-dimensional lattices, by eliminating redundant gauge degrees of freedom, while scarifying fully or partially locality [examples are loop-string-hadrons~\cite{raychowdhury2020loop,kadam2023loop,kadam2025loop,davoudi2021search}, local irrep basis~\cite{klco20202,ciavarella2021trailhead}, q-deformed theories~\cite{zache2023quantum,hayata2023q}, and dual or magnetic basis~\cite{mathur2016lattice,haase2021resource,bauer2023efficient,d2024new,grabowska2025fully,burbano2024gauge}]. Group-element basis with discrete subgroups~\cite{lamm2019general,alexandru2019gluon,ji2020gluon,alexandru2022spectrum} or other digitization of the group manifold are also explored~\cite{jakobs2023canonical,romiti2024digitizing}. Other approaches bypass the need for a spatial lattice and form other variants of finite-dimensional gauge theories that aim to restore the target infinite-dimensional gauge theory in certain limits [examples are the quantum-link and qubitization models~\cite{chandrasekharan1997quantum,brower1999qcd,alexandru2023qubitization,chandrasekharan2025qubit,liu2025phases}, orbifold lattices~\cite{kaplan2003supersymmetry,buser2021quantum,bergner2024toward,Halimeh:2024bth,Hanada:2025yzx,Halimeh:2025ivn,bergner2025exponential}, and light-front forms~\cite{kreshchuk2022quantum,kreshchuk2021simulating}]. It is not presently clear which formulation provides the most optimal representation of a gauge theory from quantum-computational perspectives, but limited studies have started to provide comparative analyses~\cite{nguyen2022digital,kan2021lattice,Davoudi:2022xmb,gustafson2024primitive,rhodes2024exponential}. 

Another thrust concerns algorithmic advances, which aim to develop or apply algorithms with more optimal asymptotic scaling in system size, evolution time, and accuracy~\cite{rhodes2024exponential,rajput2022hybridized}. Perhaps more exciting are new algorithmic and experimental paradigms in qudit-based~\cite{wang2020qudits,ringbauer2022universal}, fermionic~\cite{bravyi2002fermionic,o2018majorana}, and bosonic~\cite{grimsmo2020quantum,chabaud2023resources,liu2024hybrid,cai2021bosonic} quantum computing. Qudits are $d$-level generalizations of the qubit, and are naturally present in several atomic, superconducting, semiconducting, and optical quantum simulators~\cite{altman2021quantum}. In fact, the first qudit-based quantum-simulation experiment of gauge-theory dynamics was recently performed in Ref.~\cite{meth2023simulating} for a one-plaquette U$(1)$ LGT coupled to fermions: high-dimensional internal levels of trapped ions were used to encode the gauge degree of freedom, whereas qubits were used to encode fermions. Various theoretical proposals have also emerged in recent years highlighting the utility of qudits in quantum simulations of LGTs \cite{ciavarella2022conceptualaspectsoperatordesign,popov2024variational,calajo2024digital,kurkccuoglu2024qudit,illa2024qu8its,Ballini2025symmetry,gaz2025quantumsimulationnonabelianlattice,jiang2025nonabeliandynamicscubeimproving,joshi2025efficientquditcircuitquench,joshi2025probinghadronscatteringlattice,zache2023fermion}. Bosons can also be directly used to encode gauge degrees of freedom, hence removing the error incurred~\cite{jordan2012quantum,tong2022provably,ciavarella2025truncation} when truncating the gauge-field Hilbert space in qub(d)it-based simulations. Theoretical proposals for simulating quantum field theories on various platforms~\cite{yang2016analog,lamata2014efficient,marshall2015quantum,davoudi2021toward,jha2023toward,ale2024quantum,briceno2023toward,crane2024hybrid} have now led to the first experimental implementations in small systems~\cite{zhang2016fermion,than2024phase,saner2025real}. Last but not least, fermionic quantum computers, which can be realized in ultracold fermion lattices and fermionic Rydberg arrays, or in semiconductor platforms, will eliminate fermion-to-qubit encoding overhead, and will naturally implement Fermi statistics~\cite{gonzalez2023fermionic,zache2023fermion,schuckert2024fermion,ott2025error}.
 
Regardless of hardware architecture and encoding paradigm, accurate and precise predictions for nonequilibrium gauge-theory dynamics requires quantum supercomputers that work in the fault-tolerant regime, i.e., when errors can be corrected on-the-go and do not accumulate in the end of the computation. Fault tolerance, and the theory and practice of quantum error correction, is one of the most vibrant areas of research and development in quantum computing and quantum information theory at present times, with increasingly more efficient and hardware-aware error-correction protocols being generated and tested~\cite{egan2021fault,ryan2021realization,krinner2022realizing,ryan2022implementing,google2023suppressing,acharya2024quantum,eickbusch2024demonstrating,bluvstein2024logical,campbell2024series,lacroix2024scaling}. Gauge theories, due to their intricate Hilbert space and local redundancies, present two opportunities in error correction. First, their implementation can be combined with various error-correcting codes to reduce the encoding overheads, via leveraging Gauss's laws~\cite{rajput2023quantum,carena2024quantum}. Second, one may leverage deeper connections between error-correction codes and physics. For example, when the code states of the Kitaev surface code~\cite{kitaev1997quantum,kitaev2003fault,bravyi1998quantum} are treated as ground states of the code Hamiltonian, these states realize a topological phase of matter that also exists in a pure $\mathbb{Z}_2$ LGT in $(2+1)$D~\cite{wegner1971duality}. There is an ongoing effort in understating the gauge theory underlying, as well as topological aspects of, other error-correcting codes; see, e.g., Refs.~\cite{sang2024mixed,rispler2024random,de2025low}. It would be exciting to explore links between various non-Abelian gauge theories and potentially new error-correction schemes, and leverage such connections for more efficient quantum simulation of gauge theories.

Despite the ultimate need for fault-tolerant quantum computing, the present Review makes it clear that studies of nonequilibrium dynamics in simpler models, and in the presence of hardware imperfections, are still greatly valuable: They reveal not-previously-explored rich phenomenology of nonequilibrium gauge theories, and deepen our understanding of fundamental, and often universal, mechanisms in quantum many-body physics and quantum field theories. They further build many fascinating linkages between energy spectra, equilibrium phases and phase transitions, rich vacuum structure of field theories, entanglement properties, and various thermodynamics quantities on one hand, and the emergence of particles and equilibrium phases out of nonequilibrium conditions, and even mechanisms hindering such equilibration, on the other hand. It is conceivable that more linkages, and possibly surprises, result from such real-time explorations in increasingly more complex models and scenarios, both using classical tools such as tensor networks, and ultimately using large-scale powerful quantum simulators. Equally intriguing is quantifying the computational complexity of simulating gauge-theory dynamics under controlled nonequlibrium conditions. A potential outcome of this endeavor could be to identify a rigorous case for quantum advantage in quantum computing.

In summary, with this Review, we strive to bring to the spotlight an ongoing vibrant effort in leveraging quantum technologies in studying nonequilibrium physics of gauge theories, for applications in nuclear and high-energy physics and beyond. This program has brought together, closer than ever, multiple scientific disciplines, has generated new ideas and perspectives, and will likely get expanded and invigorated in the upcoming years. Future quantum technologies are set to solve problems of great practical values, and an immediate target will be enlightening the intriguing world of out-of-equilibrium phenomena in gauge theories.

%%%%%%%%%%%%%%%%%%%%%%%%%%%%%%%%%%%%%%%%%%%%%%%%%%%%%%%%%%%%%%%%%%%%%%%%%%%%%%%%%%%%%%%%%%%%%%%%%%%%%%%%%%%%%%%%%%%%%%%%%%%%%%%%%%%%%%%%%%%%%%%%%%%%%%%%%%%%%%%
\footnotesize
\begin{acknowledgements}    
    The authors thank Jürgen Berges, Fabian Grusdt, and Philipp Hauke for stimulating discussions in the early stages of this work.

    J.C.H.~acknowledges funding by the Max Planck Society, the Deutsche Forschungsgemeinschaft (DFG, German Research Foundation) under Germany’s Excellence Strategy – EXC-2111 – 390814868, and the European Research Council (ERC) under the European Union’s Horizon Europe research and innovation program (Grant Agreement No.~101165667)—ERC Starting Grant QuSiGauge. Views and opinions expressed are however those of the author(s) only and do not necessarily reflect those of the European Union or the European Research Council Executive Agency. Neither the European Union nor the granting authority can be held responsible for them. This work is part of the Quantum Computing for High-Energy Physics (QC4HEP) working group. 
    
    N.M.~acknowledges funding, during early stages of this work, by the U.S. Department of Energy (DOE), Office of Science, Office of Nuclear Physics, Inqubator for Quantum Simulation (\url{https://iqus.uw.edu}), via the program on Quantum Horizons: QIS Research and Innovation for Nuclear Science (Award no. DE-SC0020970). 

    J.K.~thanks the hospitality of the Aspen Center for Physics, which is supported by National Science Foundation grant PHY-2210452. J.K.~acknowledges support from the Deutsche Forschungsgemeinschaft (DFG, German Research Foundation) under Germany’s Excellence Strategy– EXC–2111–390814868 and DFG Grants No.~KN1254/1-2, KN1254/2-1 and TRR 360 - 492547816, as well as the Munich Quantum Valley, which is supported by the Bavarian state government with funds from the Hightech Agenda Bayern Plus. J.K.~further acknowledges support from the Imperial-TUM flagship partnership.
    
    Z.P.~acknowledges support by the Leverhulme Trust Research Leadership Award RL-2019-015 and EPSRC Grant EP/Z533634/1. Statement of compliance with EPSRC policy framework on research data: This publication is theoretical work that does not require supporting research data.  This research was supported in part by grant NSF PHY-2309135 to the Kavli Institute for Theoretical Physics (KITP). Z.P.~acknowledges support by the Erwin Schr\"odinger International Institute for Mathematics and Physics.
    
    Z.D.~acknowledges support from the U.S. DOE, Office of Science, Early Career Award (award no.~DESC0020271); the National Science Foundation (NSF) Quantum Leap Challenge Institutes (QLCI) (award no.~OMA2120757); the U.S.~DOE, Office of Science, Office of Nuclear Physics, via the program on Quantum Horizons: QIS Research and Innovation for Nuclear Science (award no.  DESC0023710); the U.S.~DOE, Office of Science, Office of Advanced Scientific Computing Research, Accelerated Research in Quantum Computing program: Fundamental Algorithmic Research toward Quantum Utility (FARQu); the Simons Foundation through the Emmy Noether Fellows Program at the Perimeter Institute for Theoretical Physics; and the Department of Physics, Maryland Center for Fundamental Physics, and College of Computer, Mathematical, and Natural Sciences at the University of Maryland, College Park. She is further grateful to the hospitality of the Excellence Cluster ORIGINS at the Technical University of Munich and the Ludwig Maximilian University of Munich where parts of this Review were completed. Visit to ORIGINS was enabled by the Deutsche Forschungsgemeinschaft (DFG, German Research Foundation) under Germany’s Excellence Strategy (EXC-2094—390783311).
\end{acknowledgements}
\normalsize
\bibliography{references}

\end{document}